\documentclass[12pt,epsf]{article}
\input epsf
\usepackage{amssymb}
\newcommand{\nc}{\newcommand}
\nc{\beq}{\begin{equation}}
\nc{\eeq}{\end{equation}}
\nc{\beqa}{\begin{eqnarray}}
\nc{\eeqa}{\end{eqnarray}}
\nc{\lra}{\leftrightarrow}
\def\sfrac#1#2{{\textstyle\frac#1#2}}

\nc{\sss}{\scriptscriptstyle}
{\nc{\lrp}{\mbox{\raisebox{-.0ex}{$\stackrel{\leftrightarrow}{\partial_\mu}$}}}

\def\intdp{\int {d^{\,4}p\over(2\pi)^4}}
\def\intdqeps{\int {d^{\,4+2\epsilon}q\over(2\pi)^{4+2\epsilon}}}
\def\intdq{\int {d^{\,4}q\over(2\pi)^4}}
\def\intdk{{d^{\,4}k}}
\def\intdx{\int {d^{\,4}x}}
\def\intdy{\int {d^{\,4}y}\,}
\def\intdz{\int {d^{\,4}z\,}}
\def\intdr{{d^{\,4}r\over(2\pi)^4}}
\def\pmie{p^2-m^2+i\varepsilon}
\def\p2m2{p^2-m^2}
\def\pE{p_{\sss E}}
\def\intdpe{\int {d^{\,4}\pE\over(2\pi)^4}}
\def\intdped{\int {d^{\,d}\pE\over(2\pi)^d}}
\def\intdpd{\int {d^{\,d}p\over(2\pi)^d}}
\def\tr{{\rm tr}}
\def\phm{\phantom{-}}
\def\dsl{\hbox{/\kern-.5500em$\partial$}}
\def\psl{\hbox{/\kern-.5800em$p$}}
\def\ksl{\hbox{/\kern-.5800em$k$}}
\def\qsl{\hbox{/\kern-.5800em$q$}}
\def\asl{\hbox{/\kern-.5800em$a$}}
\def\bsl{\hbox{/\kern-.5800em$b$}}
\def\csl{\hbox{/\kern-.5800em$c$}}
\def\Asl{\hbox{/\kern-.6800em$A$}}
\def\Dsl{\hbox{/\kern-.6800em$D$}}
\def\ncdot{\!\cdot\!}
\def\fd#1#2{{\delta #1\over \delta #2}}
\def\fdd#1#2#3{{\delta^2 #1\over \delta #2\delta #3}}
\def\fddd#1#2#3#4{{\delta^3 #1\over \delta #2\delta #3 \delta #4}}
\newcommand\lsim{\mathrel{\rlap{\lower4pt\hbox{\hskip1pt$\sim$}}
    \raise1pt\hbox{$<$}}}
\newcommand\gsim{\mathrel{\rlap{\lower4pt\hbox{\hskip1pt$\sim$}}
    \raise1pt\hbox{$>$}}}

\begin{document}

\input epsf.tex

\vspace{0.2in}
\centerline{\Large\bf  Advanced Concepts in Quantum Field Theory}
\vspace{0.15in}
\centerline{\bf With Exercises} 
\vspace{0.3in}

\centerline{\large 
	James M.~Cline\footnote{e-mail: 
	jcline@physics.mcgill.ca}}

\medskip

\centerline{\it Dept. of Physics, McGill University}
\centerline{\it 3600 University St.}
\centerline{\it Montreal, PQ H3A 2T8 Canada}

\bigskip
\bigskip

\section*{Preface}

These notes represent the second half of a year-long quantum field 
theory course that was given at McGill University.  It assumes the
reader understands the basics of field theory at the tree level.
I start with the loop expansion in scalar field
theory to illustrate the procedure of renormalization, and then
extend this to QED and other gauge theories.  My goal is to introduce
 the most important concepts and developments in QFT, without
necessarily treating them all in depth, but allowing you to 
learn the basic ideas.  The topics to be covered  include:

\begin{itemize}

\item Perturbation theory: the loop expansion; regularization; 
dimensional regularization; Wick rotation; momentum cutoff; 
$\lambda\phi^4$ theory;
renormalization; renormalization group equation; Wilsonian viewpoint;
the epsilon expansion; 
relevant, irrelevant and marginal operators; Callan-Symanzik equation;
running couplings; beta function; anomalous dimensions; IR and UV
fixed points; asymptotic freedom; triviality; Landau pole

\item The effective action: generating functional; connected diagrams;
one-particle-irreducible diagrams; Legendre transform 

\item
 Gauge theories: QED; QCD; anomalies; gauge invariance and
unitarity; gauge fixing; Faddeev-Popov procedure; ghosts; unitary gauge; covariant gauges; 
Ward Identities; BRS transformation; vacuum structure of
QCD; instantons; tunneling; theta vacuua; superselection sectors;
the strong CP problem.
\end{itemize}

I would like to thank Balasubramanian Ananthanarayan 
(IISc, Bangalore) for his encouragement to publish these notes.

\newpage

\tableofcontents
\newpage

\section{Introduction}
\label{intro}

Physics, like all sciences, is based upon experimental observations.
It's therefore a good thing to remind ourselves: what are the major
experimental observables relevant for particle physics?  These are the
masses, lifetimes, and scattering cross sections of particles.
Poles of propagators, and scattering and decay amplitudes are the 
quantities which are related to these observables:
\beqa
	\hbox{pole\ of\ } {1\over p^2-m^2} &\leftrightarrow&
\hbox{(mass)}^2\\
	{\cal T}_{a\to bc} &\leftrightarrow& \hbox{decay\ rate}\\
	{\cal T}_{ab\to cd} &\leftrightarrow& \hbox{scattering\ cross\
section}
\eeqa
These observables, which are components of the {\it S-matrix}
(scattering matrix) are the main goals of computation in 
quantum field theory.  To be precise, ${\cal T}$ is the transition
matrix, which is related to the S-matrix by eq.\ (7.14) of \cite{harry}:
\beq
	{S}_{fi} = {\bf 1}_{fi} + (2\pi)^4 i 
	\delta^{(4)}(p_f-p_i)  {\cal T}_{fi}
\eeq

Toward the end of 198-610A you learned about the connection between
Green's functions and amplitudes.  The recipe, known as the LSZ
reduction procedure (after Lehmann, Szymanzik and Zimmerman)
\cite{BD,IZ}, is the
following.  For a physical process involving $n$ incoming and $m$ 
outgoing particles, compute the corresponding Green's function.
Let's consider a scalar field theory for simplicity:
\beq
	G_{(n+m)}(x_1,\dots,x_n,y_1,\dots,y_m) = 
	\langle 0_{\rm out} |T^*[\phi(x_1)\cdots\phi(x_n)\phi(y_1)\dots
	\phi(y_m)]| 0_{\rm in} \rangle
\eeq
The blob represents the possibly complicated physics occurring in
the scattering region, while the lines represent the free propagation
of the particles as they are traveling to or from the scattering
region.  It is useful to go to Fourier space:
\beq
	\tilde G_{(n+m)}(p_1,\dots,p_n,q_1,\dots,q_m) = i^{n+m}
	{(2\pi)^4 \delta^{(4)}(\sum p_i-\sum q_i)
	 \Gamma_{n+m}(p_1,\dots,p_n,q_1,\dots,q_m)\over
	(p_1^2-m^2)\dots(p_n^2-m^2)(q_1^2-m^2)\dots(q_m^2-m^2)}
\eeq
The delta function arises because we have translational invariance
in space and time, so momentum and energy are conserved by the
process.  This equation makes the picture explicit.  We see in the
denominator the product of all the propagators for the free propagation.
In the numerator we have the function $\Gamma_{n+m}$, called the {\it
proper vertex function}, which represents the blob in the picture.
This function contains all the interesting physics, since we are
interested in the interactions between the particles and not the
free propagation.\\
\medskip
\centerline{\epsfxsize=2.5in\epsfbox{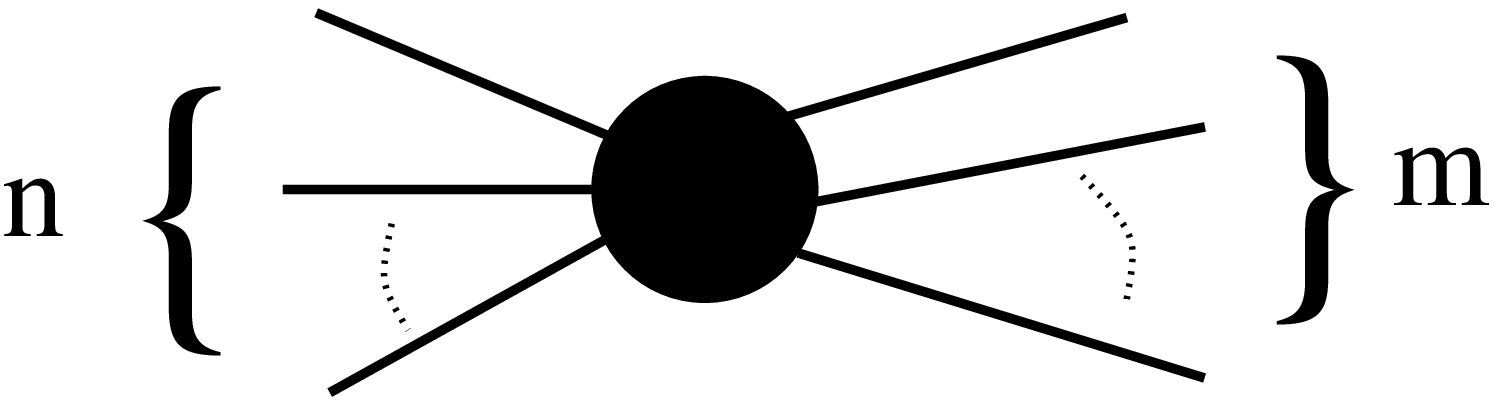}}
\centerline{\small
Figure 1.$n\to m$ scattering process.}
\medskip

Now we can state the LSZ procedure: to convert the Green's function to
a transition amplitude, truncate (omit) all the external line
propagators.  In other words, the proper vertex $\Gamma_{n+m}$ is the 
T-matrix element we are interested in.  I will come back to the
nontrivial proof of this statement later.  First, we would like to  use
it to get some concrete results.  The problem is that, in general,
there is no analytic way to compute $\Gamma_{n+m}$ if it is nontrivial.
In free field theory, the only nonvanishing vertex function is 
$\Gamma_{2} =i(p^2-m^2)$, the inverse propagator.  Once we introduce
interactions, we get all the vertex functions, but they can't be
computed exactly.  We have to resort to some kind of approximation.
Since we know how to compute for free fields, the most straightforward
approximation is that in where the interactions are weak and can
be treated perturbatively.  In nature this is a good approximation for
QED and the weak interactions, and also for the strong interaction
at sufficiently high energies.

However, these realistic theories are a bit complicated to start with.
It is easier to learn the basics using a toy model field theory.
The simplest theory with interactions which has a stable vacuum is
$\lambda\phi^4$.  We simply add this term to the Klein-Gordon 
Lagrangian for a real scalar field:\footnote{I use the metric 
convention $p_\mu p^\mu = E^2 - \vec p^{\,2}$.}
\beq
	{\cal L} = \frac12 \left(\partial_\mu\phi\partial^\mu\phi
	- (m^2-i\varepsilon)\phi^2\right) - {\lambda\over 4!}\phi^4.
\eeq
The factor of $1/4!$ is merely for convenience, as will become
apparent. There are several things to notice.  (1) The $i\varepsilon$
is to remind us of how to define the pole in the propagator so as to
get physical (Feynman) boundary conditions.  We always take
$i\varepsilon\to 0$ finally, so that (2) the Lagrangian is
real-valued.  The latter is necessary in order for $e^{-iS/\hbar}$ to
be a pure phase.  Violation of this condition will lead to loss of
unitarity, {\it i.e.}, probability will not be conserved.  (3) The
$\lambda\phi^4$ interaction comes with a $-$ sign: the Lagrangian is
kinetic minus potential energy. The $-$ sign is necessary so that the
potential energy is bounded from below.  This is the reason we consider
$\lambda\phi^4$ rather than $\mu\phi^3$ as the simplest realistic
scalar field potential.  Although  $\mu\phi^3$ would be simpler,
simpler, it does not have a stable minimum---the field would like to
run off to $-\infty$.

The above Lagrangian does not describe any real particles known in
nature, but it is similar to that of the Higgs field which we shall
study later on when we get to the standard model.  The coupling
constant $\lambda$ is a dimensionless number since $\phi$ has
dimensions of mass.  We will be able to treat the interaction as a
perturbation if $\lambda$ is sufficiently small; to determine how
small, we should compute the first few terms in the perturbation series
and see when the corrections start to become as important as the
leading term.

The tool which I find most convenient for developing perturbation
theory is the Feynman path integral for the Green's functions.  Let's
consider the generating function for Green's functions:
\beq
	Z[J] = \int {\cal D}\phi\, e^{iS[J]/\hbar}
\eeq
where
\beq
	S[J] = \int d^{\,4}x \left( {\cal L} + J(x)\phi(x) \right)
\eeq
Recall the {\it raison d'\^etre} of $Z[j]$: the Green's functions
can be derived from it by taking functional derivatives.  For example,
the four-point function is
\beqa
	G_4(w,x,y,z) &=& \left.{1\over i^4 Z}\,{\delta^4Z[J]\over \delta J(x_1)\delta
J(x_2\delta J(y_1)\delta J(y_2) }\right|_{J=0}\\
	&=& {1\over Z[0]}\int {\cal D}\phi\, e^{iS[0]/\hbar} \phi(w)
	\phi(x)\phi(y)\phi(z) 
\eeqa
If we had no interactions, this Green's function could be computed
using Wick's theorem to make contractions of all possible pairs of
$\phi$'s as shown in figure 2.  This is not very interesting: it just
describes the free propagation of two independent particles.  The 
corresponding Feynman diagram is called {\it disconnected} since the
lines remain separate.  The disconnected process is not very 
interesting experimentally.  It corresponds to two particles in a
collision missing each other and going down the beam pipe without
any deflection.  These events are not observed (since the detector
is not placed in the path of the beam).

\medskip\bigskip
\centerline{\epsfxsize=1.25in\epsfbox{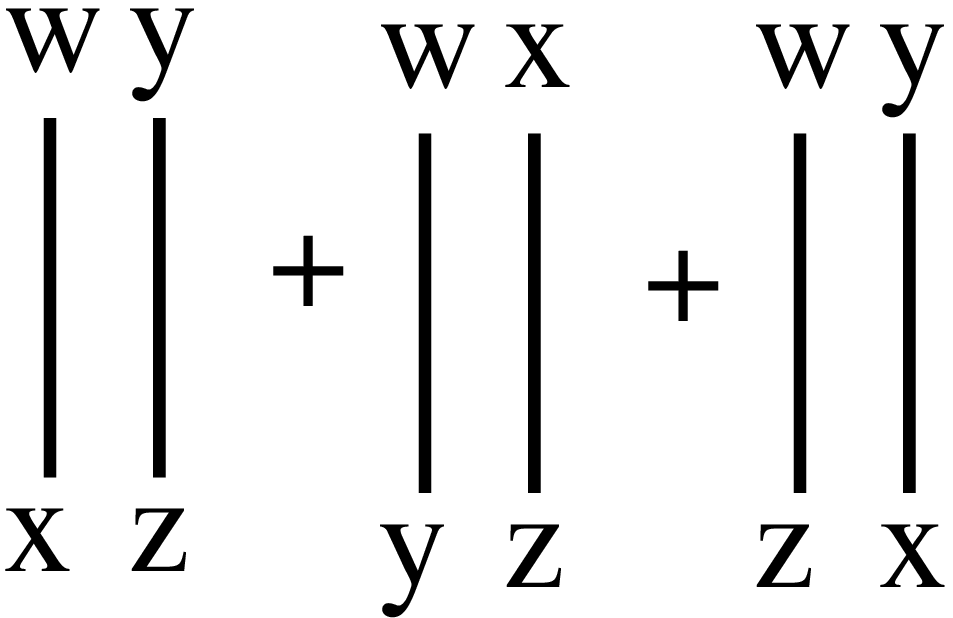}}
\centerline{\small
Figure 2. 4-point function in the absence of interactions.}
\medskip
 
But when we include the interaction we get scattering between the 
particles.  This can be seen by expanding the exponential to first
order in $\lambda$:
\beq
	G_4(w,x,y,z) \cong {1\over Z[0]}\int {\cal D}\phi\, e^{iS[0]/\hbar} \phi(w)
	\phi(x)\phi(y)\phi(z) \int d^{\,4}x'  
	\left(-{i\over\hbar}{\lambda\over 4!}\phi^4 \right).
\eeq

Before evaluating this path integral, I would like to digress for a moment
to discuss its much simpler analog, the ordinary integral
\beq
	Z= \int_{-\infty}^\infty {dx\over\sqrt{2\pi}} e^{\frac{i}2 a x^2}
\eeq
This is related to the well-known trigonometric ones,
the Fresnel integrals, and they
can be computed using contour integration \cite{complex} along
the contour shown in fig.\ 2.5:
\beq
	Z = 2 \int_0^\infty {dx\over\sqrt{2\pi}} e^{\frac{i}2 a x^2}
	= -2e^{i\pi/4}\int_\infty^0 {dy\over\sqrt{2\pi}} e^{-\frac{1}2 a y^2}
	= {e^{i\pi/4}\over\sqrt{a}}
\eeq
The second equality is obtained by using the fact that the integral
around the full contour vanishes, as well as that along the circular
arc (at $\infty$).  Therefore we have shown that the oscillatory
Gaussian integral is related to the real one.  The integral $Z$ is
analogous to the field theory generating functional $Z[0]$.   
And the analogy to the 2-point function (the propagator) is
\beq
{1\over Z}\int_{-\infty}^\infty {dx\over\sqrt{2\pi}} e^{\frac{i}2 a x^2} x^2
= {1\over Z}{2\over i}{dZ\over da} = {i\over a}
\eeq
If we carry out the analogous procedure in field theory, we obtain
the momentum-space propagator $\tilde G_2 = i/(p^2-m^2)$.  This
little exercise shows you where the factor of $i$ is coming from. It
also belies the statement you will sometimes hear, that the Feynman
path integral only rigorously exists in Euclidean space. 

\medskip\bigskip
\centerline{\epsfxsize=1.5in\epsfbox{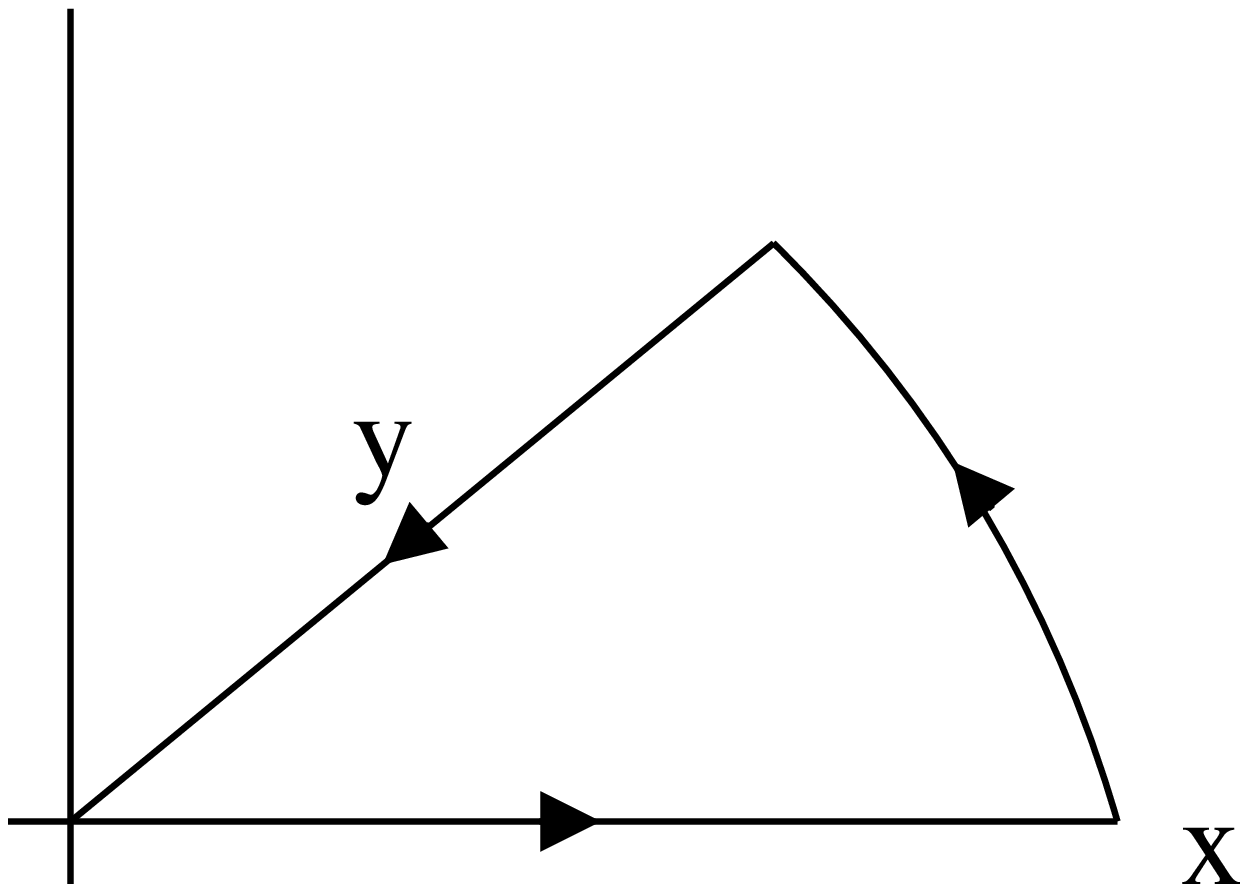}}
\centerline{\small
Figure 2.5. Complex contour for evaluating complex Gaussian integral.}
\medskip

Now let's return to the 4-point function.  
When we do the contractions, we have the possibility of contracting
the external fields with the fields from the interaction vertex.  This
new {\it connected} contribution (figure 3) comes {\it in addition} to the 
disconnected one shown in figure 4.  In the latter, we contract the
$\phi$'s within the interaction only with themselves.  This results
in a {\it loop} diagram, in fact a two-loop diagram.  There is also
another kind of disconnected diagram where one of the particles is
freely propagating, while the other feels the effect of the interaction,
figure 5.  We are going to focus on the connected contribution, fig.\ 3,
for right now.  This kind of diagram is called a {\it tree diagram}
because of its stick-like construction, to distinguish it from
loop diagrams, such as the figure-eight appearing in fig.\ 4.

\medskip\bigskip
\centerline{\epsfxsize=0.8in\epsfbox{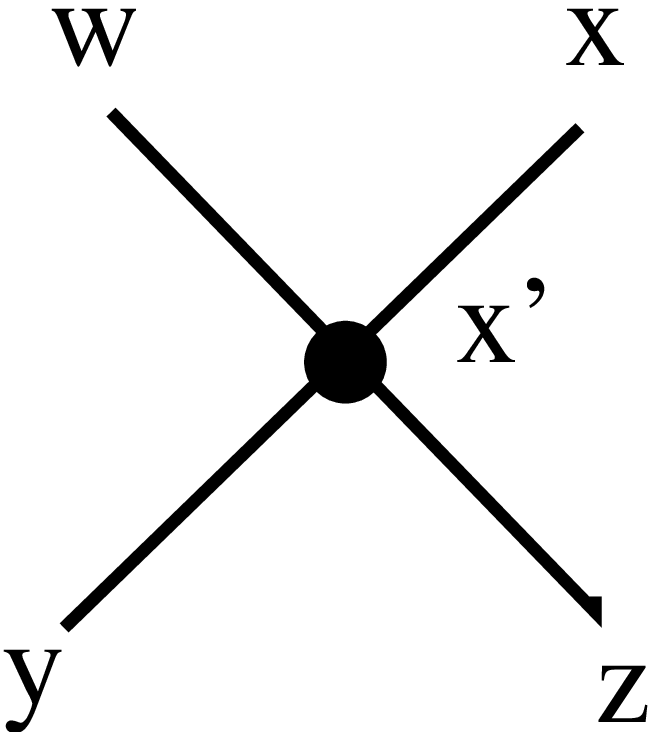}}
\centerline{\small
Figure 3. Connected 4-point function at linear order in $\lambda$:
a tree diagram.}
\medskip

\medskip\bigskip
\centerline{\epsfxsize=3.in\epsfbox{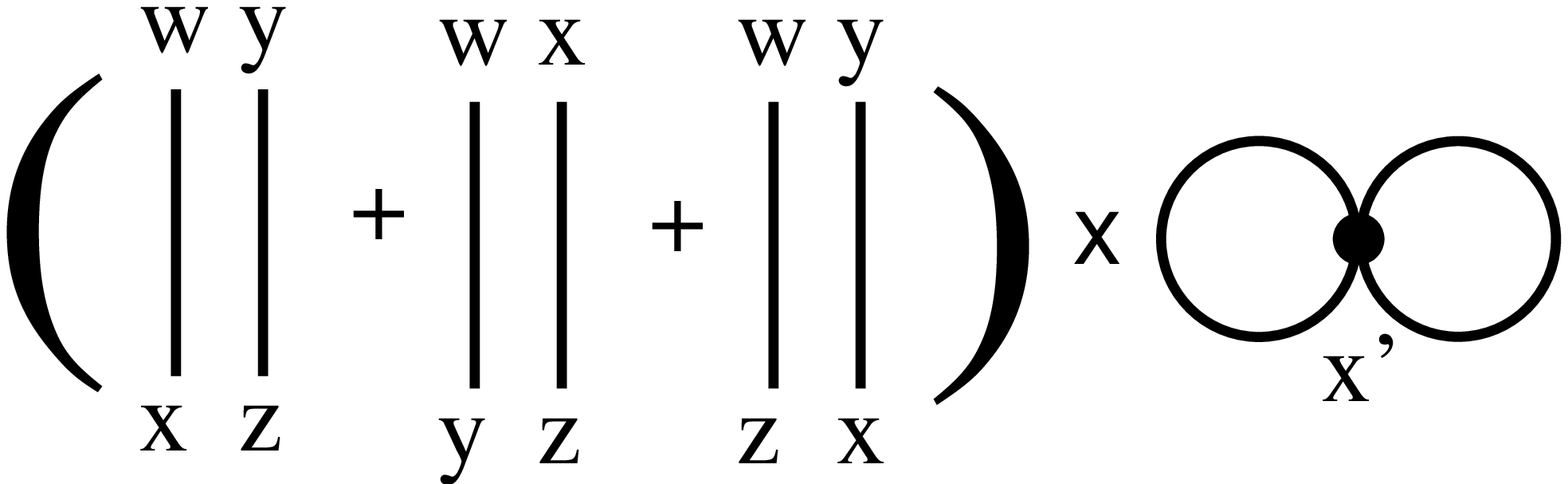}}
\centerline{\small
Figure 4. Contribution to disconnected 4-point function at linear order in $\lambda$.}
\medskip

\medskip\bigskip
\centerline{\epsfxsize=2in\epsfbox{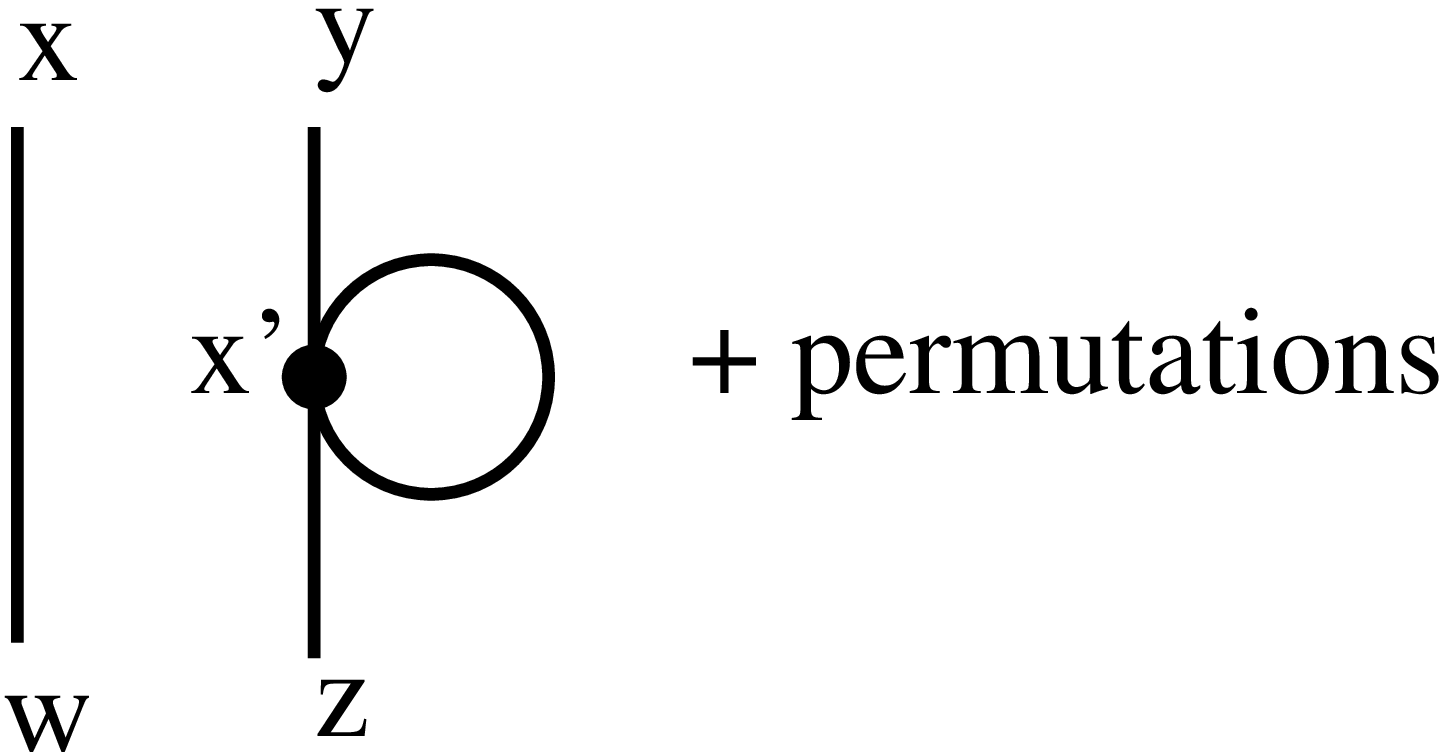}}
\centerline{\small
Figure 5. Another contribution to the disconnected 4-point function at 
linear order in $\lambda$.}
\medskip

In evaluating the connected diagram, we have to take into account all
the possible ways of Wick-contracting.  Start with $\phi(w)$: it has 4
possibilities among the 4 fields of $\phi(x')^4$; let us choose one of
them.  Then $\phi(x)$ has 3 remaining possibilities.  $\phi(y)$ has 2
and $\phi(z)$ is left with 1.  Thus there are $4\cdot 3\cdot 2\cdot 1 =
4!$ contractions which all give the same result.  This is called the
{\it statistical factor} of the Feynman diagram, and it is the reason
for including $1/4!$ in the normalization of the coupling: the $4!$
factors cancel out in the physical amplitude, making the final result
look nicer.  
\beq
	G_4(w,x,y,z) = -i\lambda \int d^{\,4}x'
	G_2(w,x') G_2(x,x') G_2(y,x') G_2(z,x'),
\eeq 
where $G^{(2)}(x,x')$ is the Klein-Gordon propagator,
\beq
	G_2(x,x') = \int {d^{\,4}p\over (2\pi)^4} {e^{-ip(x-x')}
	\over (p^2-m^2 + i\varepsilon)}
\eeq
When we carry out the LSZ procedure to find the proper
4-point vertex, it is very simple: 
\beq
\label{pv}
	\Gamma_4 = -i\lambda 
\eeq
Note that I have slipped into the habit of setting $\hbar=1$ here.
We will work in these units except briefly below when we will want 
to count powers of $\hbar$ to distinguish quantum mechanical from
classical contributions to amplitudes.

Let's pause to give the physical application: this is the transition
matrix element which leads to the cross section for two-body scattering,
by eq.\ (7.25) of \cite{harry}.  The formula for the cross section
can also be found in section 34 of the very useful Review of Particle 
Properties, available on the web \cite{PDG}.  The differential
cross-section is given in eq.\ (34.30) of that reference as
\beq
	{d\sigma\over dt} = {1\over 64 \pi s}\, {1\over |{\bf p}_{1\rm
	cm}|^2 } |{\cal M}|^2
\eeq
where the Mandelstam invariants are $s=(p_1+p_2)^2$, $t = (p_1-p_3)^2$
in terms of the 4-momenta of the incoming ($p_1$, $p_2$) and outgoing
($p_3$, $p_4$) particles.  The matrix element ${\cal M}$ is just another
name for the proper vertex (\ref{pv}).
If the scattering occurs at energies much
larger than the mass of the particle, we can approximate it as being
massless.  Then $s = 4 E^2$ if each particle has energy $E$, and
$|{\bf p}_{1\rm	cm}|^2 = E^2$.  Furthermore, the total cross section
in the massless limit is given by
\beq
\label{xsect}
	\sigma = \int_{-s}^0 dt\, {d\sigma\over dt} = {\lambda^2\over	
	64\pi E^2}
\eeq
in the center of mass frame.  Suppose that $\lambda$ is of order unity.
For particles whose energy is 100 GeV,
near the limiting energy of the LEP accelerator, this is a cross 
section of order $5\times 10^{-7}$ GeV$^{-2}\times (0.197$ GeV-fm)$^2=
10^{-8}$ fm$^2 = 10^{-38}$ m$^2 = 10^{-10}$ barn $= 100$ pb.  For
comparison, the cross section for $e^+e^-\to Z$ at the $Z$ resonance,
measured by LEP, is 30 nb, some 300 times larger. 

\section{The Loop Expansion}
\label{sec:loop}

In the previous example of the 4-point function, the first connected
diagram arose at linear order in $\lambda$.  However, if we consider
the 2-point function, we already have a connected piece at zeroth order:
it is simply the propagator, $
	\widetilde G_2 = {i/(p^2 - m^2)}.$
When we compute the first order correction to this, we get the diagram
of figure 6.  There are 4 ways of doing the first contraction and
3 of the second, so the statistical factor is $4 \cdot 3/4! = 1/2$.
In position space, this diagram is a correction to the 2-point function
given by
\beq
	\delta G_2(x,y) = -{i\lambda\over 2}\int d^{\,4}x'
	G_2(x',x') G_2(x,x') G_2(y,x') 
\eeq
When we transform to momentum space and remove the overall factor
of $(2\pi)^4\delta^{(4)}(p_1-p_2)$ for 4-momentum conservation,
we are left with 
\beq
	\delta \widetilde G_2 = -{i\lambda\over 2}G_2(x',x')(\widetilde G_2)^2  = 
	-{i\lambda\over 2} (\widetilde G_2)^2 \intdp {i\over \pmie}	
\eeq 
What is the physical meaning of this correction?  This is most easily
seen by expanding the following expression in $\delta m^2$:
\beq
\label{expand}
	{i\over  p^2-m^2- \delta m^2}  = 
	{i\over  p^2-m^2} + {i \delta m^2\over (p^2-m^2)^2} +\cdots
\eeq
This expansion has exactly the same form as 
$\widetilde G_2 + \delta\widetilde G_2$.  We can therefore see that
the loop diagram of figure 6 is nothing more nor less than a correction
to the mass of the particle:
\beqa
\label{mass_shift}
	m^2_{\rm phys} &=& m^2 + \delta m^2;\quad
	\delta m^2 = i{\lambda\over 2} \intdp {1\over \pmie}
\eeqa
\medskip\bigskip
\centerline{\epsfxsize=1.5in\epsfbox{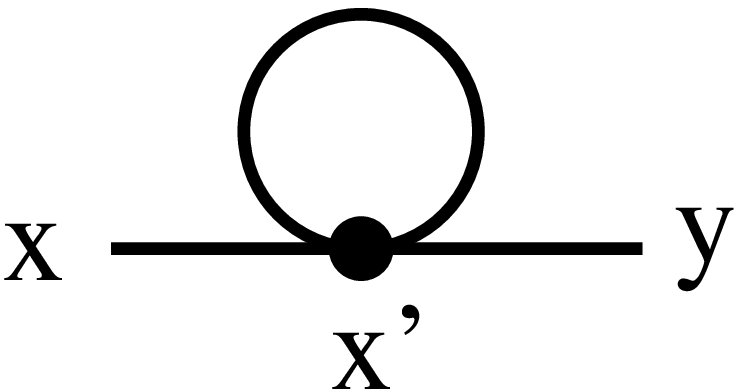}\hfil\small}
Figure 6. Correction to the 2-point function.
\medskip

This interpretation is not merely a consequence of expanding to first
order in $\lambda$.  We could continue the procedure of expanding eq.\
(\ref{expand}) to arbitrary orders.  It corresponds to the series of
diagrams in figure 7.  This is not to say that fig.\ 7 is the complete
answer for the corrections to the mass.  There are other corrections
starting at two loops which come in addition to the one we have
calculated.  Notice that the extra diagrams in fig.\ 7 relative to
fig.\ 6 have no effect on our expression (\ref{mass_shift}) for the
mass shift.  What we have computed is the one-loop contribution to the
mass of the particle.  The extra diagrams in fig.\ 7 which contain two
or more loops are just an iteration of our one-loop result when we
expand (\ref{expand}) to higher order in $\delta m^2$.  An example of a
higher order diagram which gives an intrinsically two-loop
contribution to the mass is fig.\ 8, called the ``setting sun''
diagram.

\medskip\bigskip
\centerline{\epsfxsize=5.5in\epsfbox{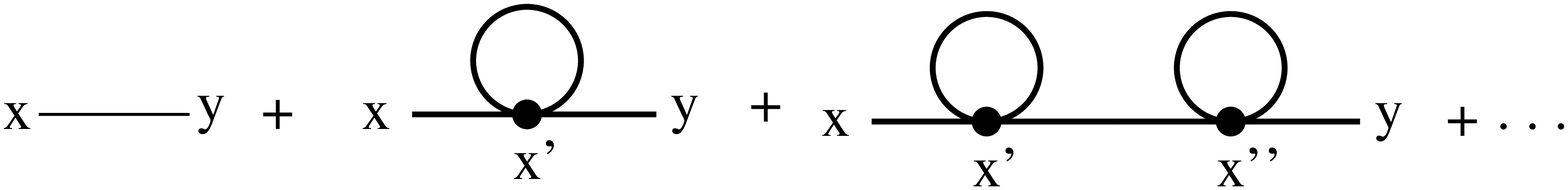}}
\centerline{\small
Figure 7. A series of corrections to the 2-point function.}
\medskip

\medskip\bigskip
\centerline{\epsfxsize=2.5in\epsfbox{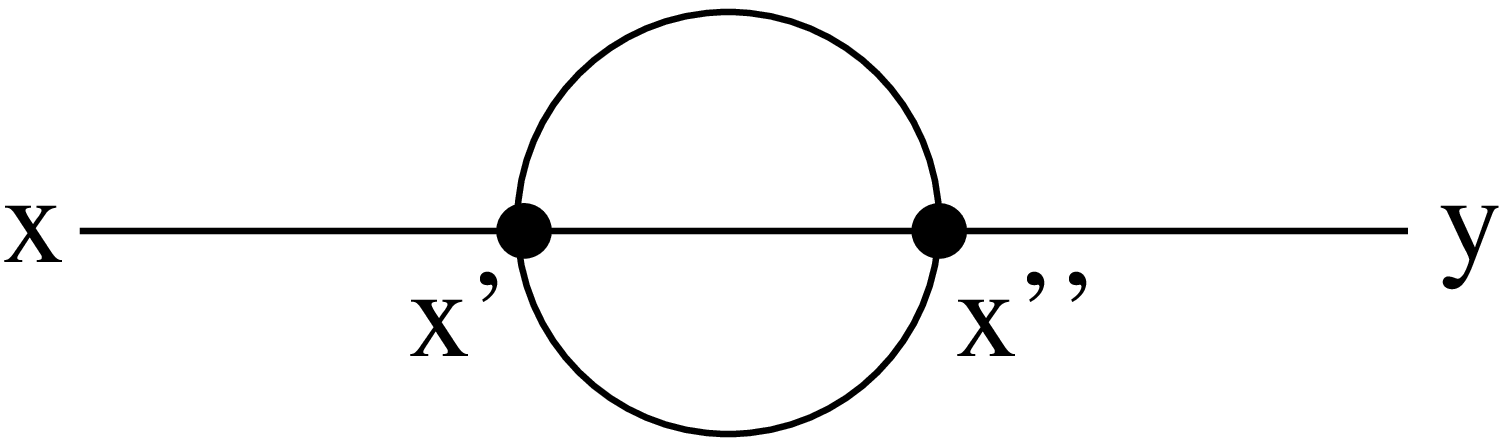}}
\centerline{\small
Figure 8. A two-loop correction to the 2-point function.}
\medskip

At some point above, I started using the units $\hbar=1$.  However, 
it is enlightening to restore the $\hbar$'s for a moment to see how
they enter the loop expansion.  Recall that each power of $\lambda$ is
accompanied by $1/\hbar$.   On the other hand, each propagator, since
it is like the inverse of the action, comes with $\hbar$ in the
numerator.  The series for the propagator can be written as
\beq
  {i\hbar\over \p2m2} + {i \hbar^2 \delta m^2\over (p^2-m^2)^2} +\cdots
\eeq
We see that the tree diagram comes with one power of $\hbar$, the
one-loop diagram has $\hbar^2$, {\it etc.}.  If we were to imagine
taking $\hbar\to 0$, the tree diagram would be the leading contribution.
It is easy to see that the same is true not just for two-point
functions, but for any Green's function.   The important conclusion is that
{\it tree diagrams represent the classical contributions to a given
process, while loops are quantum corrections.}  We can check this
on another example, the 4-point function, whose first loop correction
is shown in fig.\ 9.  Let us now truncate the external legs for 
simplicity, {\it \`a la} LSZ:
\beq
	\Gamma_4 = -i{\lambda\over\hbar} + O\left({\lambda^2\over\hbar^2}
	\left({\hbar\over \p2m2}\right)^2\right)
\eeq
Again, the loop contribution is suppressed by an additional power of
$\hbar$ relative to the tree diagram.  

\medskip\bigskip
\centerline{\epsfxsize=6.5in\epsfbox{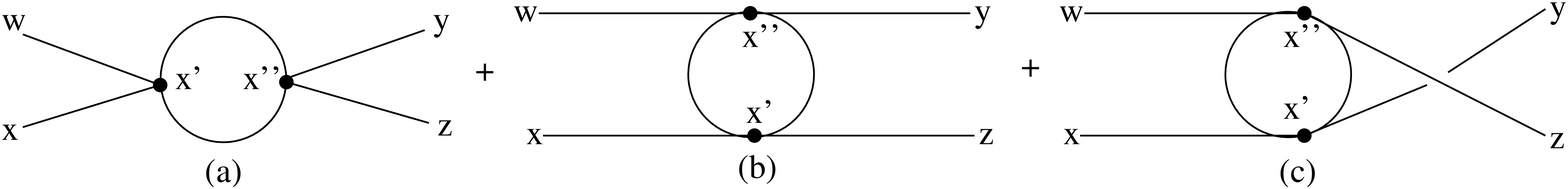}}
\centerline{\small
Figure 9. One-loop corrections to the 4-point function.}
\medskip

\section{The Feynman Rules}
\label{sec:frules}

While we are discussing this diagram, we might as well work out its
value.  We have to expand the interaction in $e^{iS}$ to second order,
so that
\beqa
	\delta G_4 &=& {1\over Z[0]}\int {\cal D}\phi\, e^{iS[0]/\hbar} \phi(w)
	\phi(x)\phi(y)\phi(z) \frac12 \left[ \int d^{\,4}x'  
	\left(-{i\over\hbar}{\lambda\over 4!}\phi^4 \right)\right]^2.\\
&=& {1\over Z[0]}\int {\cal D}\phi\, e^{iS[0]/\hbar} \phi(w)
	\phi(x)\phi(y)\phi(z) \frac12 \left[ \int d^{\,4}x'  
	\left(-{i\over\hbar}{\lambda\over 4!}\phi^4(x') \right)\right]
\left[ \int d^{\,4}x''
	\left(-{i\over\hbar}{\lambda\over 4!}\phi^4(x'') \right)\right].
	\nonumber	
\eeqa
Notice that there are three topologically distinct ways of contracting
the external legs with the vertices.  This will be more clear when we
go to momentum space (figure 10).  The distinction occurs because I
have already decided that positions $w$ and $x$ will correspond to
the incoming particles, and $y$ and $z$ will be for the outgoing ones.
Let's first treat diagram 9(a); the other two will then follow
in a straightforward way.

First the statistical factor:  To connect $\phi(w)$, we can choose
either the $x'$ or the $x''$ vertex.  That gives a factor of 2.
Let's choose $x'$.  We have 4 choices of contraction here.
Next connect $\phi(x)$ to one of the remaining legs on the $x'$
vertex; this is a factor of 3.  We know that $\phi(y)$ must go with
the $x''$ vertex, and there are 4 ways of doing this.  That leaves
3 ways of connected $\phi(z)$.  Finally, there are 2 ways of connecting
the remaining legs between the vertices.  So the statistical factor
is $(2\cdot 3\cdot 4/4!)^2/2= 1/2$.  In position space, we have
\beq
	\delta G_4^{(a)} = {(-i\lambda)^2\over 2} \int d^{\,4}x'
	\int d^{\,4}x''\,
	G_2(w,x') G_2(x,x') G_2(x',x'') G_2(x',x'') G_2(x'',y)
	G_2(x'',z)
\eeq
Next we want to Fourier tranform.  Suppose the momentum going into
and out of the graph is $q = p_1 + p_2$.  There is also momentum
going around the loop, $p$.  Momentum is conserved at each vertex,
as shown in figure 9, so one internal line carries momentum $p$,
the other has $p+q$.  When we Fourier transform the external 
legs, it just undoes the momentum integrals for those propagators,
and replaces the $1/(q_i^2-m^2)$ factors with $1/(p_i^2-m^2)$,
where $q_i$ is the integration variable appearing in the position
space propagator, and 
$p_i$ is the external momentum.  For the two internal progators,
one of the momentum integrals is for overall 4-momentum conservation.
The remaining one is the only one that is left. When we
remove the external legs and the overall delta function,  we get
\beq
\label{dg4}	
	\delta\Gamma_4^{(a)}(q) = {(-i\lambda)^2\over 2}
	\intdp {i\over \pmie}\, {i\over (p+q)^2 - m^2+i\varepsilon}
\eeq
The contributions (b) and (c) are the same but with external momenta
$r$ and $l$ replacing $q$.  The three respective graphs are called 
the $s$-, $t$- and $u$-channels, after the Mandelstam invariants
$s=q^2$, $t=r^2$ and $u=l^2$.

\medskip\bigskip
\centerline{\epsfxsize=6.5in\epsfbox{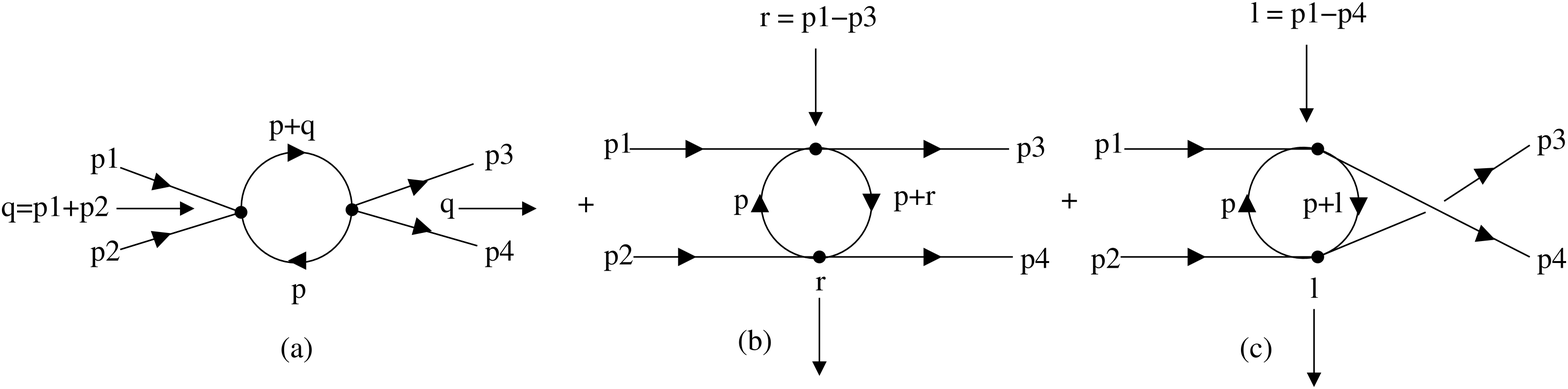}}
\centerline{\small
Figure 10. Momentum space version of figure 9.}
\medskip

After going through this procedure enough times, one comes to realize
that there is no need to start in position space and then Fourier
transform.  One can write down directly the expression for the truncated
loop diagram by associating the proper factors with each element:
\begin{itemize}
\item for each vertex a factor $-i\lambda$;
\item for each internal line, a propagator $\displaystyle {i\over \pmie}$;
\item for each closed loop, a momentum integral $\displaystyle\intdp$
\end{itemize}
In addition, one should 
\begin{itemize}
\item conserve momentum at each vertex
\item compute the statistical factor
\end{itemize}
These are the {\it Feynman rules} for the $\lambda\phi^4$ theory,
which make it relatively easy to set up the computation for a given
loop diagram.  Every theory (QED, QCD, {\it etc.}) will have its own
Feynman rules corresponding to the detailed way in which the particles
couple to each other.

\section{Evaluation of diagrams; regularization}
\label{sec:reg}

We now have formal expressions for the loop diagrams, but we still
have to evaluate them.  Let's start with the easiest diagram, fig.\
6, eq.\ (\ref{mass_shift}).   The first thing we have to do is
to make sense out of the pole in the propagator.  This is what the $i\varepsilon$ is
for.  The procedure is called the {\it Wick rotation}.  We will do
the integral over $p_0$ separately, and use the techniques of contour 
integration.
Notice that the poles in the complex $p_0$ plane are located at
\beq
	p_0 = \pm( \sqrt{\vec p^{\,2} + m^2} -i\varepsilon)
\eeq
Their positions are indicated by the x's in fig.\ 11.  According to
the theory of complex variables, we can deform the integration contour
without changing the value of the integral as long as we don't move
it though the poles.  This defines the sense in which we are allowed to
rotate the integration contour from being along the real $p_0$ axis 
to the imaginary one, as shown by the arrow.  When we do the Wick
rotation, we are making the replacement
\beq
	p_0 \to i p_0 
\eeq
everywhere in the integral.  This is equivalent to going from Minkowski
to Euclidean space, so we can also denote the new momentum
variable by $\pE$, for the Euclidean momentum.  Notice that because
I am using the mostly $-$ convention for the Minkowski space metric,
the square of the 4-momentum vector transforms as
\beq
	p^2 \to -\pE^2
\eeq
when Euclideanizing. We thus obtain
\beq
\label{mass_shift2}
	\delta m^2 = -i^2{\lambda\over 2} \intdpe {1\over \pE^2+m^2}
\eeq
Notice that the $i\varepsilon$ prescription is no longer necessary
once we have Wick-rotated.  Also, the factor of $i$ coming from the
integration measure is just what we needed to cancel the explicit one
that was already present, so that our mass shift is real-valued.  

\medskip\bigskip
\centerline{\epsfxsize=2.5in\epsfbox{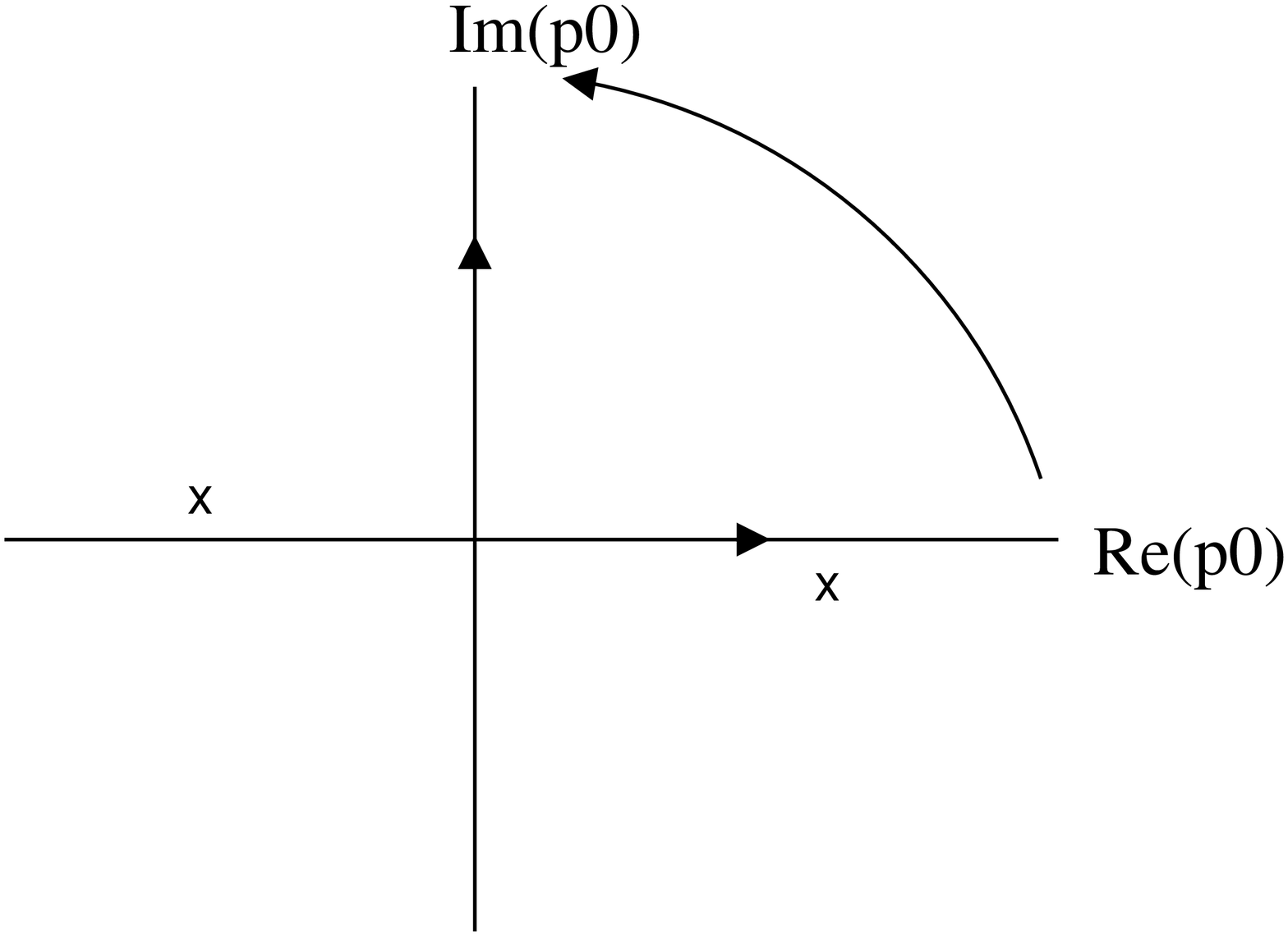}}
\centerline{\small
Figure 11. Integration contours for the Wick rotation.}
\medskip

Now we have made sense out of the pole in the propagator, so the
integral is looking a bit nicer, but there is just one problem:
it diverges quadratically as $p\to\infty$!  This is one of the famous
ultraviolet divergences of quantum field theory, which comes from the
fact that we have infinitely many degrees of freedom:  virtual 
pairs of $\phi$ particles with arbitrarily large momentum are
contributing to the quantum correction to the mass.  We have to cut
off this divergence somehow.  

We don't really believe this theory is a correct description of
nature up to arbitrarily high energies and momenta.  Let us suppose
we only think it is true up to some very high scale $\Lambda$.  
Then it
makes sense to cut off the integral at this scale.  This gives us a
way of defining the correction.  The integral can be factored into
angular and radial parts by going to spherical coordinates in
4 dimensions:
\beqa
	p_1 &=& p \sin\theta_1 \sin\theta_2 \cos\phi\nonumber\\
	p_2 &=& p \sin\theta_1 \sin\theta_2 \sin\phi\nonumber\\
	p_3 &=& p \sin\theta_1 \cos\theta_2 \nonumber\\
	p_4 &=& p \cos\theta_1
\eeqa
Then we get 
\beq
\label{mass_shift3}
	\delta m^2 = {\lambda \over 2(2\pi)^4} \int d\Omega_3
	\int_0^\Lambda  d\pE \pE^3{1\over \pE^2+m^2}
\eeq
where the volume of the 3-sphere is given by
\beq
	\int d\Omega_3 = \int_0^\pi 
	\sin^2\theta_2
	d\theta_2 \int_0^\pi \sin\theta_1
	d\theta_1 \int_0^{2\pi} d\phi  =  2\pi^2
\eeq
The radial integral can be rewritten using $u=\pE^2$ as
\beqa
	\frac12 \int_0^{\Lambda^2} du {u\over u+m^2} &=&
	\frac12 \int_{m^2}^{\Lambda^2+m^2} du {(u-m^2)\over u}\\
	&=& \frac12 \left(u -{m^2\ln u}\,\right|^{\Lambda^2+m^2}_{m^2}\\
	&=& \frac12 \left(\Lambda^2 - m^2\ln(1 + \Lambda^2/m^2)
	\right)
\eeqa
so that finally
\beq
	\delta m^2 = {\lambda\over 32\pi^2}
	\left(\Lambda^2 - m^2\ln(1 + \Lambda^2/m^2)\right)
\eeq

The procedure of rendering the Feynman diagram finite is called
{\it regularization}.  There are many ways of doing it; we have
used the {\it momentum space cutoff} method here.   A more elegant
but less physically transparent method, dimensional regularization,
will be introduced a little later on. 

The result is that the mass which is physically measured is only
approximately given by the original parameter $m^2$ which was in
the Lagrangian.  We refer to this parameter as the {\it bare}
mass  and we rename it $m_0^2$ to emphasize that it is not
yet corrected by perturbation theory.  The physical mass is
$m_0^2 + \delta m^2$, to linear order of accuracy in perturbation
theory.

\section{Renormalization}
\label{sec:renorm}

The physical mass is something that is experimentally measured,
whereas the cutoff is an arbitrary parameter which is not very
well defined.  It might by 10 TeV or it might be $10^{19}$ GeV.  The
experimentally measured values must not depend on our arbitrary
choice of $\Lambda$.  This tells us that the bare parameter $m^2_0$
must really be a function of $\Lambda$.  Defining $\hat\lambda=\lambda
/(32\pi^2)$,
\beqa
	m^2_0(\Lambda)  &=&  m^2_r-{\hat\lambda}
	\left(\Lambda^2 - m_0^2\ln(\Lambda^2/\mu^2)\right)
	\nonumber\\ 
\label{counterterm}
	\longrightarrow \quad m^2_0(\Lambda)  &=& 
	{m_r^2- \hat\lambda \Lambda^2\over 1 -
	\hat\lambda\ln\Lambda^2/\mu^2} \ \cong \ m_r^2(1 +
	\hat\lambda\ln\Lambda^2/\mu^2) - \hat\lambda \Lambda^2
\eeqa
where $m_r$ and $\mu$ are parameters which do not depend on
$\Lambda$.  The second equality in (\ref{counterterm}) takes account
of the fact that we are only working to first order in $\lambda$.
Now when we combine the bare mass and the correction to get the
physical value, we obtain
\beq
	m^2_{\rm phys} = m_r^2(1+\hat\lambda\ln m_r^2/\mu^2),
\eeq
which does not depend on the cutoff.  The terms we added to $m^2_0$
which diverge as $\Lambda\to\infty$ are called {\it counterterms}. 
If the original Lagrangian was naively considered to have $m^2_r$ as
the mass parameter, then the one-loop  counterterms are what we need
to add to it so that physical mass is finite at one loop order. Of
course, we should continue the process to higher orders in $\lambda$
and higher loops to be more accurate.  

The procedure we have carried out here is known as {\it
renormalization}.  If $m_0^2$ was the bare mass, then $m_0^2+ \delta
m^2$ is the renormalized one, at least up to finite corrections.   So
far it looks like this business has been somewhat pointless. We did
all this work to find corrections to the mass, yet it did not allow
us to predict anything, since in the end we simply fixed the
parameters of the theory by comparing to experiment.  However, when
we continue along these lines for all the other relevant amplitudes,
we will see that renormalization allows us to make very detailed
predictions relating masses and scattering cross sections, and
telling us how these quantities scale as we change the energy scale
at which they are being probed.  To appreciate this, we should next
turn to the 4-point function, which controls 2-body scattering. 

Before doing that however, we can already draw a very interesting
conclusion from our computation of the mass.  Suppose that we were
talking about the Higgs boson, even though this isn't exactly the
right theory for it.  It is known that the Higgs mass is close to
or above about 115 GeV, since LEP II finished running with a
suggestive but not overwhelmingly convincing hint of a signal at
that mass.  On the other hand, our theory might be cut off at a 
scale many orders of magnitude greater, perhaps even the Planck
mass at $\sim 10^{19}$ GeV.  If this is the case, then there had
to be a very mysterious conspiracy between the bare parameters
and the quantum corrections: they had to cancel each other to a 
part in $10^{17}$ or so.  This does not look natural at all, leading
us to believe that there must be some physical mechanism at work.
This mystery is known as the {\it gauge hierarchy problem}.
Either there is some approximate symmetry like supersymmetry which
explains why the quantum corrections aren't really as large as our
theory predicts, or else there is some other new physics which
invalidates our theory at energy scales which are not that much
greater than the 100 GeV scale where we are presuming that it 
works.

Now to continue our discussion of renormalization to the 4-point
function, recall that 
we computed the scattering cross section at tree level in eq.\
(\ref{xsect}), which is simply related to the tree level value of
the 4-point proper vertex, $\Gamma_4 = -i\lambda$.  To correct
this, we should add the one-loop contribution $\delta\Gamma_4$
from eq.\ (\ref{dg4}).  Evaluating this kind of integral is harder,
and Feynman has invented a very useful trick for helping to do so.
The trick is to combine the two denominator factors into a single
denominator, using the identity\footnote{the more general
expression, useful for more complicated diagrams, is
\beq
	\prod_{i=1}^k {1\over D_i^{a_i}} 
	= {\Gamma(\sum_i^k a_i)\over \prod_{i=1}^k \Gamma(a_i)}
	\int_0^1 dx_1 \cdots \int_0^1 dx_k\ 
	{\delta\left(1 - \sum_i^k x_i\right) \prod_{i=1}^k x_i^{a_i-1}
	\over \left(\sum_i^k D_i x_i \right)^{\sum_i^k a_i}}
\eeq
in terms of gamma functions.}
\beq
	{1\over AB} = \int_0^1 dx\, {1\over (xA+(1-x)B)^2}
\eeq
The integral over $x$ is known as the Feynman parameter integral.
When we apply this to (\ref{dg4}), it becomes
\beqa
	\delta\Gamma_4^{(a)}(q) &=& {\lambda^2\over 2}
\intdp {1\over (p^2 + 2xp\cdot q + x q^2 -
m^2+i\varepsilon)^2}\nonumber\\
	&=& {\lambda^2\over 2}
\intdp {1\over ((p+xq)^2 + x(1-x) q^2 - m^2+i\varepsilon)^2}\nonumber\\
\label{dg4a}	
	&=& {\lambda^2\over 2}
\intdp {1\over (p^2 + x(1-x) q^2 - m^2+i\varepsilon)^2}\\
\label{dg4b}
	&=& i{\lambda^2\over 2}
\intdpe {1\over \left(\pE^2 - x(1-x) q^2 + m^2\right)^2}
\eeqa
and the crossed diagrams (b,c) are the same but with $q\to r$ and
$q\to l$.
In (\ref{dg4a}) we assumed that it is permissible to shift the
variable of integration, and in (\ref{dg4b}) we did the Wick
rotation on $p$, but not on $q$---the latter is still a Minkowski
4-momentum.  Notice that the shift of integration variable might
be considered somewhat of a cheat, if we were thinking that the
original $p$ variable was cut off at $\Lambda$.  That cutoff will
get shifted when we change variables.  However, this 
only introduces an error of order $q^2/\Lambda^2$, which vanishes
in the limit $\Lambda\to\infty$.
 Now we can do the integral as we did 
before.  Defining $M^2(q) = m^2 - x(1-x)q^2$, 
\beqa 
	\delta\Gamma_4^{(a)}(q) &=& i{\lambda^2\over 32\pi^2}
	\int_0^1 dx \int_0^{\Lambda^2} du {u\over (u+M^2(q))^2}\\
&=& i{\lambda^2\over 32\pi^2}
	\int_0^1 dx \int_{M^2(q)}^{\Lambda^2+M^2(q)} du {u- M^2(q)
\over u^2}\\
&=& i{\lambda^2\over 32\pi^2}
	\int_0^1 dx \left[ \ln(1+\Lambda^2/M^2(q))
	 -(1+M^2(q)/\Lambda^2)^{-1} \right]\\
\label{fourpt}
&\cong& i{\lambda^2\over 32\pi^2}
	\int_0^1 dx \left[ \ln(\Lambda^2/M^2(q))
	 -1 \right]
\eeqa

This result is much richer than our computation of the mass shift:
here we have found the the quartic coupling gets renormalized not
just by a constant shift, but in fact a function of the external
momentum.  Let's defer the study of this finite part for later, since
we are discussing the renormalization of the theory.  First, notice
that this result is less divergent than the mass shift; this is only
log divergent rather than quadratic.  Second, there is no momentum
dependence in the divergent part, which is important for
renormalizability.  The divergence can be removed by defining the
{\it bare coupling} $\lambda_0$ as
\beq
\label{bare_c}
	\lambda_0 = \lambda_r + {3\lambda_r^2\over 16\pi^2}
	\ln \Lambda/\mu
\eeq
where, again, $\lambda_r$ and $\mu$ are independent of $\Lambda$.
(The factor of 3 comes from adding the crossed contributions.)
If we had found that the nontrivial function of external momenta
was multiplied by a divergent coefficient, we would be in trouble.
In that case we would have to modify the original (bare) theory
with a counterterm that did not resemble the simple operators
in the action we started with.  In fact, a theory with a nontrivial
function of momentum in the action would be {\it nonlocal}.  If
you Fourier transform a generic function $f$ of momentum back to position
space, you would generally need arbitrarily large numbers of 
derivatives.  For example, by Taylor-expanding $f$,
\beqa
	f(q_1+q_2) \tilde\phi(q_1) \tilde\phi(q_2)
	\tilde\phi(q_3) \tilde\phi(q_4) &=& 
	\sum_n f_n (q_1+q_2)^{2n} \tilde\phi_1 \tilde\phi_2
	\tilde\phi_3 \tilde\phi_4 \\
	&\to& \sum_n f_n 
	\partial_\mu\partial^\mu( \phi_1 \phi_2 )
	\phi_3 \phi_4,
\eeqa
which would be a rather ugly term to add to a Lagrangian.  We say
that an interaction with arbitrarily large number of derivatives 
is nonlocal because we know that
\beqa
	\phi(x)\phi(x+a) &=& \phi(x) e^{a {d\over dx}} \phi(x) \\
	&=& \phi(x)\sum_n {1\over n!} \left(a {d\over dx}\right)^n \phi(x)
\eeqa
for example.   This would be something like a theory with
instantaneous action at a distance (the distance being $a$).  Such
an interaction contradicts our cherished belief that all interactions
are fundamentally local in nature, occuring at a specific point in
spacetime, at least down to the (100 GeV)$^{-1}$ distance scales
which have been experimentally explored.

Fortunately, we are able to absorb the new infinity simply into the
coupling constant $\lambda$. In this way, the scattering cross
section computed at one loop is independent of $\Lambda$, just like
the mass shift was.  Similarly to that example, we must determine an
unknown combination  $\lambda_r - {\lambda_r^2\over 16\pi^2}\ln\mu$
by comparing the theory to the measured value of the cross section.

Again, we come to the question: where is the predictive power of 
the theory if we are merely fixing the unknown parameters in
the Lagrangian by comparing to measured quantities?  The answer is
that by fixing just these {\it two} quantities, the mass and the 
coupling, we are able to compute everything else, and thus make
predictions for an arbitrarily large number of other physical
processes.  This is the crux of {\it renormalizability}: a theory
is said to be renormalizable if only a finite number of parameters
need to be shifted in order to remove all the infinities.

For example, we could consider an inelastic scattering process, in
which two particles collide to create four particles.  Some of the
tree diagrams for this process are shown in fig.\ 12. Now consider a
one-loop correction, fig.\ 13.  For simplicity, let's assume that the
net external momentum entering and leaving the graph is zero.  This
assumption can't be true for a real process in which all the
external particles are on their mass shells ({\it on-shell}, meaning
each particle satisfies $p^2 = m^2$), since in that case we should
have at least $4m$ of energy entering the diagram in order to 
produce four particles at rest.  However, the point I want to make
concerns the contribution to this diagram from very high-momentum
virtual particles, which is insensitive to the small external
momentum.  The Feynman rules tell us that (after Wick-rotating)
\beq
\label{g6est}
	\Gamma_6 \sim \intdpe {1\over ({\pE^2+m^2})^3}
\eeq
which is a convergent integral.  Similarly, it is easy to see that
higher-point functions are even more convergent.  At one loop order,
the only divergent quantities are $\Gamma_2$ and $\Gamma_4$, and this
is why we can renormalize the theory using only two parameters,
$m^2$ and $\lambda$.

\medskip\bigskip
\centerline{\epsfxsize=3.5in\epsfbox{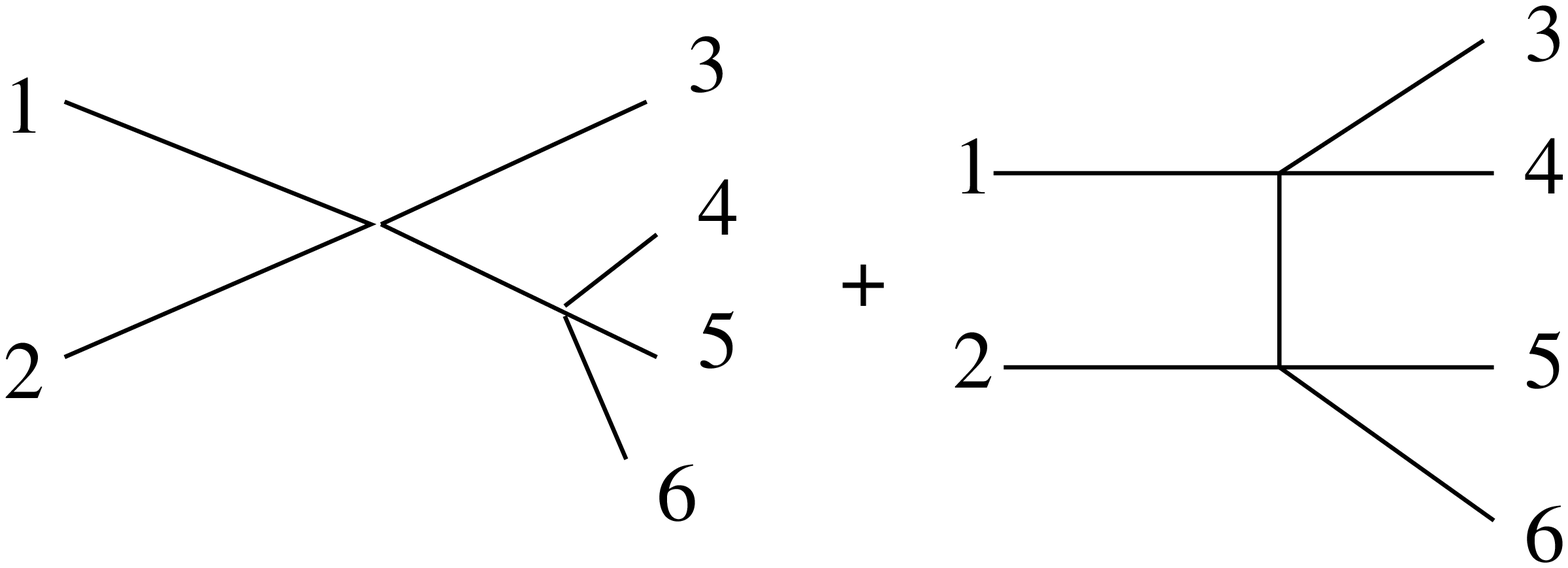}}
\centerline{\small
Figure 12. Tree-level contributions to $2\to 4$ scattering.}
\medskip
 
\medskip\bigskip
\centerline{\epsfxsize=2.5in\epsfbox{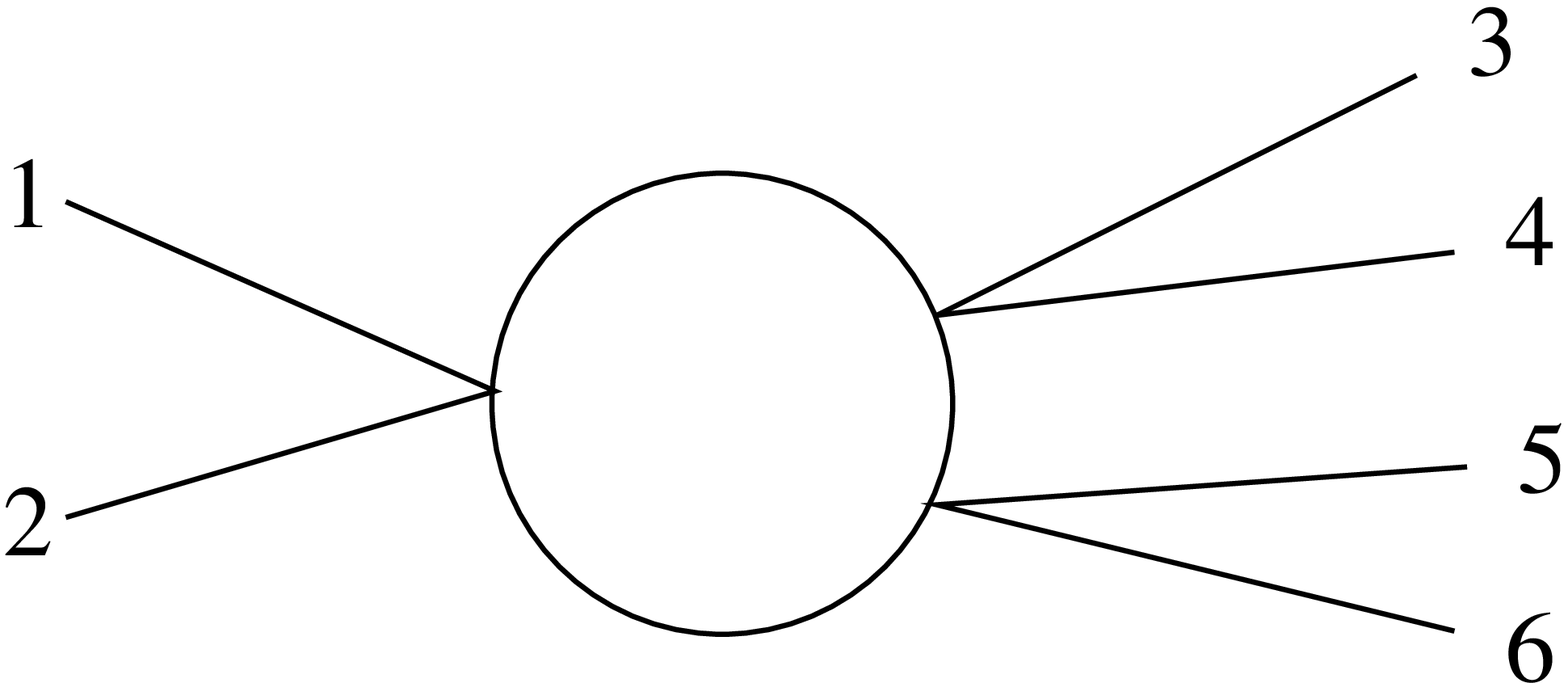}}
\centerline{\small
Figure 13. One-loop contribution to $2\to 4$ scattering.}
\medskip
 
The estimate (\ref{g6est}) is an example which leads the way to a
general {\it power-counting} argument of the type which was first
introduced by S.\ Weinberg, for establishing the renormalizability of
a theory.  The argument goes like this: let's consider an arbitrarily
complicated Feynman diagram, which has $L$ loops, $V$ vertices, and
$E$ external legs.  We would like to know which diagrams in the 
theory are divergent, to extend the concept of renormalization beyond
just one loop.  We know that each loop will give us an integral of
the form $\intdp$; how badly these integrals can diverge in the UV
(ultraviolet, high momentum) depends on how many propagators there
are.  To count the number of propagators, we need the number of 
internal lines, $I$.  Before doing contractions, there are $4V - E$
lines available to contract, and afterwards there are
\beq
	I= 2V-E/2
\eeq
internal lines.  There is also a relationship between these 
quantities and the number of loops.  Each internal line has a
momentum associated with it, and each vertex must conserve
momentum.   The total number of internal lines minus vertices is 
related to the
number of independent momenta, which is also the number of loops: 
\beq
	L = I - V + 1 = V - E/2 + 1
\eeq
By dimensional analysis, the diagram must be divergent if
$4L$ exceeds $2I$ (since each propagator goes like $1/p^2$), 
and we call this difference the {\it superficial
degree of divergence}, $D$:
\beq
	D = 4L - 2I = 4V - 2E + 4 - 4V + E = 4 - E
\eeq
$D$ is the minimum power of $\Lambda$ with which a diagram with 
$E$ external legs can be expected to scale: $E=2$ gives the quadtratic
divergence $D=2$, and $E=4$, $D=0$ is log divergent.  We did not yet
discuss the vacuum diagrams, with no vertices, as in fig.\ 14, 
but it is easy to see that they also obey this rule, being
quartically divergent.  (Vacuum diagrams give contributions to the
energy density of the vacuum, {\it i.e.,} the cosmological constant.)

\medskip\bigskip
\centerline{\epsfxsize=1.in\epsfbox{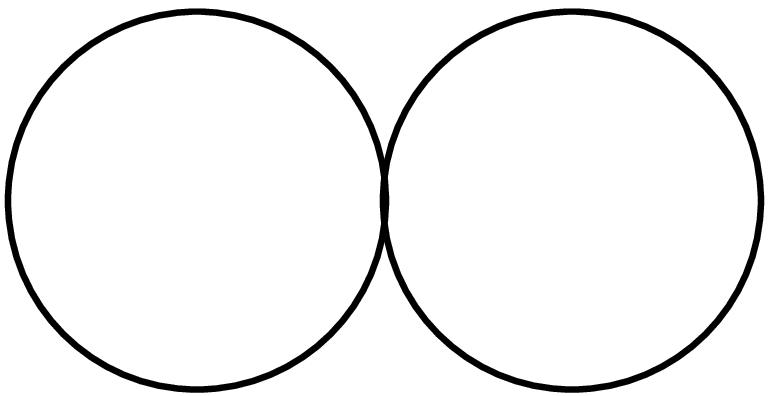}}
\centerline{\small
Figure 14. ``Figure eight'' vacuum diagram, which is quartically
divergent.}
\medskip

According to this superficial counting, one might expect all the
higher point functions to be finite, as indeed our one-loop examples
proved to be.  However, one can always embed one of the divergences
from $\Gamma_2$ or $\Gamma_4$ inside a diagram with an arbitrary 
number of external legs, rendering it divergent, which is why this
kind of power-counting criterion {\it is} superficial.  For example,
insertion of a mass correction on the external legs of the 6-point
function produces a quadratic divergence, while the other
diagram shown (b) is log divergent (fig.\ 15).  However, it should
not be too surprising that these divergences, which are not 
accounted for by the simple power-counting argument, do not pose
any new problems---in fact, they are already automatically taken 
care of by the renormalization of the mass and coupling which we 
have already done!  

\medskip\bigskip
\centerline{\epsfxsize=3.5in\epsfbox{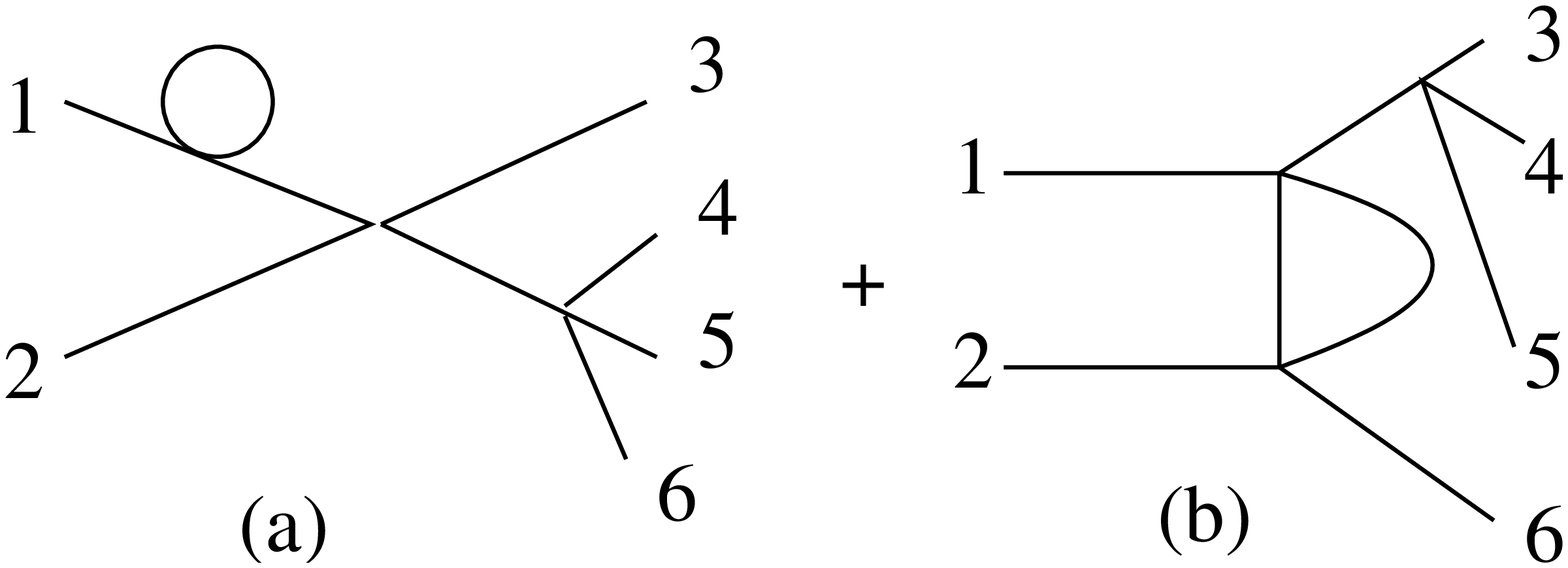}}
\centerline{\small
Figure 15. Divergent contributions to 6-point function.}
\medskip

For example, fig.\ 15(a) corresponds to correcting the mass of one of
the external legs.  Really, we should correct the masses of
all the external particles, not just one of them.  Mass
renormalization is somewhat special in this respect.  To correct the
propagator, we had to imagine summing up the entire geometric
series of diagrams which were of the form of fig.\ 7.  As the great
theorist and pedagogue of field theory Sidney Coleman put it, this
is the one nonperturbative step you must always do in perturbation
theory.  In practical terms, it means that we don't really have to
write down the diagrams like fig.\ 15(a); we merely keep in mind that
masses will get renormalized.  

What about the log divergence in fig.\ 15(b)?  It is pretty obvious
that this gets canceled by the counterterm (divergent part)
from the corresponding vertex of the tree diagram shown in fig.\ 16.

\medskip\bigskip
\centerline{\epsfxsize=1.5in\epsfbox{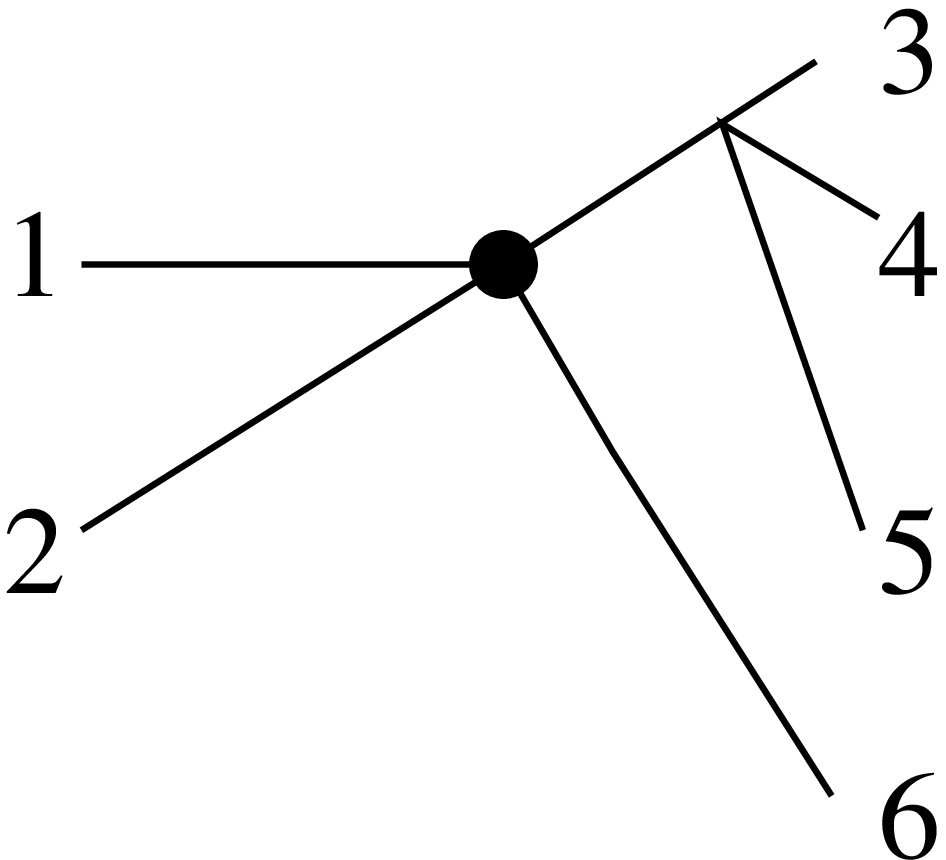}}
{\small
Figure 16. Divergent counterterm contribution to 6-point function,
which cancels the divergent part of fig.\ 15(b).}
\medskip

We can make a distinction between the kinds of diagrams which are
intrinsically important for renormalization, versus those for
which renormalization automatically works because of taking care
of the first kind.  Notice that the diagrams in fig.\ 15 can be
chopped up into smaller diagrams by severing just a single line.
These are call {\it one-particle reducible} graphs, abbreviated
1PR.  On the other hand, the diagrams like figs.\ 6, 8, 9 cannot
be split into two separate graphs by cutting a single line.  They
are called {\it one-particle irreducible}, 1PI.   The latter are the
important ones from the point of view of renormalization; they
are the {\it primitively divergent} diagrams, if they are divergent.
Any other divergences that occur can be considered to come from
the embedding of the {\it primitively divergent} diagrams inside
a more complicated diagram.

So far we have concentrated mainly on one-loop effects.  Does anything
qualitatively new happen at two loops?  In fact, yes.  Notice that
our counterterms so far only include two of the three operators of
the original Lagrangian:
\beq 
	{\cal L}_{\rm c.t.} = -\frac12\hat\lambda(m^2_r\ln\Lambda^2/\mu^2
	-\Lambda^2)\phi^2 - {3\lambda_r^2\over 4!\,16\pi^2}\ln(\Lambda/\mu)
	\phi^4
\eeq
These are the divergent parts of the bare Lagrangian.  But we might
have also expected to obtain a term of the form
\beq
\label{wfr}
	Z_\phi(\Lambda) \partial_\mu\phi \partial^\mu\phi
\eeq
where $Z_\phi$ is some divergent function of $\Lambda$.  This is the 
normal situation, but in the present theory, it happens to first
arise at two loops, through the setting sun diagram, fig.\ 8.  The
thing that distinguishes this from the mass shift, fig.\ 6, is
that the divergent part of fig.\ 8 does have momentum dependence.
It gives a contribution 
\beq
	\delta \Gamma_2(q) = {(-i\lambda)^2\over 6} \intdp \intdr
	{i\over ((p-r)^2-m^2)} {i\over ((p+q)^2-m^2)} 
	{i\over (r^2-m^2)} 
\eeq
which we can Taylor-expand in powers of the external momentum $q$.  
Notice that odd powers of $q$ must have vanishing coefficients because
$\delta \Gamma_2(q)$ is a Lorentz scalar, and there is no 
4-vector to contract with other than $q_\mu$.  Let us write
\beq
	\delta \Gamma_2(q) = \sum_{n=1} q^{2n} \delta\Gamma_2^{(n)}(\Lambda)
\eeq
By counting powers, it is easy to see that the only divergent terms
in this expansion are 
\beqa
	\delta\Gamma_2^{(0)} &=& \delta\Gamma_2(0) \ \sim \ \Lambda^2 \\
	\delta\Gamma_2^{(1)}&\sim& \ln\Lambda
\eeqa
The first term, with no momentum dependence, is a new, two-loop
contribution to the mass correction.  The second term, since it
goes like $q^2$, is the Fourier transform of the operator 
$\partial_\mu\phi \partial^\mu\phi$.  All the higher terms in 
the expansion are convergent without the cutoff and do not require
renormalization.  Let us suppose that $\delta\Gamma_2^{(1)}$ has
the form 
\beq
	\delta\Gamma_2^{(1)} = a + b\ln\Lambda
\eeq
Then if we add a term 
\beq
	{\cal L}_{\rm c.t.} = \hbox{previous terms}\ + 	
	\frac12\delta Z_\phi\, \partial_\mu\phi \partial^\mu\phi
\eeq
to the counterterm Lagrangian, with 
\beq
	\delta Z_\phi = -(b\ln\Lambda/\mu)
\eeq
for some $\mu$, then the divergent part of the new contribution will
be canceled.  Let's verify this by treating  ${\cal L}_{\rm c.t.}$
as a perturbation to the 2-point function.  In the path integral,
it gives
\beq
	\int{\cal D}\phi\, e^{iS} \phi(x) \phi(y)\ 
	{i\over 2}\int d^{\,4} x'\, \delta Z_\phi\, \partial_\mu\phi \partial^\mu\phi
\eeq
which contributes
\beq
	\delta\Gamma_2(q) =  q^2 \delta Z_\phi +\dots
\eeq
to $\delta\Gamma_2(q)$.  The factor of $1/2$ is
canceled by the number
of ways of doing the contractions.  This result 
therefore cancels
the divergent part of fig.\ 17.

\medskip\bigskip
\centerline{\epsfxsize=2.5in\epsfbox{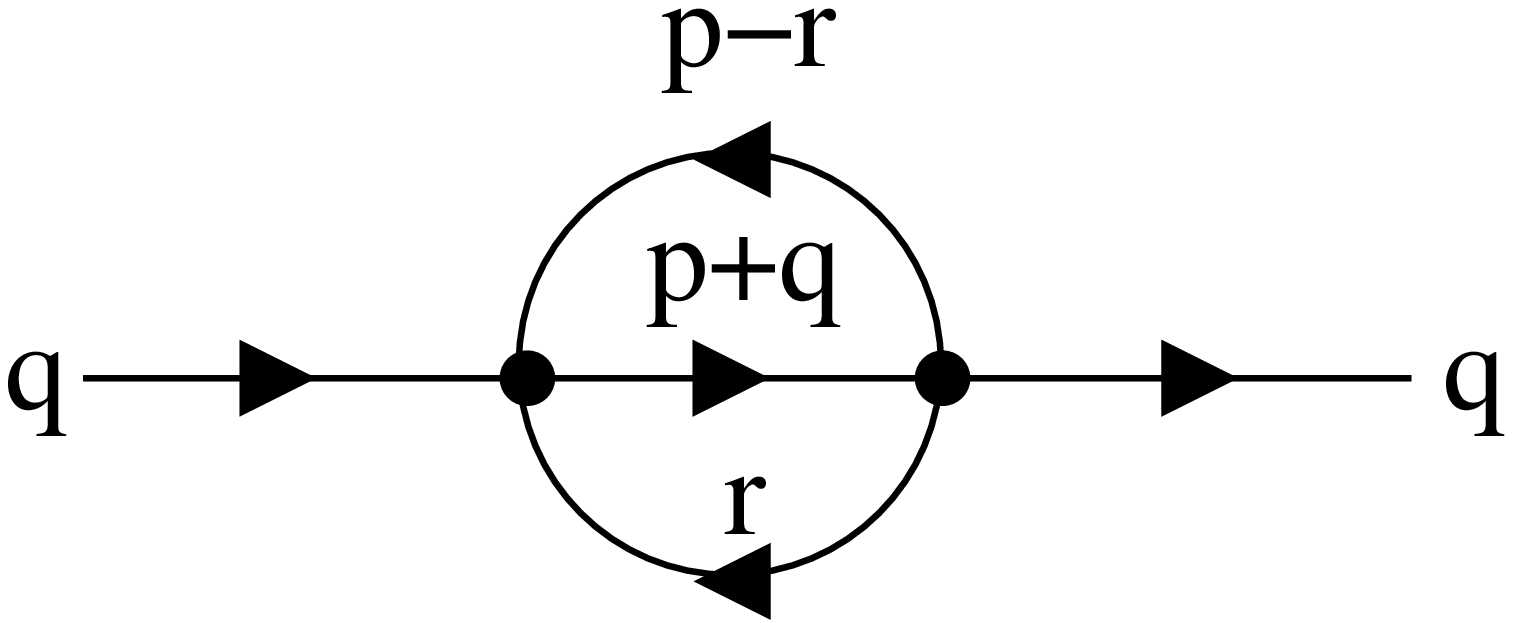}}
\centerline{\small
Figure 17. Momentum-space version of setting sun diagram.}
\medskip

This last kind of correction is called {\it wave function
renormalization}, because it can be absorbed into a rescaling
of the field.  Indeed, the renormalized Lagrangian has the form
\beq
{\cal L}_{\rm ren} = 	\frac12\left(1+\delta Z_\phi(\Lambda)\right)\partial_\mu\phi \partial^\mu\phi
	-\frac12 \tilde m_0^2(\Lambda)\phi^2 -{\tilde\lambda_0(\Lambda)\over 4!}
	\phi^4
\eeq
I have used tildes on $\tilde m_0^2$ and $\tilde\lambda_0$ to distinguish them
from the final renormalized couplings which we will now define.
If we rescale the field by $\phi = \phi_0/\sqrt{Z_\phi}$, where
$Z_\phi = 1+\delta Z_\phi$, then
\beq
\label{orig}
	{\cal L}_{\rm ren} = \frac12\partial_\mu\phi_0
	 \partial^\mu\phi_0
	-\frac12 {\tilde m_0^2\over Z_\phi}\,(\Lambda)\phi_0^2 
	-{1\over 4! }{\tilde \lambda_0\over Z_\phi^2}(\Lambda)
	\phi_0^4
\eeq
In light of this, we should redefine the bare mass and coupling to
include the effect of wave function renormalization:
\beq
\label{Lren}
	{\cal L}_{\rm ren} \to \frac12\partial_\mu\phi_0
	 \partial^\mu\phi_0
	-\frac12 {m_0^2}(\Lambda)\phi_0^2 
	-{\lambda_0\over 4!}(\Lambda)
	\phi_0^4
\eeq
where 
\beq	m_0^2 = {\tilde m_0^2\over Z_\phi}; \qquad 
	\lambda_0 = {\tilde \lambda_0\over Z_\phi^2}
\eeq
The final form (\ref{Lren}) is how the renormalized Lagrangian
is conventionally defined: the kinetic term of the bare field
is canonically normalized.
The divergences appearing in the newly defined 
counterterms for the mass 
and the coupling no longer exactly cancel the divergences of 
the diagrams like fig.\ 6 and fig.\ 9 and their higher-loop 
generalizations; they also include the effects of wave function
renormalization. 

For example, the divergent $\delta Z$ terms we have absorbed into $\lambda$
are such as to partially cancel out the corresponding divergences in fig.\
18.   However, we truncate the external lines, including the setting sun
diagrams, when we implement the LSZ procedure, so the extra factors of
$1+\delta Z$ in $\lambda_0$ have to be taken out again to get a finite 
1PI amplitude if we are computing it from the renormalized Lagrangian
(\ref{Lren}).  For the 4-point function, we write the renormalized amplitude
(which is finite) in terms of the bare one (the one computed from (\ref{Lren})
\beqa
\label{wfr2}
	\Gamma_4^{(r)} &=& \sqrt{Z_\phi}^{\,\,4} \Gamma_4^{(0)}\\
	&=& (1+\delta Z_\phi)^{2} \Gamma_4^{(0)}\\
	&\cong& (1+2\delta Z_\phi) \Gamma_4^{(0)}
\eeqa
where we expanded perturbatively in the last approximation.  The
factor of $2=4/2$ is there because there are 4 external legs in the
diagram.  Notice that the amplitudes in question are the 1PI vertices.
Also, in the above description, $\Gamma_4^{(0)}$ is the proper
vertex that was computed from the renormalized Lagrangian
(\ref{Lren}), including all the loop corrections to the relevant 
order.  But we call it the unrenormalized vertex since it is
not finite until w.f.r.\ is applied.

\medskip
\centerline{\epsfxsize=6.in\epsfbox{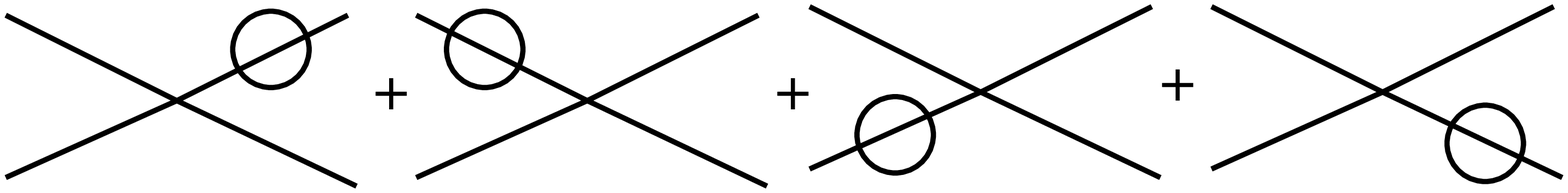}}
{\small
Figure 18. Two-loop divergent contributions to the 1PR Green's
functions, which are partially canceled by the $\delta Z$ factors hidden in
$\lambda_0$.}
\medskip

Another example of how wave function renormalization works is provided by fig.\
19. Four of the eight $\delta Z/2$ factors hidden in the vertices cancel  the
divergence associated with w.f.r.\ in the internal lines.  However, the
remaining four would like to do the same for the external lines, which are not
present in the proper vertex.   So again eq.\ (\ref{wfr}) must be applied.

In this discussion we have stressed the removal of divergences, but
you should keep in mind that w.f.r.\ would still be necessary even
if $\delta Z$ was completely finite.  Otherwise the amplitudes we
compute would not be correctly normalized.  A field $\phi$ is canonically
normalized if its creation and annihilation operators give the actual
number of particles in a state (the number operator) when acting upon that
state:
$	a^\dagger a|\Phi\rangle  = N|\Phi\rangle$.
Such a field has a propagator $i/(p^2-m^2)$, not $iZ_\phi/(p^2-m^2)$.
But in general it is necessary to start with a tree-level propagator
of the latter form so that when we add the loop corrections to it,
the physical propagator has the correct normalization.  

{\it N.B. when you solve problem 4(c), compute the renormalized 3-point
function, so you can get some practice with implementing w.f.r.}

\medskip
\centerline{\epsfxsize=4.in\epsfbox{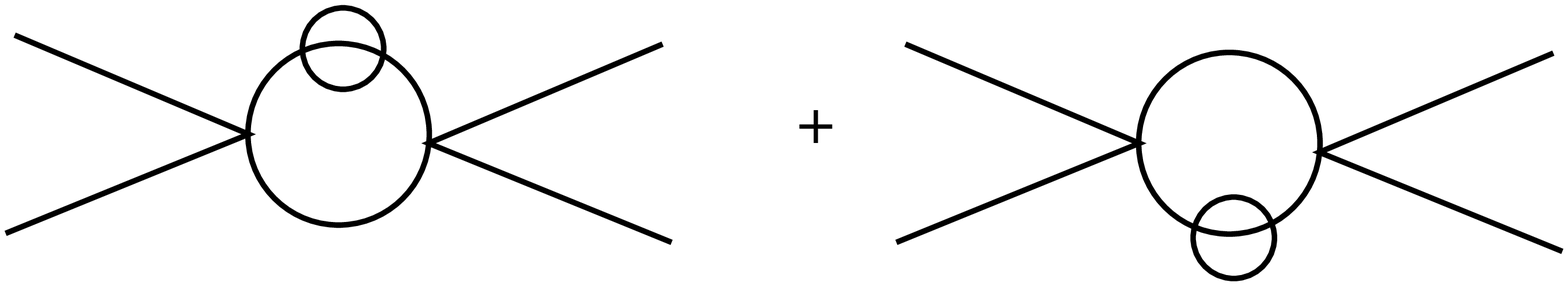}}
{\small
Figure 19. Three-loop divergent contributions to the 1PR Green's
functions, which are only partially canceled by the $\delta Z$ factors hidden in
$\lambda_0$.}
\medskip

Let's take a moment to clarify some potentially confusing terminology.  
The original Lagrangian we write down before worrying
about renormalization is the {\it bare Lagrangian}, while (\ref{Lren})
is the renormalized one, even though it is written in terms of
the bare parameters and the bare field.  We can split ${\cal
L}_{\rm ren}$ into two pieces, a finite part and a counterterm part:
\beq
	{\cal L}_{\rm ren}  = {\cal L} + {\cal L}_{\rm c.t.},
\eeq
where ${\cal L}_{\rm c.t.}$ contains all the pieces which diverge
as $\Lambda\to\infty$.   The parameters which appear in ${\cal L}$
are closely related to what we call the {\it renormalized parameters},
and to the physical measured values:
\beq
	\hbox{physical parameter} \ = \ \left(\hbox{parameter in\ } 
	{\cal L}\right)
	\ +\  \left(\hbox{finite part of loop calculations}\right)
\eeq
There is another way of specifying what we mean by the renormalized
parameters.  It is always possible to write the bare parameters as
\beqa
	m^2_0(\Lambda) &=& Z_m(\Lambda) m^2_r;\\
	\lambda_0(\Lambda) &=& Z_\lambda(\Lambda) \lambda_r,
\eeqa
where $m^2_r$ and $\lambda_r$ are finite and
$\Lambda$-independent---these are the renormalized values---while all
the $\Lambda$-dependence is contained in the factors $Z_m$ and
$Z_\lambda$.   It is important to notice that this definition does
not uniquely specify the values of the renormalized parameters,
however.  The reason is the ambiguity in separating divergent from
convergent contributions.  Since $\Lambda$ has dimensions of mass,
we cannot simply subtract $\ln\Lambda$; we have to subtract 
$\ln\Lambda/\mu$, where $\mu$ is some arbitrarily chosen mass scale.
This arbitrariness will play a crucial role in defining the
concept of the {\it renormalization group}.

It is remarkable that our renormalized Lagrangian has exactly the
same form as the bare one.  In fact this is the definition of 
a renormalizable theory: no new operators are required to be added
to the
theory to make it finite.  This did not have to be the case by
any means: renormalizable theories are a set of measure zero in the
space of all possible theories.  Consider the result of adding an
interaction term of the form
\beq
\label{higher_n}
	{\cal L}_{\rm int} = -{g_n\over M^{n-4}} \phi^{n}
\eeq
to the Lagrangian.  We can repeat the power counting argument we did
previously to find out which diagrams are superficially divergent.
To be even more general, let's suppose we are working in $d$ spacetime
dimensions instead of 4.  Then
\beqa
	I &=& n V/ 2 - E/2;\\
	L &=& I - V + 1 = (n/2-1)V -E/2 + 1;\\
\label{SDD}
	D &=& dL-2I = (d(n/2-1)-n)V - (d/2-1)E + d\\
	  &=& (n-4)V - E + 4 \hbox{\ \ for \ } d=4
\eeqa
This gives us the important result that when $d=4$ and $n>4$,  an
$E$-point function becomes more and more divergent as we  add more
vertices (hence loops). Then in addition to the 0-, 2- and 4-point
functions being divergent, as we had in $\lambda\phi^4$ theory, we
get divergences in amplitudes with arbitrarily large numbers of
external legs, provided we go to high enough order in perturbation
theory.  To remove the infinities from this theory, we have to add
counterterms of the form (\ref{higher_n}) to all orders, {\it i.e.,}
infinitely many of them.  But to fix the values of the finite parts
of the counterterms, we would have to match to the measured values of
the cross sections for $2\to n$ scattering processes, for all $n$. 
Hence there is no predictive power in such a theory---it requires
infinitely many experimental inputs.  

For this reason,  nonrenormalizable theories were considered to be
unphysical in the early days of quantum field theory---and even until
relatively recent times.  This viewpoint has changed since the
crucial contributions of K.G.\ Wilson to the theory of
renormalization.  We now recognize that operators of the form
(\ref{higher_n}) {\it will be} present in any realistic theory, which
does not pretend to be accurate to arbitrarily high energies.  The
reason the new operators are not a problem is that their coefficients
are small, because the mass scales which suppress them are large. For 
example if we have a heavy particle $\psi$ of mass $M$, coupled to
$\phi$ with the interaction $g\phi^3\psi$, then the virtual exchange
of a $\psi$ particle as shown in fig.\ 20 produces the effective interaction
\beq
\label{effint}
	-{g^2\over M^2} \phi^6
\eeq
at low external momenta, $p^2\ll M^2$.  But at large momenta, one sees
that the factor of $1/M^2$ is really coming from the internal propagator,
\beq
	-{g^2\over M^2} \phi^6 \to {g^2\over p^2- M^2} \phi^6
\eeq
In this way, what looked like a nonrenormalizable operator at low energy
can be seen to originate from a renormalizable theory, that contained an
additional heavy particle.  We say that we have {\it integrated out} the
heavy particle (in the sense of having done the path integral over $\psi$)
to obtain the effective interaction (\ref{effint}).

\medskip
\centerline{\epsfxsize=3.5in\epsfbox{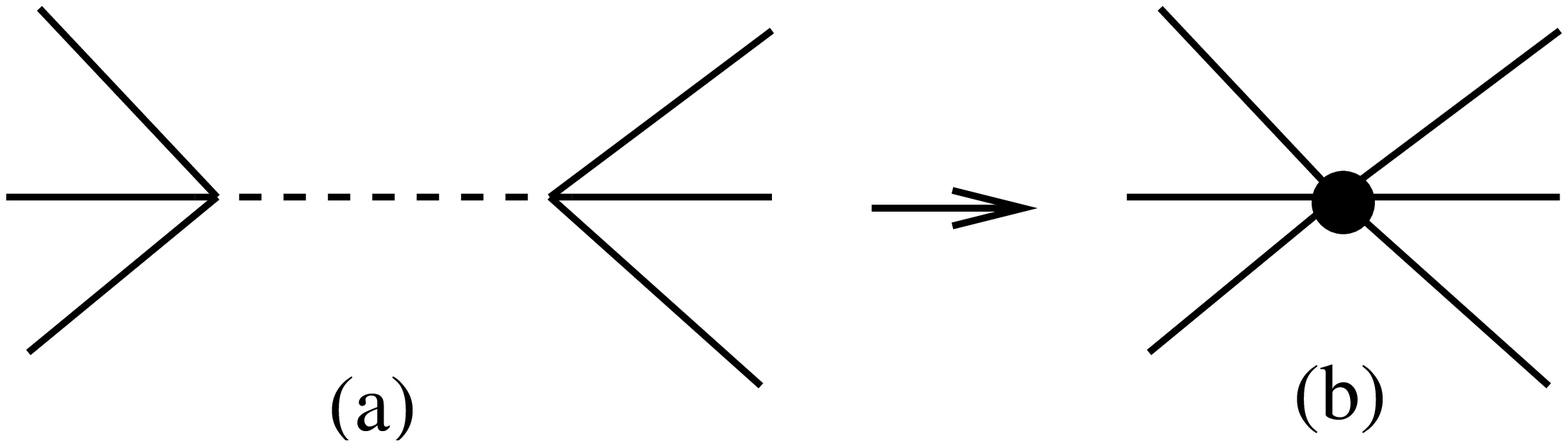}}
{\small
Figure 20. (a) Virtual exchange of a heavy particle (dashed line) giving
rise to (b) an effective $\phi^6$ interaction.}
\medskip

From (\ref{SDD}) we can see which theories are renormalizable in other
dimensions.  In $d=6$, $\phi^3$ is the highest power that is allowed.
For larger $d$, any $n$ greater than $2$ gives a nonrenormalizable 
theory.  On the other hand, in $d=2$, {\it all} operators are
allowed.

\section{Running couplings and the renormalization group}
\label{sec:running}

So far we have introduced renormalization merely as a trick for
removing the ultraviolet divergences from a quantum field theory.
However, it has much deeper implications than that.   The major
concept that will be revealed is that physical parameters like
the coupling constant are not really constants, but they
depend on the scale of energy at which the experiment is performed.
The couplings are said to {\it run} in response to changes in the energy
scale. 

We can see this behavior in $\lambda\phi^4$ theory if we return to
the one-loop correction to the 4-point function, which was partially 
calculated in eq.\ (\ref{fourpt}).   Let's finish the computation
now.  All that remains is to do the Feynman parameter integration.
If we write $M^2(q) = m^2(1 + (4/a)(x^2-x))$ where $a = 4m^2/q^2$,
then \cite{ramond}
\beqa
\label{phi4loop}
	\delta\Gamma_4^{(a)} &=& i{\lambda^2\over 32\pi^2}\int_0^1 dx\left[
	\ln(\Lambda^2/m^2) - 1 - \ln(1 + (4/a)(x^2-x))\right]\\
\label{phi4loop2}
	&=& i{\lambda^2\over 32\pi^2}\left[ \ln(\Lambda^2/m^2) + 1-
	\sqrt{1-a} \ln\left({\sqrt{1-a}+1 \over\sqrt{1-a}-1}
	\right) \right]
\eeqa
For $q^2\gg m^2$ ($a\ll 1$), this becomes
\beq
	\delta\Gamma_4^{(a)} = i{\lambda^2\over 32\pi^2}
	\left[ \ln(\Lambda^2/q^2) + 1\right]
\eeq
and, by adding to this the counterterm, ({\it i.e.,} by replacing
$\lambda$ by $\lambda_0$), and the crossed-channel contributions,
the complete 4-point function at one loop is given by
\beqa
	\Gamma_4 &=& -i \lambda_r \left( 1 + {\lambda_r\over 32\pi^2}
	\sum_{q^2=s,t,u}\ln{ q^2\over \mu^2} \right)\\
	&\longrightarrow& \lambda_{\rm eff}(E) \cong \lambda_r 
	\left( 1 + {3\lambda_r\over 32\pi^2}
	\ln{ 4E^2\over \mu^2} \right)
\eeqa
The last approximation is assuming that $s$, $t$ and $u$ are of 
the same order of magnitude, which is true for generic scattering
angles.\footnote{You might worry about the fact that $t$ is negative,
so that we have the logarithm of a negative number for the $t$-channel
contribution.  This is not an intrinsically bad thing since 
$\Gamma_4$ is allowed to be complex.  We will discuss the analytic
behavior of this amplitude in more detail later on.  Suffice it to
say that for large $|t|$, the term in parentheses is dominated by its
real part, which is gotten from taking the absolute value of the 
argument of the log.}

This result has the important implication that the effective value of
$\lambda$ is energy-dependent, and it increases logarithmically as we go
to higher energies.  (Recall that in the center of mass frame, $q^2
=s = 4E^2$.) This is our first example of a running coupling, and it
gives a hint about how renormalization determines the running. 
Notice that the dependence on energy is tracked by the dependence on
the arbitrary scale $\mu$ which was introduced in the process of
renormalization.  This $\mu$-dependence in turn goes with the 
dependence on the cutoff $\Lambda$.  The divergent part of the diagram
was much easier to compute than the finite part, yet it appears that
we could possibly have deduced the dependence on 
$q^2$ (when $q^2\gg m^2$) just by knowing the divergent part.  This
is just what the renormalization group does for us, as I shall now
try to explain.

It was mentioned earlier that there is a fundamental ambiguity in 
choosing the counterterms because there is no unique definition of
the divergent part of a diagram.  This ambiguity manifests itself
in the fact that the value of $\mu$ is arbitrary.  Suppose we measure
the cross section at some energy $E_0$.  From this we deduce the 
value of $\lambda_{\rm eff}(E_0)$.  This {\it effective coupling}
is the value of $\lambda$ that gives the measured value for the cross
section when we just do a tree-level calculation.  However, this only fixes some
combination of the two parameters $\lambda_r$ and $\mu$.  We can 
always choose a different value of $\mu$ and compensate by shifting
the value of $\lambda_r$.  Since the value of $\mu$ is arbitrary,
no physical quantity can depend upon it.  Thus, similar to the
way in which $\lambda_0$ had to be a function of $\Lambda$, 
$\lambda_r$ must be regarded as function of $\mu$:
\beq
\label{lameff}
\lambda_{\rm eff}(E) = \lambda_r(\mu)\left(
	 1 + {3\lambda_r(\mu)\over 32\pi^2}
	\ln{ 4E^2\over \mu^2} \right)	
\eeq
$\mu$ is known as the {\it renormalization scale}.   It is closely
related to the {\it subtraction point}.  The latter is another name
for the value of the external momenta at which the experiment was done,
 that was
used to fix the value of the renormalized coupling.  (It is the scale
at which the divergent parts of the amplitude were subtracted, hence
the name.) If we choose $\mu = 2 E_0$, the renormalized coupling is
equal to the effective coupling, which is rather nice.  

A technical point: it is also permissible to choose the subtraction
point to be an unphysical value of the  external momenta, that is,
where the vertex is off-shell.  For example, it is sometimes
mathematically simpler to renormalize for vanishing external momenta,
$p_i=0$.  The subtraction point can be different for each amplitude
$\Gamma_n$.  For $\Gamma_4$ one might choose  $p_i\cdot p_j =
M^2(\delta_{ij} - 1/4)$ since this gives $s=t=u=M^2$.  However it is
done, the renormalized couplings will then be simply related to an
amplitude which is analytically continued away from its physical
value.  One could always translate this back into a more physical
subtraction scheme, since it is similar to a change in the choice of
$\mu$, upon which no physics can depend.

Now let's explore the consequences of demanding that no physical
quantity depends on $\mu$.  In the present discussion, $\lambda_{\rm
eff}$ has direct physical meaning, being defined in terms of the
measured value of the cross section.  Thus
\beq
	\mu{\partial\over\partial\mu}\lambda_{\rm eff}(E) \ =\  0 
	\ = \ \mu{\partial\lambda_{r}\over\partial\mu} - 
	{3\lambda_r^2\over 16\pi^2} + 
	O\left(\lambda_r\, \mu{\partial\lambda_{r}\over\partial\mu}\right)
\eeq
which implies that 
\beq
	\beta(\lambda_r) \equiv 
	\mu{\partial\lambda_{r}\over\partial\mu} = 
	{3\lambda_r^2\over 16\pi^2} + O(\lambda_r^3)
\eeq
This is known as the {\it beta function} for the coupling $\lambda$,
and it will play a very important role in the theory of
renormalization. We can integrate $\beta(\lambda_r)$ between 
$\mu_0$ and $\mu$ to find the $\mu$-dependence of $\lambda_r$:
\beqa
\label{l_exact}
	\lambda_r(\mu) &=& {\lambda_r(\mu_0) \over 1 - 
	{3\lambda_r(\mu)\over 16\pi^2}
	\ln {\mu \over\mu_0}}\\
\label{l_approx}
	 &=& \lambda_r(\mu_0)\left( 1 + 
	{3\lambda_r(\mu_0)\over 16\pi^2}
	\ln {\mu \over \mu_0} \right) + O(\lambda_r^3)
\eeqa
The expansion (\ref{l_approx}) is in recognition of the fact that we
have not yet computed higher order terms in the perturbation series,
so we cannot trust (\ref{l_exact}) beyond order $\lambda_r^2$ (be
warned however that I am going to partially take back this statement
in a moment!). We see that $\lambda_r(\mu)$ depends on $\mu$ in
exactly the same way as $\lambda_{\rm eff}(E)$ depends on $E$ (on
$2E$, to be precise---due to the fact that the total energy entering
the diagram is $2E$).  Thus the renormalized coupling is very closely
related to the physically measured coupling, and the way it runs does
tell us about the energy-dependence of the latter.  We quickly become
accustomed to thinking about the two quantities interchangeably. 
This is the real power of renormalization.

Moreover, there is a way of making the connection between 
$\lambda_{\rm eff}$ and $\lambda_r$ stronger.  Since we have the
freedom to choose the value of $\mu$ at will, we can ask whether some
choices are more convenient than others.  Suppose we are doing
measurements near some fixed energy scale, like the mass of the $Z$
boson, $E=M_Z$.  It makes sense to choose $\mu = 2 E$, because then
the correction of order $\lambda_r^2$ vanishes in eq.\
(\ref{lameff}), and we have $\lambda_{\rm eff}(E) = \lambda_r(\mu)$. 
For this value of $\mu$, we get the correct answer already from the
tree-level calculation. Of course this is neglecting higher orders in
perturbation theory, but it seems plausible that minimizing the size
of the first order corrections will also lead to the higher order
corrections being relatively small.

This point can be seen explicitly in the expansion of the integrated
result (\ref{l_exact}).  Suppose we could believe this result to all
orders in $\lambda_r$, and further suppose that we had chosen the
logarithm to be large so that the $O(\lambda_r^2)$ correction is
sizeable compared to $\lambda_r$.  Then it easily follows that all
the higher order corrections are large too, since the expansion has
the form $1+x+x^2+\cdots$.  On the other hand, we wondered if such a 
conclusion could really be drawn, since we have not computed the
$O(\lambda_r^3)$ contributions.  Interestingly, the higher terms in
our expansion do have meaning: they contain what are known as the
{\it leading logarithms} in the perturbative expansion of the 4-point
function.  Consider the effect of the 2-loop contribution to $\lambda_{\rm
eff}$ on the beta function, which could be written generically in the
form
\beq
	\beta(\lambda_r) = b_1 \lambda^2 + b_2 \lambda^3 + 
	\cdots
\eeq
If we truncate at the cubic order and integrate,  we get
\beqa
	\lambda_r(\mu) &=& {\lambda_r(\mu_0) \over
	1 - b_1\lambda_r(\mu_0)\ln{\mu\over\mu_0} + {b_2\over b_1}\lambda_r(\mu_0)
	\ln\left({1/\lambda_r(\mu) + b_2/b_1 \over 
	1/\lambda_r(\mu_0) + b_2/b_1 }\right)}\\
	&\cong& {\lambda_r(\mu_0) \over
	1 - b_1\lambda_r(\mu_0)\ln{\mu\over\mu_0} - {b_2}\lambda_r^2(\mu_0)
	\ln{\mu\over\mu_0} + \dots}
\eeqa
If we expand this again in powers of $\lambda_r(\mu_0)$, we
get
\beq
	\lambda_r(\mu) \cong \lambda_r(\mu_0) \left( 1 + 
	\left(b_1\lambda_r(\mu_0)+ b_2 \lambda_r^2(\mu_0)\right)
	\ln{\mu\over\mu_0} +
	\left(b_1\lambda_r(\mu_0)
	\ln{\mu\over\mu_0}\right)^2 + \cdots\right)
\eeq
We see that the $b_2$ term is subleading to the $b_1^2$ term if the
log is large.  This is what is meant by leading and subleading logs.
By integrating the beta function at order $\lambda_r^2$, we sum up
all the leading logs to all orders in perturbation theory.  The $b_2$ 
term gives us the next-to-leading order (NLO) logs,  $b_3$ the 
next-to-next-to leading (NNLO), and so on.  The existence of
leading and subleading logs can also be guessed from comparing some
of the higher order diagrams for the 4-point function, as in fig.\ 21.
The first of these (a) is just a product of the one loop diagram, as far
as the dependence on external momentum goes, so it has two powers of
logs.  The second (b) can have only a single log because the loop with
three propagators is UV convergent, and only the loop with the single
propagator contributes a log.  The miracle is that the renormalization
is able to tell us how diagram (a) contributes to the leading
dependence on energy of $\lambda_{\rm eff}$ even though we only
computed the one-loop contribution, fig.\ 10.  This process of 
summing up the leading logs is called the {\it Renormalization 
Group (RG) improved} perturbation expansion.

\medskip
\centerline{\epsfxsize=4.5in\epsfbox{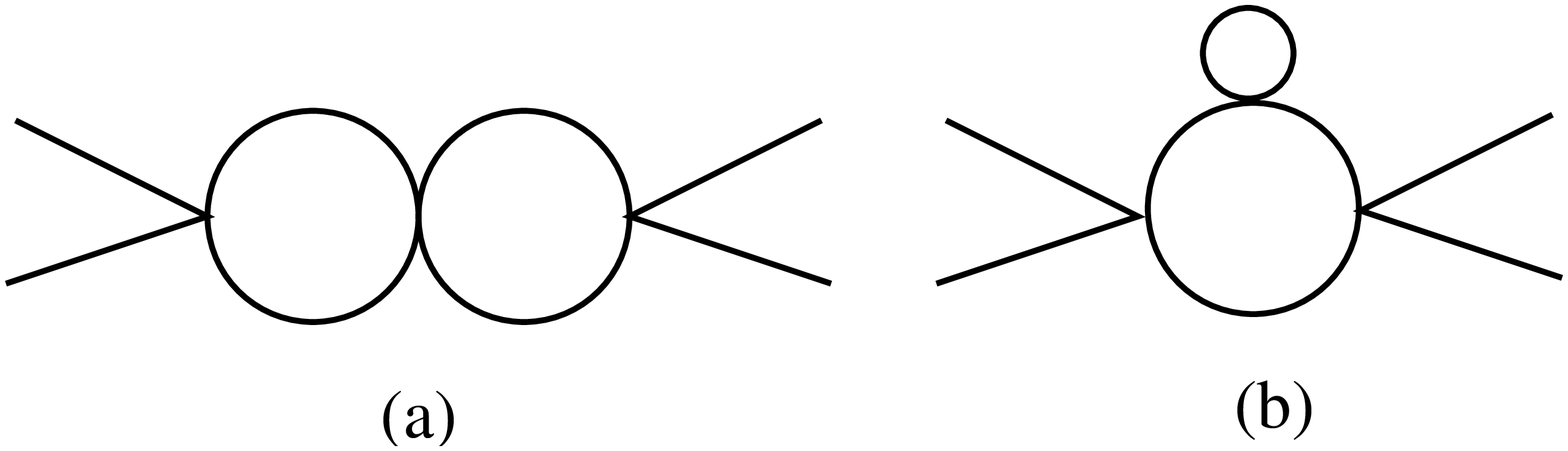}}
\centerline{\small
Figure 21. (a) Leading and (b) subleading log contributions to the
running of $\lambda$.}
\medskip

By now you must be really wondering what the Renormalization Group
{\it is}.  It is only trivially a group in the sense of group 
theory.  An element of the renormalization group is a scale
transformation of the renormalization scale \cite{gross}:
\beq 
\mu \to e^t\mu
\eeq
Two successive transformations of this kind are another
scale transformation,
\beq
	e^{t_1}e^{t_2} \mu =  e^{t_1+t_2}\mu
\eeq
so it is obviously a group, but that is not really the point.  The
goal is to deduce how Green's functions scale with the external
momenta, as we have illustrated in the case of the 4-point function.
Let's now do it more generally, for an $n$-point function.  

Imagine starting with the renormalized theory and computing the 
$n$-point proper vertex with a cutoff.  It depends on the bare
parameters,  the cutoff, and the external momenta $p_i$: 
$\Gamma_n(\lambda_0, m_0, \Lambda, p_i)$.  Recall that it is not
enough to just use the properly defined parameters $\lambda_0, m_0$
to get finite results as $\Lambda\to\infty$; we also need to do
wave function renormalization:
\beq
	\Gamma_n^{(r)} = Z_\phi^{n/2} \Gamma_n^{(0)}
\eeq  

$\Gamma_n^{(r)}$ is a physical observable, and therefore you might
guess that it must not depend on the arbitrary renormalization scale
that was introduced in the process of removing UV divergences. 
Strangely, however, it  is $\Gamma_n^{(0)}$ which is really independent
of  $\mu$.  $\Gamma_n^{(0)}$ is a function of $m^2_0$, $\lambda_0$, and
$\Lambda$, and $m^2_0$, $\lambda_0$ are actually independent of
$\mu$.  Indeed, you can see from comparing  (\ref{bare_c}) and 
(\ref{l_approx}) that $\lambda_0$ does not depend on $\mu$ because
the implicit $\mu$ dependence in $\lambda_r(\mu)$ cancels the
explicit dependence from $\ln(\Lambda/\mu)$.  Similarly, the 
implicit $\mu$ dependence in $m^2_r(\mu)$ cancels the
explicit dependence from $\ln(\Lambda/\mu)$ in eq.\ 
(\ref{counterterm}).  Hence we can write
\beq 
\mu{\partial\over\partial\mu} \Gamma_n^{(0)}(\lambda_0(\Lambda),
m^2_0(\Lambda),\Lambda) = 0
\eeq

On the other hand, $Z_\phi$ does depend explicitly on $\mu$,
with the functional form
\beq
	Z_\phi = 1 + c\lambda_r^2\ln(\Lambda/\mu)
\eeq
for some constant $c$, and it is clear from the way in which
$\lambda_r$ is known to depend on $\mu$ that the $\mu$ dependence
in $Z_\phi$ is {\it not} canceled by that of $\lambda_r$.
$\Gamma_n^{(r)}$ actually does depend on $\mu$.
The general form of the {\it renormalization group equation} (RGE) 
exploits
the $\mu$-independence of $\Gamma_n^{(0)}$ to determine just how
$\Gamma_n^{(r)}$ must depend on $\mu$:
\beqa
	\mu{\partial\over\partial\mu} \Gamma_n^{(0)} &=& 0\\
	\longrightarrow 0 &=& \mu{\partial\over\partial\mu}
	\left( Z_\phi^{-n/2} \Gamma_n^{(r)}(\lambda_r,m_r,p_i,\mu)
	\right)\\
\label{RGE}
	&=&  Z_\phi^{-n/2}\left( \mu{\partial\over\partial\mu} +
	\beta(\lambda) {\partial\over\partial\lambda} +
	\gamma_m m{\partial\over\partial m} -
	n\gamma(\lambda) \right) \Gamma_n^{(r)}(\lambda,m,p_i,\mu)
\eeqa
where the dimensionless functions are defined as
\beqa
\label{betafn}
	\beta(\lambda) &=& \mu{\partial\lambda\over\partial\mu};
	\\
	\gamma_m  &=& \mu{\partial\over\partial\mu} \ln m \\
	\gamma(\lambda) &=& \mu{\partial\over\partial\mu} \ln
	\sqrt{Z_\phi}
\eeqa
The function $\gamma(\lambda)$ is called the {\it anomalous
dimension}, for reasons that will become clear.

We want to use the renormalization group equation to deduce
how $\Gamma_n^{(r)}$ changes if we rescale all the momenta:
\beq
	p_i \to e^t p_i
\eeq
But since the procedure is a little complicated, I want to first
illustrate what is going on using the 4-point function at
one loop as an example.  We had
\beqa
	\Gamma_4^{(r)}(p_i,\lambda_r,\mu) &=& \lambda_r\left(
	 1 + \sum_{q^2=s,t,u}{\lambda_r\over 32\pi^2}
	\ln{ q^2\over \mu^2} \right)\nonumber\\
	 \to \Gamma_4^{(r)}(e^t p_i,\lambda_r,\mu) &=& 
	\lambda_r\left(
	 1 + \sum_{q^2=s,t,u}{\lambda_r\over 32\pi^2}
	\ln{ e^t q^2\over \mu^2} \right)\nonumber\\
\label{lastgam}
	&=& \Gamma_4^{(r)}(p_i,\lambda_r,e^{-t}\mu)
\eeqa
But we know that a change in $\mu$ can always be compensated by
changing the value of $\lambda_r$:\footnote{we can ignore the effect of 
w.f.r.\ here because we are only working at one loop in this
example}
\beq
	\Gamma_4^{(r)}(p_i,\lambda_r(\mu),\mu) = 
	\Gamma_4^{(r)}(p_i,\lambda_r(e^t\mu),e^t\mu).
\eeq
By simply renaming the slot for the explicit dependence on $\mu$,
this becomes
\beq
	\Gamma_4^{(r)}(p_i,\lambda_r(\mu),e^{-t}\mu) = 
	\Gamma_4^{(r)}(p_i,\lambda_r(e^t\mu),\mu).
\eeq
Putting this together with (\ref{lastgam}), we get
\beq
\label{rescale}
\Gamma_4^{(r)}(e^t p_i,\lambda_r,\mu) = 
	\Gamma_4^{(r)}( p_i,\lambda_r(e^t\mu),\mu)
\eeq	
We see (again) that going to higher energy scales for the external
particles is equivalent to changing the renormalized 
coupling constant, by rescaling $\mu$.  Recall that the $\mu$
dependence of $\lambda_r$ was determined by integrating
the beta function (\ref{betafn}).

There were three things in the above example that made it special:
(1) The 4-point function is dimensionless.  (2) It did not depend on
the mass.  (3) There is no effect of wave function renormalization at
one loop.  Concerning (1), 
when the vertex has nonvanishing dimensions, then the
first step, rescaling $p_i$ by $e^t$, cannot be undone simply by
absorbing it into $\mu$.  This is obvious with the 2-point function
(inverse propagator) at tree level:
\beq
	\Gamma_2 = p^2 \to e^{2t} p^2.
\eeq
In general, the
dimension of $\Gamma_n^{(r)}$ is (mass)$^{4-n}$, {\it i.e.,}
\beqa
	\Gamma_n^{(r)} &\sim& |p_i|^{4-n} f\left(\lambda_r(\mu), {p_i\over
	\mu},
	{m_r(\mu)\over\mu}, \mu\right)\\
	 &\to&
	e^{(4-n)t} |p_i|^{4-n}f\left(\lambda_r(\mu), {e^t p_i\over\mu},
	{m_r(\mu)\over\mu}, \mu\right)\\
	&=& e^{(4-n)t} |p_i|^{4-n}f\left(\lambda_r(\mu), {p_i\over e^{-t}\mu},
	{e^{-t}m_r(\mu)\over e^{-t}\mu}, \mu\right)
\label{rescaled}
\eeqa
so that when we do the first step, we pick up a factor of
$e^{(4-n)t}$ which wasn't there for the 4-point function.
Concerning point (2), we see that the effect of $e^t$ inside $f$
can no longer be absorbed into a redefinition of the
explicitly-appearing $\mu$, since it can now be combined with both
$m$ and with $p_i$.  And even if we could ignore the mass, point
(3) implies that the function $f$ is {\it not} invariant under
a combined change of $\mu$ and $\lambda_r$, since we have to 
take w.f.r.\ into account as well.  The final step is to use the
$\mu$-independence of the unrenormalized vertex to let $\mu\to
e^t \mu$:
\beqa
	\Gamma_n^{(r)} &\sim& e^{(4-n)t} 
	\underbrace{\left({Z_\phi(\lambda(e^t\mu),e^t\mu) \over 
	Z_\phi(\lambda(\mu),\mu) }\right)^{-n/2}}
	|p_i|^{4-n}
	f\left(\lambda_r(e^t\mu), {p_i\over \mu},
	{e^{-t}m_r(e^t\mu)\over \mu}, \mu\right)\\
	&&\qquad\qquad e^{-n\int_0^t \gamma(\lambda_r(e^t\mu))dt}\
	\nonumber
\eeqa
Putting it all together, we get
\beq
\label{rge_soln}
	\Gamma_n^{(r)}(e^tp_i,\lambda_r(\mu), m_r(\mu),\mu) = 
	e^{(4-n)t -n\int_0^t \gamma(\lambda_r(e^t\mu))dt}
	\Gamma_n^{(r)}(p_i,\lambda_r(e^t\mu),e^{-t}m_r(e^t\mu),\mu)
\eeq
This is the most general form of the solution to the RGE.  Notice
that I did not actually refer to the RGE to write it down; rather
I used the $\mu$-independence of the unrenormalized vertex.  However,
you can verify that (\ref{rge_soln}) does satisfy the RGE.  What was
the point of differentiating $Z_\phi^{-n/2}\Gamma_n^{(r)}$ with 
respect to $\mu$ and then integrating again?  Remember that by 
doing this for $\Gamma_4$, we were able to go from the simple 
one-loop result (\ref{l_approx}) for the effective coupling, to the
RG-improved version (\ref{l_exact}) which summed up the leading logs.
Integrating the RGE allows us to do likewise for the other proper
vertices.

Now we can see the origin of the name {\it anomalous dimension}.
If $\gamma(\lambda)$ was a constant, the overall rescaling factor
in front of the vertex function would be
\beq
	e^{(4-n-n\gamma)t}
\eeq
so having a nonzero value for $\gamma$ is akin to changing the 
dimension of the vertex.  It transforms as though its overall
momentum dependence was of the form $|p_i|^{4-n-n\gamma}$ instead
of $|p_i|^{4-n}$.  This is the physical consequence of wave function
renormalization.

Another point worth noting is that the presence of the mass generally
makes it much more difficult to explicitly solve the RGE's in 
realistic problems, where we typically have several coupling
constants whose RGE's are coupled to each other.  If the external
momenta are sufficiently large, then it is usually a good
approximation to ignore the dependence on $m$.  

Let's illustrate the RGE's for one more physical process, the
$2\to 4$ scattering.  How does $\Gamma_6(p_i)$ scale with changes
in the energy scale $p_i\to e^t p_i$?  Naively, we would expect the
dependence 
\beq
	\Gamma_6(e^t p_i) \sim e^{-2t} \Gamma_6(p_i)
\eeq
from the dimensionality of the process:  the diagrams of figure
12 contain one propagator, which at high energies looks like $1/p^2$.
But we know that the loop corrections to the vertices will alter this,
in particular diagrams like fig.\ 15(b) which introduce
logarithmic dependence on the external momenta.  If we ignore the
mass dependence (assuming $p^2\gg m^2$ in the virtual propagators)
and the effect of w.f.r.\ (since this is higher order in $\lambda$),
then the RGE tells us that 
\beq
	\Gamma_6(e^t p_i) \cong e^{-2t} \left({\lambda_r(e^t\mu)\over
	\lambda_r(\mu)}\right)^2
	\Gamma_6(p_i)
\eeq
since the tree diagram is of order $\lambda^2$.  The significant thing
is that the $\lambda_r(\mu)$ factors appearing here are the {\it 
renormalization group improved} ones (\ref{l_exact}), so indeed we
have summed up the leading logs in getting this expression.  If we
further improve this by including the effect of w.f.r., we will see
the anomalous dimension coming in, $e^{-2t}\to e^{-(2+\gamma)t}$
in the approximation where the anomalous dimension $\gamma$ is
relatively constant over the range of scales of interest.  Finally,
if we are working at scales where the masses are not negligible,
we will have functions like $1/(p^2-m_r^2(\mu))$ in the amplitude,
which scales like $e^{-2t}/(p^2- e^{-2t}m_r^2(e^{t}\mu))$.

\section{Other regulators}
\label{sec:reg}

The momentum space cutoff we have been using is the most intuitive
form of regularization, but it is not the most efficient one.  For
a scalar field theory which does not have any gauge symmetries to
preserve it is fine.  But in a gauge theory, the momentum space cutoff
is bad news.  Under a gauge transformation, the vector potential and
the electron field transform like
\beqa
	A_\mu(x) \to A_\mu(x)+ \partial_\mu \Omega(x);
	\qquad \psi(x) \to e^{ie\Omega(x)}\psi(x)
\eeqa
Suppose we have limited the quantum fluctuations of these fields
to having (Euclidean) momenta with $|p|< \Lambda$.  But if $\Omega(x)$
is rapidly varying in space, then the fields will have larger momenta,
exceeding the cutoff, after the gauge transformation.  In other words,
the momentum cutoff does not respect the gauge symmetry.  And breaking
the gauge symmetry leads to all sorts of technical problems in
defining the theory and insuring its consistency.

In the process of trying to define and prove the renormalizability
of the SU(2) electroweak theory of the SM, 't Hooft invented a very
elegant alternative called {\it dimensional regularization} (DR).  One
regards the dimension of spacetime as a continuous parameter, under
which physical quantities can be analytically continued.  A loop
diagram of the form
\beq
\label{inta}
	\intdpd {1\over (p^2+m^2)^\alpha}
\eeq
converges for $d<\alpha/2$.  By evaluating loop integrals
in the region of $d$ where they converge, and then analytically
continuing back to $d=4$ (or close to 4), we are able to
regularize the divergences.  The expressions so obtained will have
poles of the form $1/(d-4)$ which prevent us from taking $d\to 4$
until the divergences have been subtracted.  So in this method,
quantities which diverge as $\Lambda\to\infty$ using a cutoff
display divergences in the form of $1/(d-4)^n$.

In scalar field theory, DR is very simple to implement; we just need
to analytically continue the momentum space integral:
\beq
	{d^{\,4}p\over (2\pi)^4} \to
	{d\Omega_{d-1} p^{d-1} dp \over (2\pi)^d}
\eeq
There is a nice trick for deriving the volume $\Omega_{d-1}$ of the
$d-1$-sphere.  Evaluate the following integral (in Euclidean space) using 
either Cartesian or spherical coordinates:
\beqa
	\int {d^{\,d}p\over (2\pi)^{d/2} }e^{-(p_1^2+\cdots+p_d^2)/2}
&=& \left ( \int_{-\infty}^{\infty} {dp\over (2\pi)^{1/2}} 
	e^{-p^2/2} \right)^d
	= (1)^d \\
	&=& \int {d\Omega_{d-1} p^{d-1} dp \over (2\pi)^{d/2}}
	e^{-p^2/2} \\
	&=& {\Omega_{d-1}\over(2\pi)^{d/2}}2^{d/2-1}\int_0^\infty
	du \, u^{d/2-1} e^{-u}\\
	&=& \frac12 {\Omega_{d-1}\over\pi^{d/2}}\Gamma(d/2)
\eeqa
Therefore
\beq
	\Omega_{d-1} = {2\pi^{d/2}\over \Gamma(d/2)}
\eeq
and we thus know how to define all the integrals that will arise.

In particular, the integral (\ref{inta}) evaluates to
\beqa
\label{intb}
	\intdpd {1\over (p^2+m^2)^\alpha} &=& 
	\frac12{\Omega_{d-1}\over (2\pi)^d} \int_0^\infty du {u^{d/2-1}\over
	(u+m^2)^\alpha} \\
\eeqa
\beqa
	&=& \frac12{\Omega_{d-1}\over (2\pi)^d} (m^2)^{d/2-\alpha}
	\underbrace{\int_0^\infty dy\, y^{d/2-1} (1+y)^{-\alpha}}\\
	&& \qquad\qquad\qquad\qquad{\Gamma(d/2)
	\Gamma(\alpha-d/2)\over\Gamma(\alpha)}\\
	&=& {\Gamma(\alpha-d/2)\over (4\pi)^{d/2} \Gamma(\alpha)} m^{d-2\alpha}
\eeqa
In four dimensions, we know that the integral diverges for the common
values of $\alpha=1$ and 2.  Let us write the number of dimensions
as
\beq
	d = 4-2\epsilon
\eeq
with the limit $\epsilon\to 0$ representing the removal of the cutoff
(analogous to $\Lambda\to\infty$).  The analytic properties of the
gamma function tell us how the integral diverges.  For positive 
integer values $n$,
\beq
	\Gamma(-n+\epsilon) = {(-1)^n\over n!}\left[{1\over\epsilon}
	+ \psi(n+1) + \frac{\epsilon}2\left({\pi^2\over 3} + \psi^2(n+1)
	-\psi'(n+1)\right) + O(\epsilon^2) \right]
\eeq
where $\psi$ is the logarithmic derivative of the gamma function,
$\psi(n+1) = 1 + 1/2 + \dots + 1/n - \gamma$, $\psi'(n)=
\pi^2/6 + \sum_{k=1}^n 1/k^2$, $\psi'(1) = \pi^2/6$ and $\gamma = 
\psi(1) = -0.5772\dots$ is the Euler-Masheroni constant \cite{ramond}.
Normally one does not need to know all these finite parts; the
$1/\epsilon$ pole is the most interesting part.

There is one more subtlety with DR: the action no longer has the right
dimensions unless we adjust the coupling constant.  The renormalized 
action now reads
\beq
	S = \int d^{\,d}x\left(\frac12(\partial\phi_0)^2 - \frac12 m_0^2
	\phi_0^2 - {\lambda_0\over 4!}\phi_0^4\right)
\eeq
$\phi_0$ has mass dimension of $d/2-1$ to make the kinetic term
dimensionless; then the mass term is still fine, but $\lambda_0$
must now have mass dimension of $d-2d+4 = -d+4=2\epsilon$.  To make this 
explicit, we will keep $\lambda_0$ dimensionless, but introduce
an arbitrary mass scale $\mu$ into the quartic coupling,
\beq
	S \to \int d^{\,d}x\left(\frac12(\partial\phi_0)^2 - \frac12 m_0^2
	\phi_0^2 - {\lambda_0\over 4!}\mu^{2\epsilon}\phi_0^4\right)
\eeq

If we compute the 1-loop correction to the mass using DR, the result
turns out to be
\beqa
	\delta m^2 &=& {\lambda m^2\over 32\pi^2} \left({4\pi\mu^2\over m^2}
	\right)^{\epsilon}\Gamma(-1+\epsilon)\nonumber\\
	&=& {\lambda m^2\over 32\pi^2}
	\left(1+\epsilon\ln\left({4\pi\mu^2\over m^2}\right) +\cdots\right)
	\left(-{1\over\epsilon} +\psi(2) + \cdots\right)\nonumber\\
	&=& {\lambda m^2\over 32\pi^2} \left(-{1\over\epsilon} +\psi(2)
	-\ln\left({4\pi\mu^2\over m^2}\right) + O(\epsilon)
	\right)
\eeqa
Similarly, the $s$-channel 
correction to the 4-point can be computed, with the result
\beqa 
	\delta\Gamma_4^{(a)}(q)
	&=& i\mu^{2\epsilon}{\lambda_0^2\over 32\pi^2}\left(
	{1\over\epsilon} + \psi(1) +
	\int_0^1 dx \ln\left({4\pi\mu^2\over M^2(q)}\right)
	 + O(\epsilon) \right)
\eeqa
where $M^2(q) = m^2 - q^2 x(1-x)$ as in the momentum space cutoff
computation, eq.\ (\ref{fourpt}).  Notice that we kept one factor
of $\mu^{2\epsilon}$ in front, not expanding it in powers of $\epsilon$,
because we want $\Gamma_4$ to have the same dimension as the tree level
diagram, $-i\lambda_0 \mu^{2\epsilon}$.  This factor will disappear at the
end of any computation, after we have subracted the divergences so that
the cutoff can be removed, $\epsilon\to 0$.

In comparing the results in DR with those of the momentum space cutoff we can
notice that the $1/\epsilon$ poles correspond exactly to the factors of
$\ln\Lambda^2$ in the latter.  In fact, at one loop the poles always appear in
combination with $\ln 4\pi\mu^2$, and this quantity takes the place of
$\ln\Lambda^2$.  Recall that when we renormalized the parameters, the
$\ln\Lambda$'s always got replaced by the log of the renormalization scale. In
DR, the factors of $\ln\mu$ appear already in the loop diagrams, rather than
being put into the counterterms.  Nevertheless, the $\mu$ in DR plays exactly
the same role as the renormalization scale we introduced previously.

To reiterate, the $1/\epsilon$ poles keep track of the log divergences.  In an
$n$ loop diagram, we will get leading divergences of the form $1/\epsilon^n$,
and these will be associated with the leading logs.  This is one reason DR is
such an efficient regulator: the dominant divergences of any diagram  allow one
to immediately deduce the leading log contributions to running couplings.  But
what about the quadratic divergences, such as $\delta m^2 \sim
\lambda\Lambda^2$?  These are completely missing in DR.  The process of analytic
continuation has set them to zero.  It is as though DR automatically provides
the counterterms to exactly cancel the quadratic divergences, and only exhibits
the logarithmic ones.   This might seem like a way of avoiding the hierarchy
problem for the Higgs boson mass, but our experience with the $p$-space cutoff
tells us that the quadratic divergences must really be there, even if DR does
not see them.  

As for the finite parts of the diagrams, involving terms like $\psi(n)$, these
are not very interesting, since the counterterms also have finite parts which
could be chosen to cancel the $\psi(n)$'s.  The different schemes for choosing
finite parts for the counterterms are called {\it renormalization
prescriptions}.  There is one extremely simple prescription due to 't Hooft and
Weinberg: take the counterterms to cancel the poles exactly, and nothing else.
This is called {\it minimal subtraction} (MS).  (There is also a variant called
$\overline{\rm MS}$ in which one subtracts all the factors of $\psi(1)+\ln 4\pi$ along
with the $1/\epsilon$'s.)  The detailed form of the RGE's is
prescription-dependent, and they take a particularly simple form in the MS
scheme. 

Before leaving the subject of dimensional regularization, let's
do one more computation, since it is a bit more subtle than
the cutoff regulator.  The beta function in dimensions other than
4 gets a new contribution of
order $\epsilon$.  This is because he full bare coupling 
constant is $\lambda_{\rm bare}=\lambda_0 \mu^{2\epsilon}$,  so even
at tree level the renormalized coupling has $\mu$ dependence.  The
argument goes like this \cite{Preskill}.  When we compute the
counterterms to the coupling, we find
\beq
\label{lambdabare}
	\lambda_{\rm bare}=\mu^{2\epsilon}\lambda_r\left(
	1 + {3\lambda\over 32\pi^2\epsilon}+\hbox{\ finite\ }
	+ \hbox{\ higher order \ }\right)
\eeq
More generally we could write
\beq
	\lambda_{\rm bare} = \mu^{2\epsilon}\left(\lambda_r
	+ {a_1(\lambda)\over\epsilon} + {a_2(\lambda)\over\epsilon^2}
	+\dots \right)
\eeq 
Now $\lambda_{\rm bare}$ is regarded as being independent of $\mu$,
which only comes in when we try to fix the value of the renormalized
coupling.  Therefore
\beqa
	\mu{\partial\over\partial\mu}\lambda_{\rm bare} &=&  0
	= 2\epsilon \lambda_{\rm bare} + 
	 \mu^{2\epsilon}
	\underbrace{\mu{\partial\over\partial\mu}\lambda_r}
	\left(1 + {a_1'(\lambda)\over\epsilon} + {a_2'(\lambda)\over\epsilon^2}
	+\dots \right)\\
	&& \phantom{AAAAAAAAAA}\beta(\lambda_r) 
\eeqa
We can solve this for the beta function 
\beq
\label{WF}
	\beta(\lambda_r,\epsilon) = -2\epsilon\left( \lambda_r
	+ {a_1(\lambda_r)\over\epsilon}
	-\lambda_r{a_1'(\lambda_r)\over\epsilon} +\dots\right)
	= -2\epsilon\lambda_r +{3\lambda_r^2\over 16\pi^2} + \dots
\eeq
Usually we will be interested in the limit $\epsilon\to 0$, but below
we will show that this new term can have physical significance.  
Aside from the new term, this exercise illustrates how  one computes
the nonvanishing part of the beta function in DR.  Interestingly, the
terms of order $1/\epsilon^n$ in $\beta$ which  would blow up as
$\epsilon\to 0$ must actually vanish due to relations between the
$a_i$'s.  This harks back to our discussion of leading logs: the one
loop result fully determines the leading logs, and this is why the
higher $a_i$'s are determined by $a_1$.

Although DR is the most popular regulator on the market, there are
others which are sometimes useful.  I'll introduce a few of them
briefly.

{\it Pauli-Villars regularization} is carried out by adding fictitious
fields $\Phi_i$ with large masses $M_i$ whose function is to cancel the
divergences.  Once this is done, the limit $M_i\to\infty$ can be
taken, so that any low-energy effects of the fictitious fields
vanish.  In order to make the divergences due to the $\Phi_i$'s
cancel those of the physical fields, one needs to make them
anticommuting instead of commuting:
\beq
	\{ \Phi_i(t,\vec x), \dot\Phi_j(t,\vec y)\} = i \delta_{i,j} 
	\delta^{(3)}(\vec x-\vec y)
\eeq
When we get to the discussion of fermions, we will learn that diagrams
with a loop of anticommuting fields get an extra minus sign relative
to those with commuting fields.  Thus consider adding to the $\lambda
\phi^4$ Lagrangian the terms
\beq
	\frac12\left( (\partial\Phi)^2 - M^2\Phi^2) \right)
	-{\lambda\over 4} \phi^2 \Phi^2 
\eeq
for a single Pauli-Villars field $\Phi$.  It is not hard to verify
that, with the normalization of the coupling as given, the
combinatorics of the new diagrams are correct for cancelling the 
leading UV divergences in $\Gamma_2$ and $\Gamma_4$.  Unfortunately,
this only cures the quadratic divergences in $\Gamma_2$.  For the
log divergences, we are left with something proportional to 
$(M^2-m^2)\log(\Lambda^2)$, which means we could not get away with
just having one Pauli-Villars field.  Since there are two kinds of
divergences (quadratic and logarithmic) we need two PV fields,
whose masses $M_1$ and $M_2$ are chosen in such a way as to eliminate
both kinds.  In general one needs several fields to render all
the loop integrals finite.  Therefore this is rather messy compared
to DR.  However, it does have the virtue of preserving gauge
invariance in gauge theories, and this is why it was commonly used
before the invention of DR. 

\medskip
\centerline{\epsfxsize=4.5in\epsfbox{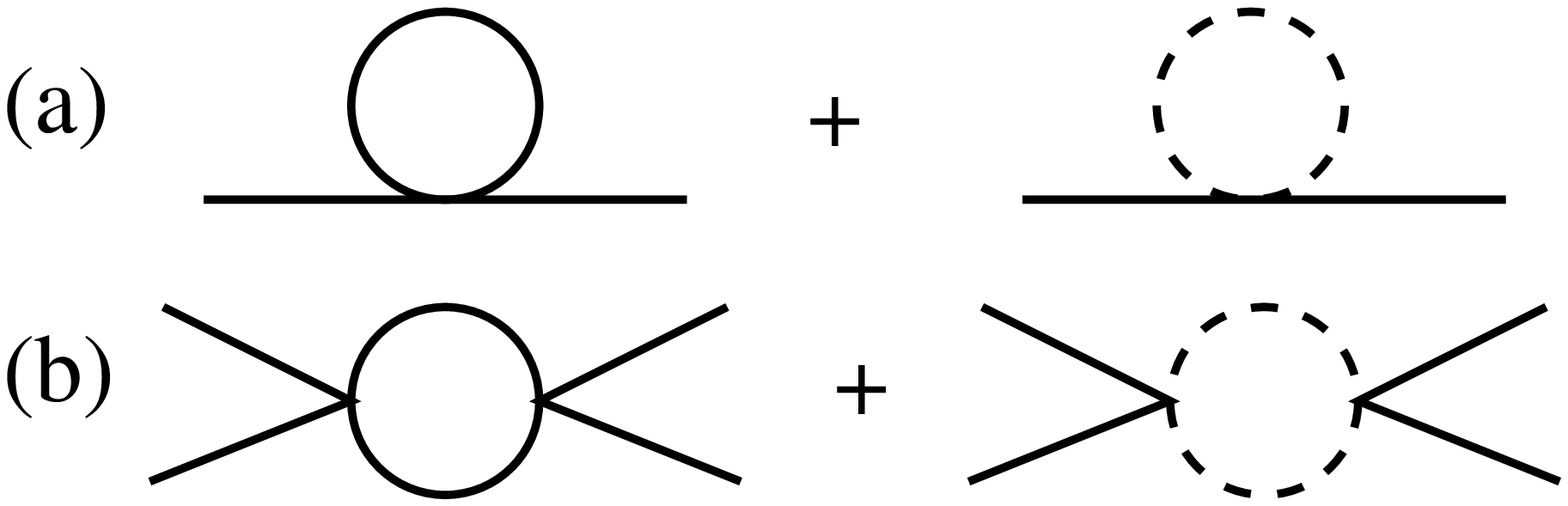}}
{\small
Figure 22. Cancellation of divergences in (a) $\Gamma_2$ and 
 (b) $\Gamma_4$ by Pauli-Villars field (dashed line).}
\medskip

A variant on the sharp momentum space cutoff is to insert a smooth
function $f(p,\Lambda)$ in the Euclidean space loop integrals, with the
property that $\lim_{p\to\infty} f(p,\Lambda)=0$ for fixed $\Lambda$
and $\lim_{\Lambda\to\infty} f(p,\Lambda)=1$ for fixed $p$.  For
example, $f = e^{-p^2/\Lambda^2}$ has this property.  

Yet another way of regularizing loop diagrams 
is to give a parametric representation
of the propagators (in Euclidean space),
\beq
	{1\over p^2+m^2} = \lim_{\Lambda\to \infty}\int_{1/\Lambda^2}^\infty d\alpha e^{-\alpha(p^2+m^2)}
\eeq
For any $\alpha>0$, the momentum integrals converge, and are easy to
do because of the Gaussian factors.  The hard work is postponed to
performing the integrals over the $\alpha$ parameters.  From these
integrals we will get
divergences of the same form as in the $p$-space cutoff method.

An extremely important form of regularization for nonperturbative
computations is the {\it lattice cutoff}, where we imagine spacetime
consists of discrete points on a hypercubic lattice with spacing
$a$ on each side.  Every point in spacetime can be labeled by a
set of integers $n_\mu$, with $mu=0,1,2,3$.
The action (in Euclidean space) is discretized by writing
\beq
	S = \sum_{n_\mu}a^4 \left( \frac 12 \left[ \sum_\nu 
	{1\over (2a)^2}(\phi_{n_\mu+e_\nu} - \phi_{n_\mu-e_\nu})^2
	-m^2\phi_{n_\mu}^2\right] + V(\phi_{n_\mu}) \right)
\eeq
where $e_\nu$ is the unit vector in the $\nu$ direction, and the
fields are defined only on the points of the lattice.  Going to
momentum space, the kinetic term in the action becomes
\beq
	S_{\rm kin} = \frac12\int {d^{\,4}p\over (2\pi)^4}\,
	\phi_p^* \left[\sum_\nu \left({e^{iap_\nu} - e^{-iap_\nu} 
	\over 2a}\right)^2 + m^2 \right] \phi_p 
\eeq
where the integration region is the hypercube defined by
\beq
	-{\pi\over a} \le p_\nu \le {\pi\over a}.
\eeq
So $1/a$ plays the role of $\Lambda$.  This method is not very 
useful for doing perturbative calculations since the propagator 
has a complicated form,
\beq
	P(p) = {1\over a^{-2}\sum_\nu\sin^2(ap_\nu) + m^2}
\eeq
which makes loop integrals very difficult to do.  Furthermore
Lorentz invariance is broken by this method, and only restored in
the continuum limit ($a\to 0$).  Nevertheless, the lattice regulator
is the only one which allows for a nonperturbative evaluation of
the Feynman path integral.  By discretizing spacetime, we have 
simply converted the functional integral into a regular finite-dimensional
multiple integral, assuming we have also put the system into a finite
box of size $L^4$.  Then there are $(L/a)^4$ degrees of freedom, and
the path integral can be evaluated on a computer, without perturbing
in the coupling constant.  This idea, applied
to QCD, was one of the great contributions of K.G.\ Wilson, to which
we shall return later in the course.

A final method called zeta function regularization deserves mention,
even though it is not useful for loop integrals.  Rather, it comes
into play when one wants to evaluate the Feynman path integral itself.
Formally we can write 
\beqa
	\int {\cal D}\phi e^{\frac{i}2\int d^{\,4}x(
	(\partial\phi)^2 - m^2\phi^2)} &=&
	\int {\cal D}\phi e^{-\frac{i}2\int d^{\,4}x 
	\phi(\partial^2 + m^2)\phi} \\
	&=& \left(\det(\partial^2 + m^2)\right)^{-1/2}
\eeqa
This comes from the finite-dimensional formula
\beqa
	\int {d^{\,n} x\over\sqrt{2\pi i}} \exp \left({\frac{i}2\sum_{a,b} x_a A_{a,b}
x_b}\right) &=& \int {d^{\,n}y\over\sqrt{2\pi i}}\exp \left({\frac{i}2\sum_{a} \lambda_{a}
y_a^2}\right)\\
	&=& \prod_{a=1}^n  \sqrt{ 1\over \lambda_a} 
	= \sqrt{ 1\over \det A }
\eeqa
To evaluate the determinant of an operator, we can use the famous
identity that
\beqa
	\ln\det A = \tr\ln A = -\left.{d\over ds}\right|_{s=0}\tr
	A^{-s} &=& \lim_{s\to 0} {d\over ds} 
	\underbrace{\sum_n \lambda_n^{-s}} \\
	&\equiv& \lim_{s\to 0} {d\over ds} \zeta_{\sss A}(s) 
\eeqa
where $\lambda_n$ is the $n$th eigenvalue of the operator $A$.
In QFT the sum is UV divergent because $A = p^2-m^2$, but for
sufficiently large values of $s$ it converges.  The nice thing
about the
zeta function $\zeta_{\sss A}(s)$ associated with this operator
is that its analytic continuation to $s=0$ exists, so the limit
is well defined.  

The function $\zeta_{\sss A}(s)$ associated with an operator
$A$ can be defined in terms of
a ``heat kernel,'' $G(x,y,\tau)$, which satisfies the diffusion-like
equation
\beq
	{\partial\over\partial\tau}G(x,y,\tau) = A_x G(x,y,\tau)
\eeq
where in the present application $A_x = \partial_x^2+m^2$.  The
solution can be written in terms of the eigenvalues $\lambda_n$
and the normalized eigenfunctions $f_n(x)$ as 
\beq
	G(x,y,\tau) = \sum_n e^{-\lambda_n\tau} f_n(x) f_n(y),
\eeq
which has the property $G(x,y,0) = \delta(x-y)$ by completeness
of the eigenfunctions.  The zeta function is defined as
\beqa
	\zeta_{\sss A}(s) &=& {1\over \Gamma(s)} \int_0^\infty d\tau\,
	\tau^{s-1}\int dx\,G(x,x,\tau)\\
	&=& {1\over \Gamma(s)} \sum_n\int_0^\infty d\tau\,
	\tau^{s-1} e^{-\lambda_n\tau} = \sum_n \lambda_n^{-s}
\eeqa
Now the difficulty is in computing $G(x,x,\tau)$.
For further details, see chapter III.5 of Ramond
\cite{ramond}.

\section{Fixed points and asymptotic freedom}
\label{sec:fp}

In our derivation of the RG-improved coupling constant
(\ref{l_exact}), we noted that $\lambda_r(\mu)$ grows with
$\mu$; in fact it blows up when $\mu$ reaches exponentially
large values:
\beq
	\lambda_r(\mu) \to \infty \hbox{\quad as\quad }
	\mu \to \mu_0 e^{16\pi^2\over 3\lambda_r(\mu_0)}
\eeq
which is called the Landau singularity---Landau observed that the
same thing happens to the electron charge in QED.  However, we 
cannot be sure that this behavior really happens unless we 
investigate the theory nonperturbatively.  Since the coupling constant
becomes large, perturbation theory is no longer reliable, and it
is conceivable that $\lambda(\mu)$ could turn around at some scale
and start decreasing again.  Lattice studies of $\phi^4$ theory have
confirmed that the Landau singular behavior is in fact what happens.
One finds that it is not possible to remove the cutoff in this
kind of theory while keeping the effective coupling nonzero at
low energies.  This can be seen by summing the leading log
contributions to the bare coupling:
\beq
\label{bare_c2}
	\lambda_0 = {\lambda_r\over 1 - {3\lambda_r\over 16\pi^2}
	\ln \Lambda/\mu}
\eeq
Unless $\lambda_r = 0$, the coupling blows up at a finite value 
of $\Lambda$.  If we insist on taking the limit $\Lambda\to\infty$,
we must at the same time take $\lambda_r\to 0$.  This means that
the interactions disappear and we are left with a free field theory.
This is what is meant by the {\it triviality} of $\phi^4$ theory.

Triviality was considered a bad thing in the old days when it was
felt that the ability to take the limit $\Lambda\to\infty$ should be
a necessary requirement for a consistent field theory.  That would be
true if we demanded the theory to be a truly fundamental description.
But if we are satisfied for it to be an effective theory, valid only
up to some large but finite cutoff, there is no need to take
$\Lambda\to\infty$.  Thus $\phi^4$ theory can still be physically
meaningful despite its triviality.

Here we see an example of a {\it fixed point} of the renormalization
group:  $\lambda_r(\mu) \to  0$ as  $\mu\to 0$.  We call it an
{\it infrared stable }(IRS) fixed point since it occurs at low (IR)
energy scales.  This example happens to be a trivial fixed point.
Much more interesting would be a nontrivial fixed point, where
$\lambda_r$ flows to some nonzero value as $\mu\to 0$.  If this
were to happen, it would lead to the startling conclusion
that $\lambda_r$ at low energies is quite insensitive to the value
of the bare coupling we put into the theory.  In this situation, the
theory itself predicts the coupling, rather than us having to fix
its value by adjusting a parameter.

There is one famous example of an IRS fixed point, due to K.G.\
Wilson and M.E.\ Fisher \cite{Wilson}, which occurs in the $\phi^4$ theory in $d<4$
dimensions.  We can find this using the $\beta$ function (\ref{WF}),
by noting that $\beta(\lambda_*)$ must vanish at an infrared
fixed point $\lambda_*$, 
\beq
	\int_{\lambda_*} {d\lambda\over\beta(\lambda)} = 
	\int_{\mu=0} d\ln\mu = \infty
\eeq
as well as at an ultraviolet one,
\beq
	\int^{\lambda_*} {d\lambda\over\beta(\lambda)} = 
	\int^{\mu=\infty} d\ln\mu = \infty
\eeq
Solving (\ref{WF}) for $\beta=0$, we obtain the usual trivial
fixed point at $\lambda_*=0$, and in addition the nontrivial
one
\beq
	\lambda_* = {32\over 3}\pi^2\epsilon
\eeq
which can be seen in fig.\ 23.  As advertised, this value is
completely insensitive to the initial value of the coupling 
at very large renormalization scales, so it seems that we
have indeed found a theory with the remarkable property that there 
are no free parameters determining the strength of the interaction
at very low energies.  We must be a little more careful however;
remember that the full coupling constant is dimensionful in
$4-2\epsilon$ dimensions:
\beq
	\lambda_{*,\rm phys} = \lambda_* \mu^{2\epsilon}
\eeq
As we move deeper into the infrared, $\lambda_{*,\rm phys}$ does
get smaller even if the fixed point has been reached, but it does
so in a way which is completely determined.  Its absolute
size is fixed by the dimensionless coupling $\lambda_*$ and 
the energy scale $\mu$ at which we are measuring the 
coupling.

\medskip
\centerline{\epsfxsize=4.5in\epsfbox{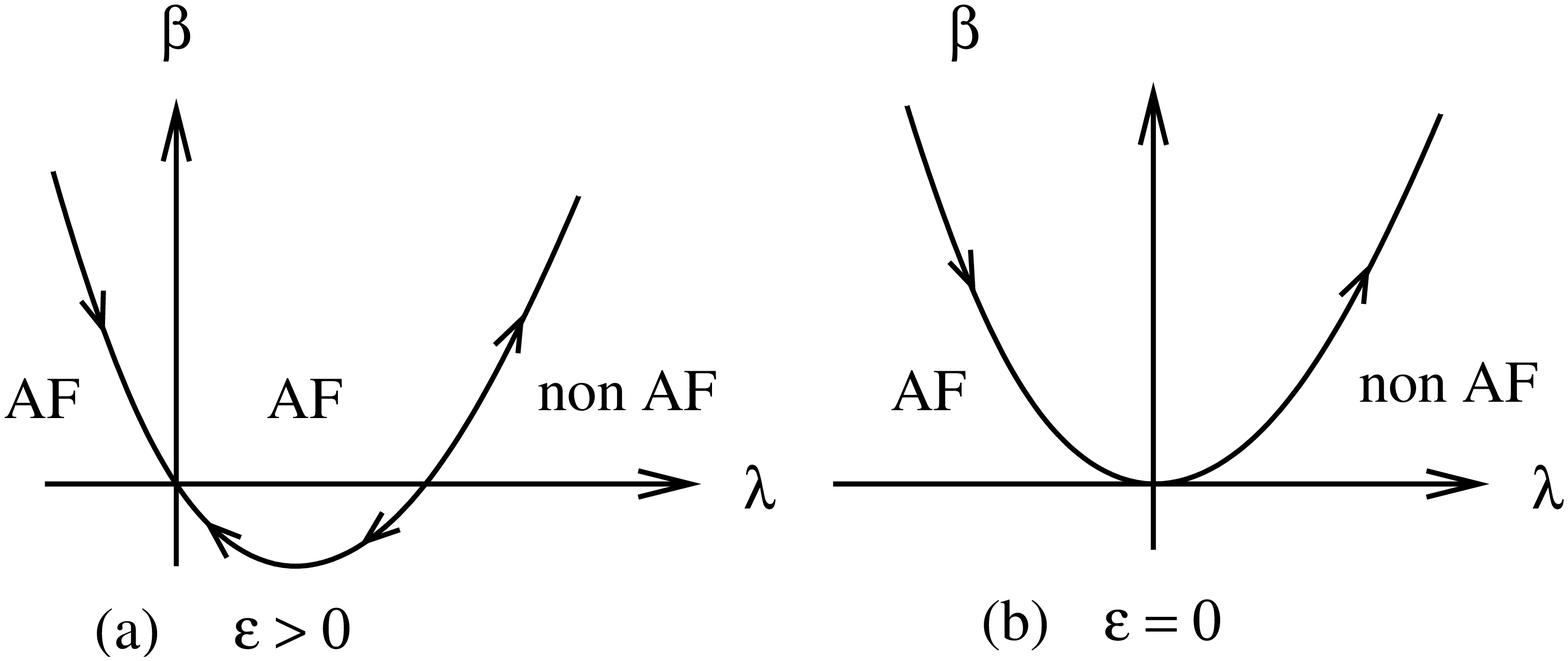}}
{\small
Figure 23. $\beta(\lambda)$ for (a) $\epsilon>0$ ($d<4$) and
(b) $\epsilon=0$ ($d=4$).  Arrows show the direction of flow of
$\lambda$ in the UV.  Asymptotically free regions are denoted by
AF.}
\medskip

How do we know whether this is a UV or an IR fixed point?  This
depends on the {sign} of $\beta(\lambda)$ for $\lambda$ near
$\lambda_*$. If $\beta(\lambda)<0$ for $\lambda<\lambda_*$ [this is
true in the region $0<\lambda<\lambda_*$ in fig.\ 23(a)], then
$\lambda(\mu)$ is a decreasing function of $\mu$.  $\lambda$  {\it
flows} away from the fixed point as $\mu$ increases, and toward it as
$\mu$ decreases.  Similarly If $\beta(\lambda)>0$ for
$\lambda>\lambda_*$ $\lambda$ flows toward  $\lambda_*$ as $\mu$
decreases.  Therefore the Wilson-Fisher fixed point is IR stable.  I
have shown the direction of flow of the  coupling as $\mu$ {\it
increases} in fig.\ 23.

For a nonvanishing physical value of $\epsilon$ like $\epsilon=1$,
the value of $\lambda_*$ is quite large, and so one cannot
necessarily trust the perturbative approximation that was used to
derive it.  However, it was shown by Wilson that the expansion in
$\epsilon$ gives surprisingly accurate results even for such large
values of $\epsilon$, especially if one includes a few higher orders
of corrections.  This method has given useful results for statistical
mechanics systems in three spatial dimensions at finite temperature.
This works because in Euclidean space, the Lagrangian looks like
a Hamiltonian, and the Feynman path integral becomes a partition
function.  After Wick rotating,
\beq
	\int{\cal D}\phi\, e^{iS} \to \int{\cal D}\phi\, e^{-H/T}
\eeq 
We can rewrite the Euclidean space Lagrangian as
\beq
	\frac12\left( (\nabla\phi)^2+m^2\phi^2\right) +\frac{1}{4!}
	\lambda\phi^4 = {1\over T}\left( 
	\frac12\left( (\nabla\tilde\phi)^2+m^2\tilde\phi^2\right)
	+\frac{1}{4!} \tilde\lambda\tilde\phi^4\right)
\eeq
where $\phi = \tilde\phi/\sqrt{T}$ and $\lambda = \tilde\lambda T$.

We have seen an example of an IR fixed point, but what about UV fixed
points?  In fig.\ 23 (a), the trivial fixed point is reached as
$\mu\to\infty$, so this is a UV fixed point.  The coupling $\lambda$
flows toward zero in the ultraviolet.  This is very different from
the behavior in 4D when $\lambda>0$, where we had $\lambda\to\infty$,
which was the Landau singularity.  It is an example of the very
important phenomenon of {\it asymptotic freedom} (AF), so called because
the theory becomes a free field theory for asymptotically large
values of $\mu$.  Asymptotic freedom is a marvelous property from the
point of view of perturbative calculability: the coupling gets weaker
as one goes to higher energies, and perturbation theory becomes more
accurate.  Since such theories make sense up to arbitrarily large
scales, they have the potential to be truly fundamental.  Furthermore,
it might seem that, just like in the case of an IR fixed point, this
could provide an example of a theory with no free parameters, since
we can generate a physical coupling at low energies starting with
an infinitesimally small one at high scales.  But this is not the
case.  Since the coupling becomes larger in the infrared, eventually
one reaches a scale $\mu = \Lambda$ where it can no longer be
treated pertubatively.  Thus we get to trade what we thought
was a dimensionless input parameter to the theory ($\lambda$) for
a dimensionful one $\Lambda$, where in this context $\Lambda$ is 
defined to be the energy scale at which $\lambda(\Lambda) \sim 1$.
This exchange of a dimensionless for a dimensionful parameter is
called {\it dimensional transmutation.}

In the case of $d=4$, $\lambda<0$, the $\phi^4$ theory is  AF, but
since the potential is unbounded from below, this is not a physically
interesting case.  Until 1974 there were no convincing examples of
asymptotic freedom in $d=4$, but then it was discovered by 't Hooft,
Politzer (working independently) and Gross and Wilczek that
nonAbelian gauge theories, like the SU(3) of QCD, have a negative
$\beta$ function at one loop, and are therefore AF.  Now that 't
Hooft has received the Nobel Prize for  proving the renormalizability
of the electroweak theory, some expect that Politzer, Gross and
Wilczek will soon be so honored for their discovery of AF.  (At the
time, Politzer was a graduate student at Harvard, and was the first
to get the sign of the $\beta$ function right after his senior
competitors Gross and Wilczek at Princeton had initially made a sign
error in their calculation.)  The important difference between
nonAbelian theories and U(1) theories like QED is the interaction of
the gauge boson with itself, as in fig.\ 24.

\medskip
\centerline{\epsfxsize=4.5in\epsfbox{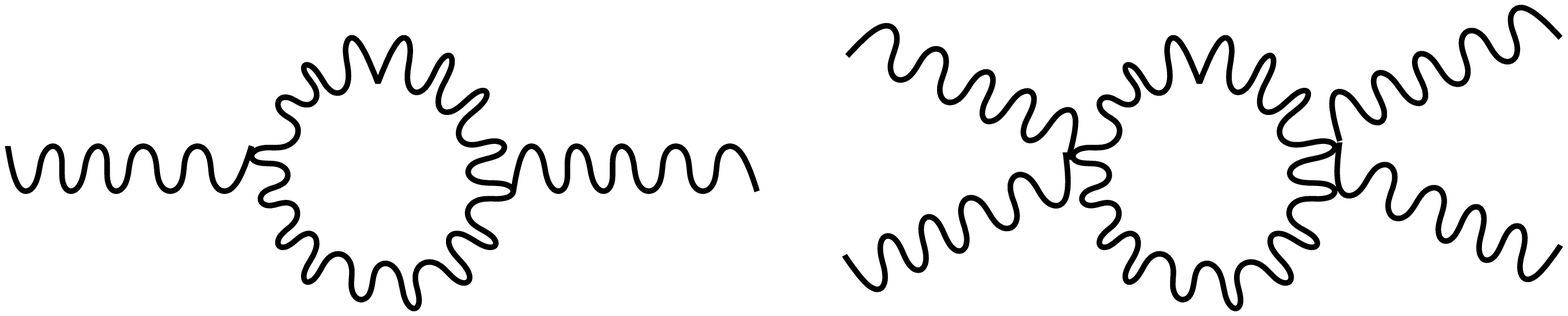}}
{\small
\centerline{
Figure 24. Some diagrams appearing in nonAbelian gauge theories.}}
\medskip

To make dimensional transmutation more explicit, the one-loop
GR-improved coupling in QCD runs like
\beq
	g^2(\mu) = {g^2(\mu_0)\over 1 + b_1
g^2(\mu_0)\ln(\mu^2/\mu_0^2)}
\eeq
and the beta function is
\beq
	\beta(g) = -b_1 g^3
\eeq
where for $n_f=6$ flavors of quarks and $N=3$ for SU(3) gauge theory,
$b_1 = (11N/3 - 2n_f/3)/(8\pi^2) = 21/(24\pi^2)$.  (The dependence
on number of flavors comes from internal quark loops as in fig.\
25, which means that actually $n_f$ also depends on $\mu$, since
a given quark will stop contributing when $\mu$ falls below its
mass $m_q$.)\  Let us define a new scale $\Lambda_{\sss QCD}$ (not to be confused
with the cutoff) such that 
\beq
	1 + 2b_1 g^2(\mu_0)\ln(\mu/\mu_0)\equiv 
	2b_1 g^2(\mu_0)\ln(\mu/\Lambda_{\sss QCD})
\eeq
and
\beq
	g^2(\mu) = {1\over 2b_1\ln(\mu/\Lambda_{\sss QCD})}
\eeq
which illustrates the claim that the renormalized coupling
becomes independent of the value $g^2(\mu_0)$ at the reference
scale.  But the dimensionful parameter $\Lambda_{\sss QCD}$
still remains: it has become the free parameter in the theory.  
It has the interpretation of being the energy scale at which the
coupling starts to become nonperturbatively large.

\medskip
\centerline{\epsfxsize=2.in\epsfbox{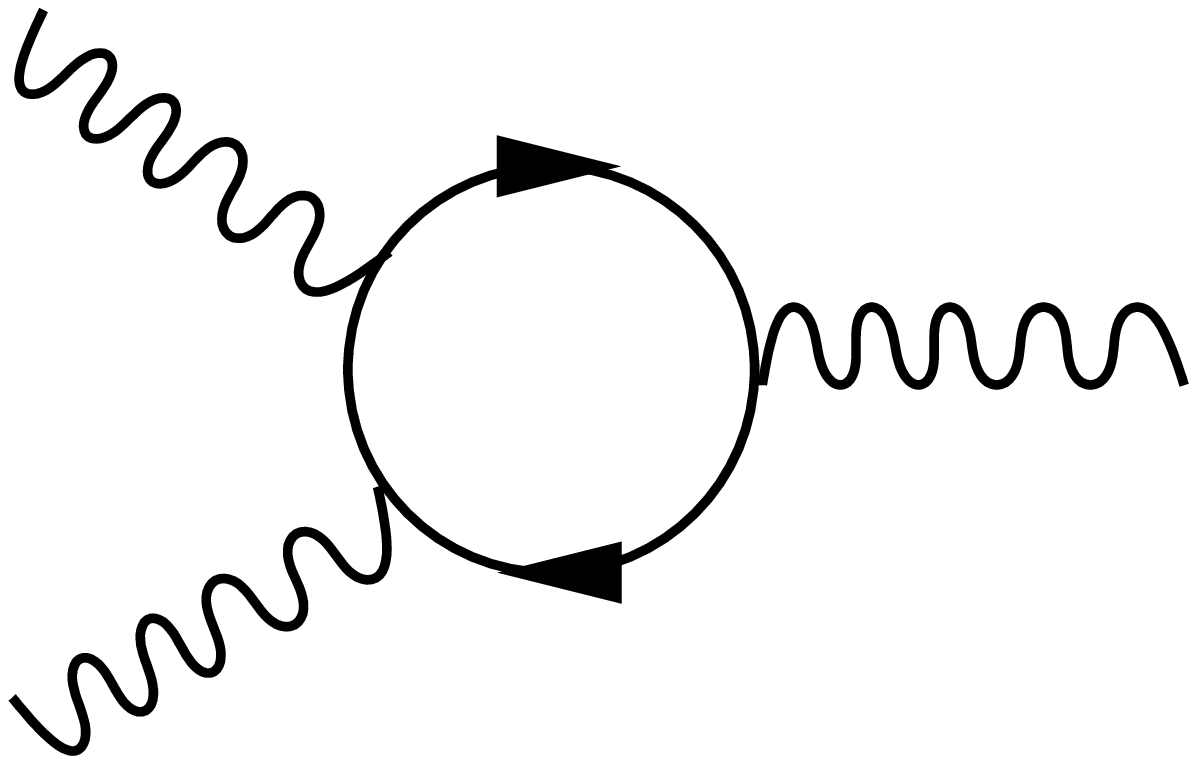}}
{\small
\centerline{
Figure 25. Quark loop contribution to running of QCD coupling.}}
\medskip 

Another way of thinking about dimensional transmutation is that
the theory has {\it spontaneously generated} a mass scale.  This
phenomenon is always associated with nonperturbative effects.
Indeed, solving for $\Lambda_{\sss QCD}$ in terms of the coupling,
we get
\beq
	\Lambda_{\sss QCD} = \mu e^{-{1\over 2b_1 g^2(\mu)}}
\eeq
which is independent of $\mu$ by construction.  It is nonperturbative
because $e^{-{1\over 2b_1 g^2(\mu)}}$ has no expansion in powers of
$g^2(\mu)$.  $\Lambda_{\sss QCD}$ cannot be predicted since it is
the experimentally determined input parameter to the theory; its
measured value is approximately 1.2 GeV, although the precise value 
depends on the renormalization scheme one chooses.

\section{The Quantum Effective Action}
\label{sec:qea}

In problem 5, you showed that $iW[J] = \ln(Z[J])$ is the generator of 
connected Green's functions.\footnote{Warning: I invert the meanings
of $W$ and $Z$ relative to Ramond.  I think this notation is more
standard than his.} These of course are much more physically useful
than the disconnected ones.  But we have seen that the 1PI vertices
are yet more important than the full set of connected diagrams.  It is
natural to wonder if there is  generating functional which can give
just the amputated 1PI diagrams. There is: it is called the {\it
effective action}, $\Gamma[\phi_c]$.  It is defined in terms of
the proper vertices $\Gamma_n$ by
\beq
\Gamma[\phi_c] = \sum_n {1\over n!}\int \left(\prod_{j=1}^n d^{\,4}x_j\,
\phi_c(x_j) \right) \Gamma_n(x_1,\dots,	x_n)
\eeq
where $\Gamma_n(x_1,\dots,x_n)$ is the 1PI amputated 
$n$-point vertex, written in position space.  Thus the $\phi_c$
factors mark the positions of the external lines in our usual
way of computing diagrams.\footnote{\bf Note: my earlier definition
of $\Gamma_n$ had the factor of $i$ from $e^{i\Gamma}$ in the
$\Gamma_n$'s, whereas here I am defining $\Gamma_n$ to be
real.  I should go back to the earlier sections and remove the
$i$'s from the $\Gamma_n$'s to be consistent.}
The physical significance of $\Gamma[\phi]$ is that we can get 
{\it all} the diagrams of the theory, both 1PI and 1PR, by thinking
of $\Gamma[\phi_c]$ as an effective action from which we 
construct tree diagrams only.  For example, the complete 6-point
amplitude has the form shown in fig.\ 26, where each external
leg is associated with a factor of $\phi_c$, and there is no
propagator for the external legs.  The heavy dots represent the
complete 4-point proper vertices and propagators summed to all 
orders in the loop expansion.  
One could imagine generating such diagrams from doing the
path integral with $e^{i\Gamma[\phi_c]}$ but only including the
tree diagrams.

\medskip
\centerline{\epsfxsize=2.5in\epsfbox{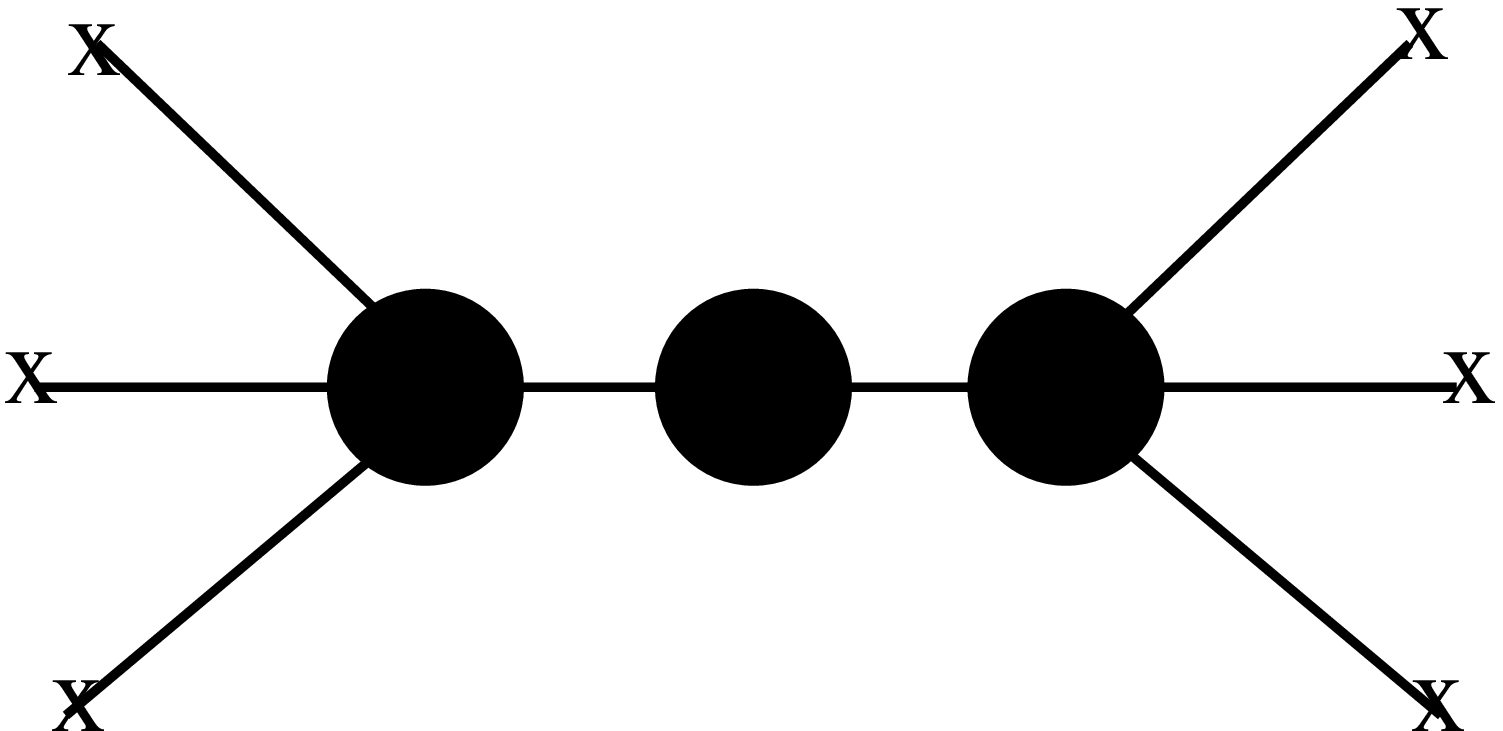}}
\centerline{\small
Figure 26. Generating all diagrams from 1PI vertices for the 6-point
function.}
\medskip

There are different ways of computing $\Gamma[\phi_c]$.  One
which is especially nice for one-loop computations is the {\it
background field} method \cite{Abbott}.
 We do it by splitting the field into two
parts, a classical background field $\phi_c$ and a 
quantum fluctuation $\phi_q$:
\beq
	e^{i\Gamma[\phi_c]} = \int{\cal D}\phi_q\,
	e^{iS[\phi_c+\phi_q]}
\eeq
Let us now
expand $S$ in powers of $\phi_q$.  At zeroth order, we get 
$S[\phi_c]$, which can be brought outside the functional integral.
At first order, we get
\beq
\label{classeq}
\phi_q\left(-\partial^2-m^2 - {\lambda\over 3!}\phi_c^2 \right)\phi_c
\equiv 0
\eeq
To perform the functional integral, it is convenient to make this term
vanish so there is no tadpole for $\phi_q$.  It does vanish provided
that $\phi_c$ satisfies its classical equation of motion, hence the 
identification of $\phi_c$ as the classical field.  Now when we 
rewrite the integrand, we get
\beq
	e^{i\Gamma[\phi_c]} = e^{iS_0[\phi_c]}\int{\cal D}\phi_q\,
	\exp\left({iS_0[\phi_q]- i\int d^{\,4}x [V(\phi_c+\phi_q) - 
	\phi_q V'(\phi_c)]}\right)
\eeq
where $S_0$ is the free part of the action.  The subtracted part
is removed due to the interaction term in the equation of motion
for $\phi_c$.  This is the key to getting rid of the 1PR diagrams:
they are diagrams constructed from vertices that have only a single
internal line.  Since all the internal lines are due to $\phi_q$, 
we eliminate such diagrams by removing the term linear in $\phi_q$
from the interaction.  

How this works at tree level is clear, but in loop diagrams it
is possible to induce mixing between $\phi_q$ and $\phi_c$
which feeds into 1PR diagrams.  In $\phi^4$ theory such
mixing starts at one loop, as shown in fig.\ 27, and it will 
give rise to corrections like fig.\ 28 which are no longer 1PI.
\medskip
\centerline{\epsfxsize=2.5in\epsfbox{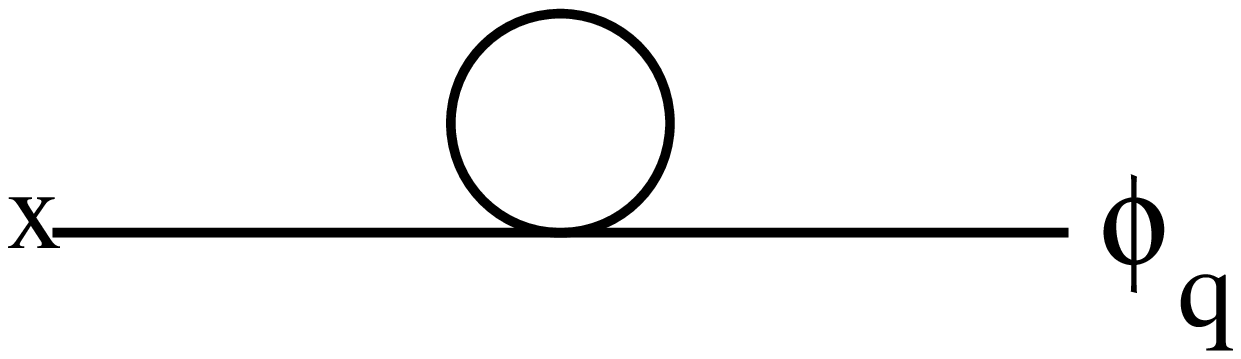}}
\centerline{\small
Figure 27. Diagram which mixes $\phi_c$ (the "x") with the quantum
flucutation $\phi_q$.}
\medskip
\centerline{\epsfxsize=2.in\epsfbox{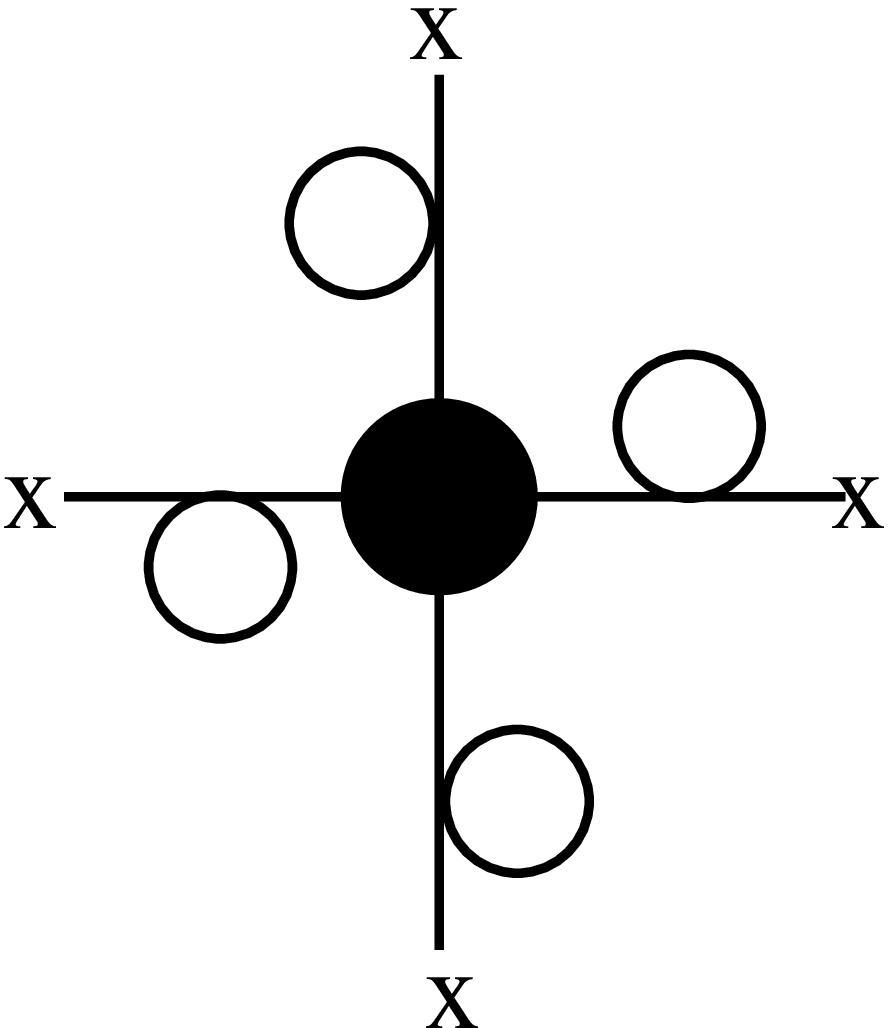}}
\centerline{\small
Figure 28. 1PR diagram induced by the mixing shown in fig.\ 28.}
\medskip
The reason is that once we introduce loops, the classical
equation will get quantum corrections; thus (\ref{classeq}) is
no longer quite right.  At one loop we get the correction
\beq
\label{classeq2}
\phi_q\left(-\partial^2-m^2 - {\lambda\over 3!}\phi_c^2 
- {\lambda\over 3!}\langle\phi_q^2 \rangle \right)\phi_c = 0
\eeq
Adding this correction will take care of the unwanted
mixing terms.   Let's now show  how things work
at lowest nontrivial order using the background field method.

I often find this to be a nice way of keeping
track of the signs of the quantum corrections to mass and coupling.
For the mass correction in $\phi^4$ theory, we can write
\beqa
	i\Gamma &\cong& \int{\cal D\phi}e^{iS_0}\int d^{\,4}x\,
	\left( -i{\lambda\over 4!} {4!\over 2! 2!}\phi_c^2\phi_q^2\right)\\
\longrightarrow	-\frac{i}{2} \delta m^2 \phi_c^2 &=&
	-{i\lambda\over 4} \phi_c^2 G_2(x,x)\\
	\longrightarrow	\delta m^2 &=& {\lambda\over 2}	
	\intdp {i\over \pmie} 
\eeqa
which agrees with our previous calculation (\ref{mass_shift}).
This way of determining the sign is more direct than our previous
reasoning.  Similarly we can obtain $\delta\lambda$:
\beqa
	i\Gamma &\cong& \int{\cal D\phi}e^{iS_0} \frac12\left(
	\int d^{\,4}x\,
	\left( -i{\lambda\over 4!} {4!\over 2! 2!}
	\phi_c^2\phi_q^2\right) \right)^2\\
\longrightarrow	-\frac{i}{4!} 
	\delta\lambda \int d^{\,4}x\,\phi_c^4 &=&
	-{\lambda^2\over 16}\int d^{\,4}x \int d^{\,4}y\,
	 \phi_c^2(x) \phi_c^2(y) G_2(x-y)^2\\
\eeqa
We can Taylor-expand $\phi^2_c(y) = \phi^2_c(x) +
(y-x)\cdot\partial\phi^2_c(x) + \cdots$.  The higher order
terms represent the momentum dependent parts of the 4-point
function, which we could keep track of if we wanted to, but if
we just want to confirm the $p$-independent part of the coupling,
we can neglect these.  Then
\beqa
	-\frac{i}{4!}\delta\lambda \int d^{\,4}x\,\phi_c^4 &=&
	-{\lambda^2\over 16}\int d^{\,4}x\,  \phi_c^4(x)
	\int d^{\,4}y\, G_2(x-y)^2\\
\longrightarrow	-\frac{i}{4!}\delta\lambda &=& 
-{\lambda^2\over 16}\intdp \left({i\over \pmie}\right)^2
\eeqa
This also agrees with the sign and magnitude of our previous
computation of the divergent part of $\delta\lambda$.

In general the effective action $\Gamma[\phi]$ is a {\it nonlocal}
functional of the classical fields: unlike the action which is of
the form $\int d^{\,4}x\, {\cal L}(x)$, which has  exclusively
local terms, $\Gamma_n(x_1,\dots,x_n)$ depends on noncoincident
positions of the fields.  This is of course not a violation of our
requirement that the fundamental interactions should be local. 
The nonlocality in $\Gamma$ is due to us having accounted for the
effects of propagation: virtual $\phi$ particles transmit
interactions over a distance, in a way which respects causality,
thanks to our proper choice of Feynman boundary conditions in the
propagators.  This would not be the case for a generic function of
the $x_i$'s we might randomly write down. That is an important
distinction between the underlying theory and the effective
theory.  

Since we can always reexpress nonlocalities in terms of higher
derivatives using Taylor expansions, another way of writing
$\Gamma$ which makes it look more like a local functional is
\beq
	 \Gamma[\phi_c] = \int d^{\,4}x\, \left[ -V_{0}(\phi_c) + 
	V_2(\phi_c) (\partial\phi_c)^2 + 
	V_4(\phi_c) (\partial\phi_c)^4 + \cdots \right]
\eeq
If we now specialize to constant fields, the derivative terms
drop out and we are left with only $V_{0}$, which is called the
{\it effective potential}.  Let's give one very famous example
\cite{CW} due to Coleman and Weinberg.  They noticed that it
is possible to sum up all the one-loop terms to all orders in
$\phi_c$ because of the fact that these terms just amount to a
field-dependent shift in the mass.  This is illustrated in 
fig.\ 29, where the $x$'s represent $\phi_c$'s (with no internal
propagator).  The diagram $\bigcirc$ is the value of $\ln(Z[0])/i$,
that is, the log of free path integral with no insertions of the 
external field:
\beqa
\bigcirc &=& -i\,\ln\left({\rm
det}^{-1/2}(-\partial^2-m^2)\right)\nonumber\\
	 &=& {i\over 2}\,\tr\ln(-\partial^2-m^2) \nonumber\\
	&=& {i\over 2}\,\intdp\langle p|\ln(p^2-m^2)|p\rangle\nonumber \\
	&=& {i\over 2}\, \langle p|p\rangle \intdp
	\ln(p^2-m^2+i\varepsilon)\nonumber\\
	&=& {i\over 2}\,\intdx \intdp
	\ln(p^2-m^2+i\varepsilon)\nonumber\\
\eeqa
Here we have used that fact that $\langle p|q\rangle  = 
\intdx e^{i(p-q)\cdot x}$ to get the factor of the volume of
spacetime.
Now the claim is that when we add up the series in fig.\ 29, we get
precisely
\beqa
\label{Vcw}
	\bigcirc\!\!\!\,\bullet &=& -\intdx\, V_{\rm\sss CW}(\phi_c) 
=  {i\over 2}\intdx\intdp \ln(p^2-m^2 - \lambda\phi_c^2/2  +i\varepsilon)\nonumber\\
	&=& -\frac12\intdx\intdpe \ln(\pE^2 + m^2 + \lambda\phi_c^2/2) + {\rm const}
\eeqa
To see that it is true, one just has to expand (\ref{Vcw}) as a power
series in $\lambda$ and compare each term to the corresponding
Feynman diagram.  The infinite imaginary constant from $\ln(-1)$ has no
physical significance since we could have avoided it by normalizing
our path integral measure appropriately.

\medskip
\centerline{\epsfxsize=4.5in\epsfbox{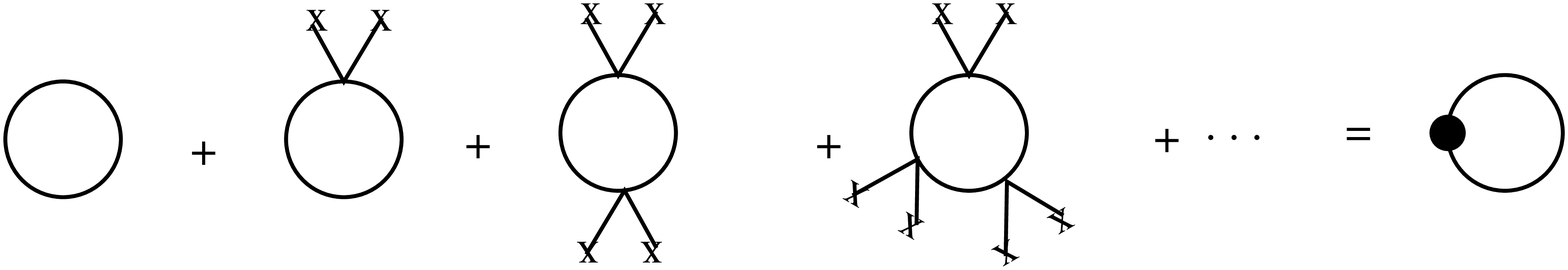}}
\centerline{\small
Figure 29. Resummation for Coleman-Weinberg potential.}
\medskip

To evaluate the Coleman-Weinberg potential, it is convenient to
first differentiate it with respect to $m^2$; 
we can then use our favorite regulator to evaluate it.  With the
momentum space cutoff, and defining $M^2 = m^2 + \lambda\phi_c^2/2$,
\beqa
	{\partial  V_{\rm\sss CW}\over \partial M^2} &=& 
	\frac12\intdpe {1\over \pE^2 + m^2 + \lambda\phi_c^2/2}\nonumber\\
	&=& {1\over 32\pi^2}\int_{M^2}^{\Lambda^2+M^2} du 
	{u-M^2\over u}\nonumber\\
	&=& {1\over 32\pi^2}
	\left(\Lambda^2 - M^2\ln(1+\Lambda^2/M^2)\right)
\eeqa
Ignoring terms which vanish as $\Lambda\to\infty$ and an irrelevant
constant, the integrated potential is
\beq
V_{\rm\sss CW} = {1\over 64\pi^2}\left(2{ \Lambda^2}\,{ M^2}
 -{{ M^4}}\,\ln  \left({\frac {{ \Lambda^2}}
{{ M^2}}}
 \right)-\frac12\,{{ M^4}} \right)
\eeq
You can easily see that the only divergent terms if we reexpand
this in powers of $\phi_c$ are the $\phi_c^2$ and $\phi_c^4$ terms,
and moreover the log divergent shift in the $\phi_c^4$ term agrees
with our previous calculations.  The quadratically divergent part
can be completely removed by renormalizing the mass (and the
cosmological constant, but we shall ignore the latter), so the
$M^4$ term is the most interesting one.  After renormalization,
and adding the one-loop result to the tree-level potential, we 
obtain
\beq
	V(\phi_c) = \frac12 m^2\phi^2 + \frac{1}{4!}\lambda\phi^4
	+ {1\over 64\pi^2}\left(m^2+\frac12\lambda\phi^2\right)^2\ln\left({
	m^2 + \frac12\lambda\phi^2\over\mu^2}\right)
\eeq
I have defined $\mu$ here in such a way as to absorb the $M^4$ which
is not logarithmic into the log.  We can see the $\mu$-dependence 
of the effective coupling constant by taking
\beq
	\lambda_{\rm eff} = \left.{\partial^4 V\over \partial\phi_c^4}
	\right|_{\phi_c=0} = \lambda + {3\lambda^2\over
32\pi^2}\left(\ln\left({m^2\over\mu^2}\right) + \frac32\right)
\eeq
Notice that since we are computing the effective potential, we have
set all external momenta to zero.  Therefore, roughly speaking,
 the coupling is evaluated at the lowest possible energy scale in the
 theory, $m^2$.  This result thus has the same basic form as our 
energy-dependent effective coupling (\ref{lameff}).  

A curiosity about the Coleman-Weinberg potential can be observed
for sufficiently large values of $\mu$: the minimum of the potential
is no longer at $\phi=0$, but at some nonzero value, as illustrated in
fig.\ 30.  If this behavior occurred, it would be an example of
{\it spontaneous symmetry breaking}.  Although the tree-level
potential has the symmetry $\phi\to-\phi$, the quantum theory 
gives rise to a nonzero {\it vacuum expectation value} (VEV),
$\langle\phi\rangle$, which breaks the symmetry.  In the present
case, we call it {\it radiative symmetry breaking} since it is
induced by radiative corrections---another term for loop corrections. 
Spontaneous symmetry breaking is an important feature of the standard
model; the Higgs field gets a VEV which breaks the SU(2)$\times$U(1)
gauge group to the U(1) of electromagnetism, and gives masses to the
fermions.  

\medskip\bigskip
\centerline{\epsfxsize=2.in\epsfbox{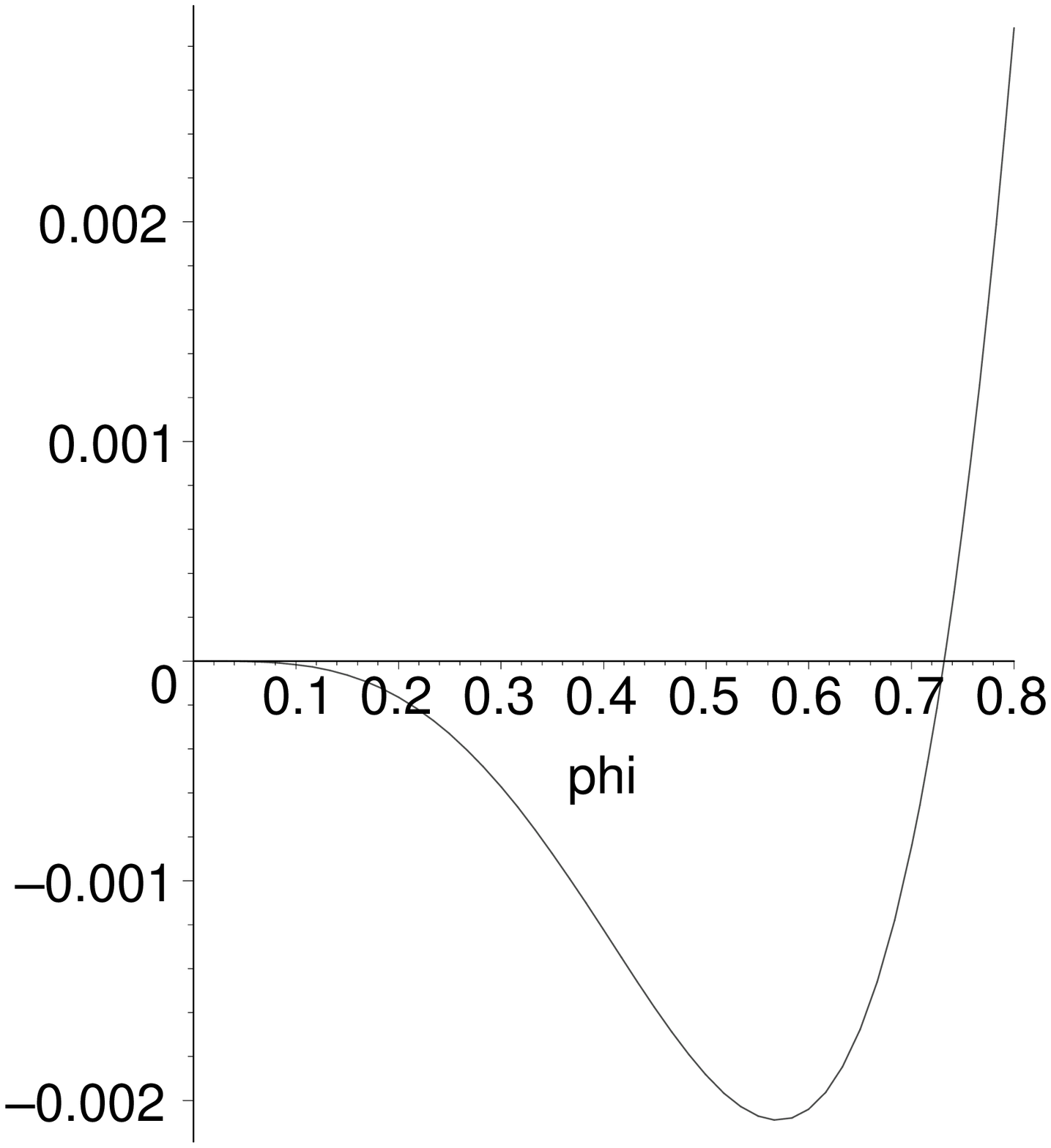}}
\centerline{\small
Figure 30. Coleman-Weinberg potential at very large $\mu$.}
\medskip

It would be interesting if a theory gave rise to spontaneous
symmetry breaking without having to put in a negative value for
$m^2$ by hand, which is the usual procedure.  Unfortunately, the
Coleman-Weinberg potential cannot be relied upon to do this since
the value of $\mu$ required is nonperturbatively large.  For
simplicity we consider the massless case:
\beq
	\mu^2 \cong \frac\lambda{2} \langle\phi\rangle^2
	\exp\left({32\pi^2\over\lambda}\right)
\eeq
For such large value of $\mu$, the loop correction is so big that
we have already reached the Landau singularity.  This can be seen
by RG-improving the one-loop result to get the log into the
denominator.  Thus the calculation cannot be trusted in this
interesting regime.  However, it is possible to get radiative
symmetry breaking in more complicated models, in a way that does
not require going beyond the regime of validity of perturbation
theory.

Now I would like to discuss a more formal definition of the effective
action, in which 
$\Gamma$ is related to $W[J]$ through
a functional Legendre transformation.  In this approach we 
define the {classical field} $\phi_c$ by 
\beq
\label{phiceq}
	\phi_c[J] = {\delta W\over \delta J}
\eeq
where we do not set $J=0$ (until the end of the calculation).  
$\phi_c$ is the expectation value of 
$\phi$ when there is a source term $+\int d^{\,4}x\, \phi J$ in the 
action; thus it satisfies the classical equation 
\beq
\left(\partial^2 +  m^2 + \frac{\lambda}{2}\phi_c^2\right)\phi_c(x) = J(x)
\eeq
plus the quantum corrections like we derived in eq.\ (\ref{classeq2}).
Hence 
\beqa
	\phi_c &=& \left(-\partial^2 - m^2 -
\frac{\lambda}{2}\phi_c^2\right)^{-1} (-J)\nonumber\\
	&=& \left[(-\partial^2 - m^2)
	\left(1 -\left(-\partial^2 - m^2\right)^{-1}
	\frac{\lambda}{2}\phi_c^2\right)\right]^{-1}(-J)\nonumber\\
\label{phic}
	&=& -\left[ G_2\cdot J + G_2\cdot\frac{\lambda}{2}\phi_c^2\cdot G_2
	\cdot J 
	+ G_2\cdot\frac{\lambda}{2}\phi_c^2\cdot G_2\cdot\frac{\lambda}{2}\phi_c^2\cdot G_2
	\cdot J + \cdots \right]
\eeqa
where I am using the notation\footnote{\bf Note: I am henceforth
defining $G_2$
to be the inverse of $(-\partial^2 - m^2)$, which in momentum space
looks like $1/(p^2-m^2)$, not $i/(p^2-m^2)$  This differs from
my previous usage.  Furthermore I am redefining the source
term in the Lagrangian to be $+J\phi$ in order to match the
convention which is more standard in the derivation of the 
effective action below.}
\beqa
	G_2\cdot J &=& \intdy G_2(x-y)J(y)\nonumber\\
	G_2\cdot\frac{\lambda}{2}\phi_c^2\cdot G_2
	\cdot J &=& \intdy \intdz G_2(x-y)\frac{\lambda}{2}\phi_c^2(y)
	G_2(y-z) J(z)\qquad \hbox{\it etc.}
\eeqa
This could be evaluated perturbatively by taking $\phi_0 = G_2\cdot J$,
substituting this into (\ref{phic}) to obtain $\phi_1 = \phi_0 + 
G_2\cdot\frac{\lambda}{2}\phi_0^2\cdot G_2 \cdot J$, and continuing
this process to the desired order in $\lambda$.  However we won't 
need an explicit solution in order to carry out the proof of the 
following claim: the effective action is related to $W[J]$ by
\beq
	\Gamma[\phi_c] = W[J] - J\cdot\phi_c
\eeq

To prove this, we will start functionally differentiating 
$\Gamma[\phi]$	
(let's drop the subscript $c$ and replace it with $\phi_c(x) = 
\phi_x$ when necessary) with respect to $\phi$:
\beq	
	\fd\Gamma\phi = \fd{W}{J}\cdot \fd{J}\phi - \fd{J}\phi
\cdot\phi - J = -J
\eeq
The latter equality follows from the fact that $\fd{W}{J}=\phi_c$.
In passing, we note that $\fd{\Gamma}{J}=0$ when $J=0$; similarly
$\fd{W}{\phi}=0$ when $\phi=0$.

Next let's differentiate again:
\beq
\label{Jphi}
	\fdd{\Gamma}{\phi_x}{\phi_y} = -\fd{J_x}{\phi_y} 
	= - \left[\fd{\phi_x}{J_y}\right]^{-1}
	= - \left[\fdd{W}{J_x}{J_y}\right]^{-1}
\eeq
If we recall problem 5,\footnote{Note the important
generalization that 
\beq
\label{Wrel}
 {\delta^n W\over \delta J_1\cdots \delta J_n} = 
	-i^{n+1}\langle T(\phi_1\cdots\phi_n)\rangle_{\rm con}
\eeq
which follows from the fact that $W = -i\ln Z$, so
${\delta^n W\over \delta J_1\cdots \delta J_n} = -i
Z^{-1}{\delta^n Z\over \delta J_1\cdots \delta J_n}+$
(terms that remove the disconnected parts) $ = 
-i \langle T(i\phi_1\cdots i\phi_n)\rangle_{\rm con}$}
\beqa
\fdd{W}{J_x}{J_y} &=& -i \left\langle (i\phi_x) (i\phi_y)
 \right\rangle_{\rm con}
\nonumber\\	&=& G_2^{\rm con}(x-y)
\eeqa
Here $G_2^{con}(x-y)$ is the connected part of the full propagator, 
to all
orders in perturbation theory.  Because $\Gamma_2$ is the full 
propagator with both legs multiplied by inverse propagators, we see
that 
\beq
	\fdd{\Gamma}{\phi_x}{\phi_y} = \Gamma_2[\phi]
\eeq
as promised.  Notice that the inverse propagator is automatically
1PI, in analogy to the fact that $(1+x+x^2+\cdots)^{-1} = 1-x$.

We can continue this procedure to find higher derivatives of
$\Gamma[\phi]$ and inductively prove the desired relation.
Things get a little more complicated as we go to higher point
functions.  To obtain $\Gamma_3$ (which does not exist in $\phi^4$
theory, but it does in $\phi^3$ theory), it is useful to differentiate
relation (\ref{phiceq}) with respect to $\phi$ \cite{AL}:
\beq
\label{fdd1}
	\fdd{W}{J_x}{J_y}\cdot\fd{J_y}{\phi_z} = \delta_{x,z}
\eeq	
Using (\ref{Jphi}) we can substitute for $\fd{J_y}{\phi_z}$
to write this as
\beq
\label{fdd2}
\fdd{W}{J_x}{J_y}\cdot\fdd{\Gamma}{\phi_y}{\phi_z} = -\delta_{x,z}
\eeq
and take a derivative with respect to $J$:
\beq
\fddd{W}{J_w}{J_x}{J_y}\cdot\fdd{\Gamma}{\phi_y}{\phi_z}
	+
\fdd{W}{J_x}{J_y}\cdot\fd{\phi_y}{J_u}\cdot
\fddd{\Gamma}{\phi_u}{\phi_y}{\phi_z}
 = 0
\eeq
Now using (\ref{fdd1}) and (\ref{fdd2}) this can be solved for
$\fddd{\Gamma}{\phi_u}{\phi_y}{\phi_z}$:
\beq
\label{G3rel}
	\fddd{\Gamma}{\phi_u}{\phi_v}{\phi_w} = 
  -\intdx\intdy\intdz \fddd{W}{J_x}{J_y}{J_z} 
	\fdd{\Gamma}{\phi_x}{\phi_u}\fdd{\Gamma}{\phi_y}{\phi_v}
	\fdd{\Gamma}{\phi_z}{\phi_w}
\eeq
To see why the $-$ sign should be there, note that eq.\ (\ref{Wrel})
implies that $\fddd{W}{J_x}{J_y}{J_z} = -\langle T(\phi_x \phi_y 
\phi_z)\rangle_{\rm con}$.  For the $\frac{1}{3!}\mu\phi^3$ interaction,
we would thus get (in momentum space)
\beqa
&&\!\!\!\!\!\!\!\!\!\!\!\!\!\!\!\!\!\!\!	\fddd{W}{J_1}{J_2}{J_3} \fdd{\Gamma}{\phi_1}{\phi_1}
\fdd{\Gamma}{\phi_2}{\phi_2}\fdd{\Gamma}{\phi_3}{\phi_3}\nonumber\\
&& = -(-i\mu){i\over p_1^2-m^2}\,
{i\over p_2^2-m^2}\,{i\over p_3^2-m^2}\times  (p_1^2-m^2)
(p_2^2-m^2) (p_3^2-m^2)  = \mu
\eeqa
whereas we know that $\fddd{\Gamma}{\phi_1}{\phi_1}{\phi_3} = -\mu$
(since the action is the kinetic term minus the potential).
Therefore the right hand side of (\ref{G3rel}) is indeed
the connected 3-point function with external propagators
removed by the $\fdd{\Gamma}{\phi}{\phi}$ factors, which is
the same as the proper vertex $\Gamma_3$.
I will leave as an exercise for you to show how things work for the
4-point function.

In the previous manipulations we have been referring to the effective
action at all orders in the loop expansion.  However, it is also
possible to truncate it, and work with the 1-loop or 2-loop effective
action.  To reiterate: the complete result for any amplitude at this
order will be given by the tree diagrams constructed from $\Gamma$.

I will quote one other interesting result from Ramond \cite{ramond}
without deriving it.  It can be shown that the equation satisfied
by the classical field is 
\beq
\label{classeq3}
	(\partial^2+m^2)\phi_x - J_x = - {1\over Z[J]}
	{\partial V\over
	\partial\phi}\left(-i{\delta\over\delta J_x}\right)Z[J]
\eeq
where the argument of $V'$ instead of being $\phi_x$ is the
operator $-i{\delta\over\delta J_x}$.  This is a general result;
specializing to $\phi^4$ theory, we get
\beq
\label{classeq4}
	(\partial^2+m^2)\phi_x - J_x = -{\lambda\over 3!}\phi_x^3 + 
 i{\lambda\over 4}\fd{\phi^2_x}{J_x} + {\lambda\over 3!}
	\fdd{\phi_x}{J_x}{J_x} 
\eeq
By restoring factors
of $\hbar$, one can show that the three terms on the r.h.s.\ of
(\ref{classeq4}) are of order $\hbar^0$, $\hbar^1$ and $\hbar^2$
respectively.  Therefore the last term arises at two loops.  It can
be seen that the quantum correction we derived in (\ref{classeq2})
does correspond to the one-loop term in (\ref{classeq4}).

To conclude this section, I would like to mention another kind of
effective action which is due to Wilson \cite{Wilson}.  
Suppose we have computed the renormalized Lagrangian ${\cal L}$
for some particular value of the cutoff $\Lambda$.  We can define
the effective value of ${\cal L}$ at a lower scale, $\Lambda'$,
by splitting the field in Fourier space into two pieces:
\beq
	\tilde\phi(p) = \left\{\begin{array}{ll}
	\tilde\phi_{\sss IR},& 0<|p|<\Lambda'\\
	\tilde\phi_{\sss UV},& \Lambda'<|p|<\Lambda\end{array}\right.
\eeq
The corresponding fields in position space are obtained by doing
the inverse Fourier transform.
It is easiest to define what we mean by $|p|$ if we are working
in Euclidean space.  The Wilsonian effective action, at the scale
$\Lambda'$, is given by integrating over the UV degrees of freedom
only:
\beq
	e^{-S'[\phi_{\sss IR};\,\Lambda']} = 
	\int{\cal D}\phi_{\sss UV} 
	\,e^{-S[\phi_{\sss IR}+\phi_{\sss UV};\,\Lambda]}
\eeq
In practice, this is accomplished by integrating all Feynman
diagrams over the region $\Lambda'<|p|<\Lambda$ instead of 
$0<|p|<\Lambda$.  Notice that this way of splitting the field into
UV and IR parts respects our desire to have no mixing between 
$\phi_{\sss IR}$ and $\phi_{\sss UV}$ in the kinetic
term, since it is diagonal in the momentum basis.  Thus 
$\phi_{\sss IR}$ is similar to $\phi_c$ and $\phi_{\sss UV}$ is
like $\phi_q$ in the above discussion. 

This process of {\it integrating out} the dynamical degrees of
freedom above a certain scale is Wilson's way of defining a RG
transformation.  Suppose the original Euclidean action had been
\beq
	S[\phi;\,\Lambda] = \frac12\left((\partial\phi)^2 +
	m^2(\Lambda)\phi^2\right)
	+{\lambda(\Lambda')\over 4!}\phi^4;
\eeq
then we expect the effective action to have the form
\beq
	S'[\phi;\,\Lambda'] = \frac12\left(\left(1+\delta Z
	\left({\phi\over\Lambda'}\right)\right)(\partial\phi)^2 +
	\hat m^2(\Lambda')\phi^2\right) + 
	O\left({(\partial\phi)^4\over\Lambda'^{\,4}}\right)
	+{\hat\lambda(\Lambda')\over 4!}\phi^4 + 
	O\left({\phi^6\over\Lambda'^2}\right)
\eeq
To make this look more like the original Lagrangian, we should
renormalize $\phi = \phi'/\sqrt{1+\delta Z}$:
\beq
	S'[\phi';\,\Lambda'] = \frac12\left((\partial\phi')^2 +
	m^2(\Lambda')\phi'^2\right) + 
	O\left({(\partial\phi')^4\over\Lambda'^{\,4}}\right)
	+{\lambda(\Lambda')\over 4!}\phi^4 + 
	O\left({\phi'^{\,6}\over\Lambda'^2}\right)
\eeq

In this approach, the cutoff is now playing the role of the
renormalization scale.  The parameters like $\lambda'$ appearing in 
$S'$ are telling us what physics looks like at the scale $\Lambda'$.
It is not hard to see why when considering loop diagrams like
fig.\ 9: if all the external legs have large momenta, then it is
impossible to get contributions from the loops where the internal
lines have small momenta.  There is only a set of measure zero which
conserves momentum at the vertices while restricting $|p|<\Lambda'$.
So we can use just the tree diagrams generated by 
$S'[\phi';\,\Lambda']$ to compute physical amplitudes at the scale
of the cutoff.  This is similar in spirit though different in detail
to the way the quantum effective action works.  In particular,
the Wilsonian effective action {\it does} contain the effects of 1PR
diagrams, whereas these have to be constructed from the proper
vertices of the quantum effective action.

Wilson does one further step in his way of defining 
$S'[\phi';\,\Lambda']$.  Notice that even in the case of
free field theory, $S'$ does not keep exactly the same form as
$S$ in the way we have so far defined it, since $\Lambda$
appears in the upper limit of momentum integrals:
\beq
	S_0[\phi;\, \Lambda] = \frac12\int_0^\Lambda {
	d^{\,d}p\over (2\pi)^4}\tilde\phi_{\sss-p}(p^2+m^2\phi^2)\tilde\phi_{p}
\eeq
To make $S_0$ appear really independent of the cutoff, we can
define a dimensionless momentum $q=p/\Lambda$:
\beq
	S_0[\phi;\, \Lambda] = \Lambda^d\frac12\int_0^1 {
	d^{\,d}q\over (2\pi)^4}\tilde\phi_{\sss -q}
(\Lambda^2 q^2 + m^2)\tilde\phi_q
\eeq
and rescale the
fields by $\phi\to \zeta_0\phi$ where $\zeta_0 = \Lambda^{-d/2-1}$:
\beq
	S_0[\phi;\, \Lambda] \to \frac12\int_0^1 {
	d^{\,d}q\over (2\pi)^4}\tilde\phi_{\sss -q}
	(q^2 + (m^2/\Lambda^2))\tilde\phi_{q}
\eeq
This is equivalent to taking the unit of energy to be $\Lambda$
at any stage in the renormalization procedure.  When we integrate
out momenta in the range $\Lambda'<|p|<\Lambda$, we will pick up
an additional factor of $(\Lambda'/\Lambda)^{-d/2-1}$ in the step
where we renormalize the fields.  One notices that the effective
mass $m^2/\Lambda^2$ gets continously bigger as one integrates out 
more fluctuations.  We call $\phi^2$ a {\it relevant operator} because
its coefficient gets bigger as we go deeper into the infrared.
On the other hand, consider the coefficient of 
an operator of the form $\phi^n$:
\beq
	S_{\rm int} = c_n\prod_{i=1}^n \left(\int_0^\Lambda {d^{\,d}p\over(2\pi)^4}
	\tilde\phi_{p_i}\right)(2\pi)^d\delta^d\left(\sum_i^n p_i\right)
\eeq
scales like $\Lambda^{(n-1)d}\zeta^n
\sim \Lambda^{-nd/2 + d + n}$.  For $d=4$ this is $\Lambda^{-d + n}$,
which we recognize to be simply the dimension of the operator.  An
operator which does not scale with $\Lambda$, such as $\lambda\phi^4$
in 4D, is called {\it marginal}.  Its effects neither become very
strong nor very weak in the infrared.  Of course this naive argument
does not take into account the running due to
quantum effects.  If the coupling runs logarithmically to larger 
values in the IR, as is the case in QCD, then it is called {\it
marginally relevant}, and if it runs to smaller values, as it does
for $\lambda\phi^4$ in 4D, it is {\it marginally irrelevant}.
Finally, we see that higher-dimension operators run according to
a power law, $\Lambda^{n-d}$, and these are called {\it irrelevant}.
The term is used in a technical, not a literal sense.  One might
find it surprising that a diagram like fig.\ 31 which contributes
to the 6-point function is considered to be irrelevant in the 
IR---after all, the propagator for the internal line gets {\it bigger}
at low energies!  However, we are assuming there is a mass in the 
theory, so when we go really deeply into the IR, the propagator can
no longer grow.  

\medskip
\centerline{\epsfxsize=2.in\epsfbox{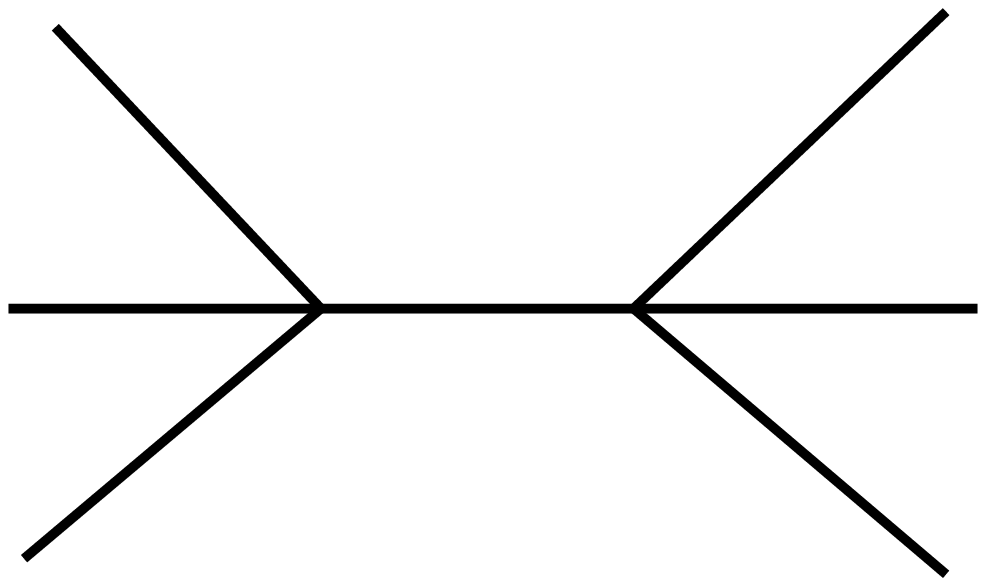}}
\centerline{\small
Figure 31. Contribution to the 6-point function in Wilsonian
effective action.}
\medskip

From this discussion we can get more insight into our calculation
of the Wilson-Fisher fixed point (\ref{WF}).  We only get the
interesting term $-2\epsilon\lambda$ because we defined $\lambda$
to be dimensionless.  This is precisely what we have done to all the
couplings when we went to the dimensionless momentum variable in
Wilson's approach.  We are taking the running cutoff to be the 
unit of energy---the renormalization scale played the same role in
the previous calculation.  It should now be clear that the cutoff
in Wilson's language is really the same thing as what we called the
renormalization scale in our first approach to renormalization.

We have touched only briefly on the profound implications of
Wilson's viewpoint.  One of the important concepts which is
closely tied to his work is that of {\it universality}.  As a result
of the irrelevant nature of higher dimensional operators, we can
modify the Lagrangian at high scales in an infinite variety of ways
without affecting the low energy physics at all, provided the 
system approaches the IR fixed point.  The microscopic Lagrangian
may look very different, yet the IR limit is insensitive to this.  Any
two theories which flow to the same IR fixed point are said to belong
to the same {\it universality class}.  

Before leaving Wilson's contributions, let us note the relation
between the fixed point and critical phenomena, namely second
order phase transitions \cite{Preskill}.  Recall the scaling behavior 
of renormalized amplitudes with energy,
\beq
	G_n^{(r)}(E,\theta_i,\lambda(\mu), m(\mu), \mu) = 
	E^D \left[{Z(E)\over Z(\mu)}\right]^{n/2} f(\theta_i,
	\lambda(E), m^2(E)/E^2)
\eeq
where $\theta_i$ are dimensionless kinematic variables, and $D$
is the dimension of the operator giving rise to the $n$-point
amplitude.  Near the fixed point, we can write
\beq
\label{fixedpoint}
	m^2(\mu) \sim C \mu^{2\gamma_m^*};\qquad
	Z^{1/2}(\mu) \sim C' \mu^{\gamma^*}
\eeq 
Therefore
\beq
	G_n^{(r)}\sim
	E^{D+n\gamma^*} f(C E^{2(\gamma_m^*-1)})
\eeq
In a second order phase transition, the correlation length
$\xi$ diverges:
\beq
	\xi^{-1} = \lim_{r\to\infty}{G_2(r)'\over G_2(r)} = 
	\lim_{r\to\infty}{(e^{-mr})'\over e^{-mr}}
	= m
\eeq
That is, the mass goes to zero.  We see that as long as $\gamma_m^*>0$,
the behavior of $m(\mu)$ in (\ref{fixedpoint}) is such that $m\to 0$
as $\mu\to 0$, corroborating our claim that the fixed point corresponds
to a phase transition.  In statistical physics, the correlation
length  diverges as a power of the deviation of the temperature from
its critical value in a second order phase transition:
\beq
	\xi  \sim (T - T_c)^{-\nu}
\eeq
The relation between the critical exponent $\nu$ and $\gamma^*_m$
is obtained from rewriting the amplitude in terms of distance
scales $L\sim 1/E$,
\beq
	G_n^{(r)} \sim L^{-D-n\gamma^*} f(m^2(L)L^2)
\eeq
and the fact that $m^2$ vanishes analytically with $(T - T_c)$:
\beq
	C \sim T- T_c
\eeq
Therefore
\beq
	f = f( (T- T_c) L^{2 - 2\gamma^*_m} ) = f(L/\xi)
\eeq
which means that 
\beq
	\xi \sim (T- T_c)^{1/(2 - 2\gamma^*_m)}
\eeq
and the critical exponent is given by
\beq
	\nu^{-1}  = 2 - 2\gamma^*_m = 2 - \epsilon/3 + O(\epsilon^2)
\eeq
This exponent has been computed to order $\epsilon^3$ in an expansion
in powers of $\epsilon$, which is then evaluated at $\epsilon=1$,
corresponding to 3 dimensions (I have switched to $d = 4-\epsilon$
in this subsection.  This expansion gives surprisingly good results;
the Ising model (in the same universality class as $\phi^4$ theory)
is measured to have $\nu = 0.642\pm 0.003$, while the $\epsilon$
expansion gives $0.626$.  There are other critical exponents as well;
the power of $L$ in front of the two-point function is
\beq
	\eta = 2\gamma^* = {\epsilon^2\over 54} + O(\epsilon^3)
\eeq
which gives 0.037 in the $\epsilon$
expansion, compared to $0.055\pm 0.010$ for the Ising model.

\section{Fermions}
\label{sec:fermions}

Up to now we have been dealing with scalars only, but in the real
world all known matter particles (as opposed to gauge bosons) are
fermions.  Let's recall the Lagrangians for spin $1/2$ particles:
\beqa
	\hbox{left-handed Weyl:}&\qquad& \psi_L^\dagger(
i\bar\sigma\!\cdot\!
	\partial)\psi_L\\
	\hbox{right-handed Weyl:}&\qquad& \psi_R^\dagger(
i\sigma\!\cdot\!
	\partial)\psi_R\\
	\hbox{Dirac:}&\qquad& \bar\psi( i\dsl
	 - m)\psi\\
	\hbox{Majorana:}&\qquad& \frac12\bar\psi( i\dsl
	 - m)\psi
\eeqa
where $\sigma^\mu = (1,\sigma^i)$, and 
$\bar\sigma^\mu = (1,-\sigma^i)$.  Weyl spinors are two component.
Recall why the equation of motion $i\bar\sigma\!\cdot\! \partial\,\psi =0$
implies a left-handed fermion: we can write it as
\beqa
	i(\bar\sigma^\mu\partial_\mu) e^{-ip_\nu x^\nu}
	 u(p) &=& (\bar\sigma^\mu p_\mu)
	e^{-ip_\nu x^\nu}u = (E - \sigma^i p_i) 
	e^{-ip_\nu x^\nu}u\nonumber\\
	&=& (E + \sigma^i p^i) 
	e^{-ip_\nu x^\nu}u = (E + \vec\sigma\ncdot \vec p) 
	e^{-ip_\nu x^\nu}u
\eeqa
Therefore, since $|\vec p|=E$ for a massless particle,
\beq
	\vec\sigma\cdot\hat p\, u(p) = - {E\over |\vec p|}\, u(p)
	= - u(p),
\eeq
hence $u(p)$ is an eigenstate of helicity with eigenvalue $-1$.
In the massless limit, helicity and chirality coincide, and the
spin being anti-aligned with the momentum is what we mean by
left-handed.

Dirac spinors are four-component, made from two Weyl spinors
by
\beq
	\psi = \left(\begin{array}{cc} \psi_R\\ \psi_L
\end{array}\right)
\eeq
using the chiral representation of the gamma matrices, where
\beq
	\gamma^0 = \left(\begin{array}{ll} 0 & 1 \\ 1 & 0 \end{array}
	\right);\qquad \gamma^i = \left(\begin{array}{ll} 0 & -\sigma^i
	 \\ \sigma^i & \phm 0 \end{array}
	\right);\qquad \gamma_5 = \left(\begin{array}{ll} 1& \phm 0 \\ 0 &
-1 \end{array}
	\right);
\eeq
they obey the anticommutation relations
\beq
	\{\gamma^\mu,\, \gamma^\nu\} = 2 g^{\mu\nu}
\eeq
A Majorana spinor is made from a single Weyl spinor using the
fact that the charge conjugate of $\psi_L$, $\sigma_2\psi_L^*$,
behaves in the same way as $\psi_R$ under Lorentz transformations:
\beq
	\psi = \left(\begin{array}{cc}  \sigma_2\psi_L^*
\\\psi_L\end{array}\right)
\eeq
If one writes out the Majorana Lagrangian in terms of
two-component spinors, it becomes clear why the factor of 
$\frac12$ is needed:
\beq
	{\cal L}_{\rm Dirac} = 	 \psi_L^\dagger(
i\bar\sigma\!\cdot\!
	\partial)\psi_L + \psi_R^\dagger(
i\sigma\!\cdot\!
	\partial)\psi_R  -m\psi^\dagger_L\psi_R 
-m\psi^\dagger_R\psi_L 
\eeq
whereas for the Majorana case, the term coming from $\psi_R^\dagger(
i\sigma\!\cdot\!\partial)\psi_R$ would be a double-counting of
$\psi_L^\dagger(i\bar\sigma\!\cdot\!\partial)\psi_L$ without
the extra factor of $\frac12$.

You have already learned that the Feynman propagator for a Dirac
fermion is
\beqa
	S_F(x',x)_{\beta\alpha} &=& -i \langle 0|T(\psi_\beta(x')
	\bar\psi_\alpha(x))|0\rangle \\ &=& 
	\intdp {e^{-ip\cdot(x'-x)}\over \psl - m + i\varepsilon}
	= 
	\intdp e^{-ip\cdot(x'-x)}{ \psl+m\over p^2 - m^2 + i\varepsilon}
\eeqa
For Weyl fermions, we can write similar expresssions, for example
\beqa
	S_F(x',x)_{\beta\alpha} &=& -i \langle 0|T(\psi_\beta(x')
	\bar\psi_\alpha(x))|0\rangle \\ &=& 
	\intdp {e^{-ip\cdot(x'-x)}\over \sigma\cdot p + i\varepsilon}
	= 
	\intdp e^{-ip\cdot(x'-x)}{ \bar\sigma\cdot p\over p^2  + i\varepsilon}
\eeqa
Here I used the identity $\sigma\!\cdot\! p \,\bar\sigma\!\cdot p\! = p^2$.
However, it is usually more convenient to work in terms of
four-component fields using the Majorana or Dirac 
form of the Lagrangian.  
In this way we can always use the familiar Dirac algebra
instead of having to deal with the $2\times 2$ sigma matrices.
For massless neutrinos, there is no harm in pretending there is
an additional right-handed neutrino in the theory, as long as
it has no interactions with real particles.  This is why we can
use the Dirac propagator in calculations where the neutrino mass
can be ignored.  If we are considering a Weyl field which has
a mass, one must use the Majorana form of the Lagrangian.  In this
case one has to be more careful when computing Feynman diagrams
because there are more ways of Wick-contracting the fermion fields,
as I shall explain below.

Note: there is another kind of mass term one can write down
for Dirac or Majorana fermions:
\beq
	m\bar\psi \psi + im_5 \bar\psi\gamma_5\psi
\eeq
To solve $(\psl - m-im_5\gamma_5)u(p)=0$, consider the form
$u(p) = (\psl + m-im_5\gamma_5)v(p)$ where $v(p)$ is an
arbitrary spinor.  The equation of motion gives
\beq
	p^2 - m^2 - m_5^2 = 0
\eeq
so $m^2 + m_5^2$ is the complete mass squared.  In fact it is
possible to tranform the usual Dirac mass term into this form
by doing a field redefinition
\beq
	\psi \to e^{i\theta\gamma_5}\psi = 
	(\cos\theta + i\sin\theta\gamma_5)\psi
\eeq
Then 
\beq
	m \to e^{i\theta\gamma_5}m e^{i\theta\gamma_5}
	= (\cos2\theta + i\sin2\theta\gamma_5)m
\eeq
showing that one can freely transform between the usual form of 
the mass and the parity-violating form.  Of course, this will
also change the form of the interactions of $\psi$, so it is
usually most convenient to keep the mass in the usual form.

In analogy with bosons, the Feynman rule for a fermion line is
given by
\beq
	{\epsfxsize=0.5in\epsfbox{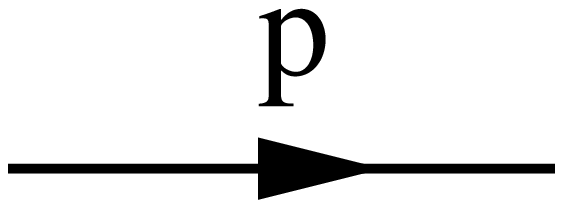}} = {i\over \psl - m
	+i\varepsilon}
\eeq
Unlike bosons, fermions have a directionality attached to their
lines because of the fact that the propagator is not symmetric under
$p\to -p$.  So we need to pay attention to the direction of flow
of momentum.

In addition to the rule for internal lines, we need to include
spinors for the external lines.  From the canonical formalism,
it is straightforward to derive:
\beqa
	u(p,s) &\quad& \hbox{for each particle entering graph} \nonumber\\
	v(p,s) &\quad& \hbox{for each antiparticle entering graph}\nonumber\\
	\bar u(p,s) &\quad& \hbox{for each particle leaving graph} \nonumber\\
	\bar v(p,s) &\quad& \hbox{for each antiparticle leaving
graph}
\eeqa
As an example of using these rules, let's compute the cross
section for charged-current scattering of a neutrino on a neutron
into electron plus
proton.  At low energies, the interaction Lagrangian has
the form
\beq
\label{4F}
	\sqrt{2}G_F \, [ \bar p (c_v-c_a\gamma_5)\gamma_\mu n]\,
	    [ \bar e \gamma^\mu P_L \nu ]+ {\rm h.c.}
\eeq
where the Fermi constant $G_F$ arises from the internal propagator
of the $W$ boson, which we have integrated out to obtain the 
nonrenormalizable effective interaction (\ref{4F}).  The 
nucleon vector and axial vector couplings are given by $c_v=1$
(which can be derived on theoretical grounds) and $c_a = -1.25$,
which is experimentally measured.

The amplitude for $\nu n\to e p$ is 
\beq
	{\cal M} = \sqrt{2}G_F \, [ \bar u_p (c_v-c_a\gamma_5)
	\gamma_\mu u_n]\,
	    [ \bar u_e \gamma^\mu P_L u_\nu ]
\eeq
and the amplitude squared is
\beq
	|{\cal M}|^2 = {2}G_F^2 \, [ \bar u_p (c_v-c_a\gamma_5)
	\gamma_\mu u_n]\,
	    [ \bar u_e \gamma^\alpha P_L u_\nu ]\,
[ \bar u_n (c_v-c_a\gamma_5)
	\gamma_\alpha u_p]\,
	    [ \bar u_\nu \gamma^\mu P_L u_e ]\,
\eeq
To get this, one uses the result that
\beq
	(\bar u_1 X u_2)^* =  u_2^\dagger X^\dagger
\gamma_0^\dagger u_1
\eeq
and
\beq
	{\gamma^\mu}^\dagger{\gamma^0}^\dagger
	= \gamma^0 \gamma^\mu;\qquad 
	{\gamma_5}^\dagger{\gamma^0}^\dagger
	= -\gamma^0 \gamma_5
\eeq
Next we sum over final polarizations and average over initial ones
using the formulas
\beq
	\sum_s u(p,s) \bar u(p,s) = \psl + m;\qquad
	\sum_s v(p,s) \bar v(p,s) = \psl - m
\eeq
to obtain
\beq
	\langle |{\cal M}|^2\rangle = \frac14 \sum |{\cal M}|^2 = 
	\frac12 G_F^2\,\tr\left[
	P_L(\psl_e+m_e)\gamma^\alpha P_L \psl_\nu \gamma^\mu
	\right]
	\cdot \tr\left[(\psl_n+m_n)X\gamma_\alpha (\psl_p+m_p)X
	\gamma_\mu\right]
\eeq
where now $X = c_v - c_a\gamma_5$.  Thus, ignoring the electron
mass,
\beqa
	\langle |{\cal M}|^2\rangle &=&  \frac12 G_F^2\,\tr\left[
	\psl_e\gamma^\alpha \psl_\nu\gamma^\mu \frac12(1+\gamma_5)
	\right]\ncdot \left( \tr\left[\psl_n X \gamma_\alpha \psl_p
	X\gamma_\mu\right] + m_nm_p \tr\left[X\gamma_\alpha
	X\gamma_\mu\right] \right)\nonumber\\
	&=&\frac44 G_F^2\left( p_e^\mu p_\nu^\alpha +  
	p_e^\alpha p_\nu^\mu - p_e\ncdot p_\nu
	\eta^{\mu\alpha} + i\epsilon^{\rho\mu\sigma\alpha}
	p_{e,\rho} p_{\nu,\sigma} \right) \nonumber\\
	&&\ncdot \left( \tr\left[\gamma_\alpha \psl_n \gamma_\mu
	\psl_p
	(c_v-c_a\gamma_5)^2 \right] + m_n m_ptr\left[\gamma_\mu
	(c_v+c_a\gamma_5)(c_v-c_a\gamma_5)\gamma_\alpha\right]
	\right)\nonumber\\
	&=& \frac{16}{4}G_F^2\left( p_e^\mu p_\nu^\alpha 
	+ p_e^\alpha p_\nu^\mu- p_e\ncdot p_\nu
	\eta^{\mu\alpha} + i\epsilon^{\rho\mu\sigma\alpha}
	p_{e\rho} p_{\nu\sigma} \right) 
	\left( [p_{n,\alpha}
	p_{p,\mu} + p_{p,\alpha}p_{n,\mu} - p_n\ncdot p_p
	\eta_{\alpha\mu}]
	(c_v^2 + c_a^2) \right.\nonumber\\ 
	&&- \left. 2ic_v c_a \epsilon_{\alpha\rho'\mu\sigma'}
	p_{n}^{\rho'}p_{p}^{\sigma'} +
	m_n m_p (c_v^2-c_a^2) \eta_{\alpha\mu}
	\right)\nonumber\\
	&=& 4 G_F^2\left( 2(p_e\ncdot p_p \, p_\nu\ncdot p_n + 
	p_e\ncdot p_n\, p_\nu\ncdot p_p)(c_v^2+c_a^2)
	+  m_n m_p (c_v^2-c_a^2)
	( 2p_e\ncdot p_\nu - 4p_e\ncdot p_\nu )\right.\nonumber\\
	&& +\left. 2c_v c_a (2\delta_{\rho'\sigma'}^{\rho\sigma})
	p_{e,\rho}p_{\nu,\sigma}p_{n}^{\rho'}p_{p}^{\sigma'}\right)
	\nonumber\\
	&=& 8 G_F^2\left( p_e\ncdot p_p \, p_\nu\ncdot p_n
	(c_v-c_a)^2 + p_e\ncdot p_n\, p_\nu\ncdot p_p (c_v+c_a)^2
	- m_n m_p p_e\ncdot p_\nu (c_v^2 - c_a^2)  \right)
\eeqa
Note that I used the identity $\epsilon_{\alpha\rho'\mu\sigma'}
\epsilon^{\alpha\rho\mu\sigma} = 
2\delta_{\rho'\sigma'}^{\rho\sigma}$
To simplify this, let's go to the low energy limit where the
proton and neutron are at rest, and the neutrino energy is much
smaller than the nucleon mass.  Then
\beqa
	\langle |{\cal M}|^2\rangle &\cong& 8 G_F^2 m_n m_p
	\left( 2 E_e E_\nu (c_v^2+c_a^2) - p_e\ncdot p_\nu
	 (c_v^2-c_a^2) \right) \nonumber\\
	&\cong& \frac{8}{16} G_F^2 m_n m_p ( 91 E_e E_\nu - 9 \vec p_e\ncdot
	\vec p_\nu)
\eeqa
Recall that the differential cross section is given by	
\beq
	{d\sigma\over dt} = {1\over 64\pi s}\, {1\over
|p_{1,\rm cm}|^2} \langle |{\cal M}|^2\rangle
\eeq
where $p_{1,\rm cm} = p_{1,\rm lab}m_n/\sqrt{s} = E_\nu
m_n/\sqrt{s}$.  Ignoring the small $\vec p_e\ncdot \vec p_\nu$ term, we
can then estimate
\beqa
	{d\sigma\over dt} &\cong& {1\over 64\pi}\, {1\over
E_\nu^2 m_n^2} \langle |{\cal M}|^2\rangle\nonumber\\
	&\cong& \frac{91}{2}G_F^2 m_n m_p E_\nu E_e\nonumber\\
	&\cong& {46\over 64\pi} G_F^2
\eeqa
We used that fact that, by energy conservation, since 
$E_\nu\ll m_n$, $E_e\cong E_\nu$.  The maximum momentum 
transfer is $\sqrt{-t} = 2 E_\nu$, when the electron comes out in the
opposite direction to the neutrino, thus the total scattering
cross section is
\beq
	\sigma = \int_{-(2E_nu)^2}^0 {d\sigma\over dt}
	\cong  {46\over 16\pi} G_F^2 E_\nu^2
\eeq
The fact that the $\vec p_e\ncdot \vec p_\nu$ term was small meant
that the differential scattering cross section is not very
directional.  If we were to obtain a result which was closer
to the form $p_e\ncdot p_\nu$, the electron would be
preferentially emitted in the same direction as the neutrino
originally had.  This is the principle which allows neutrino
detectors like SuperKamiokande to have some sensitivity to the
direction of the neutrinos.  This makes me suspect there is an
error in the above calculation, since there is no value of 
$c_a/c_v$ (which can vary for different nuclei) which makes the
$p_e\ncdot p_\nu$ term more important that the isotropic $E_e
E_\nu$ term.

Now that we have reviewed fermions at tree level, we can look at
how they behave in loop diagrams.  For this purpose we would like
to introduce interactions.  The 4-fermion Lagrangian we just
studied is not renormalizable, since by power counting diagrams
like those in fig.\ 32 are divergent.  On the other hand, 
interactions with scalars can be renormalizable:
\beq
	\mu^\epsilon \phi\,\bar\psi(g_1 + ig_2\gamma_5)\psi
\eeq
These are called Yukawa couplings; $g_1$ is called a scalar coupling
and $g_2$ is a pseudoscalar coupling, meaning that under
parity, if $\phi$ was even and $g_2=0$, the interaction would
be invariant, whereas if $\phi$ was odd and $g_1=0$, it would
also be invariant.  In the first case, $\phi$ is a scalar under
parity, and in the second it is a pseudoscalar, hence the 
terminology.  In either case, the dimension of the operator in
$d=4-2\epsilon$ dimensions requires the factor of $\mu^\epsilon$.
For simplicity, let's consider the case of the
scalar coupling, and denote $g_1=g$.  Then there are a few divergent 
diagrams we can consider, shown in fig.\ 33.

\medskip
\centerline{\epsfxsize=2.5in\epsfbox{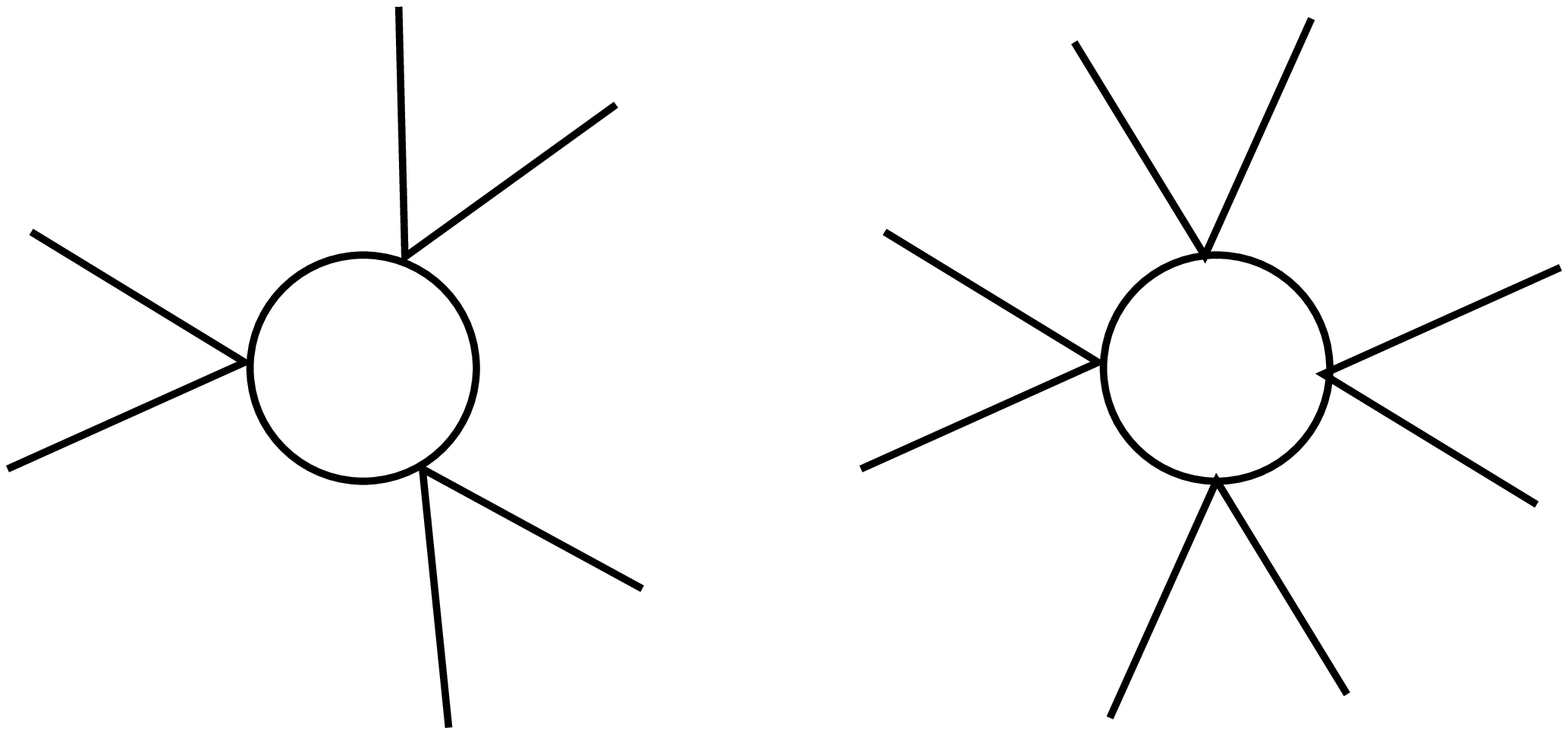}}
\centerline{\small
Figure 32.  Divergent loop diagrams in 4-fermion theory.}
\medskip

\medskip
\centerline{\epsfxsize=4.5in\epsfbox{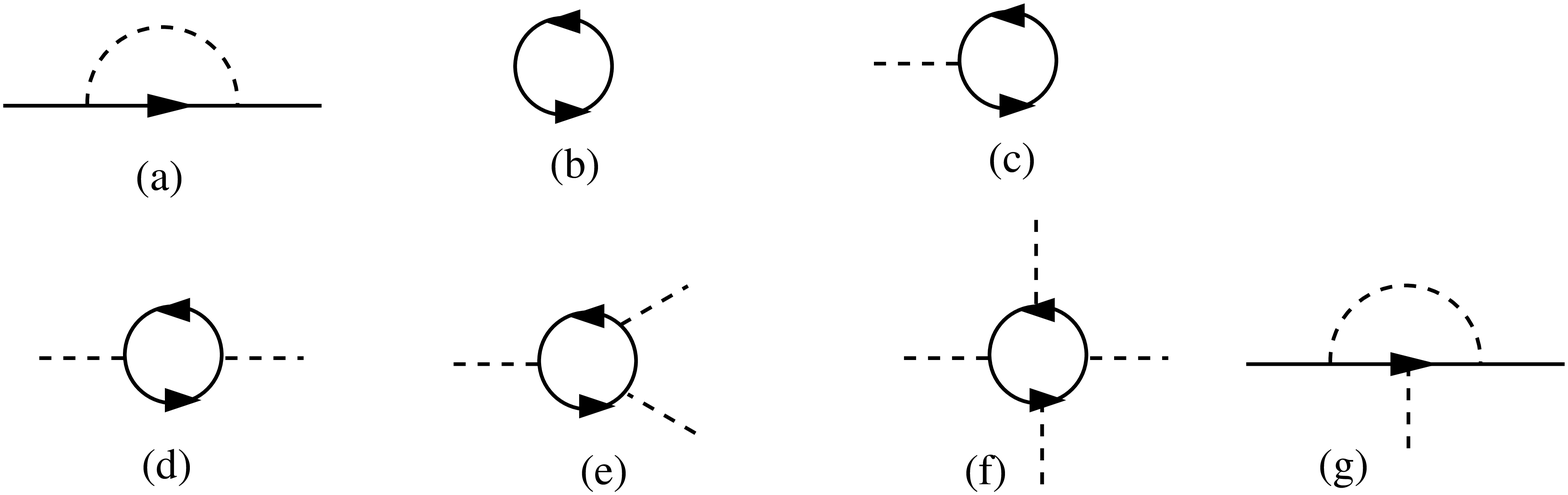}}
{\small
Figure 33.  Divergent loop diagrams with Yukawa couplings
to a scalar field (dashed lines).}
\medskip

From now on I will leave detailed computations to be presented
in the lecture and just give an outline here of what we will 
cover.  We start with fig.\ 31(d), the correction to the scalar
inverse propagator.  The important points are the following. (1)
the fermion loop gets an extra minus sign relative to boson
loops; this occurs because one has to anticommute one pair of
fermion fields in order for them to be in the right order to
give the propagator when the Wick contractions are done.
(2) There is a trace over the Dirac indices associated with
each fermion loop.  When we use dimensional regularization, we
have to know how Dirac matrices generalize to higher dimensions.
Here are some relevant formulas:
\beqa
	\gamma_\mu \gamma^\mu &=& d \\		
	\tr\, 1 &=& d \\
	\tr\, \gamma_\mu \gamma_\alpha &=& d\eta_{\mu\alpha}\\
	\tr\, \gamma_\mu \gamma_\alpha \gamma_\nu\gamma_\beta
	&=& d(\eta_{\mu\alpha}\eta_{\nu\beta} + 
	\eta_{\mu\beta}\eta_{\nu\alpha}-
	\eta_{\mu\nu}\eta_{\alpha\beta} \\
	\gamma_\mu \gamma_\alpha \gamma^\mu &=&
	- \gamma_\mu \gamma^\mu \gamma_\alpha
	+ 2\gamma_\mu \delta_{\alpha}^\mu = (2-d)\gamma_\alpha
	\\
	\gamma_\mu \,\asl\,\bsl\, \gamma^\mu &=& 4a\ncdot b +
	(d-4)\asl\,\bsl\\
	\gamma_\mu \,\asl\,\bsl\, \csl\,\gamma^\mu &=& -2
	\csl\,\bsl\,\asl + (4-d) \asl\,\bsl\, \csl 
\eeqa	
As for formulas involving $\gamma_5$, there is no natural way 
to continue them to ogher dimensions, so we are forced to
use the 4D formulas.  This turns to out be consistent for
many purposes, but we will see later that it leads to some problems
in rigorously defining a theory with chiral fermions with gauge
interactions.	

From the result of evaluating the diagram of fig.\ 33(d), we
can verify that the divergent part of the correction to the
mass has the opposite sign to what we found in $\phi^4$ theory:
\beq
	\delta m^2_{\rm scalar} = {g^2\over 16\pi^2}\,
	{12\over \epsilon}\, m^2_{\rm fermion}
\eeq
In fact the divergences of the fermion and scalar loops cancel
each other if the Yukawa and quartic couplings are related by
\beq
	g^2 = \frac{\lambda}{4!}
\eeq
and if the masses of the fermion and scalar are equal to each
other.
If there was no symmetry in the theory to enforce this kind of
relationship, these would be a kind of fine-tuning.  However,
supersymmetry (which we will hopefully have time to study toward
the end of this course) enforces precisely this kind of 
relationship.  In SUSY theories, this kind of cancellation 
also leaves the corrections to the scalar quartic coupling finite;
notice that fig.\ 33(f) goes like $g^4/\epsilon$, whereas fig.\
9 (the scalar loop contribution to the quartic coupling) goes
like $\lambda^2/\epsilon$, so these do have the correct form
to cancel each other.  

On the other hand if we look at wave function renormalization,
diagram 33(f) gives 
\beqa
-\!\!-\!\!\!\bigcirc\!\!\!-\!\!- &\sim&	i{ g^2\over 16\pi^2} \, 
{2\over\epsilon}\, p^2
\\
\left[ \hbox{compare to}\ -\!\!-\!\!-\!\!- \phantom{p^2\!\!\!\!}\right. 
&=& \left. i(p^2 - m^2) \quad \right]
\eeqa
whereas we know that the setting sun diagram of scalar field
theory goes like $\lambda^2/\epsilon$.  Thus these two diagrams
do not have the right form to cancel each other.  This shows that
while SUSY can prevent divergences in masses and couplings, it
does not do so for wave function renormalization.

Another consequence of SUSY is that the vacuum diagrams of the
form $\bigcirc$ must cancel.  We would like to see how the
diagram 33(b) gets the opposite sign to that of the scalar loop.
This comes about because of the properties of the fermionic
form of the path integral:
\beq
	\bigcirc = \Gamma = {1\over i}\ln\left(
	\int {\cal D}\psi {\cal D}\bar\psi e^{iS}\right)
\eeq
Consider the Grassmann integral
\beq
	\int \prod_i d\eta^*_i \, d \eta_i\,
	e^{i \eta^*_i M_{ij}\eta_j}= \det M
\eeq
This is in contrast to the bosonic path integral which gives
$(\det M)^{-1/2}$.  The important thing here is the sign of
the exponent.  When we take the log, this explains why the
fermionic loop has the opposite sign to the bosonic one.  As for
the factor of $1/2$ which we had for bosons, this was due to the
fact that a real scalar field is just one degree of freedom,
whereas a Weyl fermion has two spin components, and therefore
is two degrees of freedom.  If we consider the path integral
over a complex scalar field $\phi = (\phi_1 + i\phi_2)/\sqrt{2}$,
we would obtain $(\det M)^{-1}$.   We now do a detailed
computation of this determinant for fermions to show that indeed
the cancellation is exact between a Weyl fermion and a complex
scalar if they have the same mass, or between a Dirac fermion and
two such complex scalars.  In SUSY theories, scalars are always
complex, and their corresponding fermionic partners
are always Weyl.

Just like for bosons, we can construct the Coleman-Weinberg
potential for a scalar field due to the fermion loop, and the
result is the fermion determinant with the mass of the fermion
evaluated using the background scalar field:
\beq
	V_{\rm CW} = -\frac{N}{64\pi^2} (m+g\phi)^4
	\ln\left( {(m+g\phi)^2\over\mu^2} \right)
\eeq
where $N=4$ for a Dirac and 2 for a Majorana or Weyl fermion.  This
follows from the result that 
\beq
\Gamma = -i\ln \det(i\dsl - m) = -i {N\over 2} \intdx \intdp \ln(p^2-m^2)
\eeq
Notice that when we Wick-rotate, $\Gamma$ gets a positive contribution
from fermions.  This means the vacuum energy density gets a negative
contribution, since action $\sim (-$ potential).

We return to the diagrams of fig.\ 33.  The tadpole is similar
to that which arises in $\phi^3$ theory and must be renormalized
similarly, by introducing a counterterm which is linear in $\phi$.
Let us focus on diagram (a) instead; we call this a contribution
to the self-energy of the fermion.  To help get the sign of
this contribution right, I will use the background field method.
This gives
\beqa	
	i\Gamma_{\rm 1 loop} &=& \frac1{2!}(-ig \mu^{\epsilon})^2
	\left\langle \intdx\left(\bar\psi_c \phi_q \psi_q + 
	\bar\psi_q\phi_q\psi_c\right) \right. 
	 \left. \intdy\left(\bar\psi_c \phi_q \psi_q + 
	\bar\psi_q\phi_q\psi_c\right) \right\rangle \nonumber\\
	&=& (-ig \mu^{\epsilon})^2 \intdx\intdy 
	\bar\psi_c(x) \,\langle \phi_q(x) \phi_q(y) \rangle
	\,\langle \psi_q(x) \bar\psi_q(y) \rangle \,\psi_c(y) 
	\nonumber\\
	&=& -g^2 \mu^{2\epsilon} \intdp \bar\psi_c(p) \,
	\intdq {i\over \psl+\qsl-m_f}\, {i\over q^2-m_s^2} \psi_c(p)
\eeqa
since there are two ways of contracting the quantum fermion fields.
This expression is to be compared to the tree-level one,
\beq
	\Gamma_{\rm tree} = \intdp \bar\psi_c(p)(\psl-m_f)\psi_c(p)
\eeq
We can in this way identify the corrections to the mass and wave function
renormalization.  The correction to the fermion inverse propagator is
often called the {\it self-energy}, $\Sigma(p)$.  The full inverse
propagator is then $\psl-m + \Sigma$.  Comparing the above
equations, we see that
\beq
	\Sigma = i g^2\mu^{2\epsilon} \intdq {\psl+\qsl+m_f\over
	(p+q)^2-m_f^2}\, {1\over q^2 - m_s^2}
\eeq
Evaluating it in the usual way, we find the result
\beq
	\Sigma = g^2\mu^{2\epsilon} \int_0^1 dx\, {\Gamma(\epsilon)\over
	(4\pi)^{2-\epsilon} \Gamma(2)}\, {\psl(1-x) + m_f\over
(M^2)^\epsilon}
\eeq
where $M^2 = x m_f^2 + (1-x) m_s^2 - p^2 x(1-x)$.  The most divergent
part is
\beq
{\epsfxsize=0.75in\epsfbox{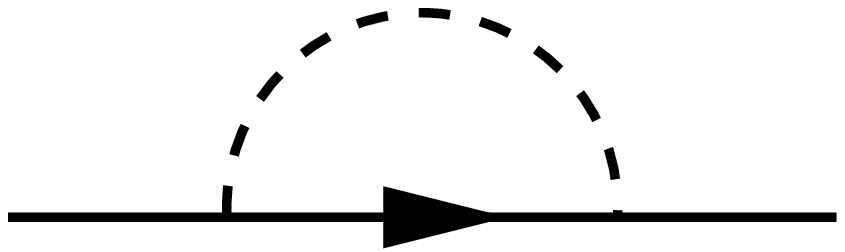}} = 
\Sigma \sim {g^2\over 16\pi^2\epsilon}	\left(\frac12\psl +
m_f\right)
\eeq

Finally, let's evaluate the remaining divergent diagram, fig.\ 33 (g),
which is called the {\it vertex correction}.
To simplify things, let's evaluate it at vanishing external momenta, since
this is all we need for getting the divergent contribution to the Yukawa
coupling.  Evaluating the diagram directly, we find the combinatoric factor
is 1, hence
\beqa
	i\Gamma_{\rm vertex} &=&(-ig\mu^\epsilon)^3 \intdq
	{i\over \qsl-m_f}{i\over \qsl-m_f}{i\over q^2-m_s^2}
	\nonumber\\ 
	&=&(g\mu^\epsilon)^3 \intdq {(\qsl+m_f)^2\over
	(q^2-m_f^2)^2(q^2-m_s^2) } \nonumber\\
	&=& (g\mu^\epsilon)^3 \intdq \left({1\over
	(q^2-m_f^2)(q^2-m_s^2) } +\hbox{convergent piece}\right)\nonumber\\	
	&=& i(g\mu^\epsilon)^3\int_0^1 dx\, {\Gamma(\epsilon)\over
(4\pi)^{2-\epsilon}(M^2)^\epsilon}\qquad 
	\left(M^2 \equiv xm_f^2 +(1-x)m_s^2\right)\nonumber\\
	&\sim& i{(g\mu^\epsilon)^3\over 16\pi^2\epsilon}
\eeqa
which should be compared to the tree level result $i\Gamma = -ig\mu^\epsilon$.

Now we can put together the above results to do the renormalization of
this theory and find the beta function for the Yukawa coupling.  Let us
write the tree plus 1-loop, unrenormalized effective action in the form
\beqa
	\Gamma &=& \intdp\left[ \frac12\phi_{-p}\left(p^2-m_s^2 + {a\over\epsilon}p^2
	-{b\over\epsilon}m_s^2\right)\phi_p +	
	\bar\psi_p\left(\psl-m_f + {c\over\epsilon}\psl - {d\over\epsilon}m_f
	\right)\psi_p \right]\nonumber\\
	&-& g\left(1+{e\over\epsilon}\right)\intdx\, \bar\psi\phi\psi
\eeqa
Comparing with our previous results, and defining $\hat g = g\mu^\epsilon/4\pi$,
we can identify
\beqa
	a = 2 \hat g^2; \quad b = 12{m^2_f\over m^2_s}\hat g^2;\quad
	c = \frac12 \hat g^2;\quad d = -\hat g^2;\quad
	e = -\hat g^2.
\eeqa
The first step is to renormalize the fields to absorb the $a$ and $c$
terms.  We define
\beq
	\psi = \sqrt{Z_\psi}\psi_r \cong \left(1-{c\over 2\epsilon}
\right)\psi_r ;\quad \phi = \sqrt{Z_\phi}\phi_r \cong 
\left(1-{a\over 2\epsilon}
\right)\phi_r
\eeq
The effective action becomes
\beqa
\label{Gammaren}
	\Gamma &=& \intdp\left[ \frac12\phi_{r,-p}\left(p^2-m_s^2\left(1+ 
	{b-a\over\epsilon}\right)\right)\phi_{r,p} +	
	\bar\psi_{r,p}\left(\psl-m_f\left(1+ {d-c\over\epsilon}\right)
	\right)\psi_{r,p} \right]\nonumber\\
	&-& g\mu^\epsilon\left(1+{e-c-a/2\over\epsilon}\right)\intdx\, 
\bar\psi_r\phi_r\psi_r
\eeqa
It is now clear how to define the bare parameters in order to absorb the divergences:
\beq
	m^2_{s,0} = \left(1- 
	{b-a\over\epsilon}\right) m^2_{s,r};\quad m_{f,0} 
	= \left(1- {d-c\over\epsilon}\right) m_{f,r};\quad
	g_0 = \left(1-{e-c-a/2\over\epsilon}\right)g_r
\eeq
Using the determined values for $e$, $c$ and $a$, and remembering that $g_0$ has 
dimensions
given by $\mu^\epsilon$, the full bare coupling is (see the parallel discussion 
starting at (\ref{lambdabare}) for the quartic scalar coupling)
\beq
	g_{\rm bare} = \mu^\epsilon\left( 1 + {5\hat g_r^2 \over 2\epsilon}\right)g_r
	\equiv \mu^\epsilon\left( g + {a_1(g)\over \epsilon}\right)
\eeq
Now we can compute the $\beta$ function for the Yukawa coupling, 
\beqa
	0&=& \mu{\partial\over\partial\mu}g_{\rm bare} = \epsilon 
	\left( g + {a_1(g)\over \epsilon}\right) + 
	\beta(g)\left(1 + {a_1'\over \epsilon}\right)\nonumber\\
	\longrightarrow \quad\beta(g) &\cong& \epsilon 
	\left( g + {a_1(g)\over \epsilon}\right)
	\left(1 - {a_1'\over \epsilon}\right)\nonumber\\
	&=& \epsilon g + {5g^3\over 16\pi^2}
\eeqa
Thus, like the quartic coupling, the Yukawa coupling is not asymptotically free;
it flows to larger values in the ultraviolet.

I must confess that this way of computing the beta function from dimensional
regularization still seems rather abstract and unintuitive to me.  Here is another
way which is more down-to-earth.  Go back to the 1-loop effective action
(\ref{Gammaren}) which has been wave-function-renormalized, and imagine what the
3-point function would look like if we had computed it using a momentum-space
cutoff, and at nonvanishing external momentum.  Remembering that $1/\epsilon$
corresponds to $\ln\Lambda^2$, and taking the external momentum to be $p$,
the effective coupling would have the
form
\beq
	g_{\rm eff} = g\left(1 - {5g^2\over 32\pi^2}\ln(\Lambda^2/p^2)\right)
\eeq
When we renormalize, this will become
\beq
	g_{\rm eff} = g(\mu)\left(1 - {5g^2(\mu)\over 32\pi^2}\ln(\mu^2/p^2)\right)
\eeq
Notice that the energy-dependent coupling increases with $p^2$ as expected 
from the sign of the beta function.  We can now compute the beta function by
demanding that the physical coupling $g_{\rm eff}$ be independent of $\mu$:
\beq
	\mu{\partial\over\partial\mu}
g_{\rm eff} = 0 = \beta\left(1 +O(g^2)\right)
	- {5g^3\over 16\pi^2}
\eeq
which gives the same result as DR to the order we are computing.  You can see the
parallels between this way of doing it and the DR method.  I tend to think that the
DR procedure is a set of manipulations that are mathematically equivalent
to using a momentum space cutoff, but the latter gives a more concrete
way of understanding what we are actually doing.

To conclude this section about fermions, I want to discuss the differences between
Majorana and Dirac particles.  The previous computation for the beta function
was for a Dirac fermion.   Implicit in this discussion was the fact that only
Green's functions like $\langle \psi \bar\psi \rangle$ were nonzero, whereas
$\langle \psi \psi \rangle$ was assumed to be zero.  Here I would like to show that
this is true for Dirac fermions, but not Majorana ones.  To put the discussion in 
context, let's suppose that we are doing the Wick contractions for a combination
of fields like
\beq
\label{bilinears}
	\langle (\bar\psi X \psi)_x (\bar\psi Y \psi)_y \rangle_{\rm con}
\eeq
where $X$ and $Y$ are matrices such as $1$, $\gamma_\mu$, $\gamma_5$, 
$\gamma_\mu\gamma_5$, {\it etc.},
and we are only interested in the connected contributions.  As we have noted 
before, one possible contraction gives
\beq
	-\tr(X\langle \psi_x \bar\psi_y\rangle Y \langle \psi_y \bar\psi_x \rangle)
\eeq
To write the other possible contraction involving $\langle\psi_x\psi_y\rangle
\langle\bar\psi_x\bar\psi_y\rangle$ in a convenient way, let's first rewrite
(\ref{bilinears}) by making use of the fact that
\beqa
\label{CC}
	(\bar\psi X  \psi) &=& - \psi^T X^T\gamma_0\psi^* \nonumber\\
	&=&  - \psi^T ({-\gamma^2}^\dagger\gamma_0)({-\gamma^2}^\dagger\gamma_0)^{-1} 
	X^T\gamma_0(-\gamma^2)^{-1}(-\gamma^2)\psi^* \nonumber\\
	&\equiv& \bar\psi^c X^c \psi^c
\eeqa
where we have defined the charge-conjugated spinor
\beq
	\psi^c = -\gamma^2 \psi^*
\eeq
and
\beq
\label{CCX}
	X^c = -\gamma_0(-{\gamma^2}^\dagger)^{-1} X^T \gamma_0 \gamma^2
	= \gamma_0 \gamma^2 X^T \gamma_0 \gamma^2
\eeq
Notice that $\gamma^2$ is antihermitian in the representation we are using.
The leading minus signs in (\ref{CC}) and (\ref{CCX}) come from anticommuting the two fields.
When we rewrite $(\bar\psi X  \psi)_x$ in this way, we see that the other possible
contraction of the fields takes the form
\beq
-\tr(X^c\langle \psi^c_x \bar\psi_y\rangle Y \langle \psi_y \bar\psi^c_x \rangle)
\eeq
The fact that $\langle \psi^c_x \bar\psi_y\rangle$ vanishes for a Dirac fermion
follows simply from the fact that fermion number is a conserved quantity in the
Dirac Lagrangian.  The latter is symmetric under the transformations
\beq
	\psi\to e^{i\theta} \psi,\qquad \bar\psi \to e^{-i\theta}\bar\psi
\eeq
However, $\psi^c \to e^{-i\theta}\psi^c$ under this symmetry, so that 
$\langle \psi^c_x \bar\psi_y\rangle \to e^{-2i\theta}
 \langle \psi^c_x \bar\psi_y\rangle$.  A nonvanishing VEV for $\psi^c_x \bar\psi_y$
would therefore break the symmetry, in contradiction to the fact that the
symmetry does exist in the underlying Lagrangian.  You can see this explicitly by
considering the path integral, and changing variables from $\psi$ to $\psi'
 = e^{i\theta} \psi$:
\beqa
	\langle \psi^c_x \bar\psi_y\rangle &=&
	\int {\cal D}\psi {\cal D}\bar\psi e^{iS(\bar\psi,\psi)} \psi^c_x \bar\psi_y
	\nonumber\\
	&=& 	\int {\cal D}{\psi'} {\cal D}{\bar\psi'}
e^{iS({\bar\psi'},\psi')} {\psi'}^c_x {\bar\psi'}_y\qquad
	\left(\psi' = e^{i\theta} \psi\right)\nonumber\\
	&=& \int {\cal D}\psi {\cal D}\bar\psi e^{iS(\bar\psi,\psi)}
	{\psi'}^c_x {\bar\psi'}_y \nonumber\\
	&=& e^{2i\theta}\langle \psi^c_x \bar\psi_y\rangle 
\eeqa
Both the action and the path integral measure are invariant under this phase
transformation.  For the action this was obvious; for the measure, we note that
the determinant of the Jacobian matrix is unity:
\beq
	\left|{\partial(\psi', \bar\psi')\over\partial(\psi,\bar\psi)}\right|
	= \left|\begin{array}{cc} e^{i\theta} & 0 \\ 0 & e^{-i\theta}
	\end{array} \right| = 1
\eeq
Since $\langle \psi^c_x \bar\psi_y\rangle = 
e^{2i\theta}\langle \psi^c_x \bar\psi_y\rangle$, it must vanish.

This argument does not work for massive Majorana fermions however, since their
action is not invariant under the phase transformation.  More precisely, it does not
make sense to perform such a phase transformation on a Majorana spinor since it
is defined as
\beq
	\psi = \left(\begin{array}{cc}  \sigma_2\psi_L^*
\\\psi_L\end{array}\right)
\eeq
which obeys the identity
\beq
	\psi^c = \psi
\eeq
Clearly, a spinor which satisfied this property (being self-conjugate) would no longer
do so if we multiplied it by a phase.  Therefore we cannot define a symmetry 
operation which forbids Green's functions of the form 
$\langle \psi^c_x \bar\psi_y\rangle$ for Majorana fermions.  In fact, since
$\psi^c = \psi$, we see that $\langle \psi^c_x \bar\psi_y\rangle = \langle 
\psi_x \bar\psi_y\rangle$.

Now we can see how to compute Feynman diagrams with Majorana fermions.  We have to
compute all possible contractions of the fermi fields, by charge-conjugating one
of the bilinears in order to put terms like $\psi\psi^T$ into the more familiar
form $\psi\bar\psi^c = \psi\bar\psi$.  One has to know how the matrix $X$ transforms
under charge conjugation.  For example, it is easy to show that
\beq
	1^c = 1;\quad \gamma_5^c = +\gamma_5;
	\quad (\gamma^\mu)^c = -\gamma^\mu;\quad 
	(\gamma_5\gamma^\mu)^c = -\gamma_5\gamma^\mu
\eeq
The fact that $\bar\psi\gamma_\mu\psi$ changes sign under charge
conjugation is intuitively clear: for electrons this is the electric current,
which should change sign when transforming from electrons to positrons.

We can now redo the previous calculations (fig.\ 33) in a theory where the fermion
is Majorana.  In this case, it is convenient to normalize the interaction term with
an additional factor of $\frac12$, as we do for the kinetic term, since this 
preserves the relationship between the Yukawa coupling and the mass if we give a 
classical expectation value of the scalar field $\phi$:
\beq
\label{LMaj}
	{\cal L} = \frac12 \bar\psi\left(i\dsl-m - g\phi\right)\psi
\eeq
When we redo the diagrams of fig.\ 33, we find that the results are identical 
to those of the Dirac fermion for all diagrams where the fermions are in the
external states, because the new factor of $\frac12$ in the coupling in (\ref{LMaj})
is exactly compensated by the number of new contractions that can be done with
the fermion fields.  However in the vacuum polarization diagram 33(d), there are
only two ways of contracting the fermions, while there are two factors of $\frac12$.
This makes sense: there are only half as many fermions circulating in the loop for
the Majorana case, relative to the Dirac one.  Therefore the only change in computing
the beta function is that the factor $a$ in (\ref{Gammaren}) is divided by 2:
\beqa
	\beta &=& {4g^3\over 16\pi^2}\qquad\hbox{(Majorana case)}
\eeqa

\section{The Axial Anomaly}
\label{sec:anomaly}

Sometimes a symmetry which seems to be present in a theory at the classical
level can be spoiled by quantum (loop) effects.  We have already seen such an
example without knowing it.  A theory with no massive parameters such as
\beq
	S = \intdx\left(\frac12(\partial\phi)^2 + \bar\psi\dsl\psi - 
	\frac{\lambda}{4!} - g\bar\psi\phi\psi\right)
\eeq
has a dilatation or scale transformation symmetry
\beq
	\phi\to e^a\phi,\quad \psi\to e^{3a/2}\psi,\quad x^\mu \to e^{-a} x^\mu
\eeq
since there is no mass scale in the Lagrangian.  However, the Coleman-Weinberg
potential breaks this symmetry due to factors like $\ln(\lambda\phi^2/\mu^2)$
and $\ln(g^2\phi^2/\mu^2)$ which are introduced through renormalization.  It is
impossible to avoid introducing the massive scale $\mu$.  A symmetry broken
by quantum effects is said to be anomalous.

When we introduced dimensional regularization, one of our motivations was
that it was important to try to preserve the symmetries which are present in
the theory.  In the previous section, we noted an example of one of these 
symmetries, the transformation $\psi\to e^{i\theta}\psi$ for Dirac fermions.
Recall that for any symmetry of a Lagrangian, there is a conserved (N\"other)
current which is constructed by promoting $\theta$ from a global constant to
a locally varying function $\theta(x)$.  The N\"other procedure is to find the
variation of the action under and infinitesimal transformation of this kind. 
If it is a symmetry, then the result can always be written in the form
\beq
	\delta{\cal L} = \intdx\, \partial_\mu\theta J^\mu = 
	- \intdx\,\theta \partial_\mu J^\mu = 0
\eeq
where $J^\mu$ is the conserved current corresponding to the symmetry.  In the
case of $\psi\to e^{i\theta}\psi$, the current is
\beq
\label{veccur}
	J^\mu = \bar\psi\gamma^\mu\psi
\eeq
If $\psi$ was the electron field, then this would be simply the electric current
(modulo the charge $e$).  For a generic fermion field, it is the particle current,
and its conservation implies that the net number of fermions of that kind does
not change in any interaction.   The current $J^\mu$ in (\ref{veccur}) 
is called the {\it vector current} for the field $\psi$, since it transforms like
a vector under the Lorentz group.  This symmetry is sometimes denoted by
U(1)$_V$.

We noted that for Majorana fermions, one cannot define a symmetry transformation
$\psi\to e^{i\theta}\psi$; however, it does make sense to consider the axial
or {\it chiral}
transformation $\psi\to e^{i\theta\gamma_5}\psi$, which is denoted by
U(1)$_A$.  In this case we have
\beq
	\left(\begin{array}{cc}  \sigma_2\psi_L^*
\\\psi_L\end{array}\right) \to \left(\begin{array}{cc}  e^{i\theta}\sigma_2\psi_L^*
\\	e^{-i\theta}\psi_L\end{array}\right)
\eeq	
which is self-consistent.  Similarly, we could do the same transformation for
Dirac fermions,
\beq
	\psi = \left(\begin{array}{cc} \psi_R\\ \psi_L
\end{array}\right) \to \psi = \left(\begin{array}{cc} e^{i\theta}\psi_R\\ e^{-i\theta}\psi_L
\end{array}\right)
\eeq
In either case, the corresponding current is given by the {\it axial vector} current,
\beq
	J_5^\mu = \bar\psi\gamma_5\gamma^\mu\psi
\eeq
The charge associated with this current, $Q_5 = \int d^{\,3}x\,J_5^0$, counts the
number of right-handed fermions {\it minus} the number of left-handed ones.  But
in a massive theory, there is not a conserved quantity because the mass term
$m(\psi_L^\dagger\psi_R + \psi_R^\dagger\psi_L)$ can convert left-handed states into
right-handed ones and {\it vice versa}.  For Majorana particles, the corresponding
statement is that the mass term can flip a left-handed particle into a right-handed
antiparticle, so the notion of a conserved particle number is spoiled.  Only if
the theory is massless is the axial rotation a symmetry.  We can see this using the
classical equation of motion:
\beq
	i\partial_\mu J_5^\mu = \bar\psi \gamma_5(i{\overleftarrow\dsl} +
	i{\overrightarrow\dsl})\psi = \bar\psi\gamma_5(m + m)\psi
\eeq
Only if $m=0$ is the axial symmetry a good one for the classical Lagrangian.

Now for the surprising subject of this section: even though the vector and
axial-vector currents may be conserved in the classical theory, it may
be impossible to preserve both of these symmetries once loop effects are
introduced.  In the present case, there is no regularization scheme that respects
both symmetries.  Dimensional regularization does not work because there is no
generalization of $\gamma_5$ to higher dimensions.  Pauli-Villars fails because
the masses of the regulator fields do not respect chiral symmetry.  We shall now
show that in the case of a single Dirac fermion, it is possible to preserve one
or the other of the two symmetries, but not both \cite{Preskill}. 

The quantity where the problem shows up is in the Green's functions of composite
operators which are products of three currents.  Define
\beq
	-i\Gamma_{\mu\nu\lambda} = \langle 0|T^*(J_{5\mu}(x_3) J_\nu(x_1)
	J_\lambda(x_2)|0\rangle
\eeq
On the basis of the classical symmetries, we would expect $\Gamma_{\mu\nu\lambda}$
to satisfy the three {\it Ward identities} for conservation of the currents:
\beq
\label{Gdiv}
	\partial_{x_3^\mu} \Gamma^{\mu\nu\lambda} = 
	\partial_{x_1^\nu} \Gamma^{\mu\nu\lambda} =
	\partial_{x_2^\lambda} \Gamma^{\mu\nu\lambda} = 0
\eeq
We will show that, due to the contribution of the {\it triangle diagrams} in 
fig.\ 34, it is impossible to satisfy all of the above relations.  

\centerline{\epsfxsize=5.5in\epsfbox{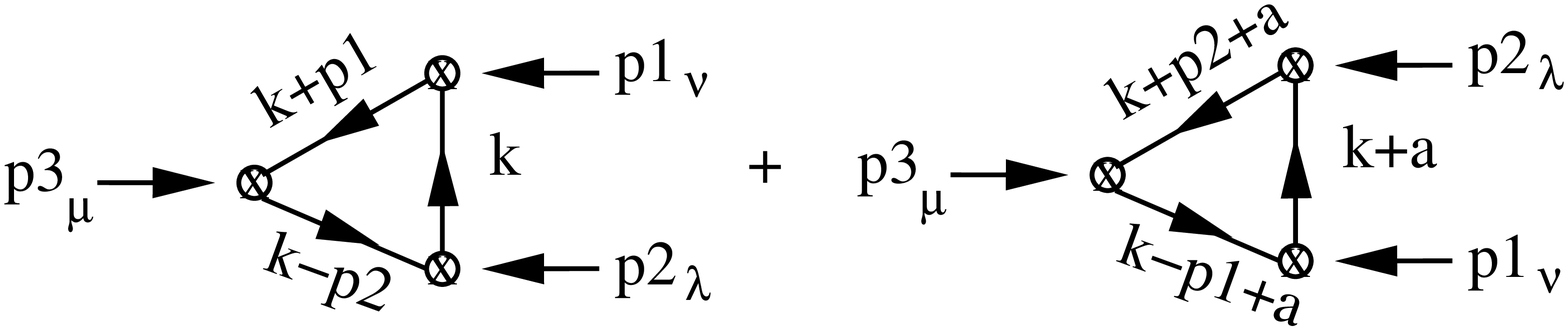}}
\centerline{\small
Figure 34. Triangle diagrams for the axial anomaly.}
\smallskip

\noindent There are
two diagrams because of the two possible ways of contracting the fermion fields.
In the figure we have gone to momentum space, and in the second diagram we have 
added an arbitrary shift $a$ to the momentum in the internal line, which correponds
to a change of integration variables $k\to k+a$ in the integral for the second
diagram.  If everything was well-behaved, the choice of $a$ should have no effect
on the final answer.  However, we will find that because the diagrams are UV
divergent, the choice of $a$ {\it does} matter.  The result of evaluating the
diagrams (details to be presented in the lecture) is that the divergences in 
(\ref{Gdiv}) are given by
\beqa
\label{axial}
	p_3^\mu \Gamma_{\mu\nu\lambda} &=& -{1\over 8\pi^2}\epsilon_{\nu\lambda\alpha
	\beta} a^\alpha p_3^\beta \\
\label{vector1}
	p_1^\nu \Gamma_{\mu\nu\lambda} &=& -{1\over 8\pi^2}\epsilon_{\lambda\mu\alpha
	\beta} (a+2p_2)^\alpha p_1^\beta \\
\label{vector2}
	p_2^\lambda \Gamma_{\mu\nu\lambda} &=& -{1\over 8\pi^2}\epsilon_{\mu\nu\alpha
	\beta} (a-2p_1)^\alpha p_2^\beta
\eeqa
To obtain this result, we can use a fact from vector calculus which saves us from having
to evaluate the loop integrals in great detail.  Namely, if we cut off the loop 
integral by a momentum space cutoff in Euclidean space, then it is an integral over
a finite volume, the 4-sphere.  We can write
\beqa
I(a)-I(0) &=& 	\int_{S_4} \intdk\, (f(k+a) - f(k))\nonumber\\ &=& \int_{S_4}\intdk \left(a^\mu\partial_\mu
	f(k) + O(a^2)\right) \nonumber\\
	&=& \vec a\!\cdot\! \int_{S_3} d\vec S\, f(k) +\dots  = 
	a^\mu \int_{S_3} d\Omega_k k^2 k_\mu f |_{k^2 \to \infty}
\eeqa
For an integral which is linearly divergent in $k$ before subtracting the two terms,
we can ignore the terms of higher order in $a$, because these involve more derivatives
with respect to $k$, and their correponding surface terms vanish at infinity.  For
example, if $f = k_\nu/(k^2+b^2)^2$, then  
\beqa
	I(a) - I(0) &=& a^\mu \int d\Omega_k k^2 {k_\mu k_\nu \over (k^2)^2}\nonumber\\
	&=& \frac14 a^\mu \delta_{\mu\nu} 
	\int d\Omega_k k^2 {k^2 \over (k^2)^2} = {\pi^2\over 2}a^\nu
\eeqa

The dependence on the arbitrary vector $a$ can be considered to be part of our
choice of regularization.  For a general choice of $a$, neither the vector nor
axial vector currents are conserved (although there might some linear combination
of them which {\it is} conserved).  We can make one choice which preserves the
vector symmetry however, which is the relevant one if we are talking about electrons
since we certainly want electric charge to be conserved:
\beq
	a = 2(p_1-p_2)
\eeq
With this choice, both of the vector current divergences (\ref{vector1}) and
(\ref{vector2}) vanish, and the axial anomaly takes on a definite value,
\beqa
\label{axial2}
	-i p_3^\mu \Gamma_{\mu\nu\lambda} &=& {1\over 2\pi^2}\epsilon_{\nu\lambda\alpha
	\beta} p_1^\alpha p_2^\beta
\eeqa

The axial anomaly is also known as the Adler-Bardeen-Jackiw (ABJ) anomaly after the
people who first published a paper about it.  However it was first discovered by
Jack Steinberger, then a graduate student in theoretical physics.  He was so dismayed
and perplexed by this result that he decided to quit theoretical physics and became
instead a very successful experimentalist.

In the above derivation we constructed a rather unphysical quantity (correlator
of three currents) to uncover the existence of the anomaly.  There is a more
physical context in which it arises; suppose we have a background U(1) gauge field
$A_\mu(x)$ to which the fermions are coupled via
\beq
	{\cal L} = \bar\psi(i\dsl + e\Asl)\psi
\eeq
With the gauge interaction, it is no longer necessary to consider the 3-current
Green's function; we can compute $\langle \partial_\mu J_5^\mu\rangle$ directly.
The other two vertices in the triangle diagram arise from perturbing to second
order in the interaction.  We get an extra factor of $\frac1{2!}$ from second
order perturbation theory, so (\ref{axial2}) becomes
\beqa
\label{axial3}
	p_{3\mu} J_{5}^{\mu}(p_3) &=& {e^2\over 4\pi^2}\epsilon_{\nu\lambda\alpha
	\beta} p_1^\alpha A^\nu(p_1) p_2^\beta A^\lambda(p_2) \nonumber\\
	 &=& {e^2\over 16\pi^2}\epsilon_{\nu\lambda\alpha
	\beta} \left(p_1^{\alpha\phantom{\beta}\!\!\!\!} A^\nu(p_1)-p_1^\nu A^\alpha(p_1)\right)\left(
	 p_2^\beta A^\lambda(p_2)- p_2^\lambda A^\beta(p_2)\right) 
\eeqa
Transforming back to position space, this gives 
\beq
\label{anomaly}
	\partial_\mu J_5^\mu = {e^2\over 16\pi^2}\epsilon_{\nu\lambda\alpha
	\beta}F^{\nu\lambda}F^{\alpha\beta} \equiv {e^2\over 16\pi^2}
	F^{\nu\lambda}\widetilde F_{\nu\lambda}
\eeq
The quantity $F^{\nu\lambda}\widetilde F_{\nu\lambda}$ can be shown to be the same
as $2\vec E\ncdot\vec B$.  If we integrate (\ref{anomaly}) over space, 
\beq
\label{dQ5dt}
	\int d^{\,3}x\, \partial_\mu J_5^\mu = \dot Q_5 = {e^2\over 8\pi^2}
	\int d^{\,3}x\, \vec E\ncdot\vec B
\eeq
This has a remarkable interpretation: it says that if electrons were massless, then
in the presence of a region of space with parallel electric and magnetic fields,
axial charge would be created at a constant rate given by (\ref{dQ5dt}).  In other
words, there would be a constant rate of production of pairs of
left-handed particles and right-handed antiparticles, or {\it vice versa}.  These
particles would start out with zero energy (remember that they are massless), but
they would be accelerated by the electric field thereafter.

The way this happens can be visualized in terms of the migration of single particle
states as shown in fig.\ 35.  If we put the system in a box, only discrete momentum
values will be allowed.  Due to the electric fields, the momentum of a given state
will increase linearly with time.  Let's focus on states with spin $S_z = +1/2$.  We
would call them left-handed if they are moving with negative group velocity ($p_z<0$)
and right-handed if $p_z>0$---except for the negative energy particles in the filled
Dirac sea we interpret things oppositely since these denote the absence of the
corresponding antiparticle.  When there is a mass gap $m$ and the electric field is
smaller than $m^2$, no transitions between negative and positive energy states take
place, so the chirality violation is not observable (it just corresponds to a constant
production of new LH states in the filled Dirac sea).  If $m=0$, the gap disappears,
and then the filled states of the Dirac sea jump to occupy positive energy states
regardless of how small the electric field is.  The production of chiral pairs
(particles with $S_z = +1/2$ and antiparticles with $S_z = -1/2$, moving in 
opposite directions) occurs and chirality violation is observable.  But one might
wonder what the role of the magnetic field is in all of this---these pictures make no
reference to $B$ after all.  Indeed if $B=0$, we could draw the same diagrams for
the $S_z = -1/2$ states, and these would produce an equal and opposite amount of chiral
charge to the $S_z = +1/2$ states.  However, in the presence of the magnetic field,
we have new contributions to the energy of a particle; there are Landau levels due
to the electrons circulating around the $B$ field lines, and there is also a magnetic
moment coupling to the spin, so that the dispersion relation takes the form
\beq
	E^2 = p_z^2 + (2n+1)eB + - 2eB S_z + m^2
\eeq
Therefore only the states with Landau level $n=0$ and $S_z = +1/2$ have zero energy;
the rest get an effective contribution to their mass.  This heuristic discussion makes
the physical origin of the anomaly more plausible.

\centerline{\epsfxsize=5.in\epsfbox{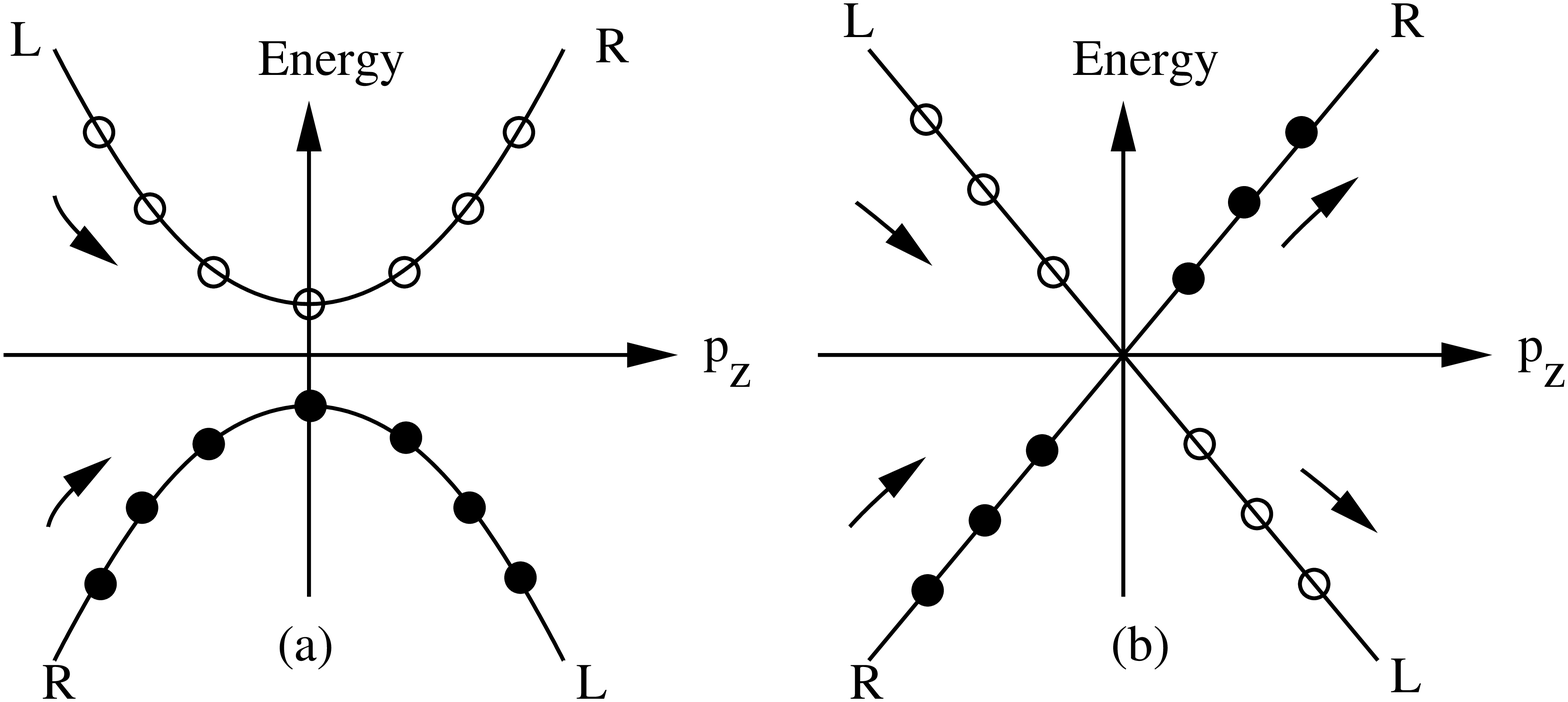}}
{\small
Figure 35. Creation of axial charge in parallel electric and magnetic fields
for (a) massive and (b) massless electrons.}
\smallskip

There are numerous ways of deriving the axial anomaly, which give equivalent
results.  One way is to regularize the axial current by point-splitting:
\beq
	J^\mu_5 = \lim_{\epsilon_\mu\to 0} \bar\psi(x+\epsilon/2) \gamma^\mu
	\gamma_5 e^{i\int_{x-\epsilon/2}^{x-\epsilon/2} dx^\nu A_\nu(x)}
	\psi(x+\epsilon/2)
\eeq
The exponential factor is required if one wants to keep the current invariant
under the electromagnetic gauge symmetry:
\beq
	\psi\to e^{ie\Omega(x)}\psi;\qquad A_\mu \to A_\mu + \partial_\mu\Omega
\eeq
The small separation vector $\epsilon_\mu$ cuts off the UV divergence of the 
momentum integral in the triangle anomaly, and gives the result we have computed
when one averages over the directions of $\epsilon_\mu$ to restore Lorentz invariance.

Yet another derivation (due to Fujikawa \cite{Fujikawa}) derives the
axial anomaly from the path integral, by showing that the path integral measure
is not invariant under the U(1)$_A$ symmetry when one carefully defines it 
in such a way as to preserve the U(1)$_V$ symmetry.   Consider a theory with
several fields which we will collectively denote by $\phi$, having action
$S[\phi]$ and some symmetry transformation $\phi\to \phi + \epsilon A\phi$, where $A$
is a matrix.  In evaluating the path integral, we can change variables from $\phi$
to $\phi' = \phi + \epsilon A\phi$,
\beqa
	\int{\cal D}\phi e^{iS[\phi]} &=& 
	\int{\cal D}\phi' e^{iS[\phi']}\nonumber\\ &=& 
	\int{\cal D}\phi' e^{iS[\phi] + i\intdx \partial_\mu\epsilon J^\mu} \nonumber\\
	&=& \int{\cal D}\phi' e^{iS[\phi]}\left(1 + i\intdx \partial_\mu\epsilon J^\mu
	\right)
\eeqa
If it is true that ${\cal D}\phi' = {\cal D}\phi$, then we see after integrating
by parts that
\beqa
\int{\cal D}\phi e^{iS[\phi]}\intdx \epsilon \partial_\mu J^\mu
 &=& 0\nonumber\\
\longrightarrow \left\langle\partial_\mu J^\mu \right\rangle = 0
\eeqa
The latter equality follows from the fact that $\epsilon(x)$ can be any function,
including a delta function with support at an arbitrary position.

The above argument can be applied to the U(1)$_A$ symmetry, and it would imply that
there is no anomaly if it was true.  The one questionable assumption we made in
the derivation was that ${\cal D}\phi' = {\cal D}\phi$.  Evidently this is not
true in the case of the U(1)$_A$ transformation.  If we define the transformation
as 
\beq
	\psi \to e^{i\epsilon\gamma_5/2}\psi,\quad
	\bar\psi \to \bar\psi e^{i\epsilon\gamma_5/2}
\eeq
then the N\"other current has the conventional normalization, and the Jacobian is
\beq
	\left|{\partial(\psi', \bar\psi')\over\partial(\psi,\bar\psi)}\right|
	= \left|\begin{array}{cc} e^{i\epsilon\gamma_5/2} & 0 \\ 0 & 
	e^{i\epsilon\gamma_5/2}
	\end{array} \right| = e^{i\epsilon\gamma_5}
\eeq	
for the field at each point $x$.  We must multiply these together for all $x$, so
the Jacobian is $\det e^{i\epsilon\gamma_5}$ in field space.  Let us compute
\beq
	\ln\det e^{i\epsilon\gamma_5} = \tr\ln(e^{i\epsilon\gamma_5})
	=  \tr\ln(1+i\epsilon\gamma_5) = i\,\tr\,\epsilon\,\gamma_5 
\eeq

Since $\tr\gamma_5$ vanishes naively, we might think that this can lead nowhere. 
However since this is a trace in field space, we have to find all the eigenvalues of
the Dirac operator and sum the expectation value of $\gamma_5$ over them.  If there
are more left-handed than right-handed solutions, we can get a nonvanishing result.
We know that the anomaly shows up most readily if we quantize in a background
U(1)$_V$ gauge field, so let us do this.  Then the explicit meaning of $\tr\,\gamma_5$
is
\beq
\label{Fuj}
{\tr}\,\epsilon\,\gamma_5 = \sum_n \intdx \,\epsilon\, \bar\psi_n \gamma_5 \psi_n
\eeq
where the functions are eigenfunctions of the Dirac operator in a background 
gauge field:
\beq
	(i\dsl +e \Asl)\psi_n = \lambda_n \psi_n
\eeq
Notice that we can rewrite the fields as $\psi(x) = \sum_n a_n \psi_n(x)$, 
$\bar\psi(x) = \sum_n b_n \bar\psi_n(x)$, where $a_n$ and $b_n$ are Grassmann 
numbers, and the path integral becomes
\beq
	\int{\cal D}\psi {\cal D}\bar\psi e^{iS} = 
	\prod\int da_n db_n\, \exp{\sum_n b_n \lambda_n a_n} = \prod \lambda_n
	 = \det(i\dsl +e \Asl)
\eeq 
Now the problem is that the sum in (\ref{Fuj}) is not well-defined without being
regulated in the ultraviolet, and we must do so in a way that respects the U(1)$_V$
symmetry.  

\def\sfrac#1#2{{\textstyle{#1\over #2}}}

\section{Abelian Gauge Theories: QED}
\label{sec:qed}

We have focused on some toy model theories so far, to keep things relatively
simple.  Although $\phi^4$ and Yukawa interactions are part of the standard model,
they appear in a somewhat more complicated way than we have treated them.  Gauge
interactions are the major ingredient we have not yet discussed.  In this section we
will treat the Abelian case, which includes QED.  

You were already exposed to the quantization of gauge fields in your previous course.
Let's review.  The gauge field can be derived from promoting the global U(1)$_V$
symmetry of a Dirac fermion $\psi\to e^{ie\theta}\psi$ to a local symmetry, 
$\psi\to e^{ie\theta(x)}\psi$.   We have now factored out the charge of the electron
$e$ in our definition of the symmetry operation, since a particle with a different
charge, like the up quark, would transform as $u\to e^{i\frac23 e\theta(x)}u$.
The new term which arises due to the spatial variation
of $\theta$ is
\beq
	\delta{\cal L} = \bar\psi(-\dsl\theta )\psi
\eeq
This can be canceled by adding the gauge interaction
\beq
	{\cal L}_{\rm int} = \bar\psi(e\Asl )\psi
\eeq
if we allow the gauge field to transform as
\beq
	A_\mu \to A+\partial_\mu\theta
\eeq
which of course is just a gauge transformation, which leaves the
field strength $F_{\mu\nu}$ unchanged.  The action for the gauge field
is
\beq
	{\cal L}_{\rm gauge} = -\frac14 F_{\mu\nu}F^{\mu\nu} = 
	\frac12\left( \vec E^2 - \vec B^2\right)
\eeq
To get more insight, let's make a choice of gauge, say Coulomb gauge, where
$\vec\nabla\cdot\vec A =0$.  The Lagrangian density can be written as
 \beqa
	\frac12\left( (\,\,\dot{\!\!\vec A} - \vec\nabla A_0)^2 -
	 (\vec\nabla\times\vec A)^2
	\right)  &=& \frac12\left(\,\, \dot{\!\!\vec A}^{\,2} + (\vec\nabla A_0)^2 -
	(\vec\nabla\times\vec A)^2 - 2 \vec\nabla A_0\cdot\dot{\!\!\vec A} 
	\right)
\nonumber\\
\hbox{integrate by parts} &\to& 
\frac12\left(-\vec A \partial_0^2 \vec A + 
	\vec A\cdot \vec\nabla^{\,2}\vec A  
- A_0 \vec\nabla^{\,2} A_0 	\right)
\eeqa
where we used the fact that $\vec\nabla\cdot\vec A =0$ to get rid of the term
$2 \vec\nabla A_0\cdot\dot{\!\!\vec A}$.  Roughly speaking, $\vec E$ looks like
the time derivative and $\vec B$ looks like the spatial derivative of the three
components of $\vec A$ (which therefore resemble three scalar fields), but we also
get the zeroth component, which is strange because it has no kinetic term and its
gradient term has the wrong sign.  If we
couple the gauge field to an external electromagnetic current $J_\mu$, the path
integral over the gauge fields is
\beq
\label{pathint}
	\int{\cal D}A_\mu e^{iS[A] - i\intdx A^\mu J_\mu}
\eeq
We can rewrite the exponent as 
\beqa
	 \intdx \left( \frac12\left[-\vec A \square \vec A + A_0\vec\nabla^2 A_0
	\right] - A_0 J_0 + \vec A\cdot \vec J\right)
	\phantom{AAAAAAAAAAAAAAA} \nonumber\\
=	\frac12\left[ 
 -\left(\vec A - \sfrac{1}{\square}\vec J\right) \square 
	\left(\vec A - \sfrac{1}{\square}\vec J\right)
+ \left(A_0 + \sfrac{1}{{\vec\nabla}^2}J_0\right) {\vec\nabla}^2 (A_0 + 
\sfrac{1}{{\vec\nabla}^2}J_0)
	\right]\nonumber\\
	+ \frac12\left[ \vec J \cdot \sfrac{1}{\square} \vec J - 
	J_0 \sfrac{1}{{\vec\nabla}^2} J_0 \right]
\eeqa
By shifting the integration variable in the  functional integral, we see the
result is
\beqa
\label{gaugePI}
	\int{\cal D}A_\mu e^{iS[A] - i\intdx A^\mu J_\mu} &=& 
	e^{\frac{i}{2}\intdx \left[ \vec J \cdot \sfrac{1}{\square} \vec J - 
	J_0 \sfrac{1}{{\vec\nabla}^2} J_0 \right]} \int{\cal D}A_\mu e^{iS[A]}
\nonumber\\	
\label{selfint}
&=& e^{-\frac{i}{2}\intdx A_\mu[J] J^\mu}
\eeqa 
where $A_\mu[J]$ is the gauge field which is induced by the external current
(see eqs.\ (4.16) and (4.17) of \cite{harry}) in Coulomb gauge:
\beq
\label{coulpot}
	A_i[J] =  \sfrac{1}{\square}J_i;\qquad 
	A_0[J] =  \sfrac{1}{{\vec\nabla}^2}J_0
\eeq
The exponent in (\ref{selfint}) is the action due to the electromagnetic interaction of the
current with itself, so the result makes sense.

However, I was sloppy in my treatment of the path integral in this derivation.
How is the constraint that $\vec\nabla\cdot\vec A =0$ implemented in the path
integral?  You have already learned about this---it is the Faddeev-Popov 
prodedure.  The problem with trying to do the path integral (\ref{pathint})
without gauge fixing is that it does not exist due to an infinite factor
associated with the gauge symmetry.  Namely, at each point in spacetime,
one can change $A_\mu$ by the gauge transformation $A_\mu\to A_\mu + \partial_\mu
\Lambda(x)$ without affecting the action.  The problem can be seen when we try to
compute the gauge field propagator.  In momentum space the action looks like
\beq
   \frac12\intdp A_\mu(-p)\left( -p^2 \eta^{\mu\nu} + p^\mu p^\nu \right) A_\nu(p)
\eeq
but the matrix $p^2 \eta^{\mu\nu} - p^\mu p^\nu$ is singular; it has an 
eigenvector $p_\mu$ with zero eigenvalue.  Therefore we cannot invert it
to find the propagator.
Of course, this eigenvector is
proportional to the Fourier transform of a pure gauge configuration,
\beqa
	A_\mu = 0 &\to& A_\mu = \partial_\mu \Lambda\nonumber\\
	A_\mu(p) = 0 &\to& A_\mu = p_\mu \Lambda(p)
\eeqa
which must have vanishing action since $E=B=0$ in this case.   We would somehow
like to avoid integrating over such gauge transformations when doing the path
integral.

If we fix the gauge, this puts one
constraint on the four fields $A_\mu$, which means that only three of them 
are independent.  The path integral measure therefore can be expressed in the
form
\beq
	{\cal D} A^\mu = {\cal D A_{\rm g.f.}^\mu} {\cal D}\Lambda
\eeq
where $A_{\rm g.f.}$ is constrained to lie in a lower-dimensional surface
in the space of fields (the gauge fixing surface in fig.\ 36) and the
gauge transformations generated by $\Lambda$ move each gauge-fixed field
configuration (points in the subspace) off this surface, along curves
called gauge orbits, along which the field strength does not change.  
Since the action does not depend on $\Lambda$, $\int {\cal D}\Lambda$
is the infinite factor which we want to remove from the path integral.
The FP procedure extracts this infinity in a well-defined way, by inserting 
a cleverly designed factor of unity into the path integral before gauge 
fixing:
\beqa
	1 &=& \int{\cal D}\Lambda\, \delta[f(A_\mu + \partial_\mu\Lambda)] 
	\det\left({\delta f(A + \partial\Lambda)_x\over \delta \Lambda_y}\right)
	\nonumber\\
	&\equiv&  \int{\cal D}\Lambda\, \delta[f(A_\mu + \partial_\mu\Lambda)] 
	\Delta_{\rm FP}[A_\mu; \Lambda]
\eeqa
When we insert this into the path integral, we can use the fact that $S[A + 
\partial\Lambda] = S[A]$ to make the shift $A\to A - \partial\Lambda$ everywhere
(since this is just shifting the integration variable), so that the path integral
becomes
\beq
	 \int{\cal D}\Lambda\,{\cal D}A_\mu e^{iS[A]}\delta[f(A_\mu)] 
	\Delta_{\rm FP}[A_\mu; 0]
\eeq
The infinite factor of the gauge group volume now comes out cleanly since
nothing in the remaining functional integral depends upon it.

\medskip
\centerline{\epsfxsize=2.5in\epsfbox{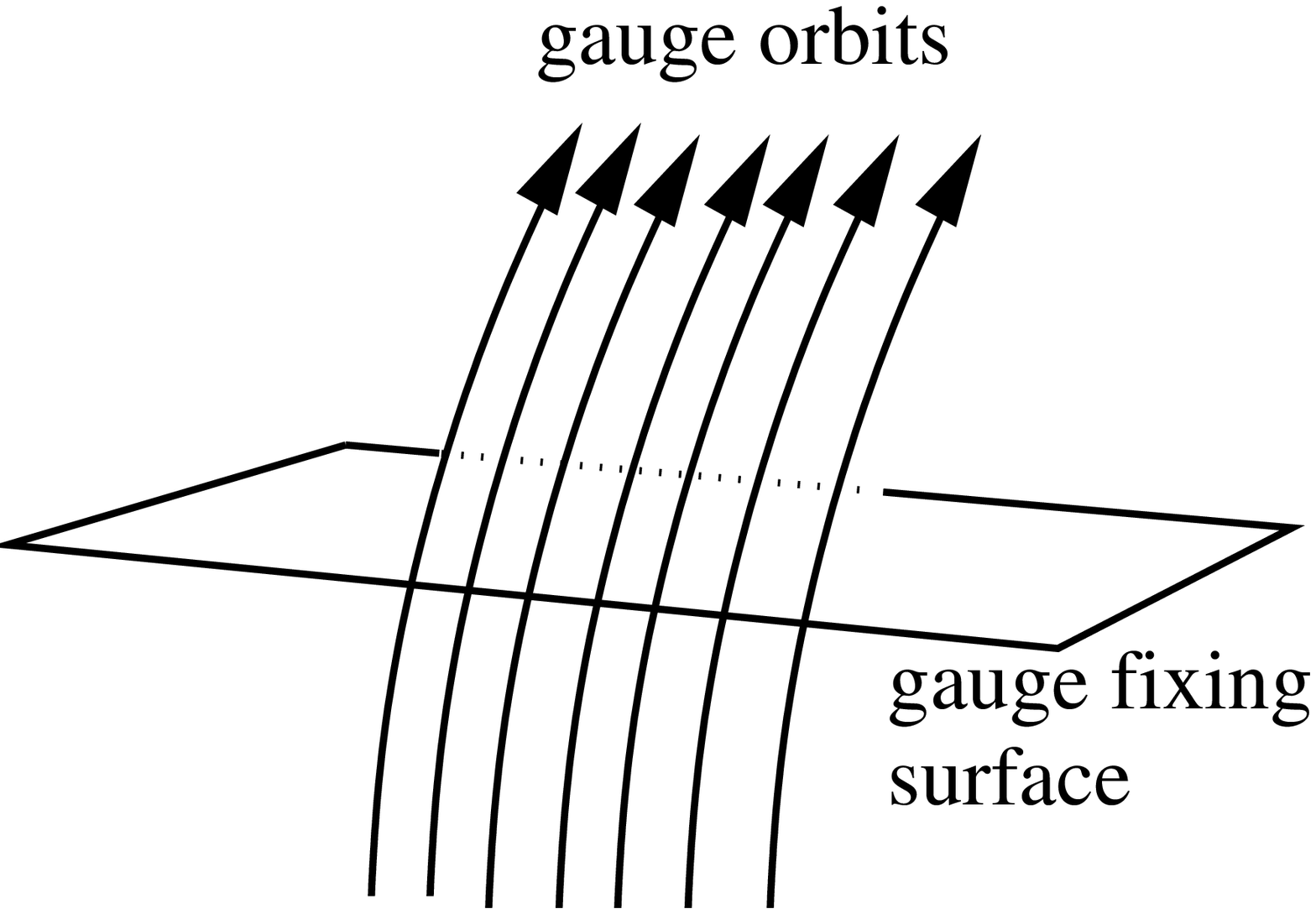}}
\centerline{\small Fig.\ 36. 
Gauge orbits piercing a gauge fixing surface.}
\medskip

The problem is now how to evaluate the remaining integral.  Let's consider
the Coulomb gauge choice again.  In that case 
\beq
	f(A) = \vec\nabla\cdot\vec A
\eeq
and the FP Jacobian matrix is 
\beq
	{\delta f(\partial\Lambda)_x\over \delta \Lambda_y}
	=
{\delta \left(\partial_\mu \eta^{\mu i}\partial_i\Lambda(x)\right)\over
\delta\Lambda(y)}
	= \vec\nabla^2 \delta(x-y)
\eeq
The determinant can be expressed using
\beq
	\ln\det \vec\nabla^2 \delta(x-y) = {\tr}\ln\vec\nabla^2 \delta(x-y)
	= \intdx\intdp \ln(\vec p^{\,2})
\eeq
similarly to our discussion of the vacuum diagrams $\bigcirc$.  However, although
this determinant is UV divergent, it does not depend on the gauge field.  It is
a harmless numerical factor which can be absorbed into the normalization of 
the path integral, and we can ignore it.  More complicated choices of gauge would
have resulted in a FP determinant which was a function of $A_\mu$; this will often
be unavoidable when we come to QCD in the next section.

We must still find a way of dealing with the delta functional which enforces
the gauge condition.  We can make things tractable by a suitable choice for
how we represent the delta function.  For example, 
\beq
	\delta(x) = \lim_{M\to\infty} \sqrt{M\over 2\pi}e^{-M^2 x^2/2}
\eeq
To express a delta functional, we need one such factor for each point of spacetime:
\beqa
	\delta[f(x)] &=& \lim_{C\to\infty} \prod_{x^\mu} 
	\sqrt{C\over 2\pi}e^{-C^2 f^2(x)/2}\nonumber\\
	&=&\lim_{C\to\infty}\det\left({C\over 2\pi}\right)e^{-\frac12
	C^2 \intdx\, f^2(x)}
\eeqa
Like the FP determinant, $\det\left({C/ 2\pi}\right)$ is another harmless
infinity we can absorb into the measure.  (Remember that the vacuum
generating functional is always normalized to be unity anyway).  Now in the 
case of $f=\vec\nabla\cdot\vec A$, it would be nice if the exponential term
looked more like part of the action.  We can accomplish this by taking
$C^2\to -i C^2$ (this is also a good representation of a delta function--the
oscillations rather than exponential decay are what
kill the integral away from the support of the delta function), so for Coulomb
gauge we get
\beq
	\delta[f(A)] = \lim_{C\to\infty}e^{i\frac12
	C^2 \intdx\, (\vec\nabla\cdot\vec A)^2}
\eeq
The action now becomes
\beq
   \frac12\intdp A_\mu(-p)\left( -p^2 \eta^{\mu\nu} + p^\mu p^\nu 
	- C^2 p_\perp^\mu p_\perp^\nu \right) A_\nu(p)
\eeq
where we defined the transverse momentum vector $p_\perp^{\mu}$ as
just the spatial components:
\beq
	p_\perp^{\mu} = p^\mu - \delta^\mu_0 p^0
\eeq
One can check that for nonzero values of $C$, the tensor 
$-p^2 \eta^{\mu\nu} + p^\mu p^\nu - C^2 p_\perp^\mu p_\perp^\nu$ is now
nonsingular and therefore invertible---the propagator exists.

What about taking the limit $C\to\infty$?  We now see that this is not really
necessary.  We can use any nonzero value of $C$ and still define the propagator.
In fact, this is generally more convenient than taking $C\to\infty$.  For example
when $C=1$, most of the terms in $-p^\mu p^\nu + C^2 p_\perp^\mu p_\perp^\nu$
cancel each other, making the propagator take a simpler form.  In terms of our
picture (fig.\ 36), it means we are no longer integrating over a thin gauge
fixing surface, but rather a fuzzy region centered about the thin slice, 
which nevertheless gives a finite result for the path integral.  In other words,
we are summing over a number of different gauge fixing surfaces, with different
weights.  There is nothing wrong with this; we only care about making the path
integral finite.

There is another way we could arrive at the same result, which is somewhat more
commonly given in textbooks.  Instead of representing the delta functional an 
exponential factor, suppose we have a true delta functional, but let the gauge
condition be
\beq
	\partial_i A_i = \omega(x)
\eeq
for an arbitrary function $\omega(x)$.  Then functionally average over different
$\omega$'s with a Gaussian weight:
\beq
	\delta[\partial_i A_i - \omega(x)] \to \int{\cal D}\omega\, 
	e^{{i C^2\over 2}\intdx\, \omega^2}
\eeq
Not only does it lead to the same result as above, but it makes even more
evident the physical picture of 
restricting the path integral to a ``fuzzy'' gauge fixing surface.

In the previous example, even with $C=1$ the propagator still has a rather
complicated form which is not Lorentz invariant.  A much more convenient choice
is the set of covariant gauges where we take $f=\partial_\mu A_\mu$ so that the
action can be written as 
\beq
\label{gfaction}
   \frac12\intdp A_\mu(-p)\left(- p^2 \eta^{\mu\nu} + p^\mu p^\nu 
	-  p^\mu p^\nu/\alpha \right) A_\nu(p)
\eeq
(we have replaced $C^2$ by $1/\alpha$).  Now when $\alpha=1$ (Feynman gauge) 
we get the simple result that the photon propagator is given by
\beq
	D_{\mu\nu}(p) = \langle A_\mu(-p) A_\nu(p) \rangle = \eta_{\mu\nu}{-i\over p^2}
\eeq
Notice the minus sign ($-i$) relative to the Feynman rule for scalar field
propagators.  It is there solely to counteract the minus sign in $\eta_{ij}$
for the spatial components.  We want the fields which correspond to the
real photon degrees of freedom (the components of $\vec A$ which are transverse
to $\vec p$) to have the correct sign for their kinetic term.  The longitudinal
component $A_0$ has the wrong sign for its kinetic term.  In a generic theory,
this would be disastrous since a wrong sign kinetic term usually implies a breakdown
of unitarity in the S matrix.  We will see that gauge invariance saves us from 
this in gauge theories.

For the case $\alpha\neq 1$, we can find the propagator using Lorentz invariance.
The propagator must have the form
\beq
	D_{\mu\nu} = A\eta_{\mu\nu} + B p_\mu p_\nu
\eeq
since these are the only Lorentz invariant tensors available.  Contracting this
with the inverse propagator, we obtain
\beq
	D_{\mu\nu} (D^{-1})^{\nu\rho} = Ap^2 \delta_\mu^{\ \rho} + 
	\left[ (1/\alpha-1)A + Bp^2 + (1/\alpha-1)B p^2 \right] = \delta_\mu^{\ \rho}
\eeq
which implies
\beq
	A = {1\over p^2};\qquad B = {\alpha-1\over p^4}
\eeq

At first it looks worrisome that the arbitrary parameter $\alpha$ is appearing in
the propagator, since we are free to choose any value we like.  Again, we will
see that gauge invariance saves us: no physical quantities can depend on gauge
parameters like $\alpha$.  Greens's functions and amplitudes will depend on
$\alpha$, but not S-matrix elements.  This can provide a useful check on complicated
calculations, if one is willing to do them in a general covariant gauge with
$\alpha$ not fixed to be the most convenient value.

The fact is that S-matrix elements, so long as they only involve the physical states
(the transverse photon polarizations, and not the longitudinal or time-like ones),
do not depend on the gauge parameter, as long as the electrons are on their mass shells
({\it i.e.,} they obey the Dirac equation $(\psl-m)\psi = 0$).  Before proving this
statement in general, let's illustrate how it works for a one-loop diagram.  The simplest
example which involves a photon propagator is the correction to the electron
self-energy, fig.\ 37.  In dimensional regularization, it is
\beq
	i\Sigma(p) = (ie\mu^\epsilon)^2 \intdq\, \gamma_\mu {i\over \psl + \qsl - m}
	\gamma_\nu {-i\over q^2} \left(\eta_{\mu\nu} + {\alpha-1\over q^2} q_\mu q_\nu
	\right) 
\eeq
Notice the factor of $\mu^\epsilon$ which is needed to keep the charge dimensionless
in $d=4-2\epsilon$ dimensions.  When we evaluate the divergent part of this diagram 
which is proportional to the gauge-fixing term, we get
\beq
\label{selfen}
	\Sigma_{\rm gauge-dep.} \sim (1-\alpha){c\over \epsilon} (\psl - m)
\eeq
This means that the mass counterterm will not depend on the gauge parameter.
In the renormalization procedure, even though one might first think that (\ref{selfen})
requires mass renormalization, we know that the true mass counterterm is determined
after doing wave function renormalization.  The latter completely absorbs the $\alpha$
dependence in (\ref{selfen}).  Notice that whereas mass and charge are associated with
physical observables, wave function renormalization is not, so it is permissible to have
$\alpha$ dependence in the latter, though not in the former.
\medskip
\centerline{\epsfxsize=2.5in\epsfbox{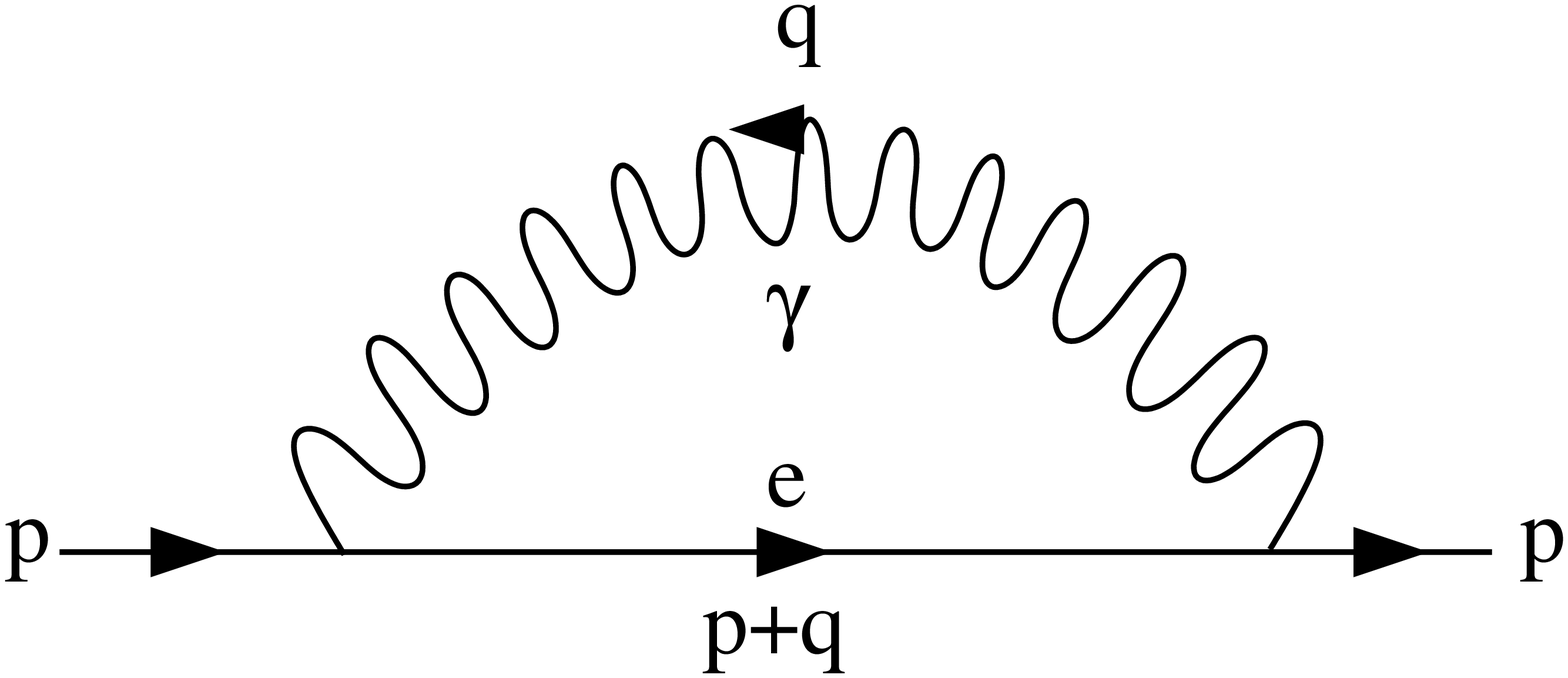}}
\centerline{\small Fig.\ 37. 
Electron self-energy in QED.}
\medskip

There are further restrictions on the form of the counterterms coming from gauge
invariance.  These come from the Ward-Takahashi identities, which are relations
between Greens' functions which we can deduce in a similar manner to Fujikawa's derivation
of the axial anomaly--though now much simpler because in the present case, the path
integral measure really is invariant under gauge transformations, by construction.
The Ward identities are most easily stated in terms of the 1PI proper vertices 
generated by the effective action.  As we did for the axial anomaly, let us first
derive the procedure for a generic collection of fields denoted by $\phi_i$, which 
transforms linearly under an infinitesimal gauge transformation,
\beq
	\delta\phi_i(x) = (A_{ij}\phi_j + A'_i)\delta\omega(x),
\eeq
where the inhomogeneous piece $A'_i$ could be a differential operator (which is the case
for the change in the gauge field).
Furthermore, suppose that the change in the action is also linear in $\delta\phi_i(x)$,
\beq
	\delta S = \intdx\, (B_i \phi_i + B')\delta\omega(x)
\eeq
We will assume that the path integral measure is invariant.  Now consider how the
path integral changes form when we change integration variables from $\phi$ to
$\phi+\delta\phi$:
\beqa
	e^{iW[J]} &=& \int{\cal D}\phi\, e^{iS[\phi]+ \intdx\,J_i \phi_i}\nonumber\\
		&=& \int{\cal D}\phi\, e^{iS[\phi]+ \intdx\,J_i \phi_i}
		\left(1 + \intdx\,\left( B_i\phi_i+B' +J_i (A_{ij}\phi_j+ A'_i) \right)\delta\omega(x)\right)
\eeqa
It follows that 
\beq
\label{ward1}
	 0 = \int{\cal D}\phi e^{iS[\phi]+ \intdx\,J_i \phi_i}
		\intdx\,\left(B_i \phi_i+B'+J_i (A_{ij}\phi_j +A'_i)\right)\delta\omega(x)
\eeq
When we take the expectation value of $\phi$ in the presence of a classical source,
this gives us precisely the classical field $\phi_c = {\delta W\over \delta J}$.
Furthermore $J = - {\delta\Gamma\over \delta\phi_c}$, so we can rewrite (\ref{ward1})
as 
\beq	
	\intdx\, {\delta\Gamma\over \delta\phi_c}\delta\phi_c = 
	\intdx\, (B \phi_c +B')\delta\omega(x) = \intdx\, \delta S[\phi_c]
\eeq
But $\intdx\, {\delta\Gamma\over \delta\phi_c}\delta\phi_c = \delta\Gamma[\phi_c]$,
the change in the effective action.  In other words, the change in the effective action
is the same as the change in the original action.  This is the general form of the
Ward identities.  We can now apply it to the specific case of QED.

For QED, we have 
\beq
	\delta A_\mu = \partial_\mu\delta\omega,\quad \delta\psi = ie\delta\omega\psi
\eeq
The action is invariant except for the gauge-fixing part,
\beq
	S_{\rm g.f.} = -{1\over 2\alpha}\intdx\, (\partial_\mu A^\mu)^2
\eeq
(note the minus sign corresponds to that in (\ref{gfaction}); we get one minus
from $\partial^2\to -p^2$ and another one from integrating by parts)
so the change in the action is given by  
\beq
	\delta S = -\frac{1}{\alpha} \intdx\, \partial_\mu A^\mu \partial^2\delta\omega
\eeq
This is the r.h.s.\ of the Ward identity.  On the l.h.s we have, after integrating
by parts,
\beq
	\intdx\, {\delta\Gamma\over \delta\phi_c}\delta\phi_c  = 
	e \intdx\,\delta\omega(x) \left( {\delta\Gamma\over \delta\psi}
	i\psi + {\delta\Gamma\over \delta\bar\psi}(-i\bar\psi) -\frac{1}{e}
	\partial_\mu {\delta\Gamma\over \delta A_\mu} \right)
\eeq
The result is true for any function $\delta\omega$, in particular a delta function, so
we can remove the integration and write
\beq
\label{bigward}
	i\left( {\delta\Gamma\over \delta\psi}
	\psi - {\delta\Gamma\over \delta\bar\psi}\bar\psi\right) = \frac{1}{e}\left(
	\partial_\mu {\delta\Gamma\over \delta A_\mu}  - 
	\frac{1}{\alpha}\partial^2\partial_\mu A^\mu \right)
\eeq
All the fields here are understood to be the classical fields, although for
clarity we have omitted the subscript.  Eq.\ (\ref{bigward}) is in fact 
the starting point for an infinite number of relations between amplitudes because
we can perform a functional Taylor expansion in the classical fields, and demand that
the equation is true at each order.  Let us first dispose of the gauge fixing term.
At order $A^1$, $\psi^0$, $\bar\psi^0$, the
only terms present are 
\beqa
\label{wardid}
	\partial^\mu \Gamma^{(2A)}_{\mu\nu} A^\nu - 
	\frac{1}{\alpha}\partial^2\partial_\nu A^\nu &=& 0\nonumber\\
\longrightarrow\quad p^\mu \Gamma^{(2A)}_{\mu\nu} A^\nu + \frac{1}{\alpha}
	p^2 p_\nu A^\nu &=& 0
\eeqa
At tree level, we know that the inverse propagator is
\beq
	\Gamma^{(2A,\,\rm tree)}_{\mu\nu} = 
   \left(- p^2 \eta^{\mu\nu} + p^\mu p^\nu 
	-  p^\mu p^\nu/\alpha \right) 
\eeq
where the superscript $2A$ means we have two gauge fields as the external particles.
Therefore the gauge-dependent term in the Ward identity exactly cancels its counterpart
in the tree-level inverse propagator.  Although the gauge-fixing term does get 
renormalized by loops, it remains true that it has no effect on the Ward identities
involving external fermions or more than two gauge bosons.

An interesting consequence of the previous result is that beyond tree level, the
vacuum polarization must be transverse:
\beq
\label{vandiv}
	p^\mu \Gamma^{(2A,\rm loops)}_{\mu\nu} = 0
\eeq
A more conventional notation instead of $\Gamma^{(2A,\rm loops)}_{\mu\nu}$ is
$\Pi_{\mu\nu}$; we define it to be the part of the inverse photon propagator that comes
from beyond tree level:
\beq
	\Gamma^{(2A)} = \frac12\intdp A_\mu(-p) \left(-p^2\eta_{\mu\nu} + (1-\alpha^{-1})
	p_\mu p_\nu + 	\Pi_{\mu\nu}(p)\right)A_\nu(p)
\eeq
Of course, eq.\ (\ref{vandiv}) corresponds to our naive expectation that the electromagnetic
current is conserved, since we can obtain $\Gamma^{(2A)}$ from the currents:
\beq
	 \Pi_{\mu\nu} = \langle J_\mu J_\nu \rangle\nonumber\\
	= \langle (\bar\psi e\gamma_\mu\psi) (\bar\psi e\gamma_\nu\psi) \rangle
\eeq
The only way this can come about is if $\Pi_{\mu\nu}$ has the tensor structure
\beq
\label{vacpolgam}
	\Pi_{\mu\nu} \propto p^2 \eta_{\mu\nu} - p_\mu p_\nu
\eeq
since these are the only Lorentz-covariant tensors around.  This is an example of
how gauge invariance provides constraints that can give a nontrivial consistency
check in practical calculations.  The form (\ref{vacpolgam}) also has an important
practical consequence: it gaurantees that the photon will not acquire a mass
through loop effects.  A mass term would have the form $\frac12 m_\gamma^2 A_\mu A^\mu$,
which would show up as  $m_\gamma^2 \eta_{\mu\nu}$ in the inverse propagator.  The
only way we could get this behavior would be if $\Pi_{\mu\nu}$ had the form
\beq
\label{phmass}
	\Pi_{\mu\nu} \propto\left( p^2 \eta_{\mu\nu} - p_\mu p_\nu\right) \left(
	{m_\gamma^2\over p^2} + \dots\right)
\eeq
This kind of behavior would imply that $\Pi_{\mu\nu}$ has an infrared divergence
as $p^2\to 0$ (notice that we can keep $p_\mu$ nonzero even though $p^2=0$
in Minkowski space).  However, at least at one loop, one does not expect any infrared
divergence because $\Pi_{\mu\nu}$ is generated by a fermion loop, and the internal
electron is massive.  In fact, even if it was massless, power counting indicates that
the diagram would still be convergent in the infrared, since it has the form
$\int d^{\,4}q/q^2$.  On the other hand, in two dimensions, the integral is indeed
IR divergent, and it is not hard to show that we indeed get a photon mass term 
as in (\ref{phmass}), with the value $m^2_\gamma = e^2/\pi$ ($e$ has dimensions of mass 
in 2D), which was shown by Schwinger \cite{Schwinger} to be an exact result.  In fact,
QED in 2D is known as the Schwinger model.

\medskip
\centerline{\epsfxsize=2.5in\epsfbox{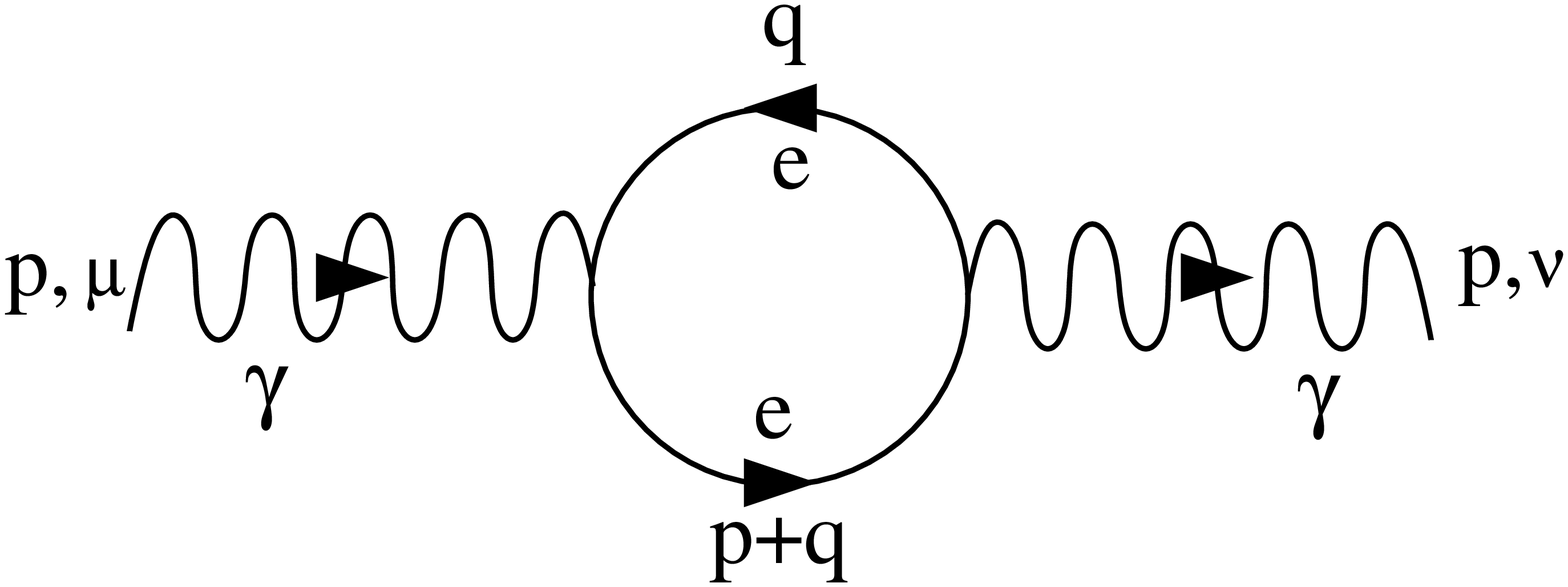}}
\centerline{\small Fig.\ 38. 
Vacuum polarization in QED.}
\medskip

To avoid generating a photon mass in 4D, it is important to use a gauge invariant
regularization.  Notice that gauge invariance forbids us from writing an explicit
mass term $\frac12 m^2_\gamma A_\mu A^\mu$ in the Lagrangian.  Therefore the massless
of the photon is guaranteed if we do not spoil gauge invariance.  This is different
from gauge fixing.  In the Faddeev-Popov procedure, we did not spoil gauge invariance
in an arbitrary way, but carefully manipulated the path integral so that the underlying
gauge symmetry is still present.  If we randomly changed the Lagrangian by adding 
nongauge invariant pieces, we can expect a photon mass to be generated by loops.  This
can be easily seen by examining the vacuum polarization diagram at zero external
momentum:
\beqa
	i\Pi_{\mu\nu}(0) &=& -(ie\mu^\epsilon)^2 \intdpd \tr\,\gamma_\mu {\psl+m\over
	p^2-m^2}\gamma_\nu{\psl+m\over 	p^2-m^2}\nonumber\\
	&=& -(ie\mu^\epsilon)^2 d\intdpd {(2p_\mu p_\nu - p^2\eta_{\mu\nu} + m^2\eta_{\mu\nu})
	\over (	p^2-m^2)^2}\nonumber\\
	&=& -(ie\mu^\epsilon)^2 d\, \eta_{\mu\nu} \intdpd {(2/d-1) p^2+ m^2
	\over (	p^2-m^2)^2}\nonumber\\
&=& -(ie\mu^\epsilon)^2 d\, \eta_{\mu\nu} \intdpd {(2/d-1)(p^2-m^2)+ (2/d)m^2
	\over (	p^2-m^2)^2}
\eeqa
If we were to regularize this using a momentum space cutoff instead of dimensional
regularization, we would get a quadratically divergent contribution to the photon mass.
But in DR, we get
\beqa
	i\Pi_{\mu\nu}(0) &=& -i(ie\mu^\epsilon)^2	\intdped\left[ {2-d\over \pE^2+m^2}
	- {2m^2\over (	\pE^2-m^2)^2}\right]\nonumber\\
	&=& 2i{(e\mu^\epsilon)^2\over (4\pi)^{2-\epsilon}}\left[(-1+\epsilon)
	\Gamma(-1+\epsilon) - \Gamma(\epsilon)\right] = 0	
\eeqa
Similarly, we could regularize using Pauli-Villars fields and also get zero, and this
was the standard approach before the invention of DR.  You can now appreciate
(having done problem 11) what an improvement DR was.

The next interesting Ward identity comes from considering the terms of order
$A^0$, $\psi^1$, $\bar\psi^1$:
\beq
		i\left( {\delta\Gamma^{(\bar\psi\psi)}\over \delta\psi(x)}
	\psi(x) - {\delta\Gamma^{(\bar\psi\psi)}\over \delta\bar\psi(x)}\bar\psi(x)\right) = 
\frac{1}{e}
	{\partial\over \partial x^\mu} {\delta\Gamma^{(\bar\psi A\psi)}\over \delta A_\mu(x)} 
\eeq
Since $\Gamma^{(\bar\psi\psi)} = \intdx\bar\psi(i\dsl-m)\psi$ and 
$\Gamma^{(\bar\psi A\psi)} = e\intdx\bar\psi\Asl\psi$ at tree level, we obtain
\beqa
	i\intdy\left[ \bar\psi(y)(i\dsl_y-m)(-\delta(x-y))\psi(x) + 
	\bar\psi(x)\delta(x-y)(i\dsl_y-m)\psi(y) \right]\nonumber\\
 = \frac{1}{e} e \intdy \partial_{\mu,x}\bar\psi(y)\delta(x-y)\gamma^\mu\psi(y)
\eeqa	
which indeed is an identity, since the mass terms cancel and the derivatives terms
match on either side of the equation.  More generally, let $S(x,y)$ be the
fermion propagator and $\Gamma_\mu(x,y,z)$ be the vertex function.  Then we can
write the Ward identity as
\beq
	{\partial\over\partial y_\mu }\Gamma_\mu(x,y,z) = -iS^{-1}(x-z)\delta(z-y)
	+iS^{-1}(x-z)\delta(x-y)
\eeq
which in momentum space is
\beq
	(p-k)_\mu \Gamma_\mu(p,p-k,k) = S^{-1}(p) - S^{-1}(k)
\eeq
At tree level, $\Gamma_\mu = i\gamma_\mu$ and $S^{-1}(p)= i(\psl-m)$.  This tells 
us nothing new yet.  

However at the loop level, things become
more interesting.  
To appreciate this, let's first look at the general structure
of the renormalized Lagrangian for QED.  Following the notation of F.\ Dyson, 
\beqa
	{\cal L}_{\rm ren} &=& Z_2 \bar\psi  i\dsl \psi - mZ_m \bar\psi\psi
	+ e\mu^\epsilon Z_1\bar\psi\Asl\psi + \frac{1}{4}Z_3F_{\mu\nu}F^{\mu\nu}
	+ \frac12 Z_\alpha (\partial_\mu A^\mu)^2 \nonumber\\
	 &=& \bar\psi_0  i\dsl \psi_0 - m_0 \bar\psi_0\psi_0
	+ e_0\bar\psi_0\Asl_0\psi_0+ \frac{1}{4}F_{0,\mu\nu}F_0^{\mu\nu}
	+ \frac1{2\alpha_0} (\partial_\mu A_0^\mu)^2
\eeqa
where we can see that the relation between bare and renormalized parameters is
\beq
\label{QEDren}
	\psi_0 = \sqrt{Z_2}\psi,\quad
	A_0^\mu = \sqrt{Z_3}A_0^\mu,\quad
	e_0 = e \mu^\epsilon {Z_1\over Z_2 \sqrt{Z_3}},\quad
	m_0 = m {Z_m\over Z_2},\quad
	\alpha_0 = {Z_3\over Z_\alpha}
\eeq
Were it not for gauge invariance, we would have to compute all three coefficients
$Z_1$, $Z_2$, and $Z_3$, to obtain the beta function for the coupling $e$.  But now
at the loop level, the Ward identity is telling us that
\beq
	Z_1(p-q)_\mu \Gamma_\mu(p,p-q,q) = Z_2(S^{-1}(p) - S^{-1}(q))
\eeq
which implies that $Z_1 = Z_2$.  

\medskip
\centerline{\epsfxsize=2.5in\epsfbox{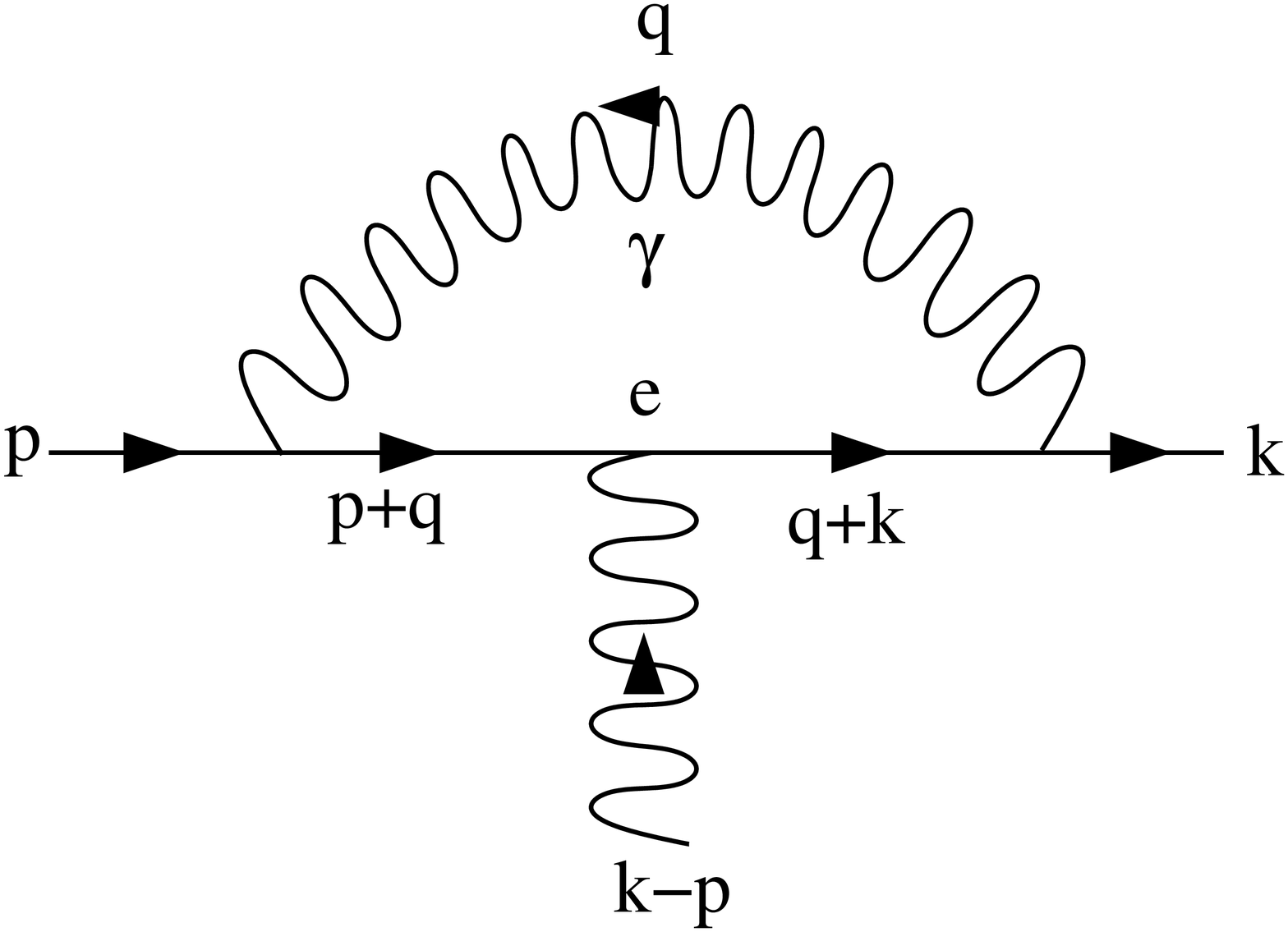}}
\centerline{\small Fig.\ 39. 
Vertex correction in QED.}
\medskip

Therefore, we can compute the relation between the
bare and renormalized charges without having to evaluate the more complicated diagram
of fig.\ 39, which determines $Z_3$.  Instead, we can just compute the vacuum
polarization diagram, fig.\ 38.  It is given by
\beq
	i\Pi_{\mu\nu} = -(i e\mu^{\epsilon})^2 \intdqeps \tr\left(\gamma_\mu 
	{i\over \psl+\qsl - m} \gamma_\nu\, {i\over\qsl - m} \right)
\eeq
If we are only interested in $Z_3$, we can evaluate this in the limit of vanishing
photon momentum.  Let's define
\beq
\label{vacpol2}
	\Pi_{\mu\nu} = \left( p^2 \eta_{\mu\nu} - p_\mu p_\nu\right)\Pi(p^2)
\eeq
We see that the effective action generated by the loop diagram has the form
\beq
	\frac14(1-\Pi(0))F_{\mu\nu}F^{\mu\nu}
\eeq
so that we should start with the renormalized Lagrangian 
\beq
	\frac14(1-\Pi(0))^{-1}F_{0,\mu\nu}F_0^{\mu\nu}
\eeq
in order to keep the photon kinetic term correctly normalized.  Hence we see that
\beq
	Z_3 = (1-\Pi(0))^{-1} \cong 1+\Pi(0)
\eeq
Now we can find the bare charge with respect to the renormalized charge, using
(\ref{QEDren}):
\beq
	e_0 = {e\mu^{\epsilon}\over \sqrt{Z_3}} \cong e\mu^{\epsilon}
	\left(1-\frac12\Pi(0)\right)
\eeq
Computation of the divergent part of the vacuum polarization gives
\beq
\label{vacpoldiv}
	\Pi(0) \sim - {e^2\over 12\pi^2\epsilon}
\eeq
The usual computation of the $\beta$ function gives 
\beq
\label{QEDbeta}
	\beta(e) = -\epsilon e + {e^3\over 12\pi^2}
\eeq
which shows that QED is not asymptotically free; the scale-dependent coupling 
runs like
\beq	
\label{betaqed}
	e^2(\mu) = {e^2(\mu_0)\over 1 - {e^2(\mu_0)\over 12\pi^2}\ln{\mu^2\over\mu^2_0}}
\eeq
and the electric charge has a Landau singularity at high scales, like the quartic 
coupling did.

The result (\ref{vacpoldiv}) also tells us how the gauge-fixing term is renormalized.
The longitudinal part of the unrenormalized one-loop inverse propagator is
\beqa
	 p_\mu p_\nu(1-\alpha^{-1}) +\Pi_{\mu\nu}  &=& 
	  (p^2\eta_{\mu\nu}{\rm\ part}) + p_\mu p_\nu(1-\alpha^{-1} -\Pi(0)) \nonumber\\
	&\to& p_\mu p_\nu(1-\alpha^{-1} -\Pi(0))/(1-\Pi(0))\nonumber\\
	&=& p_\mu p_\nu(1-\alpha^{-1}(1-\Pi(0))^{-1})\nonumber
\eeqa	
The arrow above means ``after renormalizing the gauge field wave function.'' 
We want the divergent contribution
to the effective action to be canceled by the counterterm in $\frac12 Z_\alpha
(\partial\cdot A)^2$, so we see that we should take 
\beq
	Z_\alpha = (1-\Pi(0)) = 1/Z_3
\eeq

We have touched upon the important one-loop effects in QED, figures 37--39.  But what
about diagrams with an odd number of photons, as in fig.\ 40?  
These would render the theory unrenormalizable if they were nonzero.  It is easy 
to see that the first one vanishes, however:
\beq
	\intdp {{\tr}[\gamma_\mu(\psl+m)]\over p^2 - m^2} = 0
\eeq
Each term is zero either because of Dirac algebra or antisymmetry of the integrand.
The second diagram is the triangle diagram in disguise, but without the $\gamma_5$.
It is harder to directly show that it vanishes, but a simple observation shows that
all diagrams with an odd number of external gauge bosons vanishes.  This is Furry's
theorem.  The Lagrangian is invariant under charge conjugation symmetry:
\beq
	\psi \to \psi^c = C\bar\psi^T,\qquad A_\mu \to -A_\mu
\eeq
This works because of the fact that $\bar\psi\gamma_\mu\psi  =  
- \bar\psi^c\gamma_\mu\psi^c$, so the interaction term $\bar\psi\Asl\psi$ remains
invariant.  Because of this symmetry of the original Lagrangian, the effective action
must also have the symmetry.  This includes diagrams with only external photons.  For
them, the amplitude must be invariant under simply $A_\mu \to -A_\mu$.  This insures that
if there is an odd number of external photons, the diagram vanishes.  Intuitively,
the reason the first diagram vanishes is that electrons and positrons give equal and
opposite contributions to the electric potential.  In a medium that has more electrons
than positrons (unlike the vacuum), such diagrams would no longer be forced to vanish.
Of course, this condition breaks the charge conjugation symmetry.

\medskip
\centerline{\epsfxsize=2.in\epsfbox{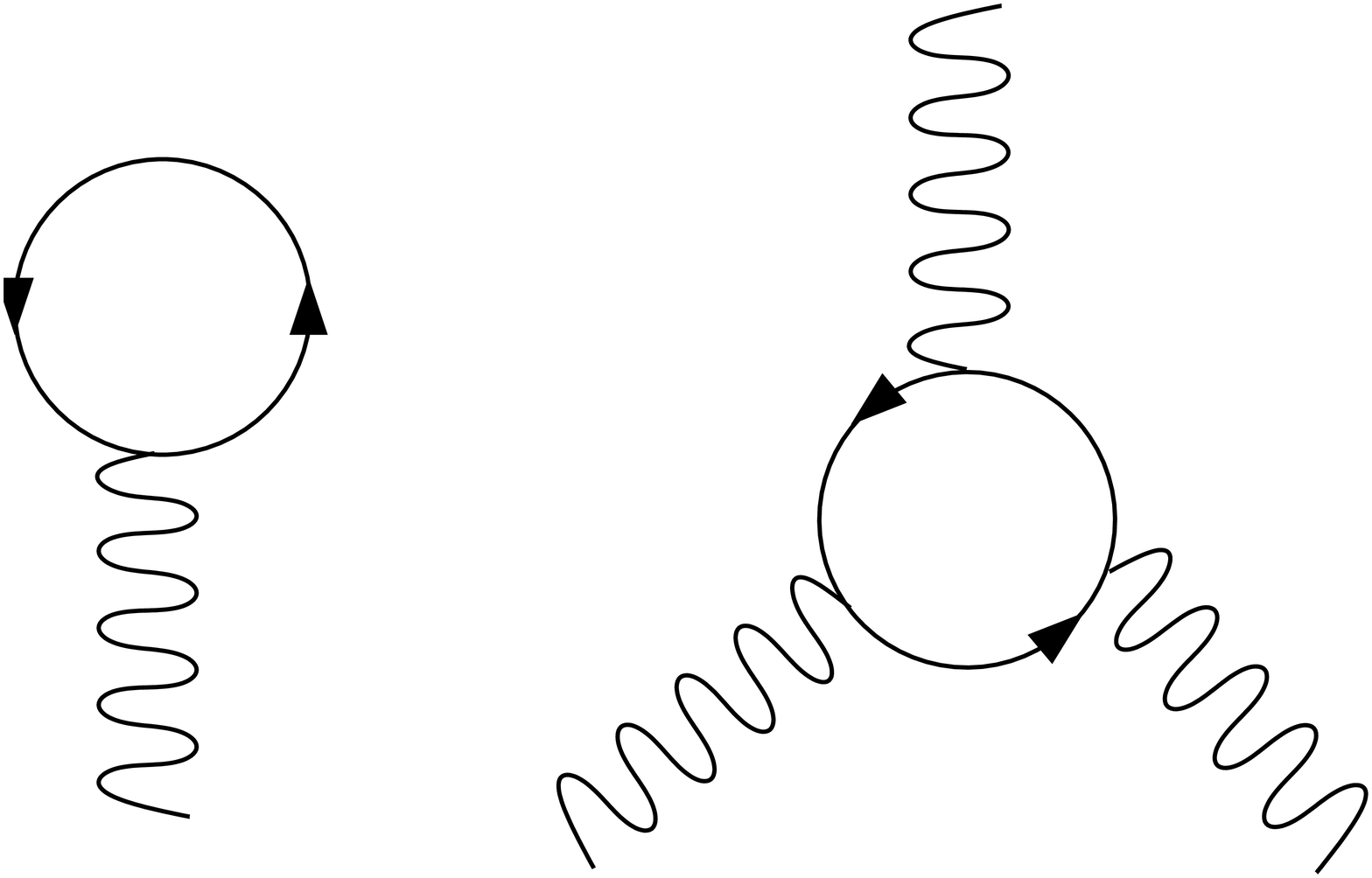}}
\centerline{\small Fig.\ 40. 
Diagrams forbidden by Furry's theorem.}
\medskip

\subsection{Applications of QED}

A new technical problem arises in QED which we have not encountered before: infrared
divergences.  For some loop diagrams, the momentum integral can diverge for low momenta
where virtual particles are going on their mass shell rather than at high momenta.
This problem typically arises when there is a massless particle in the theory.  The
vacuum polarization diagram does not suffer from this problem because the internal
electrons only go on their mass shell when the external photon momentum $p$ goes to zero,
but in this case we are saved by the explicit factors of $p^2\eta_{\mu\nu} - p_\mu p_\nu$
which must be in the numerator, thanks to gauge invariance.  But the diagrams like the
electron self-energy and vertex correction do not have this saving grace.
The divergence can be seen for example in the vertex correction even before doing any
integrals,
\beq
	i\Gamma_\mu(p,p-k,k) = (ie\mu^{\epsilon})^3 \intdq \gamma_\alpha {i\over 
	\qsl + \ksl - m} \gamma_\mu {i\over 
	\qsl + \psl - m}\gamma_\beta {-i\eta^{\alpha\beta}\over q^2}
\eeq
Suppose that one of the electrons is on its mass shell, so $p^2-m^2=0$.
Then as $q\to 0$, not only do we get a divergence from the photon propagator, but also
the electron propagator $(\psl - m)^{-1} = (\psl + m)/(p^2-m^2)$.  This would not 
happen if the photon had a mass: it is kinematically imossible for an electron
to decay into another electron and a massive particle, whereas it is possible for
an electron to emit a zero-energy photon.  This is why massless particles give
rise to infrared divergences.  In the above example, assuming the external photon
momentum is not also zero, the second electron will have to be off shell, so we won't
get an additional divergence from the other propagator.  One therefore expects a
logarithmic IR divergence, and this is what happens.

There are two ways to handle this divergence; one is to put in a small regulator mass
$m_\gamma$ for the photon, so that its propagator becomes $-i\eta^{\alpha\beta}/
(q^2-m_\gamma^2)$.  One might worry about the fact that this spoils gauge invariance, 
but it turns out that for QED, this does not lead to any disastrous consequences---it is
possible to add a mass to a U(1) gauge boson without destroying the unitarity or
renormalizability of the theory, even though gauge invariance is lost.  (The same
statement does not hold true for nonAbelian theories.)  Another way of regulating the
IR divergences is through DR: there will be additional poles in $\epsilon$ associated
with them.

However, unlike UV divergences, there is a general theorem that IR divergences do not
require further modifications of the theory like counterterms or renormalization. 
Instead, one has to carefully define what the physical observables are.  It is guaranteed
that IR divergences will cancel out of these physical quantities.  In the case of the
vertex function, the cancellation is somewhat surprising at first---it involves the
amplitude for {\it emission of an additional photon}, that is, Brehmsstrahlung.
In fig.\ 41 I have illustrated this for the case of Compton scattering, but I could
just as well have omitted the second electron and assumed the first one was scattering
off of an external electromagnetic field.
Note that we can never add two diagrams that have different particles in the initial
or final states---that would be like adding apples and oranges.  
However, when the
emitted photon has a sufficiently small energy, it will unobservable in any real
experiment.  A real experiment will not be able to distinguish an event with just an
electron in the final state from one which has an electron and a very soft photon, whose
energy is below the resolution of the detector.  Therefore, referring to the diagrams
in figure 41,  the cross section we measure
will have the form
\beqa
	{d\sigma\over d\Omega}(e\to e) + {d\sigma\over d\Omega}(e\to e\gamma)
 &\sim& |(a)+(b)|^2 + |(c)|^2 \nonumber\\
	&=& |(a)|^2 + (a)(b)^* + (a)^*(b) + |(c)|^2 + O(e^8)
\eeqa
The terms $(a)(b)^* + (a)^*(b) + |(c)|^2$ are all of the same order ($e^6$) in the 
perturbation expansion, so it makes sense to combine their effects.  Amazingly, there
is an IR divergence in $(c)$ which exactly cancels the one coming from interference of
the vertex correction with the tree diagram, $(a)(b)^* + (a)^*(b)$.  This divergence
is due to the fact that the virtual electron goes on shell (so its propagator blows up) as
the photon energy goes to zero.  The divergence comes from doing the phase space integral
for the photon.  It can also be cut off by introducing a small regulator mass for the
photon.

The specific form of the IR-divergent parts of the above differential cross sections
is (see p.\ 200 of \cite{Peskin})
\beqa
	{d\sigma\over d\Omega}(e\to e) &=& \left({d\sigma\over d\Omega}(e\to
	e)\right)_0 \left[1-{\alpha\over\pi} \log\left(-l^2\over m^2\right)
	\log\left(-l^2\over \mu^2\right) + O(\alpha^2)\right];\nonumber\\
	{d\sigma\over d\Omega}(e\to e\gamma) &=& \left({d\sigma\over d\Omega}(e\to
	e)\right)_0 \left[+{\alpha\over\pi} \log\left(-l^2\over m^2\right)
	\log\left(-l^2\over \mu^2\right) + O(\alpha^2)\right];
\eeqa
where $l=p-k$ is the momentum transfer carried by the $t$-channel photon, and these
expressions are in the limit of large $-l^2$.  Notice that the interference term
is the negative one, which had to be the case since the cross section itself
must be positive.

\medskip
\centerline{\epsfxsize=6.5in\epsfbox{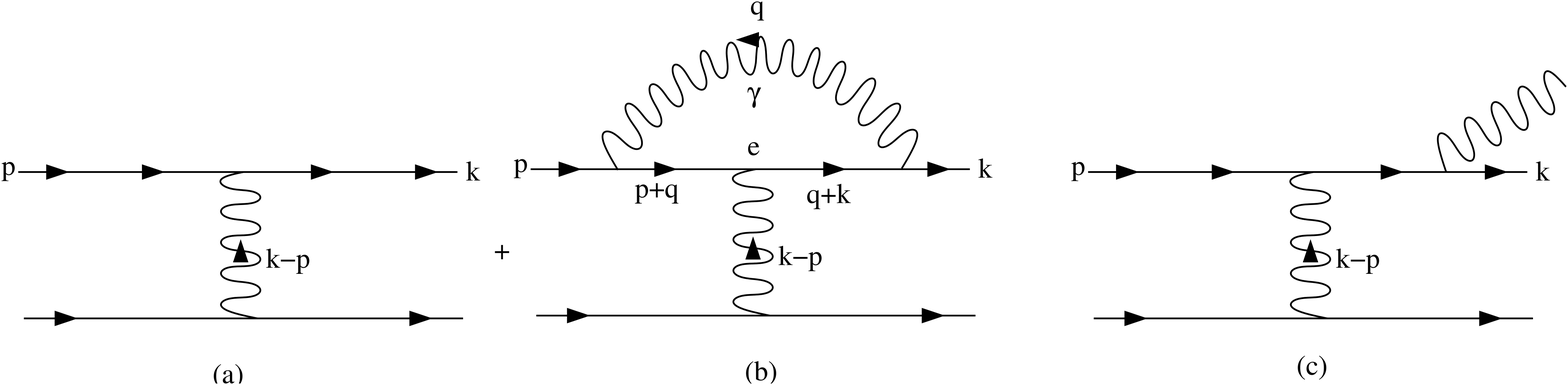}}
\centerline{\small Fig.\ 41. 
Cancellation of IR divergences between the vertex correction and soft
Brehmsstrahlung diagrams.}
\medskip

The Brehmsstrahlung cross section given above is integrated over all the emitted
photons.  To compare with a real experiment, we should only integrate over photons
up to the experimental resolution $E_r$:
\beq
		{d\sigma\over d\Omega}(e\to e\gamma_{\sss E<E_l}) =\left({d\sigma\over d\Omega}(e\to
	e)\right)_0 \left[+{\alpha\over\pi} \log\left(-l^2\over m^2\right)
	\log\left(E_l^2\over\mu^2\right) + O(\alpha^2)\right]
\eeq
So the experimentally measured cross section has the form
\beq
	{d\sigma\over d\Omega}_{\rm expt.} =\left({d\sigma\over d\Omega}(e\to
	e)\right)_0 \left[1-{\alpha\over\pi} \log\left(-l^2\over m^2\right)
	\log\left(-l^2\over E_l^2 \right) + O(\alpha^2)\right],
\eeq
still in the limit that $-l^2\gg m^2$.

Before we leave the subject of QED, it should be said that the few computations
we have touched upon don't begin to describe the fantastic successes of
the theory in explaining experimental data.  A notable example is the anomalous
magnetic moment of the electron and muon.  The electron couples
to a magnetic field via 
\beq
	{\cal L}_{\rm int} = -\vec\mu\cdot\vec B
\eeq
where the magnetic moment is related to the spin by
\beq
	\vec\mu = g\left(e\over 2m\right) \vec S
\eeq
and $g$ is the Land\'e $g$ factor.  At tree level, the Dirac equation predicts that
$g=2$.  But this prediction is altered by the finite part of the vertex correction:
\beq
	{g-2\over 2} = {\alpha\over 2\pi} = {e^2\over 8\pi^2}
\eeq
plus higher order corrections in $\alpha$.  These corrections are called anomalous
magnetic moment.  In fact they have been computed to $O(\alpha^4)$.  The
calculations are so accurate that it is not useful to go to yet higher orders
since the uncertainties due to loop diagrams with hadrons are greater than the
size of the QED corrections.  At the present, there is a small discrepancy between
the predicted and measured values of $g-2$ for the muon, which has led to a flurry
of papers exploring how supersymmetric particles in the loops might be able
to account for the result.  The discrepancy is so far only at the 3$\sigma$ level,
however.   To derive this result, it is useful to invoke the Gordon identity
for the vector current of on-shell Dirac fermions:
\beq
\label{gordon}
	\bar u(p') \gamma^\mu u(p) = \bar u(p') \left[ {p'^\mu+p^\mu\over 2m}
	+ {i\sigma^{\mu\nu}(p'_\nu-p_\nu)\over 2m} \right]
\eeq
where $-i\sigma^{\mu\nu}=\frac12[\gamma^\mu \gamma^\nu ]$.  It is the 
$\sigma^{\mu\nu}$ part of the current which couples to the magnetic field,
and one can see explicitly where the factor of $e/m$ in the electron magnetic
moment is coming from.   Eq.\ (\ref{gordon}) is the tree-level current, which 
gets corrected at the loop level to the form
\beq
	\Gamma^\mu(p',p) = \gamma^\mu F_1(q^2) +  
	{i\sigma^{\mu\nu}q_\nu\over 2m}F_2(q^2)
\eeq
where $q=p'-p$ is the photon momentum, and the functions $F_i(q^2)$ are called
form factors.  These of course are obtained from a complete computation of the
vertex function, including the finite parts.  The magnetic moment interaction is
defined in terms of a uniform magnetic field, whose momentum $q$ is therefore zero,
and so by comparing the above expressions we can deduce that
\beq
	{g - 2\over 2} = F_2(0)
\eeq

One other very famous prediction is the Lamb shift, which is the splitting between
the $2S_{1/2}$ and $2P_{1/2}$ states of the hydrogen atom.  In the Dirac theory,
these states are exactly degenerate because the spin-orbit interaction energy of
the $2P_{1/2}$ state (using the tree-level value of the Land\'e $g$-factor) is
exactly canceled by the relativistic correction to its kinetic energy, relative to
that of the $2S_{1/2}$ state. In 1947,  Willis Lamb and R.\ Retherford observed
that the  energy of the $2P_{1/2}$ state was lower than that of the $2S_{1/2}$  by
an amount corresponding to the energy of a photon with a frequency of 
$\delta\omega = 1057.77$ MHz (or Mc/sec).  This was the experimental puzzle which
motivated much of the early work on QED.  The theoretical contributions which
account for the Lamb shift consist of several parts: those from the two form
factors $F_i$, and one from the vacuum polarization.

The contribution from $F_2$ is easy to understand; this just modifies the usual
contribution of the splitting from the spin-orbit coupling in the obvious way.
This is a small contribution to the total Lamb shift: only 68 MHz.  The contribution
from the vacuum polarization is also small, $-27$ MHz, but it has an interesting
interpretation, which explains its mysterious name.  The one-loop computation of 
$\Pi_{\mu\nu}$ in DR gives
\beq	
\Pi(p) = {e^2\over 2\pi^2}
	\left[{1\over 6\epsilon}- {1\over 6\gamma} - \int_0^1dx\,
	x(1-x) \ln\left[{m^2-p^2x(1-x)\over \mu^2}\right]\right]
\eeq
To interpret it physically, we have to fix the finite part of the counterterms.
We saw that by choosing $Z_3 = (1-\Pi(0))^{-1}$, the photon wave function had
the proper normalization at one loop.  This tells us that the physically relevant
part of $\Pi(p)$ is simply $\hat\Pi(p) = \Pi(p) - \Pi(0)$:
\beq
	\hat\Pi(p) = -{e^2\over 2\pi^2}\int_0^1dx\,
	x(1-x) \ln\left[{m^2-p^2x(1-x)\over m^2}\right]
\eeq
Notice that this is the same integral which appeared in the one-loop correction
for the quartic coupling in $\lambda\phi^4$ theory.  We will come back to discuss
its significance for large values of $p^2 > 4m^2$ later.  For now, we are interested
in $p^2\ll m^2$ because the typical momenta of virtual photons in the hydrogen
atom are of order the electron momentum, which is suppressed relative to $m$ by
a factor of $\alpha$.  Furthermore the Coulomb potential is static, so its
Fourier transform depends only the spatial components of the momenta, which means
we can take $p^2\cong -\vec p^{\,2}$.  In that limit,
\beqa
	\hat\Pi(p) &=& -{e^2\over 2\pi^2}\int_0^1dx\,
	x(1-x) {\vec p^{\,2} x(1-x)\over m^2}
	= -{e^2\over 60\pi^2}{\vec p^{\,2}\over m^2}
\eeqa
We would like to see how this result modifies the Coulomb potential.  Recall our
results (\ref{gaugePI}) and (\ref{coulpot}).  In the static limit, we can
ignore the vector potential relative to the scalar one; it is the latter which gives
the Coulomb self-interaction energy:
\beqa
	-S_{\rm coul} = E_{\rm coul} &=& {1\over 2}\int d^{\,3}x \int d^{\,3}y \,
	\rho(x)\rho(y)
	\int {d^{\,3}q\over (2\pi)^3}e^{i\vec q\cdot\vec x}{-1\over
	 |\vec q|^{2}[1-\hat\Pi(-|\vec q|^{2})]}\nonumber\\
	&\cong& -{1\over 2}\int d^{\,3}x \int d^{\,3}y \,
	\rho(x)\rho(y)
	\int {d^{\,3}q\over (2\pi)^3}e^{i\vec q\cdot\vec x}{1\over
	 |\vec q|^{2}}\left[1+\hat\Pi(-|\vec q|^{2})\right] \nonumber\\
	&=& +{1\over 2}\int d^{\,3}x \int d^{\,3}y \,
	\rho(x)\rho(y) V(|\vec x -\vec y|)
\eeqa
where the modified Coulomb potential can be read off as 
\beq
	V(|\vec x|) = -{1\over 4\pi}\left({1\over |\vec x|} + {4\pi e^2
	\over 60 \pi^2} \delta^{(3)}(\vec x)\right)
\eeq
We see that the delta function contribution will lower the energy of the $S$-states 
of the hydrogen atom, but have no effect on the $P$ states since their wave
functions vanish at the origin.  Hence it gives a negative contribution to the

The extra term was evaluated in the limit $|\vec q|^2\ll m^2$. A more careful
evaluation can be done.  If we restore the
factor of $e^2$ which is conventional to put in the Coulomb potential,
the result is the Uehling potential, which is still an approximation that
assumes $r\gg 1/m$.  
\beq
V(|\vec x|) = -{\alpha\over r}\left(1 + {\alpha\over 4\sqrt{\pi}}
	\, {e^{-2mr}\over (mr)^{3/2}} \right) \equiv -{\alpha(r)\over r}
\eeq
In the limit $m\to\infty$, we recover the delta function, showing
that the latter is a good approximation for distance scales much greater than the
Compton wavelength of the electron.  We have defined an effective distance-dependent
coupling $\alpha(r)$ based on this result.  We see that $\alpha(r)$ gets larger
as one goes to shorter distances.  We already knew this from another perspective:
our calculation of the $\beta$ function gave us the analogous result for $\alpha$
as a function of the energy scale, though it was valid for large energies, hence
small distances.  The two results give the same qualitative behavior however: 
the effective electric charge increases at short distances.  This is where the
picture of vacuum polarization comes in.  We can understand this behavior if the
vacuum itself is polarized by the presence of a bare electric charge, due to the
appearance of virtual electron-positron pairs.  These pairs will behave like
dipoles which tend to screen the bare charge.  At large distances, the screening
is most effective, but as we get close to the bare charge, the effect of the
screening is reduced.  

So far we have accounted for only $68-27 = 41$ MHz of the total Lamb shift of
1058 MHz.  The biggest contribution is coming from the $F_1$ form factor in the
vertex correction.  It is the most difficult part to compute, because it contains
an infrared divergence.  In the limit of small $q^2$, it can be shown that
\beq
	F_1(q^2) = 1 + {\alpha\over 3\pi} \, {q^2\over m^2}\,
	\left[ \ln\left({m_\gamma\over m}\right) + \frac38 \right]
\eeq
(Sometimes you will see in addition to the term $\frac38$ another term
$\frac15$, which is due to the vacuum polarization.  We could push the latter
into the vertex by doing a momentum-dependent field redefinition of the photon,
but we have chosen not to do so.)  To arrive at this result, we have chosen the
finite parts of the counterterms in such a way that $F_1(0)=1$; then the
renormalized charge $e$ couples to the electric current with the conventional
normalization.  Notice that $q$ is the momentum of the photon.  If we consider
how $F_1$ corrects the Dirac equation, we get
\beq
\label{lamb}
	(i\dsl - m + e F_1(-\square) \Asl)\psi = 0
\eeq
To treat the hydrogen atom, we would normally use $eA_\mu = \eta_{0\mu}\alpha/r$, the Coulomb
potential.  The above equation is telling us that we should correct this, 
effectively replacing the Coulomb potential by
\beqa
	{\alpha\over r} &\to & \left(1 + F_1(\vec\nabla^2)\right){\alpha\over r}
	\nonumber\\
	&=& {\alpha\over r} -   {4\alpha^2\over 3m^2}\,
	\left[ \ln\left({m_\gamma\over m}\right) + \frac38 \right]\delta^{(3)}(
	\vec r)
\eeqa
We can treat the extra term as a perturbation in the Schr\"odinger equation and
compute its effect immediately for a given wave function.  Like the vacuum
polarization contribution, it has no effect on the $P$ states.  The only problem
is that it is IR divergent; how do we interpret the photon mass?

To understand what is going on, one has to go back to the original form of the
self-energy, a complicated expression, but the salient feature is that its
IR divergence is only there because of the external electrons being on shell.
In a bound state however, they are not exactly on shell; rather,
\beq
	E = \sqrt{\vec p^{\,2} + m^2} - E_b
\eeq
where $E_b \sim \alpha^2 mc^2$ is the binding energy.  This means that
\beq
	p^2 - m^2 \cong 2mE_b
\eeq
and we should expect to replace $m^2_\gamma$ by something of order 
$|2mE_b|$.  This crude procedure does in fact give a rather good estimate
of the bulk of the Lamb shift.  To do better and compute it exactly requires
special techniques of bound state computations, and many pages of calculation.

\section{Nonabelian gauge theories}
\label{sec:qcd}

\subsection{Group theory for $SU(N)$}
We have dealt with the simplest kind of gauge theory, which is based on the
gauge symmetry $U(1)$.  It is an Abelian group because any two $U(1)$ 
transformations commute with each other:
\beq
\label{U1}
	e^{i\theta_1} e^{i\theta_2} =  e^{i\theta_2} e^{i\theta_1}
\eeq
On the other hand, $U(N)$, the group of $N\times N$ unitary matrices, is
noncommutative: in general
\beq
\label{UN}
	U_1 U_2 \neq U_2 U_1
\eeq
for two such matrices.  To express this in a way that looks more like (\ref{U1}),
we use the fact that any $U(N)$ matrix can be written in the form
\beq
	U = e^{i \sum_a \theta_a T^a}
\eeq
where the matrices $T_a$ are Hermitian, and are known as the {\it generators} of the
group.  The noncommutativity of the U(N) matrices in (\ref{UN}) is due to that
of the generators.  For two matrices which are only infinitesimally different from
the unit matrix, $U_i \cong 1 + i\theta_{i,a}T^a$,
\beq
	U_1 U_2 - U_2 U_1 = i\theta_{1,a}\theta_{2,b} [T^a,T^b]
\eeq
The generators obey the {\it Lie algebra}
\beq
	 [T^a,T^b] = if^{abc}T^c
\eeq
where the {\it structure constants} $f^{abc}$ are real and antisymmetric on any
two indices.  The number of different generators is called the {\it rank} of the 
group, and is given by $N^2$ for the group U(N).  

However, the group U(N) is not {\it simple} except for the case of U(1): we
can decompose any U(N) matrix into the product of a special unitary $SU(N)$ matrix
and a U(1) phase: U(N) $=$ $SU(N)$ $\times$ U(1).  Therefore the rank of $SU(N)$ is
$N^2-1$.  You are already familiar with $SU(2)$, which is generated by the
Pauli matrices:
\beq
	T^a = \frac12 \sigma^a,\quad a=1,2,3
\eeq
The factor of $\frac12$ assures the generators are normalized in the conventional
way
\beq
	\tr T^a T^b = \frac12 \delta_{ab}
\eeq
For $SU(3)$, the generators are given by $T^a = \frac12 \lambda^a$, where 
the $\lambda^a$'s are the Gell-Mann matrices,
\beqa
\lambda_1 &=& \left(\begin{array}{ccc}0&1&0\\1&0&0\\0&0&0\end{array}\right)\quad
\lambda_2 = \left(\begin{array}{ccc}0&-i&0\\i&0&0\\0&0&0\end{array}\right)\quad
\lambda_3 = \left(\begin{array}{ccc}1&0&0\\0&-1&0\\0&0&0\end{array}\right)\nonumber\\
\lambda_4 &=& \left(\begin{array}{ccc}0&0&1\\0&0&0\\1&0&0\end{array}\right)\quad
\lambda_5 = \left(\begin{array}{ccc}0&0&-i\\0&0&0\\i&0&0\end{array}\right)\quad
\lambda_6 = \left(\begin{array}{ccc}0&0&0\\0&0&1\\0&1&0\end{array}\right)\nonumber\\
\lambda_7 &=& \left(\begin{array}{ccc}0&0&0\\0&0&-i\\0&i&0\end{array}\right)\quad
\lambda_8 = {1\over\sqrt{3}}\left(\begin{array}{ccc}1&0&0\\0&1&0\\0&0&-2\end{array}\right)\nonumber\\
\eeqa
The structure constants for $SU(3)$ are given by
\beq
\label{struct}
	\begin{array}{cc} abc & f_{abc}\\ 123&1\\147&1/2\\156&-1/2\\
	246&1/2\\257&1/2\\345&1/2\\367&-1/2\\458&\sqrt{3}/2\\
	678&\sqrt{3}/2 \end{array}
\eeq

In addition, one can consider the anticommutators,
\beq
	\{\lambda_a,\lambda_b\} = \frac43\delta_ab I + 2d_{abc}\lambda_c
\eeq
where $I$ is the $N\times N$ unit matrix.  See the particle data book for a table of the
$d_{abc}$'s for $SU(3)$.  Some other identities:
\beqa
	T^a T^b &=& {1\over 2N} \delta_{ab}I + \frac12 d_{abc} T^c +
	\frac{i}{2} f_{abc}T^c;\nonumber\\
\label{tataid}
	\sum_a T^a T^a &=& \frac{N^2-1}{2N} I\\
	\sum_a T^a_{ij} T^a_{kl} &=& \frac12\left( \delta _{il}\delta_{jk} -
	\frac{1}{N}\delta_{ij}\delta_{kl}\right)\nonumber\\
	\tr(T^a T^b T^c) &=& \frac14\left(d_{abc} + i f_{abc}\right)\nonumber\\
	\tr(T^a T^b T^c T^d) &=& \frac{1}{4N} \delta_{ab}\delta_{cd}
	+\frac18(d_{abe} + i f_{abe})(d_{cde} + i f_{cde})\nonumber\\
\label{ffid}
	f_{acd} f_{bcd} &=& N \delta_{ab}
\eeqa	

\subsection{Yang-Mills Lagrangian with fermions}

Now we would like to use this to do some physics.  A fermion $\psi_i$  which
transforms in the fundamental representation of $SU(N)$ will have an index $i$, which
in the case of QCD ($SU(3)$) we call color, and in the case of the electroweak theory
($SU(2)$) we call isospin.  It transforms as
\beq
	\psi_i\to U_{ij}\psi_j = \left(e^{i g\sum_a \omega_a T^a}\right)_{ij}\psi_j
\eeq
Here we have introduced a charge $g$ which is analogous to the electric charge.
To gauge this symmetry, we need to find a generalization of the covariant
derivative.  The ordinary derivative transforms as
\beq
	\partial_\mu\psi \to U\partial_\mu\psi + (\partial_\mu U)\psi
	= U\left( \partial_\mu + U^{-1}(\partial_\mu U)\right)\psi
\eeq
The obvious way to cancel the unwanted term $(\partial_\mu U)\psi$ is to
introduce the covariant derivative
\beq
	D_\mu\psi = (\partial_\mu - ig A_\mu^a T^a)\psi
\eeq
In other words, the gauge field has become a Hermitian matrix which can be expanded in the 
basis $T^a$.  It must transform like 
\beq
	A_\mu \equiv A_\mu^a T^a \to U A_\mu U^{-1} + 
	{i\over g}U\partial_\mu U^{-1}  = 
	U A_\mu U^{\dagger} + 
	{i\over g}U\partial_\mu U^{\dagger}
\eeq
Notice that this reduces to the usual expression in the Abelian case, where
$U = e^{ig\Lambda(x)}$ for a simple function $\Lambda$:  $A_\mu \to A_\mu +
\partial_\mu\Lambda$. 

The Feynman rules for the fermion propagator and the
interaction of the gauge field with fermions are
easy to read off from the interaction vertex.  They are just like those of
QED with the obvious changes,
\beqa
{\epsfxsize=1in\epsfbox{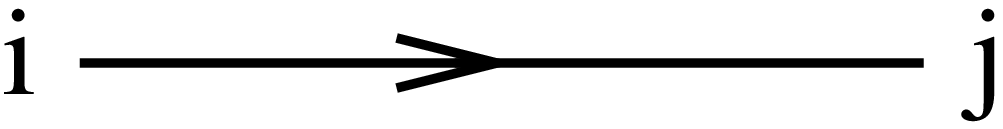}} = i{\delta_{ij}\over \psl - m};\nonumber\\
{{}\atop{\epsfxsize=1.15in\epsfbox{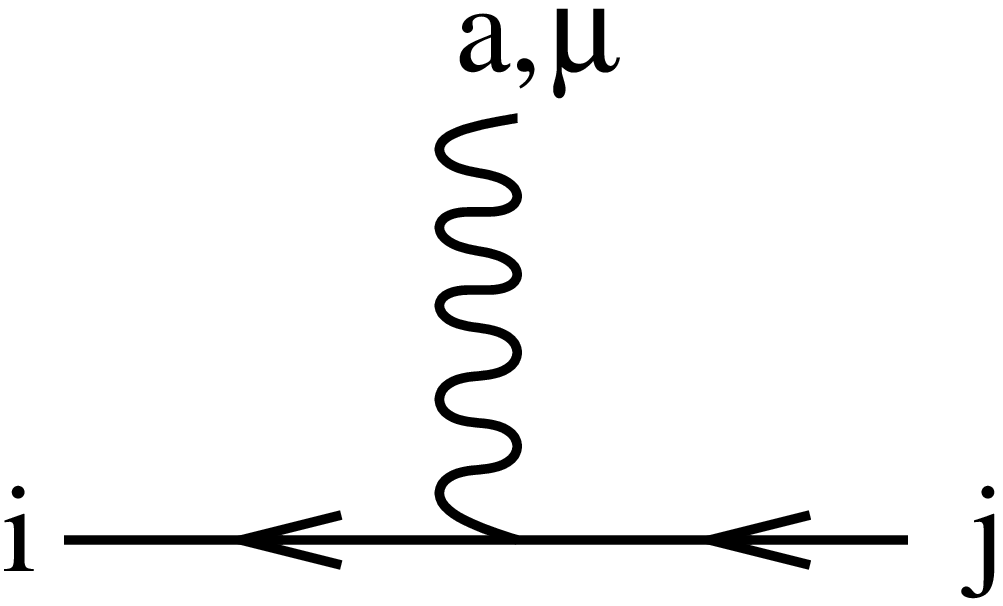}}} = ig\gamma_\mu T^a_{ij}
\eeqa

Next we need to specify the dynamics of the gauge field.  In QED this was 
given by the Lagrangian $\frac14 F_{\mu\nu} F^{\mu\nu}$.  For $SU(N)$ we expect
that the field strength will be a matrix.  If we try to define it in the same way
as in the U(1) case, $F_{\mu\nu} = \partial_\mu A_\nu - \partial_\nu A_\mu$,
then under a gauge transformation we would get
\beqa
\label{F?}
F_{\mu\nu} &{?\atop\to}& U F_{\mu\nu} U^\dagger + 
{i\over g}\partial_\mu\left(U\partial_\nu U^{\dagger}
	\right) - {i\over g}\partial_\nu\left(U\partial_\mu U^{\dagger}
	\right)\nonumber\\
	&{?\atop=}& U F_{\mu\nu} U^\dagger + {i\over g} [ \partial_\mu U\partial_\nu U^{\dagger}
	- \partial_\nu U\partial_\mu U^{\dagger} ]
\eeqa
If it weren't for the term in brackets, we could take the gauge field Lagrangian
to be
\beq
\label{F2}
	{\cal L}_{\rm gauge} =  -\frac12\tr F_{\mu\nu} F^{\mu\nu} = 
	-\frac14 \sum_a F^a_{\mu\nu} F^{a,\mu\nu}
\eeq
But this does not work with $F$ as we naively tried to define it.
To discover the right definition for the field strength, we can appeal to a different
identity which also works in the case of QED:
\beq
	F_{\mu\nu} = {i\over g}[D_\mu, D_\nu] = 
	\partial_\mu A_\nu - \partial_\nu A_\mu - ig[A_\mu,A_\nu]
\eeq
The new commutator term vanishes in the U(1) case, but under U(N) it transforms
in just the right way to cancel the unwanted terms in (\ref{F?}).  This term
is the key to all that makes nonAbelian gauge theories very different from QED;
without it, our theory would be equivalent to one with $N^2-1$ kinds of photons,
{\it i.e.} a U(1)$^{N^2-1}$ gauge theory.  When we expand (\ref{F2}), not only
do we get the inverse propagator for the gauge fields, but also a cubic and
a quartic interaction, as shown in figure 42.

\medskip
\centerline{\epsfxsize=3.5in\epsfbox{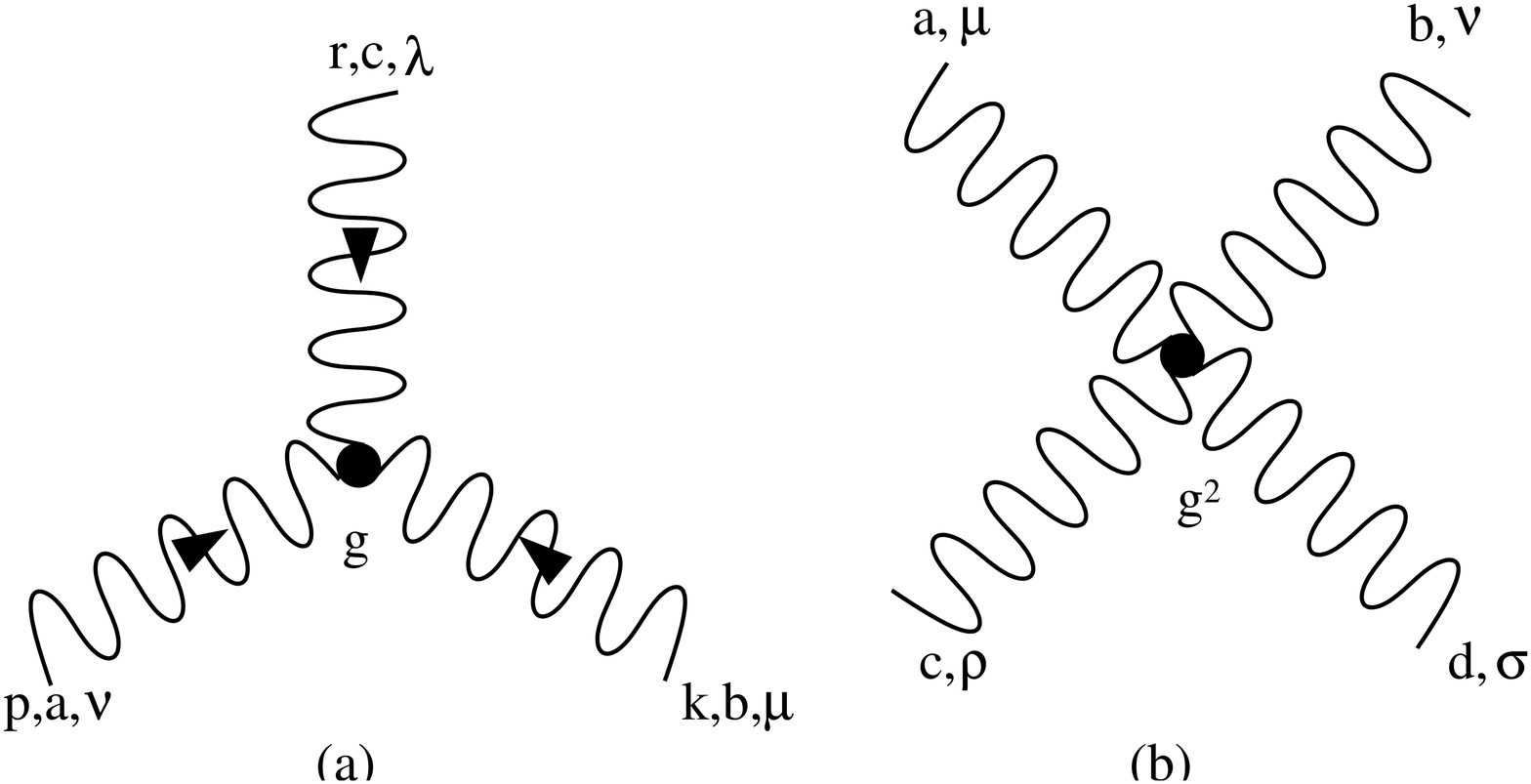}}
\centerline{\small Fig.\ 42. 
New self-interactions of the gauge bosons in Yang-Mills theory.}
\medskip

We have already studied theories with cubic and quartic interactions, with no
sign of dramatic effects like the confinement of quarks.  Yet in the case of 
nonAbelian gauge theories, these new terms completely change the character of 
the physics, giving rise to asymptotic freedom at high energies, and confinement
of quarks.  Although confinement cannot be seen in perturbation theory, we can
at least show that the gauge coupling runs to larger values in the infrared, which
is circumstantial evidence for confinement, since the interaction becomes 
increasingly strong.  To study this, we need to find the Feynman rules for the 
new interactions.  In position space, the interaction terms are
\beq
	{\cal L}_{\rm int} = -\frac{g}{2} f_{abc}A^{\mu}_b A^{\nu}_c (\partial_\mu A^a_\nu
	- \partial_\nu A^a_\mu) - \frac{g^2}{4}f_{abc}f_{ade}A^{\mu}_b A^{\nu}_c
	A^d_\mu A^e_\nu
\eeq
The first term can be simplified to $-g f_{abc}A^{\mu}_b A^{\nu}_c \partial_\mu 
A^a_\nu$ since the factor $f_{abc}A^{\mu}_b A^{\nu}$ is already antisymmetric
under $\mu\leftrightarrow\nu$.  In momentum space, we can write the corresponding
action for this term as
\beq
\label{cubic}
	S_3 = 	-g \intdp {\intdk\over(2\pi)^4} \intdr (2\pi)^4 \delta^{(4)}(p+k+r)
	f_{abc} ip_\mu 
	A^a_\nu(p) A^{\mu}_b(k) A^{\nu}_c(r) 
\eeq
To get the Feynman rule, we need to do all 6 possible contractions of the
fields in (\ref{cubic}) with the external fields whose momenta, color 
and Lorentz indices are shown in fig.\ 42 (a).  This is the same thing as
taking functional derivatives, remembering that all the momenta and indices
in (\ref{cubic}) are dummy variables:
\beqa
	\hbox{cubic Feynman rule } &=& {1\over (2\pi)^4 \delta^{(4)}(p+k+r)}\,
	i{\delta^3 S_3\over \delta A^a_\nu(p)\delta A^b_\mu(k)\delta A^c_\lambda(r)}	
	\nonumber\\
	&=& g f_{abc} \eta_{\nu\lambda}p_\mu + \hbox{permutations of }
	(a\nu p, b\mu k, c\lambda r)\nonumber\\
	&=& g f_{abc}\left[ \eta_{\nu\lambda}(p-r)_\mu +
	\eta_{\nu\mu}(k-p)_\lambda + \eta_{\mu\lambda}(r-k)_\nu \right]
\eeqa
Carrying out a similar procedure for the quartic coupling, we get
\beqa
	\hbox{quartic Feynman rule } &=& -ig^2\left[f^{abc}f^{cde}
	(\eta_{\mu\rho}\eta_{\nu\sigma} - \eta_{\mu\sigma}\eta_{\nu\rho})
	+ f^{ace}f^{bde}
	(\eta_{\mu\nu}\eta_{\rho\sigma} - \eta_{\mu\sigma}\eta_{\nu\rho})
	\right.\nonumber\\ &+& \left. f^{ade}f^{bce}
	(\eta_{\mu\nu}\eta_{\rho\sigma} - \eta_{\mu\rho}\eta_{\nu\sigma})	
	\right]
\eeqa
Notice that there are $4!$ terms as in $\phi^4$ theory, but due to the symmetry
of the interaction, each of the six
shown above appears $4$ times, which canceled the factor of $\frac14$ in the 
original interaction.

The propagator for the gauge bosons has the same structure as in QED; the only
difference is that now there are $N^2-1$ of them, so we must include a Kronecker
delta factor $\delta^{ab}$ for the group structure.  However, the gauge fixing
process gives us something new; in QED we could ignore the Faddeev-Popov determinant
because it was independent of the fields, but this will no longer be possible
(in general) in $SU(N)$.  The generalization of the FP procedure to $SU(N)$ is
to insert the following factor into the path integral:
\beqa
	1 &=& \int{\cal D}U\, \delta[f(A_U)] 
	\det\left({\delta f(A_U)_x\over \delta U_y}\right)
	\nonumber\\
	&\equiv&  \int{\cal D}U \delta[f(A_U)] 
	\Delta_{\rm FP}[A; U]
\eeqa
where $A_U =  U A_\mu U^{-1} + 	{i\over g}U\partial_\mu U^{-1}$, and ${\cal D}U$
is the functional generalization of the group invariant measure $dU$ for $SU(N)$
which has the property that $\int dU = \int d(U'U)$ for any fixed element $U'$
of $SU(N)$.  We will not have to concern ourselves with the detailed form of 
this measure since it just goes into the overall infinite factor which we
absorb into the normalization of the path integral.  But we do have to compute
the determinant, which looks formidable.  Fortunately though, we can compute it
using the infinitesimal form of the gauge transformation:
\beqa
	A_U &=&  (1+ig\omega^a T^a)A_\mu (1-ig\omega^a T^a) + {i\over g}
		(-ig)\partial_\mu\omega^a T^a \nonumber\\
	\longrightarrow \delta A &=& ig\omega^a A_\mu^b [T^a,T^b] 
	+ \partial_\mu\omega^a T^a\nonumber\\
	&=& -g f^{abc}\omega^a A_\mu^b T^c + \partial_\mu\omega^a T^a
\eeqa
Therefore
\beq
	\delta A^c_\mu =  \partial_\mu\omega^c -g f^{abc}\omega^a A_\mu^b
\eeq
Let's consider the form of the gauge condition for covariant gauges.  In 
analogy to QED, we would like to take
\beq
	f(A) = \partial^\mu A_\mu = \partial^\mu A^a_\mu T_a = 0
\eeq
We see that there are really separate conditions for each gauge boson,
so our gauge condition should have a color index as well: $f \to f^a$.  Then
the FP determinant takes the form
\beq
\Delta_{\rm FP}[A] = \det{\partial f^a\over\partial A^c_\mu}{\delta(\delta A^c_\mu)_x\over \delta(\omega^d_y)}
= \det\left|\partial_\mu( \delta_{ad}\partial_\mu - gf^{dba}A_\mu^b) \delta(x-y)\right|
\eeq
Unlike in QED, the nonAbelian nature of the theory has introduced field-dependence
into the FP determinant, so we will no longer be able to ignore it.

The problem now is how to account for this new factor in a way which we know how 
to compute.  We know from previous experience though that an anticommuting field
produces a functional determinant when we perform its path integral.  Therefore it
must be possible to rewrite the FP determinant as a path integral over some new
fictitious fields $c_a$, $\bar c_a$ which we call {\it ghosts}
\beq
	\Delta_{\rm FP}[A] = \int {\cal D }\bar c\, {\cal D}{c}\, e^{iS_{\rm ghost}}
\eeq
where
\beq
	S_{\rm ghost} = \intdx\, \bar c_a \delta_{ad}\left(-\partial^\mu\partial_\mu
	 + g f^{dba}\partial^\mu A_\mu^b\right) c_d
\eeq
Similarly to our derivation of the self interactions of the gauge bosons, we
can obtain the Feynman rules for the ghost propagator and the
ghost-ghost-gluon vertex,
\beqa
{\epsfxsize=1in\epsfbox{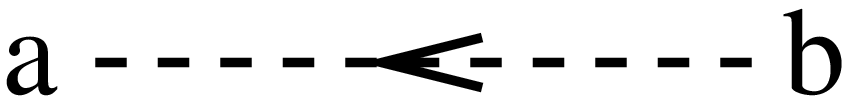}} = i{\delta_{ab}\over p^2};\nonumber\\
{{}\atop{\epsfxsize=1.15in\epsfbox{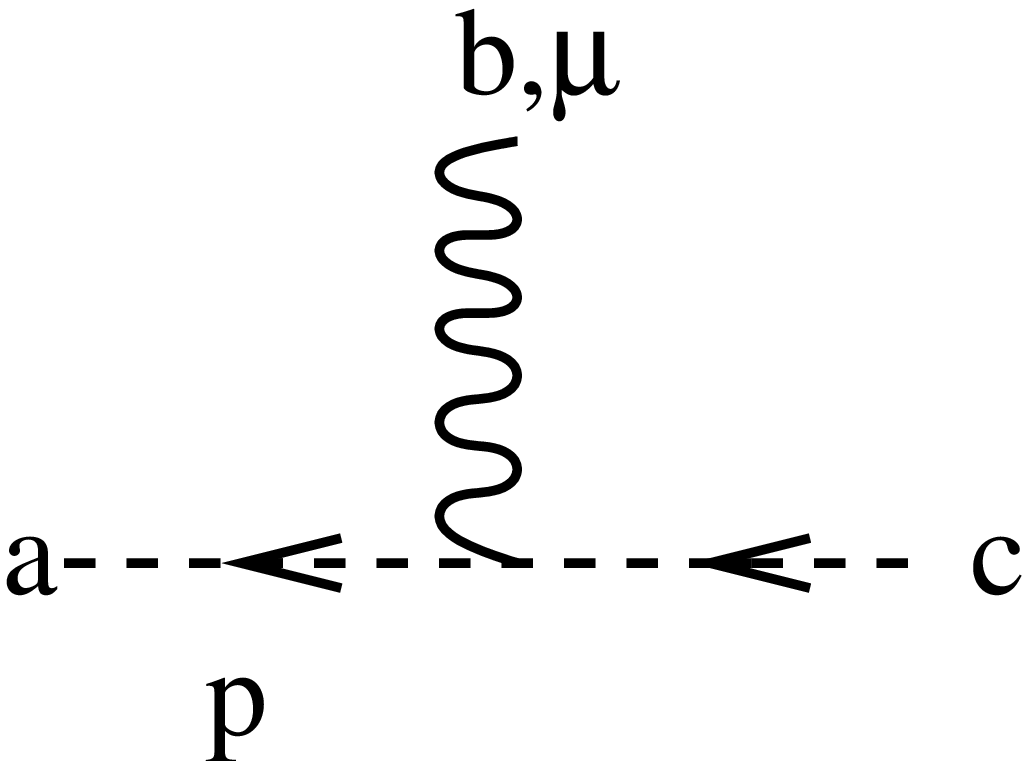}}} = -gf^{abc}p
\eeqa

The ghosts are fermionic fields, whose path integral produces the determinant
with a positive power, unlike bosons which give a negative power.  Loops of the
ghost fields will therefore come with a minus sign, just like those of fermions.
On the other
hand, their propagator looks like that of a scalar field.  Therefore the ghosts
violate the spin-statistics theorem, which says that a scalar particle should
be quantized as a boson.  However, the ghosts are not physical fields; their purpose
is to cancel out the unphysical contributions of the longitudinal and timelike
polarizations of the gauge bosons.  In QED the unphysical polarizations never
contributed to physical amplitudes in the first place, so it was not necessary to
cancel them.  But in Yang-Mills theory, the unphysical states participate in the
interaction vertices, so their effects must be explicitly counteracted.  This will
always come about through loops of the ghost fields; at tree level they are not
necessary.  To appreciate this important fact, let's look at the issue of
unitarity of the S-matrix in greater detail.  

\subsection{Unitarity of the S-matrix}

Conservation of probability in quantum mechanical scattering theory is expressed
by the unitarity of the scattering matrix,
\beq
	S^\dagger S = 1
\eeq
Let us change notation slightly from that used in \cite{harry} to that of
\cite{Peskin} to express the relationship between the $S$ matrix, the $T$
matrix, and the transition amplitude ${\cal M}$:
\beqa
	S &=& {\bf 1} + iT
\eeqa	
where $ {\bf 1}$ is the identity operator.  Then unitarity says that
\beq
\label{unitarity}
	S^\dagger S = {\bf 1} + i(T-T^\dagger) + T^\dagger T = {\bf 1}
\eeq
The matrix elements of $T$
for a $2\to n$ scattering process, for example, are related to ${\cal M}$
by
\beqa
	\langle {\bf p}_1 \cdots {\bf p}_n | T | {\bf q}_1  {\bf q}_2
	\rangle &=& (2\pi)^4 \delta^{(4)}(q_1 + q_2 - p_1 \cdots - p_n)
	{\cal M}(q_i \to p_i);\nonumber\\
	\langle {\bf q}'_1 {\bf q}'_2 | T^\dagger | {\bf p}_1\cdots  {\bf p}_n
	\rangle &=& (2\pi)^4 \delta^{(4)}(q_1' + q_2' - p_1 \cdots - p_n)
	{\cal M}^*(q_i' \to p_i)
\eeqa
One of the most common uses of unitarity is to take the matrix element of (\ref{unitarity}) between 
a pair of two-particle states (though we could equally well consider any number
of particles):
\beqa
-i\left( \langle {\bf q}'_1{\bf q}'_2 |T| {\bf q}_1  {\bf q}_2 \rangle
 - \langle {\bf q}'_1{\bf q}'_2 |T^\dagger| {\bf q}_1  {\bf q}_2 \rangle\right) &=&
\langle {\bf q}'_1{\bf q}'_2 |T^\dagger T |{\bf q}_1  {\bf q}_2 \rangle\nonumber\\
&=&
\label{unit2}
\sum_i \langle {\bf q}'_1{\bf q}'_2 |T^\dagger|i\rangle\langle i| T |{\bf q}_1 
 {\bf q}_2 \rangle
\eeqa
The explicit form for the complete set of states inserted in (\ref{unit2})
is
\beqa
	\sum_i |i\rangle\langle i| &=& \sum_n\left(\prod_{i=1}^n \int {d^3 p_i 	
\over (2\pi)^3}\,{1\over 2 E_i}\right)|{\bf p}_1 \cdots {\bf p}_n \rangle\langle 
	{\bf p}_1 \cdots {\bf p}_n|\nonumber\\
	&=& \sum_n\left(\prod_{i=1}^n \int {d^4 p_i 	
\over (2\pi)^3}\,\delta(p_i^2-m_i^2)\theta(p^0_i)\right)|{\bf p}_1 \cdots {\bf p}_n \rangle\langle 
	{\bf p}_1 \cdots {\bf p}_n|\nonumber\\
	&\equiv& \sum_f \int d\Pi_f |f\rangle\langle f|
\eeqa
When we eliminate $T$ in favor of ${\cal M}$, this equation takes the form
\beq
\label{unit3}
	-i\left( {\cal M}(q_i\to q_i') -
	{\cal M}^*(q_i'\to q_i)\right)
	= \sum_f \int d\Pi_f {\cal M}(q_i\to p_i) {\cal M}^*(q_i'\to p_i)
\eeq

Equation (\ref{unit3}) clearly requires the existence of an imaginary part of the
amplitude for the l.h.s.\ to be nonzero.  For example, at tree level in
$\lambda\phi^4$  theory, we have ${\cal M} = \lambda$, so the l.h.s.\ vanishes.
At one loop however, an imaginary part can occur at certain values of the external
momenta.  Recall the result (\ref{phi4loop2}):
\beq
{\cal M}(2\to 2) = \sum_{q^2=s,t,u} 
 {\lambda^2\over 32\pi^2}\left[ \ln(\Lambda^2/m^2) + 1-
	\sqrt{1-4m^2/q^2} \ln\left({\sqrt{1-4m^2/q^2}+1 \over\sqrt{1-4m^2/q^2}-1}
	\right) \right]
\eeq
This expression is rather ambiguous because it does not tell us which branch we
should take when the argument of the square root becomes negative.  This can be
determined by more carefully comparing to the original expressions before carrying
out the Feynman parameter integral, and paying attention to the $i\epsilon$
factors.  Nevertheless, it can be seen that the function $z\ln((z+1)/(z-1))$ is
purely real when $z$ is purely imaginary, which corresponds to $q^2 < 4 m^2$,
whereas it has both real and imaginary parts when $z$ is real, {\it i.e.}, when
$q^2 > 4 m^2$.   This function has the analytic behavior shown in fig.\ 43: a
branch cut develops in the complex $q^2$ plane at $4 m^2$.

\medskip
\centerline{\epsfxsize=2.5in\epsfbox{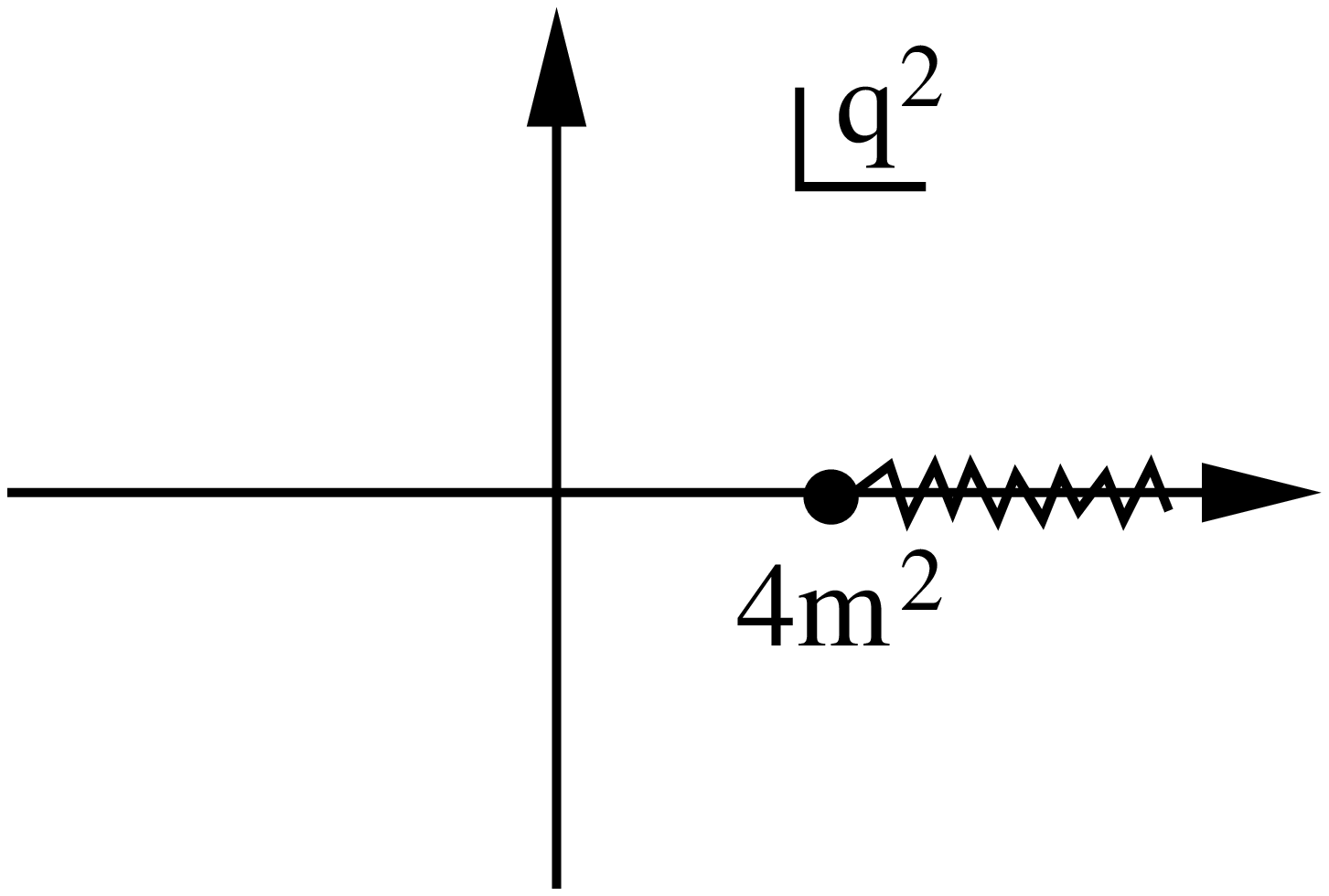}}
\centerline{\small Fig.\ 43. 
Analytic structure of the $2\to 2$ amplitude in $\lambda\phi^4$ theory in 
the complex $q^2$ plane.}
\medskip

The significance of the value $4m^2$ can be seen by considering the special case
where $q_i'=q_i$ in (\ref{unit2}).  Then ${\cal M}(q_i\to q_i)$ is the {\it 
forward scattering amplitude}, and the r.h.s.\ of (\ref{unit2}) is proportional
to the total cross section for $q_1,q_2\to$ anything, since the integrals are of
the same form as the phase space for the final states.  This is the 
{\bf optical theorem},
\beq
\label{optical}
	2{\rm Im}{\cal M}(q_i\to q_i) = \sqrt{(q_1\cdot q_2)^2 - m_1^2 m_2^2}
	\,\sigma_{\rm tot}(q_1,q_2\to{\rm\ anything})
\eeq
The factor $\sqrt{(q_1\cdot q_2)^2 - m_1^2 m_2^2}$ is often written as
$2 E_{\rm cm} p_{\rm cm} = \sqrt{s}  p_{\rm cm}$, and is related to the relative
velocity of the colliding particles by $v_{\rm rel} = 
\sqrt{(q_1\cdot q_2)^2 - m_1^2 m_2^2}/(E_1 E_2)$.  Notice that we are allowed
to evaluate the l.h.s.\ of the equation off-shell, for example at vanishing 
external momenta.  But in the case of $\lambda\phi^4$ theory, the cross
section does not exist for these unphysical values of the initial momenta, so
the r.h.s.\ is zero.  Only when we have enough energy in the initial state to
produce at least two real particles does the scattering cross section begin to
be nonzero.  This requires a minimum value of $s = (m+m)^2 = 4m^2$. 

The optical theorem (\ref{optical}) tells us something about the analytic
properties of amplitudes as a function of the kinematic variables.  For example
if we consider ${\cal M}(s)$ as a function of $s$, then in the region $s<4m^2$,
we can analytically continue ${\cal M}(s)$ away from the real axis.  Using the
identity ${\cal M}(s) = \left({\cal M}(s^*)\right)^*$ which holds in some
neighborhood of the region $s<4m^2$, and the fact that ${\cal M}(s)$ is an analytic
function of $s$ in this region, we can analytically continue to deduce that
${\cal M}(s) = \left({\cal M}(s^*)\right)^*$ holds everywhere in the complex $s$
plane.   Now if we go to the region $s>4 m^2$ and consider points just above
and below the real axis, $s \pm i\epsilon$, we see that 
\beq
	{\rm Re}{\cal M}(s+i\epsilon) = {\rm Re}{\cal M}(s-i\epsilon);\qquad
	{\rm Im}{\cal M}(s+i\epsilon) = -{\rm Im}{\cal M}(s-i\epsilon)
\eeq
There is a {\it discontinuity} in the imaginary part across the cut.

The $i\epsilon$ prescription of the Feynman propagators give us a nice way of
computing the imaginary parts of diagrams.  We use the identity
\beq
	{1\over x + i\epsilon} = {x -i\epsilon\over x^2 + \epsilon^2} 
	= P\left({1\over x}\right) - i2\pi \delta(x)
\eeq
where $P(1/x)$ denotes the principal value of $1/x$.
In the case of a bosonic propagator, this gives
\beq
\label{improp}
	{1\over p^2-m^2 + i\epsilon} 
	= P\left({1\over p^2-m^2}\right) - i2\pi \delta(p^2-m^2)
\eeq
One can show that the imaginary part of the amplitudes we have considered is
due to the second term in (\ref{improp}).  One should not be misled into
thinking that the imaginary part of the product of all propagators as per
(\ref{improp}) gives the imaginary part of the amplitude however.  The 
optical theorem tells us that {\it all} the internal lines must be
on shell.  Moreover, the delta function in (\ref{improp}) gets contributions
from both positive and negative energy states, whereas the optical theorem
tells us that the Im${\cal M}$ is associated with positive energy particles only.
Therefore it is not trivial to show the link between (\ref{improp}) and
Im${\cal M}$.  The result, which I shall not prove, but which follows from 
the optical theorem, is known as the Cutkosky rules: to find Im${\cal M}$,
make the following replacement on all the internal lines which are going on
shell:
\beq
	{i\over p^2-m^2 + i\epsilon} \to 2\pi \theta(p^0) \delta(p^2-m^2)
\eeq

Now let's relate this ideas to gauge theories.  Our concern is that unphysical
states of the gauge boson must not contribute to the imaginary part of an amplitude.
If they did, then unitarity would require them to also appear as external states
in tree-level processes.  For example, the optical theorem requires the relation
depicted in fig.\ 44 for scattering of two gauge bosons.  Notice that this
process did not arise in the Abelian theory.  Explicit calculations show that
in fact this relationship fails because the longitudinal and timelike polarizations
of the gauge fields do contribute to the imaginary part of the loop amplitude.
This is why the Faddeev-Popov ghosts are necessary in the nonAbelian theory:
their contribution subtracts the unphysical one from the loop amplitudes.

\medskip
\centerline{\epsfxsize=4.5in\epsfbox{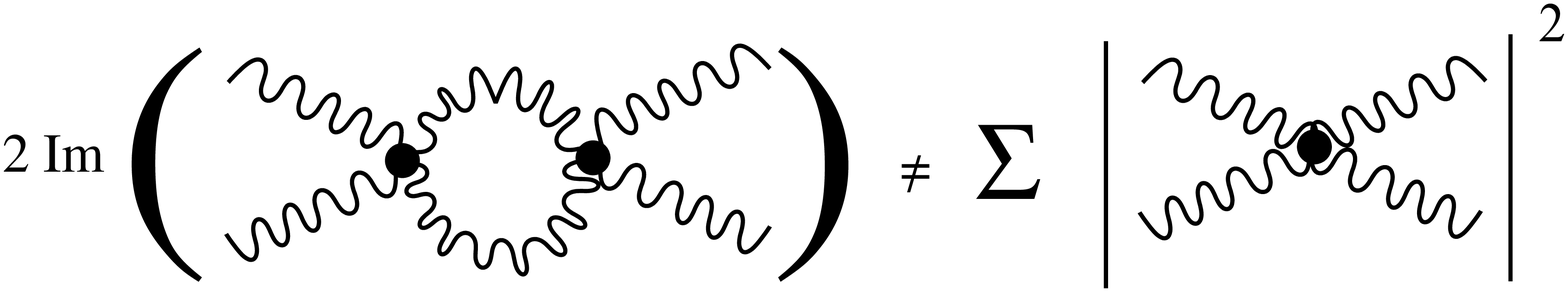}}
\centerline{\small Fig.\ 44. 
Apparent failure of unitarity in Yang-Mills theory.}
\medskip

\medskip
\centerline{\epsfxsize=5.5in\epsfbox{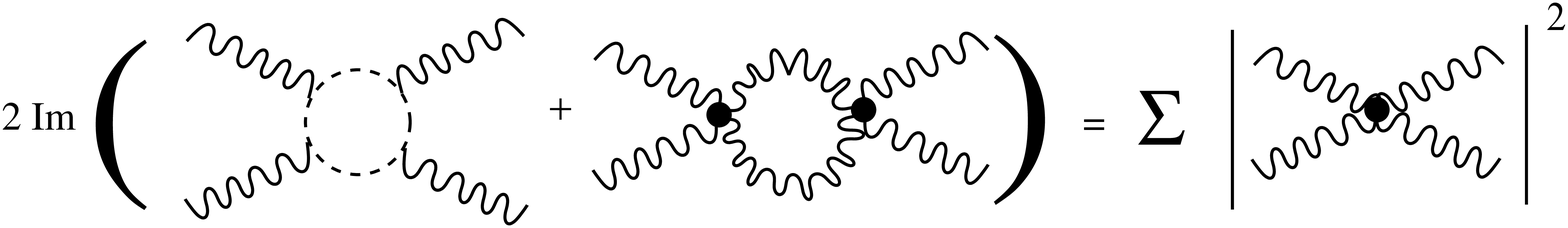}}
\centerline{\small Fig.\ 45. 
Restoration of unitarity by Faddeev-Popov ghost loop contribution.}
\medskip

\subsection{Loop structure of nonAbelian gauge theory}

Now that have shown the importance of the ghosts for unitarity, let's return
to the renormalization of the nonAbelian theory.  Now there are more
counterterms, because of the new cubic, quartic, and ghost interactions of the
gauge bosons.  These new terms also make the analog of the QED Ward identities,
called the Slavnov-Taylor identities, much more complicated than in QED.
First, let's consider the form of the renormalized Lagrangian.  If we write
out explicitly the new gauge interactions, it has the form
\beqa
	{\cal L}_{\rm ren} &=& Z_2 \bar\psi  i\dsl \psi - mZ_m \bar\psi\psi
	+ g\mu^\epsilon Z_1\bar\psi\Asl\psi + \frac{1}{4}Z_3(\partial_\mu A_\nu
	-\partial_\nu A_\mu)(\partial^\mu A^\nu
	-\partial^\nu A^\mu)\nonumber\\ &-& Z_4 g \mu^\epsilon f^{abc}A^\mu_a A^\nu_b
	\partial_\mu A^c_\nu +\frac12 Z_5 g^2 \mu^{2\epsilon}f^{abc} f^{ade}
	A_\mu^b A_\nu^c A^\mu_d A^\nu_e 
	+ \frac12 Z_\alpha (\partial_\mu A^\mu)^2 \nonumber\\
	&+& Z_6 \partial_\mu \bar c^a \partial^\mu -\frac12 Z_7 g\mu^\epsilon f^{abc}
	A^c_\mu \bar c^a{\lrp} c^b 
	-\frac12 Z_8 g\mu^\epsilon f^{abc}
	\partial^\mu A^c_\mu \bar c^a c^b
\eeqa
Note that we have suppressed color indices in some places.
In the case of QED, we were able to simplify some calculations due to the relation
$Z_1=Z_2$ which came from the Ward identities.  In QCD this is no longer possible;
the ghost Lagrangian presents a new source of gauge noninvariance when we try
to carry out the same path integral argument that led to the U(1) Ward identities.
The problem can be seen by looking at the form of the ghost Lagrangian; at tree
level it can be written as
\beq
\label{ghostlag}
	{\cal L}_{\rm ghost} = \bar c^a \partial_\mu D^\mu_{ab} c^b 
\eeq
where $D^\mu_{ab}$ is the covariant derivative in the adjoint representation.
The adjoint is the representation under which the gauge bosons themselves
transform, where the generators $T^a_{ij}$ which we introduced for the fermions
are replaced by new generators $t^a_{bc} = if^{abc}$, which are the structure
functions themselves:
\beq
D^\mu_{ab} c^b = \left(\delta_{ac}\partial_\mu + g f^{abc}A_\mu^b\right) c^b
\eeq
The expression (\ref{ghostlag}) is ``almost''
gauge invariant, but one of the derivatives is the ordinary one, not covariant.
So even if we let the ghosts transform under $SU(N)$ in the same way as the
fermions, which is suggested by the fact that they carry the same kind of index,
this term is not invariant.   Indeed, if the gauge fixing condition is
$G^a(A) = 0$, then the gauge-fixing term $\frac{1}{2\alpha}G^a G^a$ and the
ghost Lagrangian transforms to themselves plus the following piece that 
expresses the lack of invariance under an infinitesimal gauge transformation
$\omega^a$:
\beq
\label{deltagauge}
	\delta{\cal L}_{{\rm g.f.}+ {\rm ghost}} = \frac{1}{\alpha} 
	{\delta G^a\over \delta A^b_\mu} D_\mu^{bc}\omega^c 
	+\delta\bar c^a {\delta G^a\over \delta A^b_\mu}(D_\mu c)^b
	+  \bar c^a {\delta G^a\over \delta A^b_\mu}\delta(D_\mu c)^b
\eeq

Notice that we are not forced to assume that the ghosts transform in exactly the
same way as the fermions in (\ref{deltagauge}), so we have left $\delta\bar c$ and
$\delta c$ unspecified.  In fact, it is possible to choose them in a clever way
that makes $\delta{\cal L}_{{\rm g.f.}+ {\rm ghost}}$ vanish.  This was discovered
by Becchi, Rouet and Stora and around the same time by Tyutin, and is known as
the BRST symmetry.  This elegant technique is not only helpful for proving the 
renormalizability of Yang-Mills theory, but also has applications in string theory.
The symmetry requires the gauge transformation to depend on the ghost itself:
\beq
	\omega^a = \zeta c^a;\qquad \delta \bar c^a = -\frac{1}{\alpha}G^a \zeta
\eeq
where now $\zeta(x)$ must also be a Grassmann-valued field so that $\omega^a$ can be a
$c$-number (an ordinary commuting number).  The motivation for this choice
is that now the first two terms of (\ref{deltagauge}) cancel each other.  Now
we only need to define $\delta c^a$ in such a way that $\delta(D_\mu c)^b$
vanishes.  One might think that we no longer have the freedom to do this since
the transformation of $\bar c^a$ has been defined, but in fact it is kosher to
treat $\bar c$ and $c$ as independent fields in the path integral.  (This can
be proven by explicitly writing everything in terms of the real and imaginary
parts of $c$ and $\bar c$.)  It can be verified that the magic form of the
transformation of $c$ is 
\beq
	\delta c^b = \frac12 \zeta f^{bce} c^c c^e
\eeq

We now have a new continuous symmetry of the gauge-fixed Lagrangian which can
be used to find relations between amplitudes in the nonAbelian theory.  In terms
of the effective action, these relations (the Slavnov-Taylor identities) can
be written by introducing new source terms which couple to the changes that 
are induced in the fields under the BRST transformations:
\beq
	{\cal L}_{\rm source} = \tau^a_\mu (D_\mu c)^a /g + u^a f^{abc}
	c^b c^c /2 + \bar\lambda c^a T^a\psi +\bar\psi T^a c^a \lambda
\eeq
The new sources $\tau^a_\mu$, $u^a$, $\bar\lambda$ and $\lambda$ are analogous to
the source terms $J^\mu A_\mu $ for the gauge field or $J\phi$ for a scalar field.
Notice that $\lambda$ and $\bar\lambda$ are fermionic source terms since the couple
to the change in the fermionic field $\psi$.  The utility of these source terms
is that if we compute the effective action as a function of the new sources and
functionally differentiate it with respect to the sources, we get the expressions
for the changes in the classical fields with respect to the BRST transformations.
The Slavnov-Taylor identities can then be written as
\beq
	\intdx\left({\delta\Gamma\over \delta A^b_\mu }{\delta\Gamma\over \delta \tau_b^\mu }
	- {1\over \alpha g}\partial_\mu A^\mu_b 
	{\delta\Gamma\over \delta \bar c^b}-
	{\delta\Gamma\over \delta c^b }{\delta\Gamma\over \delta u^b }
	+i{\delta\Gamma\over \delta \psi}{\delta\Gamma\over \delta\bar \lambda }
	-i{\delta\Gamma\over \delta \bar\psi}{\delta\Gamma\over \delta\lambda }
	\right) = 0
\eeq
Although it may not be obvious, these identities actually reduce to the Ward
identities in the $U(1)$ case.  Essentially, each functional relation contained
in the Ward identities is also contained in the Slavnov-Taylor relations, except
multiplied by a power of the ghost fields.  So instead of looking for a term which
is of order $\psi^0$, $\bar\psi^0$, $A^1$, we would look for the same term but also
of order $c^1$ in the ghost field.  For example, ${\delta\Gamma\over \delta
\tau_b^\mu }$ gives us $\partial_\mu c^b$ at lowest order in perturbation theory,
whereas ${\delta\Gamma\over \delta \bar c^b}$ gives us $\partial^2 c^b$.
Thus the terms of order $A^1$, $c^1$ are
\beq
	\intdx\left( {\delta\Gamma\over \delta A^b_\mu}\partial_\mu c^b
	- {1\over \alpha g}\partial_\mu A^\mu_b \partial^2 c^b\right)
\eeq
which agrees with the QED Ward identity (\ref{wardid}).

These identities are still quite complicated (but they would have been even
worse had we tried to work with the broken gauge symmetry rather than the
BRST transformations).  They can be used to prove the following relations between
the renormalization constants:
\beq
\label{ymrel}
	{Z_1\over Z_2} = {Z_4\over Z_3} = {\sqrt{Z_5}\over\sqrt{Z_3}}
	= {Z_7\over Z_6} = {Z_8\over Z_6}
\eeq
In particular, we no longer have $Z_1 = Z_2$, due to the new complications from the
ghosts.  This means that we can no longer simplify the relation between the
bare and renormalized couplings,
\beq
	g_0 = g\mu^\epsilon {Z_1\over Z_2 Z_3^{1/2}}
\eeq
to get away with computing only $Z_3$ as we did in QED.  Rather, we are forced to
compute all the diagrams contributing to the fermion vertex and wave function
renormalization, in addition to the new ghost loop contributions to the vacuum
polarization.  There is one exception to this statement: in the {\it axial gauge}
\beq
	n_\mu A^\mu_b = 0
\eeq
for some fixed vector $n_\mu$, the ghosts have no interaction with the gauge bosons,
and they can be ignored once again.  This can be seen by considering the form
of the determinant:
\beqa
\Delta_{\rm FP}[A] &=& \det{\partial f^a\over\partial A^c_\mu}{\delta(\delta A^c_\mu)_x\over \delta(\omega^d_y)}
= \det\left|n_\mu( \delta_{ad}\partial_\mu - gf^{dba}A_\mu^b) \delta(x-y)\right|
\nonumber\\
&=& \det\left|n_\mu( \delta_{ad}\partial_\mu ) \delta(x-y)\right|
\eeqa
We used the fact that this determinant is multiplied by $\delta[n_\mu A^\mu_a]$
in the path integral to set $n_\mu A^\mu_a=0$.  
 The problem with this gauge is that the gauge boson
propagator is so complicated that it hardly makes life easier,
\beq
	D_{\mu\nu}^{ab}(p) = -{\delta_{ab}\over p^2}\left(\eta_{\mu\nu}
	-{1\over n\cdot p}(n_\mu p_\nu + n_\nu p_\mu) - 
	{p_\mu p_\nu\over (n\cdot p)^2} (\alpha p^2 - n^2)\right)
\eeq

There is a way we could have guessed the relations (\ref{ymrel}) without
going through all the machinery of the Slavnov-Taylor relations.  When we rewrite
the renormalized Lagrangian in terms of the bare fields, it takes the form
\beqa
\label{ymrenlag}
	{\cal L}_{\rm ren} &=& \bar\psi  i\dsl \psi - m\bar\psi\psi
	+ g_0\bar\psi\Asl\psi + \frac{1}{4}(\partial_\mu A_\nu
	-\partial_\nu A_\mu)(\partial^\mu A^\nu
	-\partial^\nu A^\mu)\nonumber\\ &-& g_0^{(1)}f^{abc}A^\mu_a A^\nu_b
	\partial_\mu A^c_\nu +\frac12 (g_0^{(2)})^2f^{abc} f^{ade}
	A_\mu^b A_\nu^c A^\mu_d A^\nu_e 
	+ \frac1{2\alpha_0} (\partial_\mu A^\mu)^2 \nonumber\\
	&+& \partial_\mu \bar c^a \partial^\mu -\frac12 g^{(3)} f^{abc}
	A^c_\mu \bar c^a{\lrp} c^b 
	-\frac12  g^{(4)} f^{abc}
	\partial^\mu A^c_\mu \bar c^a c^b
\eeqa
For clarity I have omitted the subscript ``0'' on all the bare fields in
(\ref{ymrenlag}).
The various bare couplings are related to the renormalized one by
\beqa
\label{ymcoup1}
	g_0 &=& g\mu^\epsilon {Z_1\over Z_2 Z_3^{1/2}}\\
\label{ymcoup2}
	g_0^{(1)} &=& g\mu^\epsilon {Z_4\over Z_3^{3/2}}\\
\label{ymcoup3}
	g_0^{(2)} &=& g\mu^\epsilon {Z_5^{1/2}\over Z_3}\\
\label{ymcoup4}
	g_0^{(3)} &=& g\mu^\epsilon {Z_7\over Z_6 Z_3^{1/2}}\\
\label{ymcoup5}
	g_0^{(4)} &=& g\mu^\epsilon {Z_8\over Z_6 Z_3^{1/2}}\\
\eeqa	
If all these bare couplings are equal to each other, then we can rewrite
the whole Lagrangian in terms of covariant derivatives and field strengths, 
so that its form is gauge invariant except for the gauge fixing and ghost
term.  And even though the ghost term is not invariant, it takes its original
tree level form, $\bar c^a \partial_\mu (D^\mu c)_a$.  These equalities are
insured if the renormalization constants obey the relations (\ref{ymrel}).

\subsection{Beta function of Yang-Mills theory}

If we want to show that QCD is asymptotically free, the above arguments indicate
that we have to compute the divergent parts of all of the following diagrams.
For $Z_2$, the fermion wave function renormalization, there are no new kinds
of diagrams; the self-energy diagram is the same as fig.\ 37 for QED.  The only
difference is that there is a group theory factor from the vertices:
\beqa
{{}\atop{\epsfxsize=2.0in\epsfbox{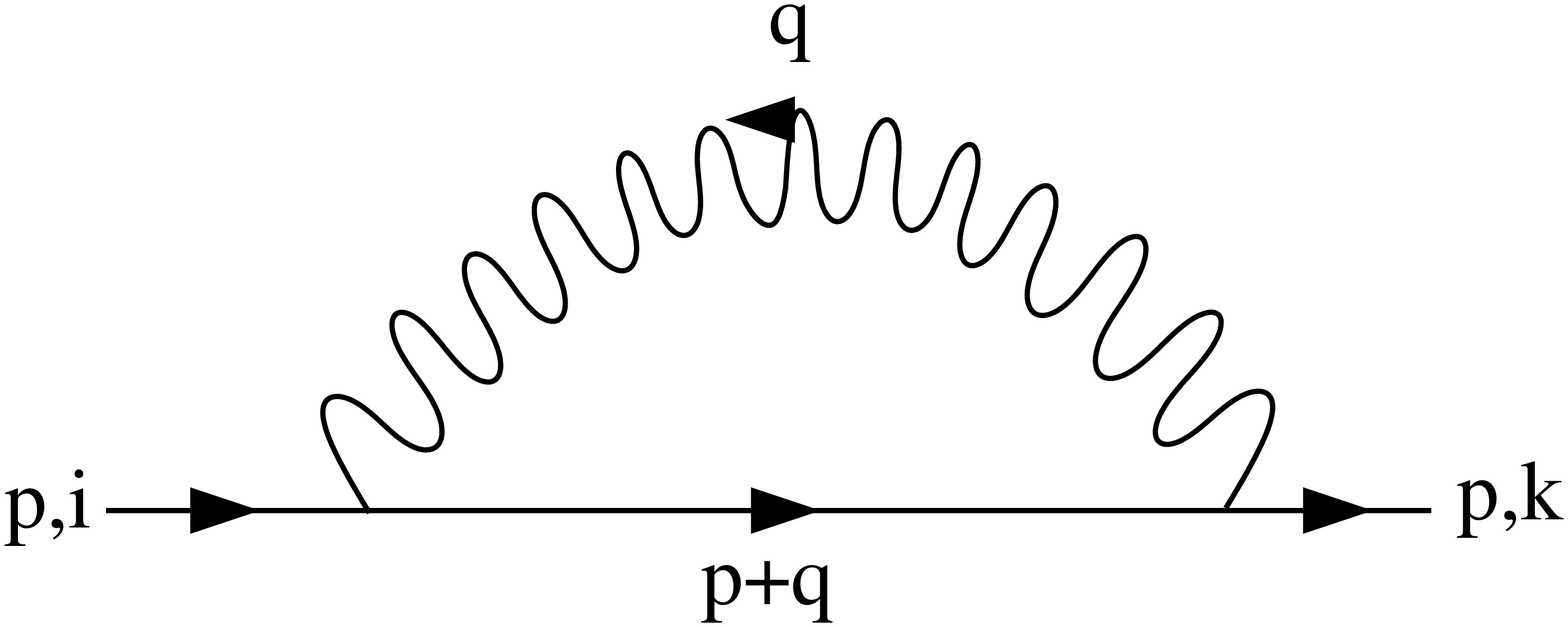}}} &=& T^a_{ij} T^a_{jk}\times
	(\hbox{QED result})\nonumber\\
	&=& {N^2-1\over 2N} \delta_{ij}\times
	(\hbox{QED result})
\eeqa
where we used the result (\ref{tataid}) for $T^a_{ij} T^a_{jk}$.

For the vertex correction, we have the diagram similar to fig.\ 39 for
QED, but in addition the new contribution of fig.\ 46. 

\medskip
\centerline{\epsfxsize=2.5in\epsfbox{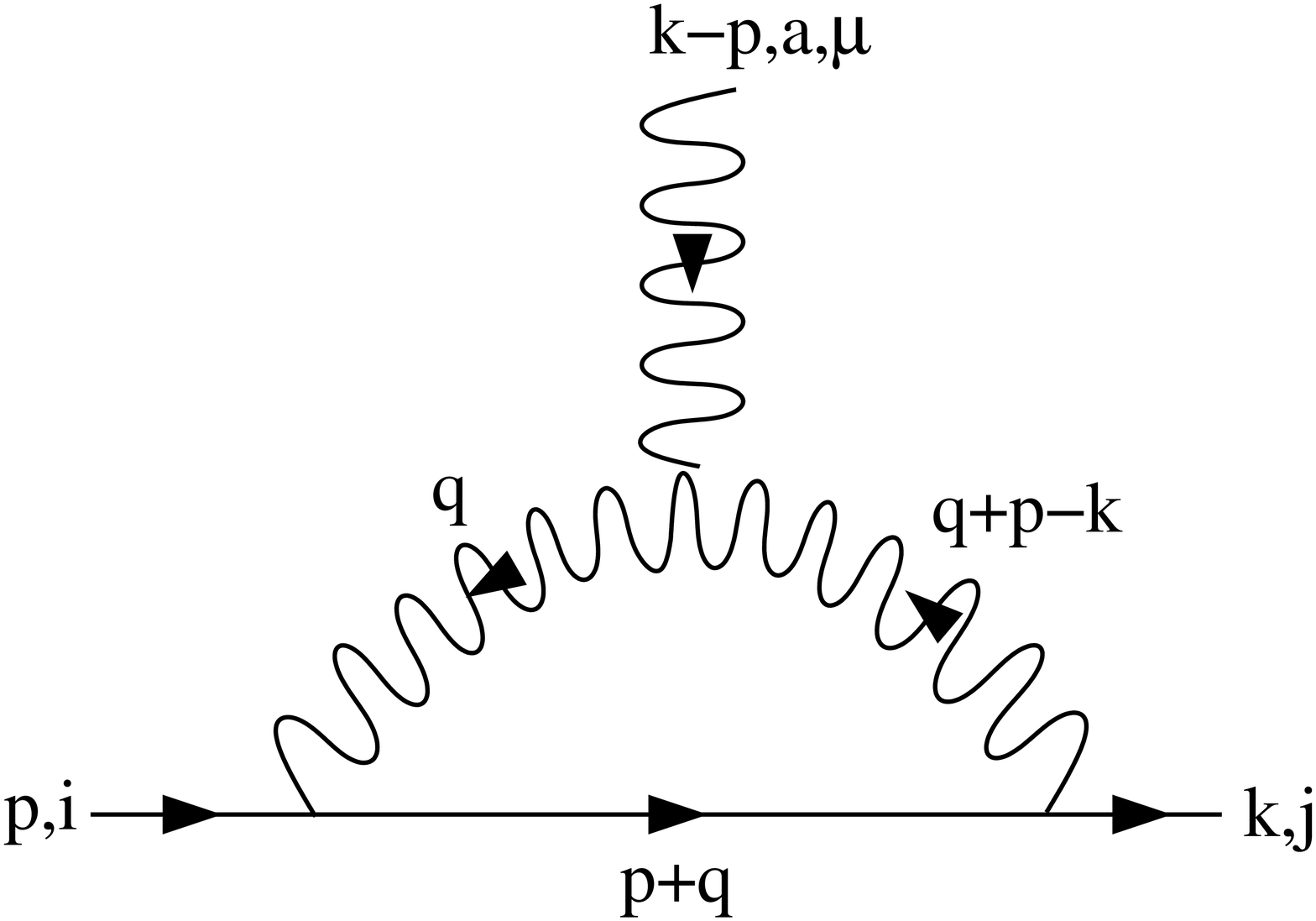}}
\centerline{\small Fig.\ 46. 
New vertex correction in nonAbelian theory.}
\medskip

I will not go through the computation of this new diagram.  However, it is rather
straightforward to compute the one which is similar to QED,
\beqa
{{}\atop{\epsfxsize=2.0in\epsfbox{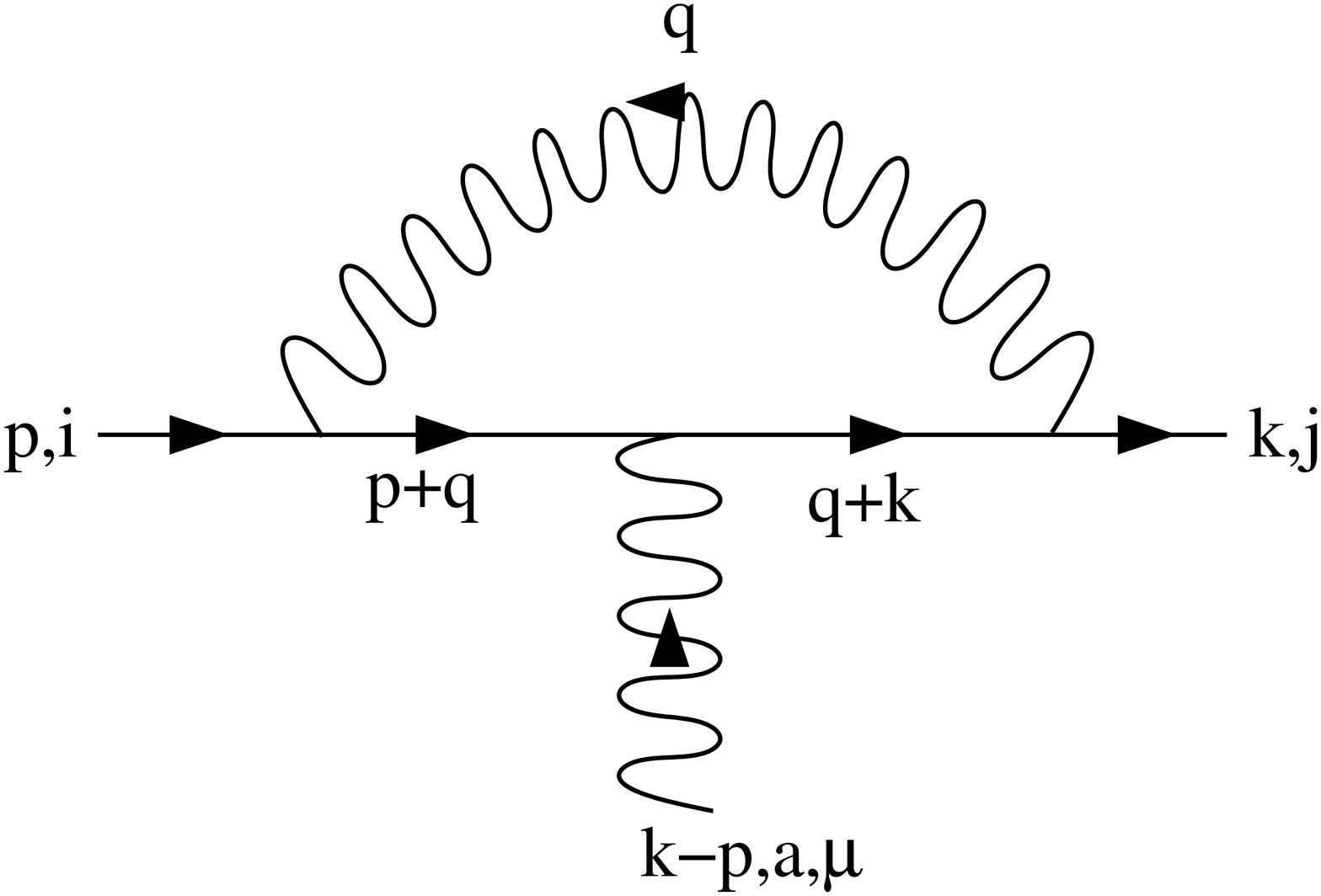}}} &=& T^b_{il} T^a_{lm} T^b_{mj}\times
	(\hbox{QED result})
\eeqa
We can write the group theory factor as
\beqa
	T^b T^a T^b = [T^b, T^a] T^b + T^a T^b T^b &=& 
	if^{bac} T^c T^b + {N^2-1\over 2N} T^a \nonumber\\
	&=& \frac{i}{2} f^{bac} [T^c, T^b] + {N^2-1\over 2N} T^a \nonumber\\
	&=& -\frac12 f^{bac} f^{cbd} T^d + {N^2-1\over 2N} T^a \nonumber\\
	&=& -\frac12 N T^a + {N^2-1\over 2N} T^a
\eeqa
We used the identity (\ref{ffid}) to obtain the last line.  (The minus sign
in $-\frac12 N T^a$ is erroneously missing in Ramond (6.52) and (6.53).)

The biggest complications in the computation of the beta function are all the
new diagrams contributing to the vacuum polarization, shown in fig.\ 47.
These have no counterpart in QED.

\medskip
\centerline{\epsfxsize=5.5in\epsfbox{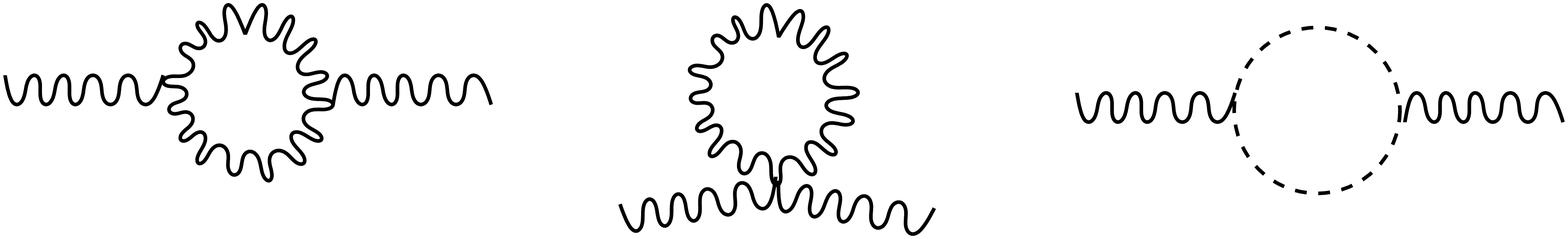}}
\centerline{\small Fig.\ 47. 
New contributions to the vacuum polarization in nonAbelian theory.}
\medskip

In the preceding description, we assumed that we would use the result
(\ref{ymcoup1}) to find the relation between the bare and the renormalized
gauge coupling; this was the procedure most analogous to QED.
But we could have used any of the other relations (\ref{ymcoup2}-\ref{ymcoup5})
to get the same result.  For example, it looks like using (\ref{ymcoup2})
or (\ref{ymcoup3}) could save some work because then we would only have to
compute two renormalization constants instead of three.  The new diagrams that
would be needed to determine $Z_4$ or $Z_5$ are the renormalization of the
cubic and quartic gauge coupling vertices, shown in figures 48 and 49.  It is
clear that, even though there are fewer constants, there are more diagrams,
so we don't save any work by doing thing this way.

\medskip
\centerline{\epsfxsize=3.0in\epsfbox{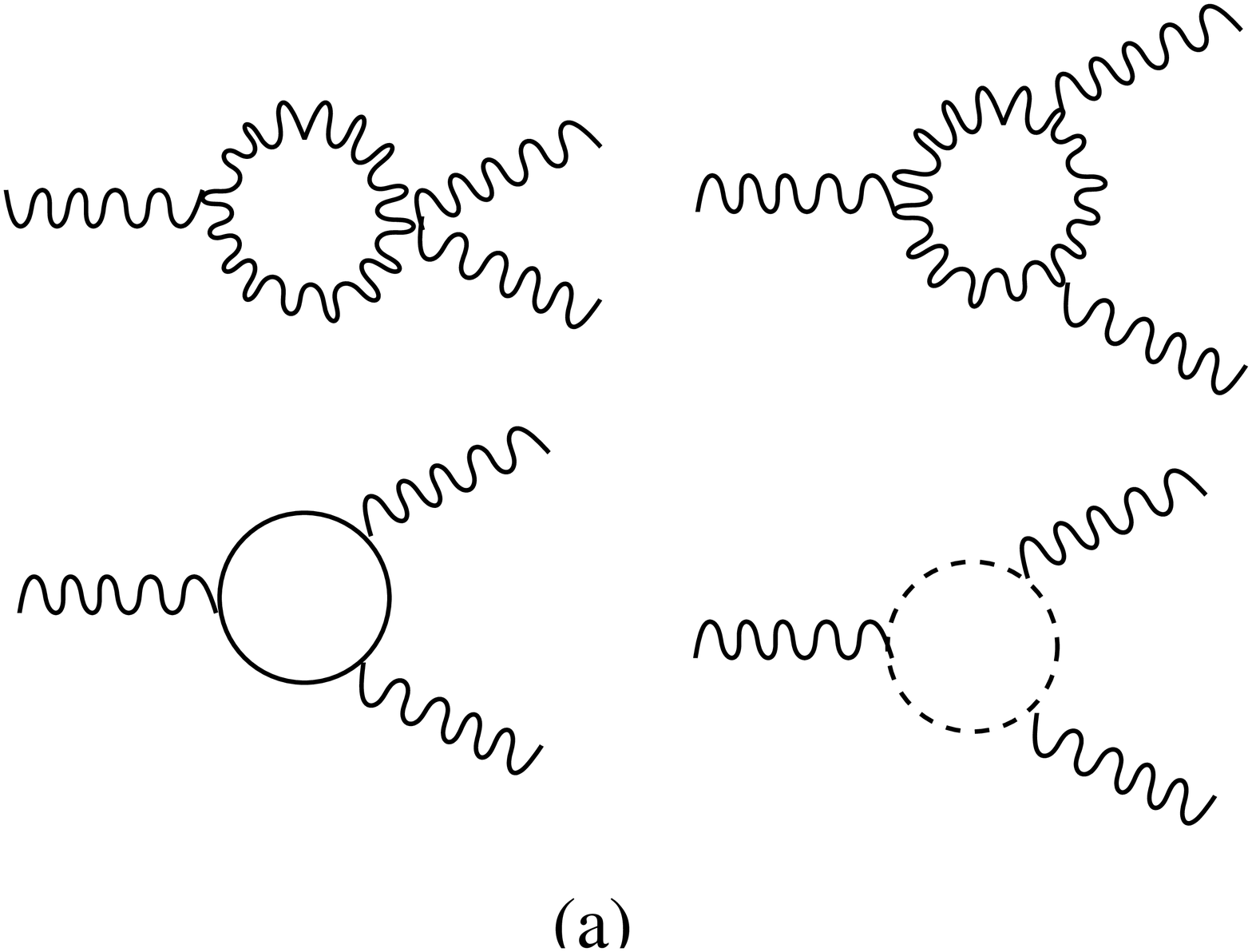}\hfill\epsfxsize=3.0in\epsfbox{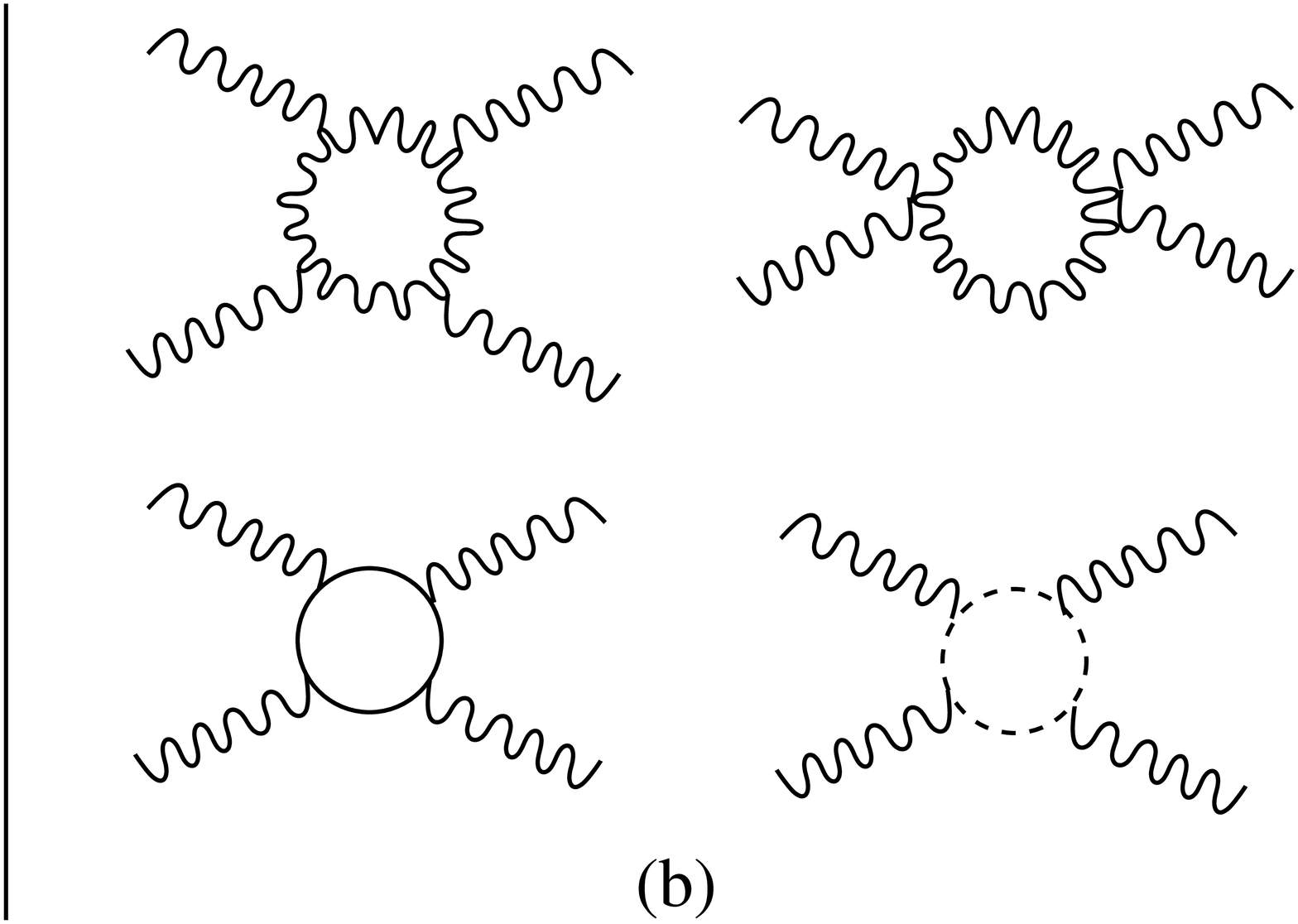}}
\centerline{\small Fig.\ 48. 
Contributions to renormalization of (a) cubic and (b) quartic gauge couplings.}
\medskip

It is remarkable that these very different and complicated ways of computing the renormalized
coupling give the same answer, thanks to the gauge symmetry.  The result for the
beta function can be expressed in a way which is applies to gauge groups which
are more general than $SU(N)$,
\beq
\label{QCDbeta}
	\beta(g) = -{g^3\over 16\pi^2}\left( {11\over 3}C_2(A)
	-\frac43 n_f C(f) - \frac13 C(s)\right)
\eeq
where $n_f$ is the number of flavors of fermions, and the numbers $C_2$ and
$C$ are group theoretic factors which are a function of 
the representation of the gauge group under which the argument transforms.
I have also shown the additional effect we would get from  enlarging our theory
to have complex scalar fields which are charged under the gauge group.  They
contribute just like fermions, except we have assumed the fermions are Dirac
particles with 4 degrees of freedom, so each fermion carries as much weight
as 4 complex scalars. 
$C_2(A)$ is the {\it quadratic Casimir invariant} for the gauge fields, which
transform in the adjoint representation.  This invariant is defined by the
identity
\beq
\label{quadcas}
	\sum_a T^a_{R} T^a_{R} \equiv C_2(R) I_R
\eeq
where $T^a_{R}$ is the group generator in the representation $R$ and
$I_R$ is the unit matrix which has the same dimensions as the generators.  Even if
you have not taken a course in group theory, you are familiar with the fact that
the matrix elements of the angular momentum operators form matrices whose
dimension is $(2l+1)\times (2l+1)$ for states with angular momentum $l$.  It
so happens that for the rotation group $SO(3)$, since there are three generators,
the adjoint representation coincides with the fundamental representation,
since 3-vectors are the fundamental objects which are transformed by $SO(3)$.
But in general the adjoint and fundamental representations are different.
From the identities (\ref{tataid}) and (\ref{ffid}) we deduce that for $SU(N)$,
\beq
	C_2(R) = {N^2-1\over 2N},\qquad C_2(A) = N
\eeq
The other group invariant is $C(R)$, called the {\it Dynkin index} for the
representation $R$, and defined by
\beq
\label{dynkin}
	\tr\left(T^a_{R} T^b_{R}\right) = C(R) \delta_{ab}
\eeq
It is easy to find a relation between $C(R)$ and $C_2(R)$; simply take the
trace over generator indices in (\ref{quadcas}) and over
$ab$ in (\ref{dynkin}), and equate the two left-hand sides.
 Since $\delta_{ab} = d(A)$, the 
dimension of the adjoint representation (which is just the number of
generators of the group) and $\tr I_R = d(R)$, the dimension of the 
representation $R$, we get
\beq
	C_2(R) d(R) = C(R) d(A)
\eeq
For fermions in the fundamental representation of $SU(N)$ this gives
\beq
	C(f) = {N^2-1\over 2N} \times N / (N^2-1) = \frac12
\eeq
Putting these together and ignoring the effect of possible scalar fields,
 we get the beta function in $SU(N)$,
\beq
\label{ymbetafn}
	\beta(g) = -{g^3\over 16\pi^2}\left( {11\over 3}C_2(A)
	-\frac43 n_f C(f) \right)=  -{g^3\over 16\pi^2}\left( 11
	-\frac23 n_f  \right)
\eeq
It is clear that the new term which makes this negative (if $n_f$ is not 
too large!) is due to the gauge bosons.  Roughly speaking, the factor
$C_2(A)$ counts how many gauge bosons are going around the new loops
due to gauge boson self-interactions, whereas $n_f C(f)$ counts the contributions
from fermion loops.  If we set $C_2(A)=0$ we can recover our old QED result by
taking $n_f =1$ and 
identifying $g = \sqrt{2}e$.  The reason for the factor of 2 is the conventional way
in which generators of $SU(N)$ are normalized: $\tr T_a T_b = \frac12 \delta_{ab}$.
In QED we take $T = 1$, not $T=1/\sqrt{2}$.  

Although the fermions weaken asymptotic freedom, in the real world of QCD there
are only 6 flavors of quarks.  We would need 17 flavors to change the sign of 
the beta function.

\subsection{Heuristic explanation of asymptotic freedom}

We see from (\ref{ymbetafn}) that there is a competition between the gauge bosons and the fermions 
for determining the sign of the beta function.  If we add enough flavors of
fermions, they win and spoil the asymptotic freedom of the theory.  We know that
fermions have this effect in QED, because their vacuum fluctuations screen the
bare charge of the electron at long distances, causing the coupling to become
weaker at low energy.  Apparently the gauge bosons in the nonAbelian theory
have the effect of {\it antiscreening} a bare color charge.  Can we understand
why this happens?  Peskin has a nice explanation (section 16.7 of \cite{Peskin}),
but I will instead give one found in \cite{Preskill}, originally found in papers
of N.K.\ Nielsen (1980) and R.\ Hughes (1981).

In the following discussion we will make use of our intuition from ordinary
electromagnetism.  In the Yang-Mills theory it is also possible to talk
about electric and magnetic fields,
\beq
	E_i^a = F_{0i}^a, \quad B_i^a = \frac12 \epsilon_{ijk} F_{jk}^a,
\eeq
but now there are $N^2-1$ components labeled by the group index.  Without
the new self-interactions of the gauge bosons, this would just be $N^2-1$
copies of QED.   So for the interaction of the gauge bosons with fermions,
at lowest order in the coupling, we can use our intuition from QED.  Of course
all the new effects will come from the gauge bosons' interactions with themselves.

We can think of the beta function as telling us about the
energy dependence of the dielectric constant for a bare charge, due to
fluctuations of the vacuum:
\beqa
	g^2(E) &=& g_0^2 / \epsilon(E)\nonumber\\
		&=& {g_0^2\over 1- 2 b_0 g_0^2 \ln(E/\Lambda)} 
\eeqa
where I am using $E$ for the renormalization scale and $\Lambda$ for the
cutoff.  The constant $b_0$ is related to the beta function by
\beq
	\beta(g) = b_0 g^3 + \cdots
\eeq
We can thus identify the ``running electric permeability'' as
\beq
	\epsilon(E) = 1- 2 b_0 g_0^2 \ln(E/\Lambda)
\eeq
In QED we know that $\epsilon(E)>1$ always, corresponding to screening.
Apparently in QCD we have antiscreening, corresponding to $\epsilon(E)<1$.

Instead of thinking about the electric polarizability of the vacuum, we could also
consider the magnetic permeability $\mu$, since
\beq
	\mu \epsilon = c^2 = 1
\eeq
Therefore, if we can understand why $\mu > 1$ for nonAbelian theories, this 
will explain why $\epsilon<1$.  In fact, the coefficient of the beta function 
is very directly related to the {\it magnetic susceptibility},
\beq
\label{chibrel}
	\chi(E) = 1 - 1/\mu(E) = -2 b_0 g_0^2 \ln(\Lambda/E)
\eeq 
Thus the sign of the beta function is the negative of the sign of the
susceptibility.

We are quite familiar with an effect that leads to $\mu>1$ in ordinary
electromagnetism: place a piece of iron inside a solenoid, and it increases
the magnetic field by a factor $\mu>1$, due to the fact that the spin magnetic
moments of the iron atoms like to align with the external field, thus increasing
the total field.  This effect is known as {\it paramagnetism}.  However, there is
another effect which leads to $\mu<1$: if a charged particle undergoes cyclotron
motion around the magnetic field lines, it creates a current loop whose magnetic
dipole moment {\it opposes} the direction of the external field and causes a 
decrease in the net field, leading to $\mu<1$.  This effect is called {\it
diamagnetism}.  Clearly, if we have quantum fluctuations of virtual fermions and
gauge bosons in the vacuum (in an external magnetic field), both effects are
possible.  It becomes a quantitative question of which one is stronger.
Apparently, diamagnetism must be the stronger effect for fermions, since otherwise
we would get $\mu > 1$ and $\epsilon<1$ even in QED.  So if this is the correct 
explanation, it must be because paramagnetism is the bigger effect for the
nonAbelian gauge bosons.   In QED they were irrelevant because of the fact that
photons don't couple to themselves, so they can't act as a source of magnetic
field.  But in QCD, a vacuum fluctuation of the gauge bosons can act as a source
of magnetic field, since the gauge field equation of motion (ignoring fermionic
sources) is
\beq
	\partial^\mu F_{\mu\nu} + ig[A^\mu,F_{\mu\nu}] = 
	\partial^0 F_{0\nu} + \partial^i F_{\mu i} + ig[A^0,F_{0\nu}]
	+ ig[A^i,F_{i\nu}] = 
	0;
\eeq
this is the generalization of Maxwell's equations to the nonAbelian case.
To find the magnetic part, take $\nu=j$,
\beq
	\partial_0 E_j + \partial^i \epsilon_{jik} B_k 
	+ ig[A^0,E_j] + ig[A^i,\epsilon_{jik} B_k] = 0
\eeq
We see that if the background gauge fields have $E_j=0$ but $B_k\neq 0$,
then a vacuum fluctuation of $A$ can give rise to a source of color
magnetic field due to the last term.

To compute the beta function from this argument, we will use another result from
classical electromagnetism: the self-energy of a magnetic field configuration in 
a magnetic medium is given by
\beq
\label{Umag}
	U = -\frac12 \chi B^2
\eeq
We will compute the correction to $U$ due to the fluctuations of the vacuum,
and from this infer the beta function.  The argument is similar in spirit to
our intuitive explanation of the axial anomaly.  We want to see how the energy
levels of the quantum states of the theory are affected by a background magnetic
field, due to the interactions of the states with the field.

First let's compute the diamagnetic contribution.  If the $B$ field is in the
$\hat z$ direction, then the energy eigenstates of a charged particle
are the Landau levels, 
\beq
\label{landau}
	E^2 = p_z^2 +m^2 + (2n+1) eB
\eeq
with $n\ge 0$, and $e=g/\sqrt{2}$, according to the observation below
(\ref{ymbetafn}).  Here we should imagine that the $B$ field is pointing
in one particular direction in the adjoint representation 
color space, say $a$, so the fermion or gauge boson 
which has the Landau levels is some 
linear combination associated with the generator $T^a$.  Because of gauge 
invariance, it should not matter which generator we choose, so for convenience
consider one of the diagonal generators, like $T_3$ for the fermions.  Then
we see that fermions with color indices $\psi_i$, $i=1,2$ couple to this external field,
while $i=3$ does not.  Since the effect we are looking for in (\ref{Umag})
is quadratic in $B$, it will not matter that $i=1$ and $i=2$ couple with different
signs.  According to this argument, the effect we are looking for will be
proportional to the trace of the square of the generator, and this is the same
for all generators, so this supports our claim that the choice of the $B$ field's
direction in color space will not affect our answer.  Note that the extra factor
of two we get here will cancel the factors of $1/\sqrt{2}$ which arise from the
relation $e=g/\sqrt{2}$, so that we could have done the whole computation ignoring
the distinction between $e$ and $g$ if we had also forgotten about counting how
many fermions the external field couples to.

Now we must compute the energy of the vacuum due to the magnetic interaction
in (\ref{landau}).  In quantum mechanics we learn that the zero-point energy
of a harmonic oscillator is $\frac12\hbar\omega$.  In quantum field theory,
each energy eigentstate of the field acts like a harmonic oscillator whose zero-point
energy is simply $\omega =\sqrt{p_z^2 + m^2+(2n+1) eB}$, and the excited states
have energy $(N+\frac12)\hbar\omega$, where the $N$ here refers to the number of
particles having energy $\omega$, not to be confused with the Landau level of a 
given eigenstate.  The zero-point energy density is given by the sum over all possible
states,
\beq
	U_{\rm dia} = \frac12 \sum_n g_n \int {dp_z\over 2\pi} \sqrt{p_z^2+m^2 + (2n+1) eB}
\eeq
where $g_n$ is the degeneracy of Landau levels per unit area of the $x$-$y$
plane.  We can easily deduce the value of $g_n$ by taking the limit $B\to
0$, since then $U_{\rm dia}$ must reduce to the usual expression in the absence
of any magnetic field,
\beq 	
	U_{\rm dia} \to \frac12 \int {d^{\,3}p\over(2\pi)^3}\sqrt{p_z^2+m^2 }
\eeq
If we identify $(2n+1) eB = p_x^2 + p_y^2 \equiv \rho^2$, and $\int dp_x
dp_y = 2\pi \int d\rho \rho$, then $2\rho\Delta\rho = 2n eB$, and $\sum_n g_n 
= 2\pi \sum \rho\Delta\rho g_n / (2\pi eB) \to \int dp_x
dp_y g_n / (2\pi eB)$.  Comparison shows that therefore 
\beq
	g_n = {eB\over 2\pi}
\eeq
and 
\beq
\label{diamag}
	U_{\rm dia}(B) = \frac12 \sum_n {eB\over 2\pi}
 \int {dp_z\over 2\pi} \sqrt{p_z^2+m^2 + (2n+1) eB}
\eeq

The sum (\ref{diamag}) is obviously quite divergent in the UV, but we don't 
care about the
most divergent part; we only care about the difference $U(B)-U(0)$
since the true vacuum contribution $U(0)$ is unobservable.  To evaluate
this difference, we can use the following identity:
\beqa
\int_{-\epsilon/2}^{(N+\frac12)\epsilon} dx\, F(x) &=&
	\sum_{n=0}^N \int_{(n-\frac12)\epsilon}^{(n+\frac12)\epsilon} 
	dx\, F(x) \nonumber\\ &=& \sum_{n=0}^N \int_{(n-\frac12)\epsilon}^{(n+\frac12)\epsilon} 
	dx\, \left[ F(n\epsilon) + (x-n\epsilon)F'(n\epsilon)+
	\frac12(x-n\epsilon)^2 F''(n\epsilon)+\dots \right]\nonumber\\
	&=& \sum_{n=0}^N \left[ \epsilon F(n\epsilon) + \frac{1}{24}  
	\epsilon^3 F''(n\epsilon)+\cdots \right]
\eeqa
therefore
\beqa
	\sum_{n=0}^N \epsilon F(n\epsilon) &=& 
	\int_{-\epsilon/2}^{(N+\frac12)\epsilon} dx\, F(x) - \frac{1}{24}  
	\sum_{n=0}^N \epsilon^3 F''(n\epsilon)+\cdots \nonumber\\
	&=& \int_{-\epsilon/2}^{(N+\frac12)\epsilon} dx\, F(x) - \frac{1}{24}
	\epsilon^2 
	\int_{-\epsilon/2}^{(N+\frac12)\epsilon} dx\,
	F''(x)+O(\epsilon^4)
	\nonumber\\ 
	&=& \int_{-\epsilon/2}^{(N+\frac12)\epsilon} dx\, F(x) - \left.\frac{1}{24}
	\epsilon^2 
	F'(x)\right|^{(N+\frac12)\epsilon}_{-\epsilon/2}+O(\epsilon^4)
\eeqa			
We can apply this to $U(B)-U(0)$ using $\epsilon= 2eB$,
$F(x) = (16\pi^2)^{-1}\int dp_z \sqrt{p_z^2+x}$, 
to obtain
\beqa
	U(B)-U(0) &=& \frac1{16\pi^2}\sum_n \int {dp_z}
	\left.\left(-\frac1{24}\right){(2 e B)^2\over 2} [p_z^2+m^2 + (2n+1) eB]^{-1/2}
	\right|^{n=\infty}_{n=0}\nonumber\\
	&=& \frac{1}{24\cdot 8\pi^2}(e B)^2 \int_{-\Lambda}^\Lambda {dp_z
	\over [p_z^2+m^2+ eB]^{1/2}} \nonumber\\
	&=& \frac{1}{96\pi^2}(e B)^2 \ln(\Lambda/m)	
\eeqa
which gives the diamagnetic contribution to the susceptilibity
\beqa
	\chi_{\rm diag} &=&- \frac{1}{48\pi^2}e^2 \ln(\Lambda/m)
\eeqa
from each spin state.  However, there is an implicit assumption in this
calculation that we are dealing with bosons.  Notice that 
\beq
\label{vacenden}
	U(0) = \frac12\int { d^{\,3}p\over (2\pi)^3}\sqrt{\vec p^2+m^2}
\eeq
is simply the contribution to the vacuum energy density which we know is 
positive for bosons and negative for fermions.  Somehow our computation above
misses the minus sign for fermions.  Let's first see how (\ref{vacenden})
relates to our previous computations of the one-loop contribution to
the vacuum energy density, which is 
\beq
	V = -\frac{i}2\intdp \ln(-p^2 + m^2 - i\epsilon) 
\eeq
It is easiest to compare $dU/dm^2$ and $dV/dm^2$.  Note that
\beq
\label{vacendender}
{dU(0)\over dm^2} = \frac14\int { d^{\,3}p\over (2\pi)^3}(\vec
p^2+m^2)^{-1/2}
\eeq
and letting $\omega = \sqrt{\vec p^2 + m^2}$, we can integrate over
$p_0$ by completing the contour in the upper half plane to get
\beqa
	{dV\over dm^2} &=&  \frac{i}2 \int { d^{\,3}p\over (2\pi)^3}
	\int {dp_0\over 2\pi} {1\over p^2 - m^2 +i\epsilon}
	\nonumber\\ &=& \frac{i}2 \int { d^{\,3}p\over (2\pi)^4}
	{1\over 2\omega} \int {dp_0\over 2\pi}
	\left( {1\over p_0 - \omega+i\epsilon} -
	{1\over p_0 + \omega -i\epsilon}
	\right)
	\nonumber\\ &=& \frac{i}2 \int { d^{\,3}p\over (2\pi)^4}
	{1\over 2\omega}(- 2\pi i)
\eeqa
This confirms (\ref{vacenden}), except that 
since we know $V$ comes with a minus sign
for fermions, it must also be the case for $U$.  

Next we must consider the paramagnetic contribution to the energy, due to 
the intrinsic magnetic moment of the particles:
\beq
	E^2 = \vec p^2 + m^2 + (2n+1)eB  - ge \vec B\cdot \vec S
		\equiv E^2_{\rm Landau} - ge \vec B\cdot \vec S
\eeq
If we expand $E$ in powers of the magnetic moment energy, we get
\beq
	E = E_{\rm Lan.}\left[1 - \frac12 {geB S_z\over E^2_{\rm Lan.}}
	- \frac18 \left({geB S_z\over E^2_{\rm Lan.}}\right)^2\right]
\eeq
and we must sum the new contributions to $\frac12\omega$ over all the Landau levels and
$p_z$'s.  Picking out the term of order $B^2$ (note that 
the linear term will give no contribution when we sum over the spin
polarizations) we get the paramagnetic
contribution to the susceptibility,
\beq
U_{\rm para}=	-\frac12\chi_{\rm para}B^2 = -\frac1{16}(g e S_z B)^2 
	\int {d^{\,3}p\over (2\pi)^3(p^2+m^2)^{3/2}}
\eeq
giving
\beqa
	\chi_{\rm para}B^2 &=& \frac1{8}(g e S_z B)^2 
	\int {d^{\,3}p\over (2\pi)^3(p^2+m^2)^{3/2}}\times ({\rm sign})\nonumber\\
	&\cong& {4\pi\over 8\cdot 8\pi^3} (g e S_z B)^2 \ln(\Lambda/m)
	\left(\begin{array}{cc} +1,&{\rm boson}\\
				-1,&{\rm fermion}\end{array}\right)
\eeqa
where we used the same reasoning for the sign.  The Land\'e $g$ factor is
always 2 for an elementary particle, at lowest order in perturbation
theory---this is true for the gluons as well as the fermions. 
Putting the two contributions (paramagnetic and diamagnetic) 
together we obtain
\beq
	\chi_{\rm total} = {1\over 16\pi^2}\left[4 S_z^2 - \frac13\right] e^2 \ln(\Lambda/m)
	\left(\begin{array}{cc} +1,&{\rm boson}\\
				-1,&{\rm fermion}\end{array}\right)
\eeq
Comparing to (\ref{chibrel}), we can read off the contributions to 
the coefficient of the beta function,
\beq
	\delta b_0 = -{1\over 32\pi^2}\left[4 S_z^2 - \frac13\right] e^2 \ln(\Lambda/m)
	\left(\begin{array}{cc} +1,&{\rm boson}\\
				-1,&{\rm fermion}\end{array}\right)\times
	(\hbox{no. of polarizations})
\eeq

First, let's check this result against QED.  In that case, the photon carries
no charge, so we get only the contribution from the fermions, which have
$S_z=1/2$.  The way we
have normalized the vacuum energy, each Dirac fermion has 4 polarization
states (antiparticles must be counted separately), so this gives a beta
function coefficient of $b_0 = +1/(12\pi^2)$, in agreement with
(\ref{QEDbeta}). 

Now let's look at the contribution from the vector gauge bosons in the
nonAbelian theory.  They have $S_z = 1$, and two polarization states.
Therefore each virtual gluon which can couple to the external color
$B$ field contributes $-(11/3)/(16\pi^2)$ to $b_0$.  Comparing to
(\ref{QCDbeta}), we see that the form seems to be correct---we understand
the factor of $11/3$, and $C_2(A)$ is counting the virtual gluons.  We
can check that $C_2(A)=N$ is the correct factor in the context of
our present computation by looking at the nonvanishing structrue
constants $f^{abc}$ given in (\ref{struct}) for the case of QCD.  Above
we argued that we should compute the trace of the square of each generator
which couples to the external field.  For example, if we take the external
field to be in the $a=1$ color direction, we see that there are three 
generators which have this index, and 
\beq
	\sum_{bc} f^{1bc}f^{1bc} = 2(1 + 1/4 + 1/4) = 3
\eeq
which agrees with our expectation.  We could of course have gotten the
same answer using any value of $a$, as the identity (\ref{ffid}) guarantees.

Thus we see that asymptotic freedom can be understood as a consequence
of the fact that the gauge bosons themselves have magnetic interactions
with an external $B$ field, and being bosons they contribute with the
opposite sign to the vacuum energy, hence the magnetic susceptibility.
The virtual gauge bosons thus do cause antiscreening of a color charge,
opposite to the screening behavior in QED.

\section{Nonperturbative aspects of $SU(N)$ gauge theory}
\label{sec:nonpert}

To conclude this course, I would like to treat one important subject
which goes beyond the realm of perturbation theory, namely the
role of instantons in gauge theories.  We will see that the vacuum
state of QCD is much more complicated than that of QED, which is 
simply described by $A^\mu=0$ or its gauge-equivalent copies.
The reason for this is topology.  

To motivate this, let's first consider a simpler model: QED in $1+1$
dimensions, where the spatial dimension is compactified on a circle.
Let's parametrize it by $\theta \in [0,2\pi]$.  A possible gauge
transformation is one that winds around the circle $n$ times:
\beq
\label{largegt}
	\Omega(\theta) = e^{in\theta}
\eeq
If we started from the vacuum state $A_\mu=0$ and performed such a 
gauge transformation on the fermions, $\psi\to \Omega\psi$, then we would
obtain a new value for the gauge field,
\beq
\label{nvacua}
	A_\mu \to -{i\over e} \Omega\partial_\mu \Omega^{-1};\quad A_0 = 0,
	\quad A_\theta = -{n\over e}
\eeq
The interesting thing about this is that $n$ has to be an integer; otherwise
the gauge transformation is not continuous around the circle.  Therefore
it is not possible to transform it back to $A_\mu=0$ using
a family of gauge transformations which are continuously connected to the
identity.   This means that we would have to pass through some field
configuration with nonvanishing field strength if we were to try to go
between the different vacuum states.

The gauge transformations (\ref{largegt}) with $n\neq 0$ are called  {\it
large} gauge transformations, and the vacuum states (\ref{nvacua}) are
called $n$-vacua.  Even though physics looks the same in any of these vacuum
states, it is not obvious that we can choose just one of them. The ``true''
vacuum state could be a superposition of the different  $n$-vacua.  An
analogous thing  happens when we consider the quantum mechanics of a double
well potential, as in fig.\ 50.  Classically,  there are two vacuum states,
at $x_0$ and $-x_0$.  But quantum mechanically, the true vacuum state is the
symmetric superposition $|+\rangle =  \frac{1}{\sqrt{2}} (|+x_0\rangle +
|-x_0\rangle)$, which has a lower energy than the antisymmetric one,
$|-\rangle = \frac{1}{\sqrt{2}} (|+x_0\rangle + |-x_0\rangle)$.   In quantum
mechanics we learn that this is related to the possibility of {\it tunneling}
between the two states.  The states $|\pm x_0\rangle$ are not eigenstates of
the Hamiltonian because the amplitude for tunneling is nonzero:
$\langle +x_0 | -x_0 \rangle \neq 0$.  Furthermore, the energy difference
between the symmetric and antisymmetric combinations is related to the
tunneling amplitude:
\beqa
	E_+ - E_-  &=& \langle+ |H|+\rangle - \langle- |H|-\rangle\nonumber\\
	&=& \frac12\left( \langle+x_0|H|-x_0\rangle + 
	\langle-x_0|H|+x_0\rangle\right) \nonumber\\
	&\cong& \langle\pm x_0|H|\pm x_0\rangle \, \langle-x_0|+x_0\rangle
\eeqa

\medskip
\centerline{\epsfxsize=2.in\epsfbox{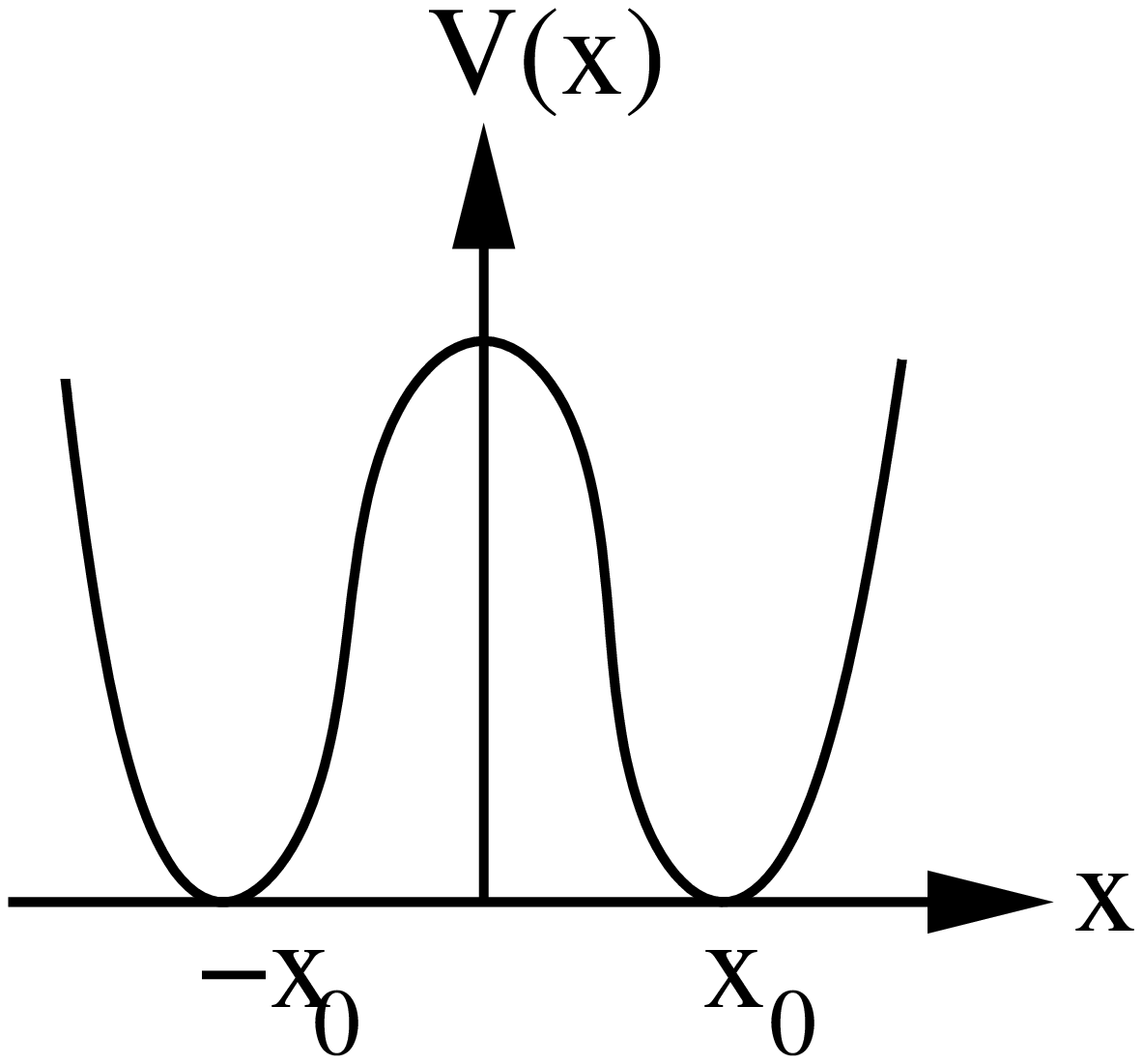}}
\centerline{\small Fig.\ 50. 
A double well potential.}
\medskip

In field theory we expect a similar phenomenon whenever there are 
degenerate vacuum states, as long as the probability to tunnel between them
is nonvanishing.  In the above example, we found degenerate vacuum states as
a consequence of topology:  the gauge transformations provided topologically
nontrivial maps from the group manifold to physical space.  In this example,
the group manifold (for $U(1)$) is a circle, since any two gauge
transformations $e^{i\Lambda}$ and $e^{i\Lambda'}$ are equivalent if $\Lambda$
and $\Lambda'$ differ by a multiple of $2\pi$.  We mapped the group manifold
onto another circle comprising the physical space.  In mathematical terms,
this mapping has nontrivial {\it homotopy}.  We could not have done the same
thing in $2+1$-D QED where the physical space is a sphere.  In this case,
a gauge transformation which wraps around the equator of the sphere can
be smoothly deformed to a trivial gauge transformation at a point.  There is
no topological barrier between large and small gauge transformations in 
this case.  

\medskip
\centerline{\epsfxsize=1.5in\epsfbox{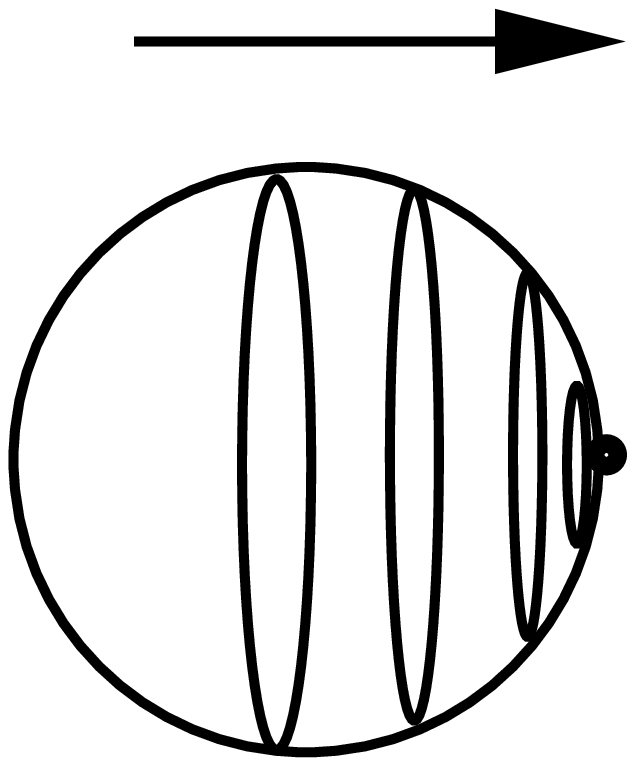}}
\centerline{\small Fig.\ 51. 
A homotopically trivial mapping.}
\medskip

The preceding example makes it clear that if we want to find the same
phenomenon in $3+1$ dimensions, we need a gauge group whose group manifold
has at least three dimensions.  A simple example is $SU(2)$: it has three
generators, and its group manifold is almost the same as that of $SO(3)$, 
a three-sphere.  To find the large gauge transformations, let us imagine
that 3-D space is compact (it could have an arbitrarily large volume),
and has the topology of a 3-sphere.  We can therefore assign coordinates
$\phi_1$, $\phi_2$, $\phi_3$ to physical space.  Denote these collectively
by $\vec\phi$. Then the large gauge $SU(2)$ transformations can be 
written as
\beq
\label{largeQCDgt}
	U_n = \exp(in\vec\phi\cdot\vec\tau)
\eeq
where $\tau_i = \frac12\sigma_i$ are the generators of $SU(2)$.  We can
find the nonAbelian gauge fields generated from the trivial vacuum by these
gauge transformations,
\beq
\label{nvac}
	A_{n,\mu} = {i\over g} U_n\partial_\mu U_n^{\dagger}
\eeq
Since they are pure gauge, they are guaranteed to have vanishing field
strengths.  As in the Abelian example, the large gauge transformations
(\ref{largeQCDgt}) cannot be continuously deformed into the trivial one
$U = 1$ because of the topological obstruction: the mapping $U_n$ winds
the group manifold around the physical space $n$ times, and $n$ must
be an integer by continuity of the fields.

We therefore expect that the true vacuum state of nonAbelian theories
will be some superposition of the $n$-vacua.  In fact, it will turn out
that the correct choice has the form $|\theta\rangle = \sum_n e^{in\theta} |n\rangle
$ where $\theta$ is an arbitrary number.  But to understand this, we have
to show that the $\theta$-vacua are the real eigenstates of the Hamiltonian,
and this requires knowing what is the amplitude for tunneling between any
two $n$-vacuua.  How do we compute tunneling probabilities in quantum
field theory?  Instantons provide the answer, at least in cases where there
is a large barrier to tunneling.  A classic reference on this subject
is \cite{Coleman}

\subsection{Instantons}

To find a transition amplitude in field theory, we can use the Feynman path
integral. We want to integrate over all field configurations which start  in
one vacuum state, say $|n\rangle$, and end in the other, say $|m\rangle$. 
If the two states were separated by a saddle point as shown in fig.\ 52,
then there will be one path  between them (the dotted line which goes over
the saddle) which minimizes the action.  This path will dominate the path
integral in a stationary phase approximation:
\beq
	\int {\cal D}\phi e^{iS} \sim e^{iS_{\rm min}}
\eeq
This approximation will be justified if $S_{\rm min}$ is large, since then
the action will have a large variation for nearby paths, and $e^{iS}$
will oscillate rapidly, giving cancelling contributions to the path
integral. 

We can find the action of this path by solving the equation of motion:
\beq
	-\ddot\phi + \nabla^2\phi  - V'(\phi) = 0
\eeq
subject to the boundary conditions $\phi(t\to-\infty) = \phi_n$,
$\phi(t\to+\infty) = \phi_m$.

\medskip
\centerline{\epsfxsize=3.in\epsfbox{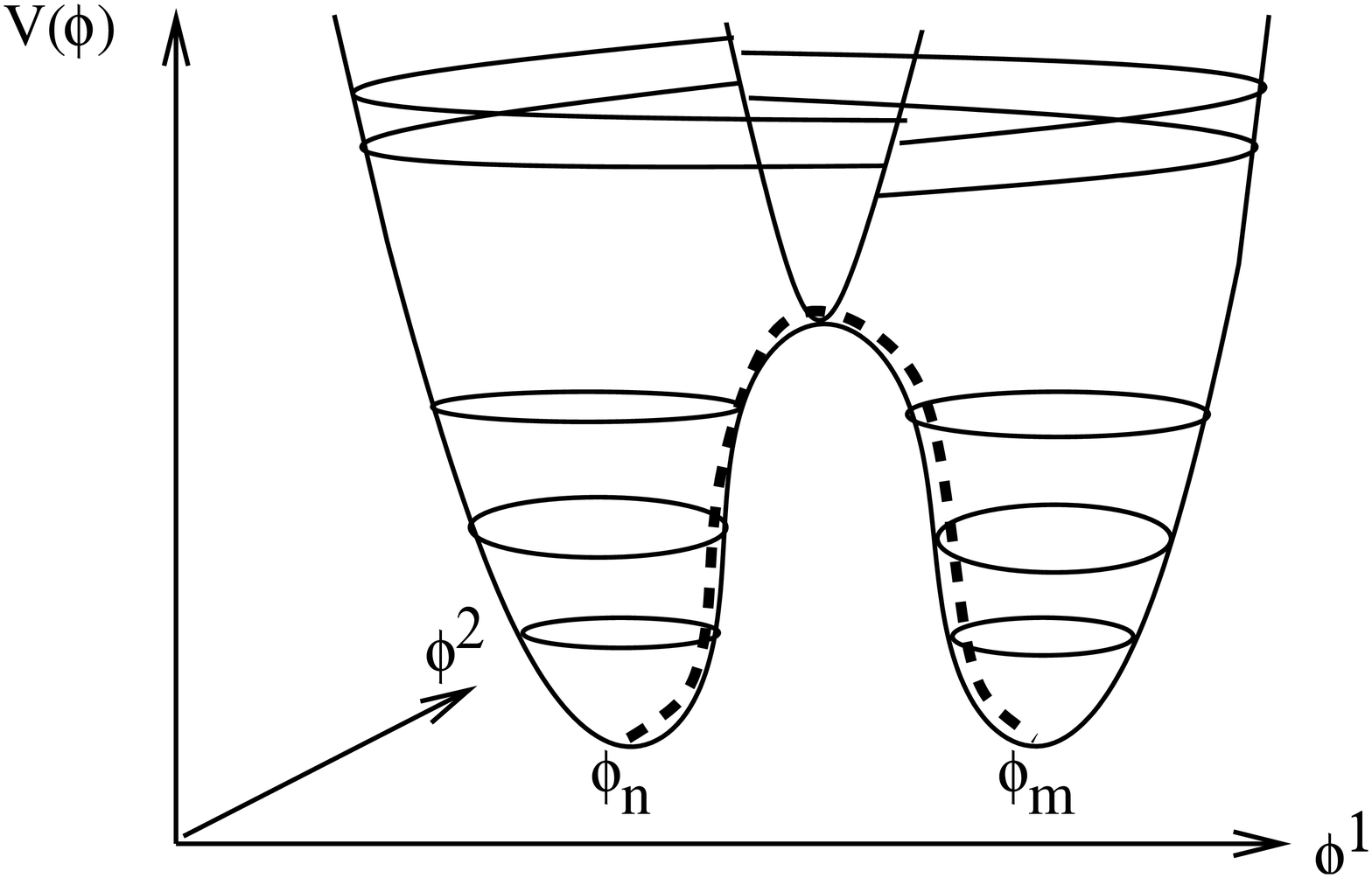}}
{\small Fig.\ 52. A potential for a field with several components,
$\phi^i$, and vacua at the values $\vec\phi_m$ and $\vec\phi_n$.
Dotted line: path of smallest action connecting the two vacua $n$ and $m$.}
\medskip

However, this method does not give a good approximation to the tunneling
amplitude.  The reason is that we should really integrate over all the
fluctuations which are close to the stationary path in order to get
a good estimate, and simply evaluating the integrand at the stationary
point does not take this into account.  We should see that the tunneling 
probability is exponentially small if the barrier is large.

There is a trick which enables us to do a very similar calculation to
the one above, which however does give a good estimate of the tunneling 
amplitude.
It is simply to go to Euclidean space before doing the path integral.
When we do this, the path integral becomes
\beq
	\int {\cal D}\phi e^{-S_E} \sim e^{-S_{\rm min}}
\eeq
We are now using the method of steepest descent rather than the
stationary phase approximation to evaluate the path integral.  Notice
that this does immediately give what we expect: an exponentially suppressed
amplitude in the case of a large barrier.

In Euclidean space, the action is given by 
\beq
	S_E = \intdx \left(\frac12\left[ (\dot\phi)^2 + (\nabla\phi)^2\right]
	+ V(\phi)\right)
\eeq
This is similar to the Minkowskian action, but with the potential reversed
in sign.  We can imagine we are still solving the Minkowskian field
equations, but with $V\to-V$ (if we also lump the gradient term into the 
potential), as in fig.\ 53.  

\medskip
\centerline{\epsfxsize=2.5in\epsfbox{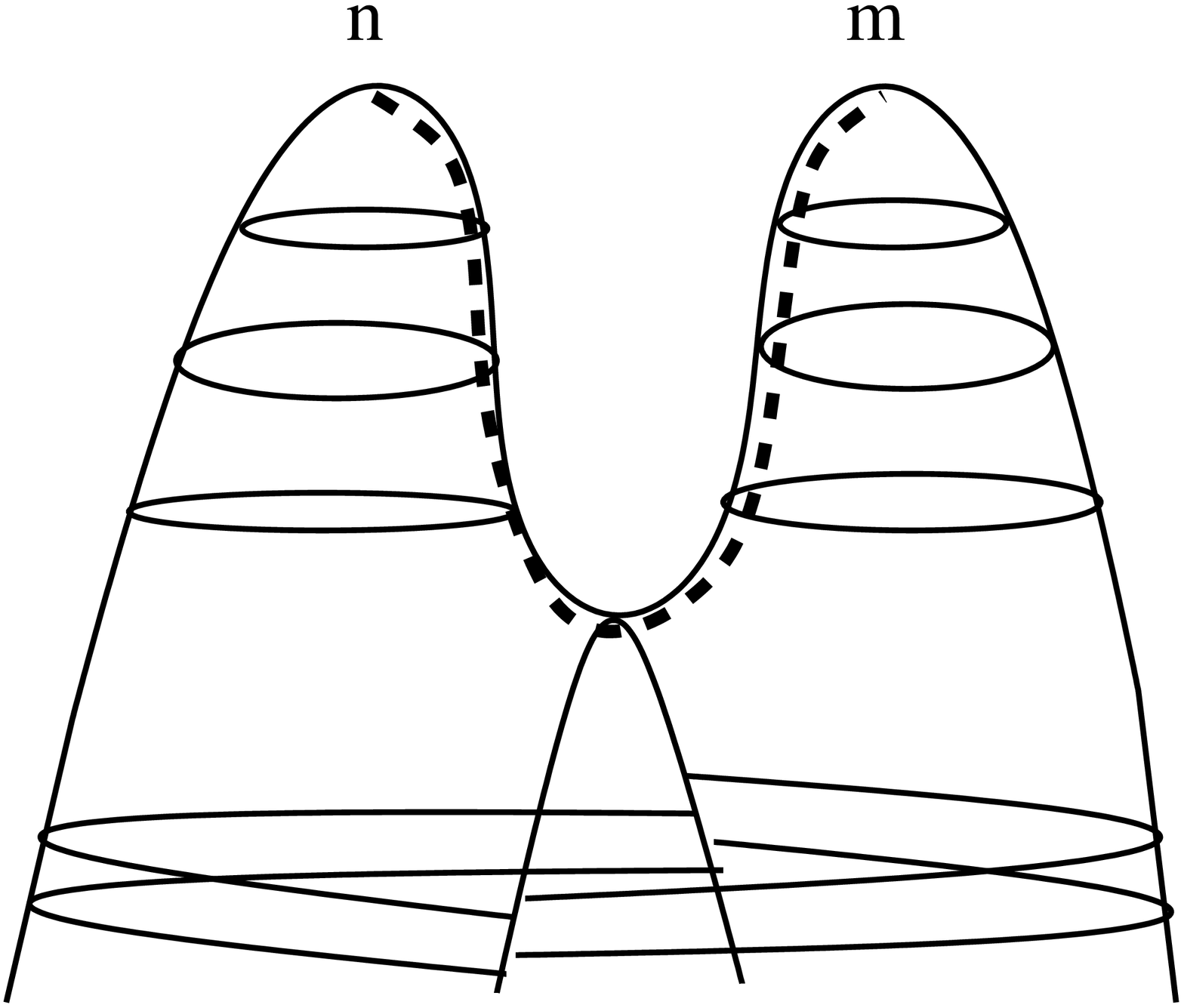}}
\centerline{\small Fig.\ 53. The same potential as in fig.\ 52, but now
in Euclidean space.}
\medskip

The equation of motion becomes
\beq
	\ddot\phi + \nabla^2\phi  - V'(\phi) = 0
\eeq
If we consider for a moment the case of quantum mechanics for a particle
in a potential, the term $\nabla^2\phi$ is absent since $\phi$ itself
represents the position of the particle.  The solution we seek is one
that starts at $n$ when $t\to-\infty$, and rolls to $m$ as $t\to+\infty$.
We can use energy conservation to write the first integral of the equation
of motion,
\beq
	E = 0 = \frac12(\dot\phi)^2 - V(\phi)
\eeq
where we chose the vacuum states to have zero energy for convenience.
The solutions are given implicitly by 
\beq
	\int^t dt = \pm \int^\phi { d\phi\over \sqrt{V(\phi)}}
\eeq
For example, if $V(\phi) = \lambda(\phi^2 - v^2)^2$, the integral can be
done, with the result
\beq
\label{phinst}
	\phi(t) = v \tanh(\sqrt{\lambda} v t)
\eeq 
The graph of this solution looks like fig.\ 54, where the width of the
transition region is $w= (\sqrt{\lambda} v)^{-1}$.  This is the amount
of (Euclidean) time it takes for the field to do most of its transition 
between the two vacua, since for times much earlier or later, it is
essentially just sitting at one of the two vacuua.  Since all the
action occurs at essentially one instant in time, 't Hooft called these
kinds of solutions {\it instantons}.  The journal Physical Review at first
refused to accept such an imaginative name, so some of the first papers
on this subject called them ``pseudoparticles,'' but the instanton name
quickly gained popularity.

\medskip
\centerline{\epsfxsize=3.in\epsfbox{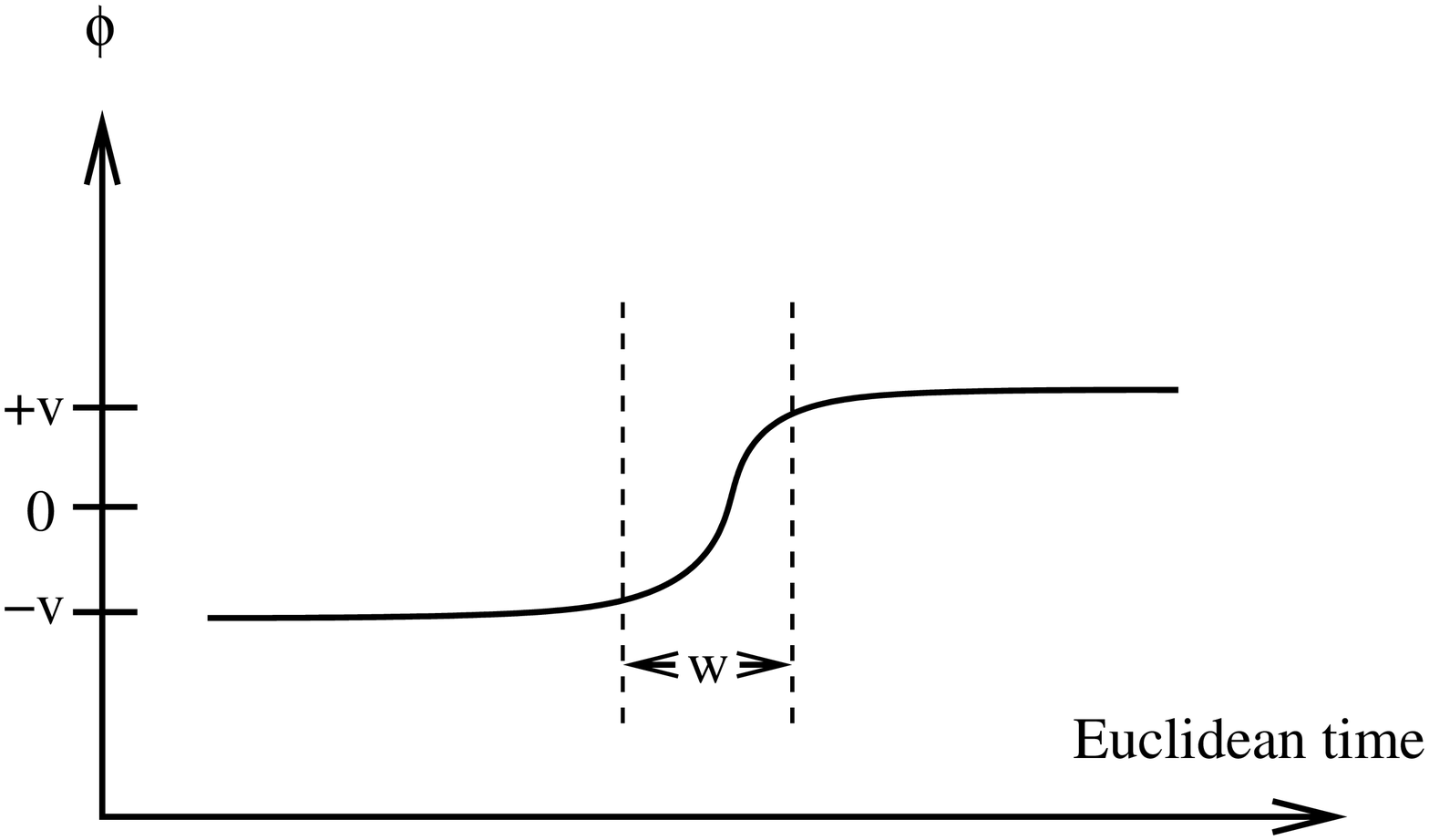}}
\centerline{\small Fig.\ 54. Graph of the instanton solution (\ref{phinst}).}
\medskip

An important feature of instantons is that they are {\it nonperturbative}.
When (\ref{phinst}) is substituted back into the Euclidean action, one finds
that 
\beq
	S_{E,\rm min} \sim {v^4\over \lambda}
\eeq
and therefore the tunneling amplitude goes like
\beq
	\langle n| m\rangle \sim e^{-c v^4/\lambda}
\eeq
This has no expansion in perturbation theory---we would never see any
sign of it even if we worked to infinite order in the perturbation expansion.
Therefore instantons provide us with our first example of a truly
nonperturbative effect in quantum field theory.

Now we are ready to apply these ideas to $SU(2)$ gauge theory.  We need
a solution to the Euclidean Yang-Mills field equations 
\beq
	\partial^\mu F_{\mu\nu} + ig[A^\mu, F_{\mu\nu}] = 0
\eeq	
which interpolates
between two of the $n$-vacua (\ref{nvac}) as $t$ goes from $-\infty$
to $+\infty$.  I will present this solution without proof, and leave it
for you (last homework problem) to verify that it is a solution:
\beq
\label{instanton}
	A_\mu(x) = -{i\over g}{ x^2\over x^2+\lambda^2} U\partial_\mu U^\dagger
\eeq
where $x^2 = x_0^2 + x_i^2$ and $\lambda$ is an arbitrary constant length
scale which characterizes the size of the instanton.  The gauge
transformation $U$ is given by
\beq
\label{largeYMgt}
	U = {1\over \sqrt{x^2}}(x_0 - i \vec x  \cdot \vec \sigma)
\eeq
Clearly, as $x^\mu\to\infty$, $A_\mu(x)$ approaches a pure gauge field.
But because of the prefactor $ x^2/( x^2+\lambda^2)$, it is not pure gauge
in the interior, so there is a barrier.  It turns out that the action of this
solution is given by
\beq
	S_I = {8\pi^2\over g^2}
\eeq
which like the scalar field example is nonperturbative.

The large gauge transformation $U$ in (\ref{largeYMgt}) can be applied
any number of times, $U_n = U^n$, to produce a map which wraps the group
manifold $n$ times around the $3$-sphere at infinity.  This means there
is a series of vacuum states labeled by integers $n$, just as we had in
the case of $(1+1)$-D QED on a circle.  They are separated from each other
by energy barriers in the field space, as shown in fig.\ 55.

\medskip
\centerline{\epsfxsize=3.in\epsfbox{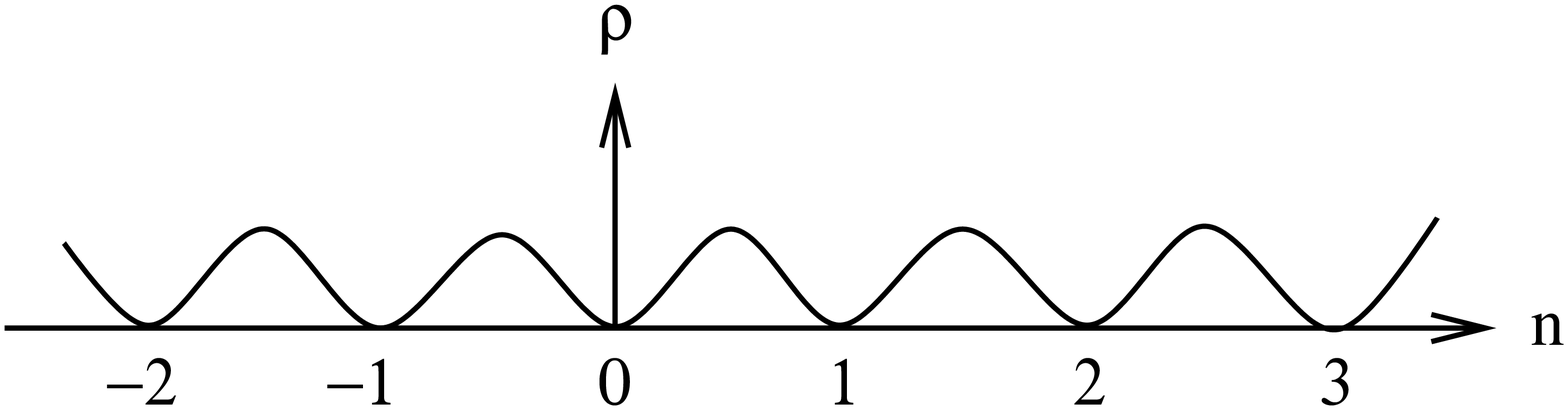}}
{\small Fig.\ 55. Energy density of field configurations
which interpolate between the $n$-vacua of Yang-Mills theory.}
\medskip

\subsection{Winding number}

Since the instantons satisfy the Euclidean field equations, they
are the field configurations which interpolate between $n$-vacuua with
the least action.  There are of course infinitely many other solutions
which will have the same topology as a given instanton, but larger action.
Is there a way in which we can compute whether a given field configuration
has a nontrivial change in topology?  I will not prove it, but one can show
that the following expressions give the winding number of a gauge
field in 2 and 4 Euclidean dimensions, for the $U(1)$ and $SU(N)$ 
gauge theories, respectively:
\beqa
\label{FFD}
	\nu &=& {g\over 4\pi}\int d^{\,2}x\, \epsilon_{\mu\nu} F_{\mu\nu};
	\nonumber\\
	\nu &=&  {g^2\over 32\pi^2}\int d^{\,4}x\, \epsilon_{\mu\nu\alpha\beta} 
	\, \tr F_{\mu\nu}  F_{\alpha\beta} \equiv {g^2\over 16\pi^2}\int d^{\,4}x\, 
	\, \tr F_{\mu\nu}  \widetilde F_{\mu\nu}                    
\eeqa
Both of these equations can be rewritten as integrals of a total divergence,
\beqa
	\nu &=& {g\over 2\pi}\int d^{\,2}x\, \partial_\mu W_\mu;\quad
	W_\mu = \epsilon_{\mu\nu} A_{\nu};
	\nonumber\\
\label{topno}
	\nu &=&  {g^2\over 8\pi^2}\int d^{\,4}x\,  \partial_\rho W_\rho;\quad
	W_\rho = \epsilon_{\rho\sigma\mu\nu} 
	\, \tr\left(A_\sigma \partial_\mu A_\nu + {2ig\over 3} A_\sigma A_\mu A_\nu
	\right)          
\eeqa
If we compactify Euclidean space by imposing a maximum radius $\sqrt{x_\mu x_\mu}
< R$ (we will take $R\to\infty$), then the divergence theorem can used to rewrite
the above integrals as surface integrals, over the $(N-1)$-sphere which bounds
the region $\sqrt{x_\mu x_\mu}< R$:
\beq
	\nu = c_{\sss N}\int d\Omega_{N-1}^\mu W_\mu
\eeq
where $c_{\sss N} = {g\over 2\pi}$ or ${g^2\over 8\pi^2}$.  Now we can argue that
any physically sensible configuration must have zero field strength at $r=R$
as $R\to\infty$, so that its action will be finite.  This means that $A_\mu$
must approach a pure gauge configuration at the boundary,
\beq
	A_{\mu} \to {i\over g} U\partial_\mu U^{\dagger} 
\eeq
Now let's parametrize the bounding $(n-1)$-sphere by the angle $\phi$ in 2D
and $\phi_1$-$\phi_3$ in 4D.  The winding number (also called the Pontryagin
index) becomes
\beqa
\label{U1case}
	\nu &=& -{i\over 2\pi}\int_0^{2\pi} d\phi\,{dU\over d\phi} U^{-1} ;
	\\
	\nu &=& {1\over 24\pi^2}\int d\phi_1 d\phi_2 d\phi_3\, \epsilon_{ijk}\,
	\tr\left({dU\over d\phi_i} U^{-1} {dU\over d\phi_j } U^{-1} 
	{dU\over d\phi_k  } U^{-1}\right)	
\eeqa
It can be shown that these expressions are unchanged by deformations of $U(\phi_i)$
which do not change the winding properties of $U$; they are topological invariants.
This is easy to see in the $U(1)$ case: suppose that $U = e^{if(\theta)}$ where
$f$ is any function which has the property that $f(2\pi) = f(0) + n2\pi$.  Then
(\ref{U1case}) gives
\beq
	\nu =  {1\over 2\pi}\int_0^{2\pi} d\phi\, {df\over d\phi} 
	= {f(2\pi) - f(0)\over 2\pi} = n
\eeq
The more complicated case of $SU(N)$ in 4D works in a similar way.  For example,
consider the instanton solution (\ref{instanton}).  We can verify its winding number
by considering the region near $\vec x = 0$ (the north pole of the 3-sphere).
In this vicinity, choose the angles $\phi_i$ to be the space coordinates
$x_i$, so 
\beq
	\partial_k U U^{-1} = i\sigma_k
\eeq
and
\beq
	\tr\left({dU\over d\phi_i} U^{-1} {dU\over d\phi_j } U^{-1} 
	{dU\over d\phi_k  } U^{-1}\right) = i^3 \tr\sigma_i\sigma_j\sigma_k
	= 2\epsilon_{ijk}	
\eeq
Then $\epsilon_{ijk}\tr(\cdots) = 12$ and we get
\beq
	\nu = {1\over 2\pi^2} \int d\phi_1 d\phi_2 d\phi_3 =1,
\eeq
since the volume of the 3-sphere (the group manifold of $SU(2)$) is $2\pi^2$.  Here
we pulled a little trick by assuming that the trace factor is a constant independent
of the angles, even though we only showed it to be the case in the vicinity of  the
north pole.  However, it can be shown that the expression for $\nu$ is invariant
under constant tranformations of the group, $U\to U U_0$, so one can do the
Faddeev-Popov procedure of inserting $1 = \int d\omega \delta(U-1)\det|\partial U_\omega
/\partial\omega|$.  Since $\int dU \cdot 1 = 2\pi^2$ for this group, we deduce that
$\int d\omega  \det|\partial U_\omega /\partial\omega|_{U=1} = 2\pi^2$, which 
enables us to evaluate $\int dU \epsilon_{ijk}\tr\left({dU\over d\phi_i} U^{-1} {dU\over d\phi_j } U^{-1} 
{dU\over d\phi_k  } U^{-1}\right)$ just by knowing the value of the integrand in
the vicinity of $U=1$ (or any other element of the group, for that matter).

Let us comment on the relation of the instanton solution (\ref{instanton}) and 
the picture we originally started with, that of tunneling in $(1+1)$-D QED
compactified on a spatial circle.   Logically, it is convenient to think about
$N$-space being compactified on an $N$-sphere, and letting time run from $-T/2$
to $+T/2$, where we will take $T\to\infty$.  Then we would look for gauge field
configurations whose winding number on the $N$-spheres changes from $n_-$ to $n_+$
as $t$ goes from $-T/2$ to $+T/2$.  The winding number of $(N+1)$-D instanton
configuration that interpolates between the two vacuua can be shown to be
\beq
	\nu = n_+ - n_-
\eeq
However, it is not very convenient to look for instanton solutions on the space
$S_N\times R$.  It is much easier to find solutions that transform simply under
$SO(N+1)$ rotations of the full Euclidean space.  (Remember, $N$ is counting only
the spatial dimensions.)  Such solutions live on the $(N+1)$-sphere rather than
$S_N\times R$.  However, in the end we are going to decompactify both spaces, so it
should not matter which one we use, as long as the instantons have the same winding
number in both pictures.  In fact, one picture should be equivalent to the other
under a gauge transformation as long as they both have the same winding number.
The two situations are depicted in fig.\ 56.

\medskip
\centerline{\epsfxsize=3.5in\epsfbox{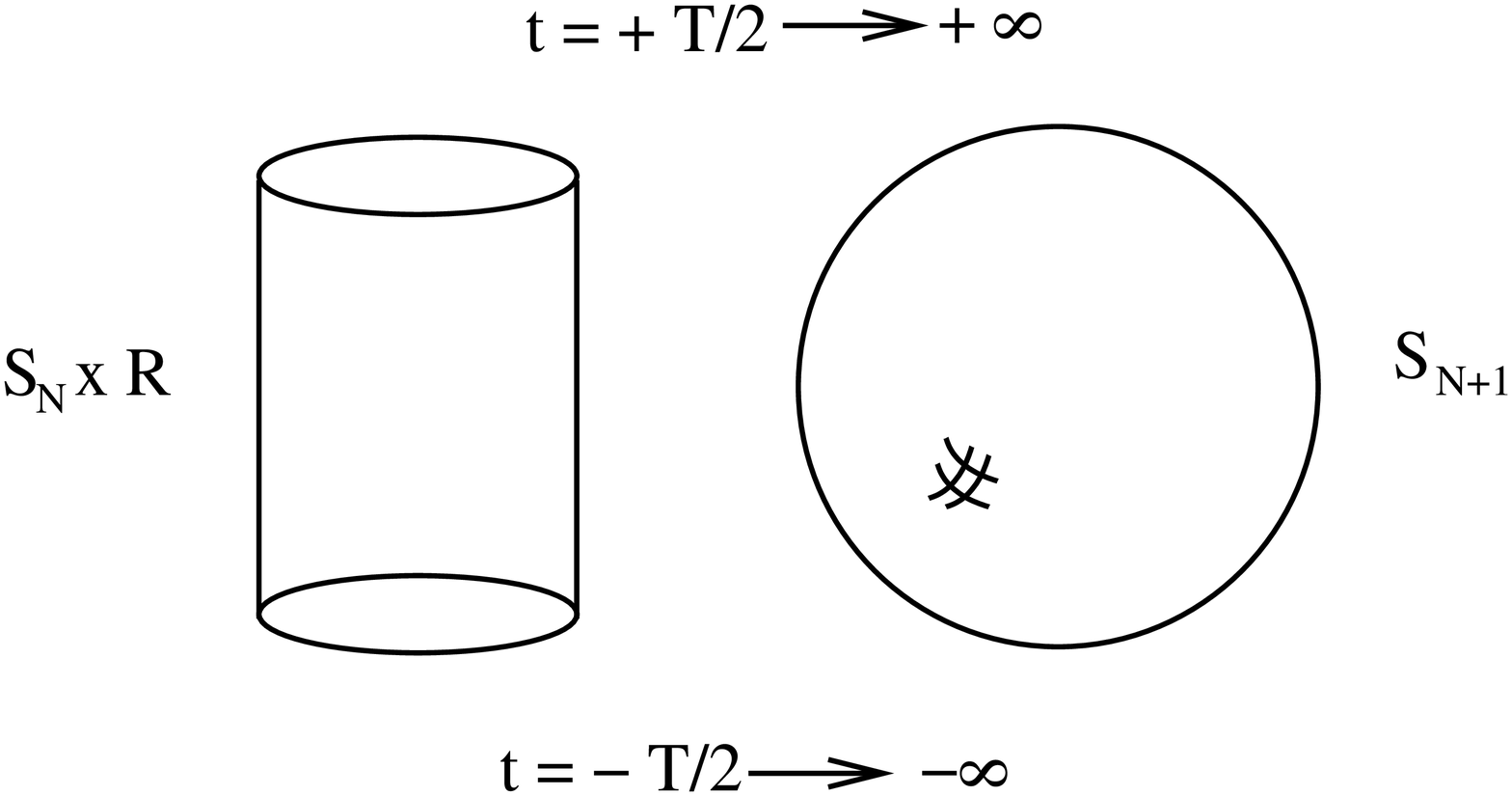}}
\centerline{\small Fig.\ 56. Two different ways of compactifying
$N+1$ dimensional Euclidean space.}
\medskip

An important note for $SU(N)$ gauge theories: although we have only displayed
the instanton solution for $SU(2)$, this is essentially all that is needed for
$SU(N)$ theories, since $SU(2)$ is a subgroup of $SU(N)$.  For example, in $SU(3)$,
we can choose one of the 3 rows and columns to transform trivially, and embed
the $SU(2)$ instanton in the other two, for example
\beq
	U_{\sss SU(3)} = \left(\begin{array}{cc} 1 & 0 \\
						0 & U_{\sss SU(2)}
	\end{array} \right)
\eeq
And any constant gauge transformation of this, $U\to U U_0$, is also a good solution.
This means that the $SU(3)$ instanton has an orientation in color space, over which
we will have to integrate when we compute the tunneling rate.

\subsection{Tunneling between $n$-vacuua}

Previously we asserted that the amplitude for tunneling goes like $e^{-S_I}$,
where the instanton action is given by $S_I = {8\pi^2\over g^2}$.  One
way of writing this is in terms of the transition amplitude between between
two neighboring $n$ vacua with winding numbers $n$ and $n+1$.  In Minkowski space
the transition amplitude would be given by 
\beq
	\langle n+1 | e^{iHT} | n\rangle
\eeq
since $e^{iHt}$ is the evolution operator, but we are doing the calculation in 
Euclidean space, so we have
\beq
	\langle n+1 | e^{-HT} | n\rangle = \int {\cal dA_\mu} e^{-S}
	\sim C e^{-S_I}
\eeq
Here $C$ is a prefactor which corrects the crude estimate $e^{-S_I}$. Coleman has
explained very nicely how to compute $C$ by expanding the action to quadratic
order about the instanton and doing the path integral over these small fluctuations
\cite{Coleman}.  Not only does one have to sum over these small fluctations, but
also over the collective coordinates of the instanton.  Namely, the instanton can
be centered around any point in spacetime, and it can have different orientiations
in color and in position space.  Moreover, we saw that the instanton has an arbitrary size
$\rho$ which must also be integrated over.  When all this is done, one gets a
tunneling amplitude of
\beq
	\langle n+1 | e^{-HT} | n\rangle \sim V T \int {d\rho\over \rho^5}
	e^{-8\pi^2/g^2(\mu)} f(\rho\mu)
\eeq
The factor $VT$ is the volume of Euclidean spacetime, which tells us that
the probability of tunneling should really be interpreted as a constant 
rate of tunneling per unit spatial volume (careful consideration of
how the $n$-vacuum states should be normalized gives a tunneling rate that
is proportional only to $VT$ and not $V^2T$).  The power of $\rho$ could
be guessed from dimensional analysis, as we will see below. 
Also $f(\rho\mu)$ is the function which arises from integrating over the small
fluctuations. This is just some functional determinant, which is UV divergent,
and requires us to renormalize, thus introducing the arbitrary renormalization
scale $\mu$.  The $\mu$-dependence of $f$ is guaranteed to cancel that of
$e^-{8\pi^2/g^2(\mu)}$.  We can choose $mu = 1/\rho$ so that $f$ becomes an
unimportant constant.  Furthermore, let us remember that
\beq
	g^2(\mu) \sim {1\over 2 b_0 \ln(\Lambda/\mu)}
\eeq
where $\Lambda$ is the scale where the coupling becomes strong, on the order of
$300$ GeV in QCD.  One then gets a tunneling rate of 
\beq
	{\Gamma\over V}\sim \int {d\rho\over \rho^5} e^{-8\pi^2/g^2(\mu)}
	\sim \Lambda^4 \int {d\rho\over \rho} (\rho\Lambda)^{11N/3-4}
\eeq
where we displayed the value of $b_0$ for pure $SU(N)$ with no quarks.  (The story
actually becomes more complicated when we include quarks; see \cite{Coleman}.)
Notice that the powers of $\rho$ are correct for giving a rate per unit time.
Although the integral diverges for large instantons, it can be argued that our
semiclassical approximation to the path integral is breaking down since their
action is becoming too small, and the only reasonable place to cut off the integral
is for instantons of size $\Lambda^{-1}$, since this is the only length scale
in the theory.  Thus one estimates finally that 
\beq
	{\Gamma\over V}\sim \Lambda^4
\eeq
which is no longer exponentially suppressed; this is a large value.  So tunneling
is a big effect, and we cannot pretend that the $n$-vacua are even good approximate
vacuum states.

\subsection{Theta vacua}

It turns out that the physically correct vacuum states are superpositions of the
$n$-vacua of the form
\beq
	|\theta\rangle = {1\over \sqrt{2\pi}}\sum_n e^{-in\theta} |n\rangle
\eeq
Exactly the same thing happens in simple quantum mechanical systems with a periodic
potential---these are the Bloch states.  Unlike $n$-vacua, 
the theta vacuua are eigenstates 
of the large gauge transformations (\ref{largeQCDgt}), $U_p$, since these 
simply increment the winding number of the $n$-vacua by $p$:
\beqa
	U_p |\theta\rangle &=& {1\over \sqrt{2\pi}}\sum_n e^{-in\theta}U_p|n\rangle\nonumber\\
	&=& {1\over \sqrt{2\pi}}\sum_n e^{-in\theta}|n+p\rangle\nonumber\\
	&=& {1\over \sqrt{2\pi}}\sum_n e^{-i(n-p)\theta}|n\rangle\nonumber\\
	&=& e^{ip\theta}|\theta\rangle
\eeqa
Furthermore, these states are orthonormal,
\beq
	\langle\theta' | \theta\rangle = {1\over 2\pi}\sum_n
	e^{-in(\theta-\theta')} = \delta(\theta-\theta')
\eeq
and we can show that there is no tunneling between two different
theta vacua:
\beqa
	\langle\theta' | e^{-HT} |\theta\rangle &=& {1\over {2\pi}}\sum_{n,m}
	e^{in\theta' - im\theta} \langle m|e^{-HT}|n\rangle	\nonumber\\
	&\equiv& {1\over {2\pi}}\sum_{n,m}
	e^{in\theta' - im\theta} A(n-m)	\nonumber\\
	&=& {1\over {2\pi}}\sum_{n}e^{in(\theta'-\theta)} \sum_{n-m}
	e^{i(n-m)\theta} A(n-m)	\nonumber\\
	&=& \delta(\theta-\theta') f(\theta)
\eeqa	
Thus if we are in a particular theta vacuum, we will stay in it forever.  This
satisfies the criteria of a good vacuum state.  One says that the theta
vacua are different {\it superselection sectors}.  Since there is no way to go
from one to the other, it makes no sense to talk about superpositions of theta vacua.

Although all theta vacua seem to be acceptable vacuum states at this point, it
is not true that the are all equivalent.  In fact, we can show that they have
different energies by computing the expectation value of $e^{-HT}$, 
\beqa
	\langle\theta|e^{-HT}|\theta\rangle &=& e^{-E(\theta)T}
	 \langle\theta|\theta\rangle 
	\cong (1  -E(\theta)T +\cdots)\langle\theta|\theta\rangle \nonumber\\
	&\cong& \delta(\theta-\theta) \sum_\nu e^{i\nu\theta}e^{-|\nu|S_i} \nonumber\\
	&\cong& VT \left(1 + 2\cos(\theta)e^{-S_i} + O(e^{-2S_i})\right)
\eeqa
We see that the vacuum energy density has the dependence
\beq
	{E(\theta)\over V} \sim {\rm const.} - 2\cos\theta e^{-S_i}
\eeq
where the factor $e^{-S_i}$ is just shorthand for the full expression which includes
the integral over instanton sizes and thus gives the actual estimate
\beq
	{E(\theta)\over V} \sim {\rm const.} - O(\Lambda^4) \cos\theta
\eeq
Even though the vacua with $\theta = 0$ and $\pi$ have the lowest energy,
there is no way to get to these vacua if we happen to be in one with higher
energy.

\subsection{Physical significance of $\theta$}

Strangely, it seems as though our discussion of the vacuum structure in $SU(N)$
gauge theory has introduced a new coupling constant $\theta$, on which physics
seems to depend.  How did we not see it when we originally formulated the theory?

In fact we can show that the $\theta$ parameter can be included at the level of 
the Lagrangian by adding the new term
\beq
\label{newL}
	{\cal L}_\theta = {\theta g^2\over 32\pi^2} \sum_a F_{\mu\nu}^a
	\widetilde F^{\mu\nu}_a
\eeq
We originally discarded this term because it can be written as a total derivative
(\ref{topno}), thus it has no effect on the equations of motion.  In fact,
the effects of such a term would vanish at any order in perturbation theory.
However, we can now see that it does have a nonperturbative effect.  Consider 
the expectation value of an arbitrary operator ${\cal O}$ in a theta vacuum.
We have to sum over all the sectors of the gauge fields which have different
winding number $\nu$:
\beqa
	\langle \theta|{\cal O}|\theta\rangle &=& 
	\sum_\nu e^{i\nu\theta} \int ({\cal D} A)_\nu e^{-S} {\cal O}\nonumber\\
	&=&  \sum_\nu  \int ({\cal D} A)_\nu 
	e^{-S} {\cal O} \exp\left(i{\theta g^2\over 32\pi^2} \sum_a F_{\mu\nu}^a
	\widetilde F^{\mu\nu}_a\right)
\eeqa
where we used the relationship between winding number $\nu$ and $F\tilde F$
from (\ref{FFD}).  We see that the factor $e^{i\nu\theta}$ which weights the
$n$ vacuua is exactly reproduced by adding the term (\ref{newL}) to the QCD
Lagrangian.

There is one big problem with this new term.  It violates parity, and it can be shown
that it would contribute enormously to the electric dipole moment of quarks and
hence neutrons.  There are stringent experimental limits which imply that 
$\theta \lsim 10^{-10}$, a rather severe fine-tuning problem, known as the
{\it strong CP problem}.  Peccei and Quinn invented an elegant solution to this 
problem:  by adding certain new fields, the theta parameter can be compenstated
by a dynamical field $a$, the axion.  The potential for the axion becomes
$E(\theta+a/f)$, where $f$ is the axion scale (the scale of breaking of a new
symmetry which is introduced), and the effective physical value of the theta
parameter becomes $\bar\theta = \theta + a/f$.  Since the axion naturally wants
to go to the minimum of its potential, this gives a dynamical explanation for
why $\bar\theta = 0$.  Axions are still an attractive idea, appearing naturally
in string theory, providing a candidate for dark matter, 
and they remain the subject of ongoing experimental searches.

\section{Exercises}
\label{sec:ex}

1.  (a) Prove that $s+t+u = m_1^2 + m_2^2 + m_3^2 + m_4^2$ 
for $2\to 2$ scattering, where the four particles might all have
different masses.  \\
(b) Compute the differential cross section $d\sigma/dt$ 
for $\psi\psi\to\psi\psi$ scattering for a
Dirac fermion of mass $m$ coupled to a scalar of mass $\mu$ via the
Yukawa interaction $g\phi\bar\psi\psi$.  Don't forget the $u$-channel
diagram.  For what scattering angle is $d\sigma/dt$ largest?  \\

2. (a) Consider a theory with $N$ Dirac fermions, labeled $\psi_i$ with 
$i=1,\dots,N$, coupled to a scalar field via the interaction Lagrangian
$g\phi\sum_i \bar\psi_i\psi_i.$  Assuming the $\phi$ mass ($\mu$) is much
greater than the fermion masses, how does the $\phi$ decay width scale with
$N$? How does the inclusive cross section for $\psi_1\bar\psi_1\to {\rm all\ }
\psi_j\bar\psi_j$ scale with $N$ near the resonance, ignoring the width of the
$\phi$?  How does the maximum value of the cross section scale with $N$, 
accounting for the width of the $\phi$? \\
  (b) Use the result you found in (a),
along with the result for a single species of fermions derived in the lectures,
to estimate the maximum value of the cross section for $Z\to {\rm all\ } q\bar
q$ ({\it i.e.,} quarks and antiquarks).  The number of decay channels for the $Z$ boson is 3 (colors) for every
flavor of quark which is kinematically allowed, plus 1 for each type of charged
lepton and neutrino.  Compare your answer to the actual value, which you can
find in the section on ``Plots of cross sections and related quantities'' of
the Particle Data Group Review of Particle Physics.  
(See http://pdg.lbl.gov/2004/reviews/contents\_sports.html)\\
(c) The comparison between the toy model prediction and the data is not very good
in part (b).  One observation that helps is that the $Z$ boson is not a scalar
particle, but rather a massive vector, with three polarization states.  How
will your prediction in (b) change if you treat each of these polarization as
though they were separate scalar particles?  \\

3.  Study eq.\ 10.47 of the section 
``Electroweak model and constraints on new physics'' in the Particle Data
Book.  (a) What would be the effective value of $g^2$ for each kind of quark and
lepton, if the $Z$ boson was a scalar particle?  Note that the partial widths
in (10.47c) for each quark are summed over 3 colors, so you should divide 
these results by 3. \\
(b) Sum all the partial widths, assuming that charmed and strange quarks
are like up and down quarks.  Do they add up to the total width
(10.48)?  Note: if you are
unsure about masses and charges of quarks and leptons, see the
summary tables at \\ 
http://pdg.lbl.gov/2004/tables/contents\_tables.html. \\

4. Consider  a model of
 two scalar fields with potential
$M^2\sigma^2/2$ + $m^2\phi^2/2 + \mu\sigma\phi^2/2$.  
Assume that the decay $\sigma\to 2\phi$ is allowed.\\
(a) Compute the decay width of the $\sigma$.\\
(b) Compute the imaginary part of the $\sigma$ self-energy.  Show that
it is related to the decay width in (a) in the
manner claimed in lecture, near the $\sigma$ mass shell.
  (This form is also mentioned at the end of the 
kinematics review in the Review of Particle Properties).\\
(c) Evaluate the most divergent terms, and determine 
the counterterms in the tree
level Lagrangian needed to cancel these divergences.\\

5.  Evaluate the self-energy correction to the fermion
$\Sigma_\psi(p)$ which was started in lecture.  First do all integrals
except the Feynman parameter integral.  For the last step, compute
only the divergent part.\\

6. Show that the series of one-particle reducible diagrams which  contribute to
the two-point function in the theory of a single scalar field with potential 
$m\phi^2/2 + \mu\phi^3/3!$ is a geometric series.  (You should see the pattern
for the statistical factor after computing it for the first few diagrams.) 
Evaluate the most divergent contribution and sum the series. Why is it
important that the series is geometric? \\ 

7. Consider a scalar field whose potential energy is $V(\phi) = A\phi^6$.\\
(a) What is the tree-level mass of the corresponding particle?  What is the
value of the cross section for $2\phi\to 2\phi$ at tree level?\\
(b) What is the dimension of the coupling constant $A$?\\
(c) Derive the invariant amplitude ${\cal M}$ for the process $2\phi\to 4\phi$
at lowest order in $A$.\\
(d) Compute the lowest order quantum correction to the mass of the particle.\\
(e)  Compute the lowest order quantum correction to the differential cross
section $d\sigma/dt$ for $2\phi\to 2\phi$.\\
(f)  Draw the connected Feynman diagrams contributing to
the two-loop correction to the 6-point function
$G_6(p_1,\dots,p_6)$.  Identify those which are 1PI or 1PR.\\

8. Consider the one-loop vertex correction $\Gamma_{\phi\bar\psi\psi}$
in the Yukawa theory ($V = -g\phi\bar\psi\psi$),
in the limit where the fermion mass vanishes, and the external momentum of
the boson vanishes, but the fermion momentum $p$ and the boson
mass $\mu$ are nonzero.  Use a momentum-space cutoff $\Lambda$ and drop all
terms which vanish as $\Lambda\to\infty$, but make no other approximations.\\
(a)
$i\Gamma_{\phi\bar\psi\psi}$ has a real and an imaginary part.  Compute
the real part (which corrects the tree-level coupling).  Note that $\int dx
\ln(Ax^2 + Bx + C)$ can be computed by completing the square in the logarithm.
For definiteness, assume that $p^2 > \mu^2$.\\
(b) Verify that the effective coupling depends on $p^2$ at large $p^2$ in the
way claimed in the lecture about the running of the coupling constant, 
$g(\mu_r)$.\\
(c) For what values of $p^2$ does $i\Gamma_{\phi\bar\psi\psi}$ develop an
imaginary part?  Show that these are the same values as those for which
the virtual particles in the loop can go on shell.  Notice the similarity
to the other situation where you have seen the loop get an imaginary part.
The imaginary part is related to unitarity of the S-matrix through the optical
theorem.\\  

9. Consider the scattering process $2\phi\to 2\phi$ in the theory with couplings
$\lambda\phi^4/4! + g\phi\bar\psi\psi$, where the external particles all have
energy $E$ which is much greater than any of the masses.\\
(a) Explain why, at such large energies, any loop corrections (including
finite part) containing 
logarithmic divergences should go like $\ln(\Lambda/E)$, rather than
$\ln(\Lambda/\mu)$, for example.\\
(b) Use the observation of (a) to deduce the energy dependence at high $E$ 
of the one-loop amplitude for $2\phi\to 2\phi$ scattering, by computing only the
most divergent contributions.  Does the effective coupling $\lambda(E)$ grow
or decrease as a function of $E$? \\ 

10. Consider the nonrenormalizable theory with potential $
M^2\phi^2/2 + \lambda M^{-2}\phi^6/6!$.  
Suppose that $M$
is some fixed mass scale which is {\it not} considered to be a function of the
cutoff, but instead is just a generic value which is parametrically 
smaller than the cutoff.  Assume that $\lambda \lsim 1$.
For this
problem think of the cutoff as being some actual physical scale, large but
not infinite.  \\
(a)  A given 1PI Feynman diagram will have ${\cal V}$ vertices and ${\cal E}$ external legs.
How many internal lines ${\cal I}$ must there be?  \\
(b) Prove that the number of loops is ${\cal L}={\cal I-V}+1$.\\
(c) Estimate the contribution of this diagram to the proper vertex with ${\cal
E}$
legs, $\Gamma_{\cal E}^{(V)}$, as a function of ${\cal V}$ and ${\cal E}$. Take the external legs
to be at zero momentum.  Assume combinatoric factors will approximately 
cancel out the factors of $1/6!$, and don't worry about minus signs, 
but keep factors of $2\pi$.   Estimate the loop integrals by power counting:
let $\prod_{i=1}^{\cal L} d^4l_i \to dl\, l^{4{\cal L}-1}$ and 
$\prod_{j=1}^{\cal I} (l_j^2 + M^2) \to (l^2+M^2)^{\cal I}$.  Will the integral
be dominated by infrared (low $l$) or ultraviolet (high $l$) contributions, for
a given value of ${\cal E}$ and ${\cal V}$?\\
(d) By summing the dominant contributions for a given ${\cal E}$, estimate the
size of full proper vertex, $\Gamma_{\cal E} = \sum_{\cal V} 
\Gamma_{\cal E}^{({\cal V})}$.\\
(e) In terms of the proper vertices, write down the form of an effective potential $V_{\rm eff}(\phi)$ which 
has the property that tree level diagrams computed from that potential reproduce
the full multiloop result.  \\
(f) Operators of dimension $d < 4$ are known as {\it relevant}, with 
$d=4$ as {\it marginal}, and with $d>4$ as {\it irrelevant}.  From
the above results, explain why this is an appropriate terminology, when the
cutoff is large.  This represents the modern understanding of what is
special about renormalizable theories (which contain only relevant and marginal
operators):  the nonrenormalizable operators have small
coefficients.\\

11.  This problem is supposed to solidify your intuitive understanding of 
renormalization, and give you practice with the Wilsonian effective action.
Consider the theory with potential $V = A\phi + \frac12 m^2\phi^2 +\frac16 \mu\phi^3$
in 6 spacetime dimensions.  \\
(a) What are the dimensions of the couplings and the field?\\
(b) Show by power counting of the loop diagrams which are generated 
 that the theory is renormalizable.\\
(c)  The tadpole term $A\phi$ can be removed by redefining
the field $\phi = \phi_0 + \delta\phi$, provided that the parameters 
satisfy a certain relation.  Find $\delta\phi$ and the relation.  Show how
the physical mass $m^2_0$ and cubic coupling $\mu_0$ depend on the original
parameters after going to the new field $\phi_0$.  
Show that the value of $\delta\phi$ and the shifted couplings could equivalently
be found by imposing that $V'(\phi_0)=0$, $m^2_{0} = V''(\phi_0)$
and $\mu_0 = V'''(\phi_0)$.\\
(d) We need to do loop integrals in 6 dimensions in the sections below.  In
$d$ dimensions, let $d^{\,d}x = d\Omega_{d-1}\, dx\, x^{d-1}$. Using the fact that
$\int_{-\infty}^\infty dx\, e^{-x^2/2} = \sqrt{2\pi}$, evaluate the integral
$\int d^{\,d}x\,  e^{-x^2/2}$ in both Cartesian and spherical coordinates to 
determine $\int d\Omega_{d-1}$.\\
(e) Let $A=0$ at the cutoff $\Lambda$.  Write down integral expressions for the
terms in $i\delta S_{\Lambda} \equiv iS_{\Lambda'} - iS{\Lambda}$
which will contribute to renormalization of the original couplings in
$S_\Lambda$ when computing Wilson's effective action at a lower cutoff
$\Lambda'<\Lambda$.  (Do not compute contributions to nonrenormalizable operators.)
By using the shorthand $\int_{\Lambda'}^\Lambda  dp = \int d^{\,d}p/(2\pi)^d$
(integrated between Euclidean momenta ${\Lambda'}<|p|<\Lambda$), $\delta_{q} = (2\pi)^d
\delta^{(d)}(\sum_i q_i)$ and $\phi_{q} = \int d^{\,d}x\, e^{iqx}\phi(x)$, write
these contributions to $S_{\Lambda'}$ in the form of integrals which do not
make any reference to the dimensionality of spacetime.  Use $q_i$ for the
external momenta and $p$ for the loop momentum in your expressions.
\\
(f) Now specializing to $d=6$, 
evaluate the most divergent parts of the integrals (from $\Lambda'$ to
$\Lambda$) in part (e).  For the self-energy diagram, Taylor expand in the
external momentum so that you can identify both the mass and the wave function
renormalization.  Why is it reasonable to neglect higher terms in the Taylor
expansion?  Assemble your results into a complete expression for these
leading contributions to $i\delta
S_{\Lambda}$, rewritten in position space rather than momentum space. \\
(g) Combine $\delta S_{\Lambda}$ with $S_{\Lambda}$ to find $S_{\Lambda'}$
and identify the corrections to the kinetic term and the couplings.  
Canonically normalize the field at the new cutoff to absorb the wave
function renormalization found in (f).  What are the new values of $A(\Lambda')$,
$m^2(\Lambda')$ and $\mu(\Lambda')$ in terms of the original parameters?
Keep in mind that you are doing a perturbative expansion.\\
(h) How does the cubic coupling run with $\Lambda'$?  By setting $\Lambda'=E$
for an experiment at energy scale $E$, decide whether scattering processes
become stronger or weaker as you go to higher energies.
Compute the beta function for this coupling,
$\beta(\mu) = \Lambda' {d\over d\Lambda'}\mu(\Lambda')$ and take note of
its sign.  Would it have been
possible to deduce this result from just one of the Feynman diagrams you 
calculated?\\
(i) Find the physical $m^2(\Lambda')$ which takes into account the shift in 
$\phi$ needed to remove the tadpole.  Then evaluate it as $\Lambda'\to 0$,
and take this to be the physical mass squared, $m_{\rm phys}^2$. 
Keep only the terms which dominate at small coupling and large cutoff.
 Assuming
$m_{\rm phys}^2\ll \Lambda$, solve for the value of $m^2(\Lambda)$ needed to  
obtain the small $m_{\rm phys}^2$, to first order in $m_{\rm phys}^2$,
and show that there is a fine-tuning (hierarchy) problem. \\
(j) Estimate the magnitude of the leading contribution to the energy density of
the vacuum ({\it i.e.,} cosmological constant) which 
is generated at the scale $\Lambda'$, assuming it was zero at the scale
$\Lambda$.\\

In the following two problems  you will solve the renormalization group equations
for the theory with potential$\frac12\mu^2\phi^2 + m\bar\psi\psi +
\lambda\phi^4/4! + g\phi\bar\psi\psi$.\\

12.  (a) Verify the result given in lecture for the anomalous
dimensions of $\phi$ and $\psi$ at one loop:
$\gamma_\phi = g^2/8\pi^2$ and $\gamma_\psi = g^2/32\pi^2$.\\
(b) Show that this gives the beta function $\beta(g) =
5g^3/16\pi^2$.\\
(c) Use the anomalous dimensions and the
vertex correction you computed in a previous assignment
 to find the contribution from the
fermion loops to the beta function of the quartic coupling.
(I was incorrect to say in lecture that this contribution 
 $\beta_\lambda$ has to be positive.)\\

13.  (a) Using Mathematica or some other programming language, numerically
integrate the beta functions to obtain the running couplings as a
function of $t=\ln(\mu_r/\mu_0)$.  (Don't forget to use the full beta
function for $\lambda$.)   Start with $\lambda=0.5,\ g=0.4$ at
$t_0=0$.  At what value of $t$ does the quartic coupling vanish?  
How many orders of magnitude increase  in the energy does this value
of $t$ represent? Plot the two couplings as a function of $t$. \\
(b) Keeping $\lambda(0)=0.5$, by trial and error find
the smallest value of $g(0)$ where a Landau pole occurs close to $t=100$. 
Find its value to 3 significant figures. 
Then show that for larger values of $g(0)$, 
it is possible to keep both couplings in the range $[0,1]$.
For what range of values of $g(0)$ is this true?
If you had to make an analytic estimate of the best value of 
$g(0)$ to take, for a fixed value of $\lambda(0)$, what would it
be?  Compare to your numerical result.
 \\
(b) Add the anomalous dimension for $\phi$ in your system of equations.
Use this together with your determination of $\lambda$ to plot the scale 
dependence of the amplitude for $\phi\phi\to\phi\phi$.  On the plot show both
the running coupling by itself, and the amplitude including the effect of the anomalous
dimension.  Does the anomalous dimension increase or decrease the
amplitude?  Do it first for the set of couplings in (a) which
give the Landau pole, but graph only the region of $t$ for which
the couplings remain perturbative.  Do it again for the second
set of couplings, graphing over the whole range of $t$.
\\




17.  Evaluate $\frac12\int d^4x\, d^4y\, J^\mu(x) D_{\mu\nu}(x-y) J^\nu(y)$
for the static charge distribution $J_0(x) = \rho(\vec x)$, $J_i=0$,
and show that it gives the self-energy of the charge distribution due to the
Coulomb potential, times an infinite integral over time (which converts the
energy into the action).
Remember to use the appropriate $i\epsilon$ prescription
to define the poles of the propagator.  Do the $q_0$ integral of the propagator
first, as a contour integral, which must be done separately for the two cases
$x_0<y_0$ and $y_0 < x_0$.  Next do the $x_0$ integral.  Then do the 
angular integral of the propagator, and finally the radial momentum integral
of the propagator.\\

18. In lecture it was claimed that the gauge fixing term does not spoil the
gauge invariance of any other terms in the effective action.  Thus the
$\alpha$-dependence which enters these terms must do so in a gauge-invariant
way.  To see how this works, compute the divergent and $\alpha$-dependent
parts  of the electron self-energy, vacuum polarization, and vertex correction
at one loop in QED (using a general Lorentz gauge), and show that they are
consistent with the Ward identities.  Verify that the beta function for the
charge is independent of $\alpha$.\\

19.  Denote the two-fermion, $n$-photon terms in the QED effective action by
\beq
	\Gamma_{\bar\psi A^n\psi} = 
	\sum_{n=0}^\infty \int \prod_{i=0}^n dp_i \,
	\bar\psi(p_{n+1})\, \Gamma_{\mu_1\dots\mu_n} \psi(p_{0}) A^{\mu_1}(p_1)
	\cdots A^{\mu_n}(p_n) \delta_{p_{n+1} - \sum p_i}
\eeq
(where $dp = {d^{\,4}p/(2\pi)^4}$ and $\delta_{p_{n+1} - \sum p_i} = 
(2\pi)^4 \delta^{(4)}(p_{n+1} - \sum_{i=0}^n p_i)$).	
Find the Ward-Takahashi identities which relate the functions 
$\Gamma_{\mu_1\dots\mu_n}$.  Draw the 1-loop diagrams corresponding
to these identities (at one-loop level) to illustrate.\\

20.  Derive the axial anomaly using the point-splitting regularization of the
axial vector current.  Use the following steps.  (a) Expand the exponential of 
the line integral to order $\epsilon_\mu$ and show that to this order,
\beq
	\partial^\mu J_{\mu 5} = i e F^{\mu\nu} \epsilon_\nu
	 \bar\psi(x+\epsilon/2)\gamma_\mu\gamma_5
	\psi(x-\epsilon/2)
\eeq
(b)  Show that the vacuum expectation value of the above quantity vanishes
at zeroth order in perturbation theory.   Then set up the calculation of the
same quantity at first order in perturbation theory in $e$ (one insertion of 
the electromagnetic interaction in a background gauge field). \\ 
(c) Let $p$ and $q$ be the momenta of the two fermion propagators, and let $l=p-q$
be the momentum of the gauge field which you have inserted.  Further, let $r=p+q$.
Show that your result is proportional to
\beq
	\int d^{\,4}l \int d^{\,4}r \, 
{e^{il\cdot x-i\epsilon\cdot r/2}\, l_\alpha r_\beta A_\nu(l) \over ((r+l)^2/4-m^2) ((r-l)^2/4-m^2)}
\eeq
Since we are interested in the $\epsilon\to 0 $ behavior, argue that we can set
$l\to 0$ when doing the $r$ integral.  You should now see how to do the $l$
integral.\\
(d)  To do the $r$ integral, note the following useful trick: rewrite $1/(r^2)^2$
as $\int_0^\infty ds\,s\,e^{-s r^2}$, and $r_\beta =
2i{\partial\over\partial\epsilon^\beta}$. \\
(e) Average over the directions of $\epsilon_\nu$ to evaluate $\epsilon_\mu
\epsilon_\nu/\epsilon^2$ and obtain the final result.\\

21. (a) In $D=2$ dimensions, we can define Dirac matrices
\beq
	\gamma_0 = \sigma_x,\quad \gamma_1 = i\sigma_y,\quad
	\gamma_5 = -\sigma_z
\eeq
Show that they satisfy the usual Dirac algebra for the anticommutators,
and also that ${\rm tr}\gamma_\mu \gamma_\nu \gamma_5 = 2\epsilon_{\mu\nu}$.\\
(b) Use the above results to do Fujikawa's derivation of the axial anomaly in 2D.
To remind you, the steps are: (1) Show that the Jacobian in the path integral
for an infinitesimal
U(1)$_A$ transformation has the form ${\rm det} (1+i\epsilon(x)\gamma_5)$ for 
either $\psi$ or $\bar\psi$, where the determinant is in the space of all functions
as well as 2D spinor indices.  
 (2) Using the identity $\ln\det = {\rm tr}\ln$, rewrite this determinant in a more explicit
form involving $\int d^{\,4}x\, \int d^{\,4}p$, and regulate the momentum space
integral in a way which respects the U(1) gauge symmetry, {\it i.e.,}
using a function $f((i\Dsl)^2/\Lambda^2)$, where $\Dsl$ is the covariant
derivative, $\Lambda$ is a UV cutoff scale, and $f$ is a function with the
properties $f(0) = 1$, $f(x)\to 0$ as $x\to\infty$.  (3) Use the result
which is analogous to $\Dsl^2 = D^2 -(i/2)F_{\mu\nu}\gamma^\mu\gamma^\nu$ 
(is this still valid in 2D?) and expand $f$ to first order in 
$F_{\mu\nu}\gamma^\mu\gamma^\nu$.  (4) Show that the first order term is sufficient
to get a nonvanishing result for the trace.  Rewrite the momentum integral in the 
form $\int dp^2 p^2 f'(p^2)$.  Use integration by parts to evaluate
the momentum integral, showing that the result for the anomaly is independent of 
the detailed form of the cutoff function $f$.  \\
(c) Based on the 2D and 4D results, guess the form of the anomaly in $2n$ dimensions.
Why is there no axial anomaly in $2n+1$ dimensions?\\

22.\ In lecture it was shown that the amplitude for $\pi^0\to \gamma\gamma$
due to the triangle anomaly is
\[ A(\pi^0\to \gamma\gamma) = {N_c e^2\over 4\pi^2 f_\pi}(Q_u^2 - Q_d^2)\,
	\epsilon_{\mu\nu\alpha\beta}\, p_1^\alpha\, p_2^\beta\, \epsilon_1^\mu\,
	\epsilon_2^\nu \]
From this compute the decay width.  Compare to the experimental value
$\Gamma = 8.0$ eV to determine the combination $N_c\,(Q_u^2 - Q_d^2)$.
Using the known values of the quark charges, find the number of colors,
rounded to the closest integer.\\

23.  (a) Compute the divergent contributions to the gluon vacuum polarization
in QCD due to diagrams involving ghosts and gluons, in Feynman gauge. 
Find the contribution to the $Z_3$ renormalization factor of QCD coming from
these  diagrams.  Notice that this is the complete answer in a theory with just
gauge bosons and no quarks.  

Hint: think about the value of a diagram which has no external momentum 
flowing through the loop, in a massless theory using dimensional regularization,
before computing group theory factors for such a diagram.

(b) To find the beta function, one must combine this result with that of other
diagrams, since it is gauge-dependent.  The simplest way to proceed for the
pure gauge theory is to evaluate  $Z_1/Z_2$ in a theory with quarks.
The result in Feynman gauge is
\beq
	Z_1/Z_2 = 1-{g^2\over{16\pi^2\epsilon}}N
\eeq
where $N=3$.   (Only in axial gauge where the ghosts decouple is $Z_1/Z_2=1$,
and then the complete result can be gotten from $Z_3$ as in QED.)
Using this result and that you found in part (a), compute the beta function for
the coupling in the pure gauge theory.


\end{document}